Supported by: CERN, CoEPP-MN-17-1

# Parton Radiation and Fragmentation from LHC to FCC-ee

Workshop Proceedings, CERN, Geneva, Nov. 22nd–23th, 2016

*Editors*

David d'Enterria (CERN), Peter Z. Skands (Monash)

*Authors*

D. Anderle (U. Manchester), F. Anulli (INFN Roma), J. Aparisi (IFIC València),
G. Bell (U. Siegen), V. Bertone (NIKHEF, VU Amsterdam), C. Bierlich (Lund Univ.),
S. Carrazza (CERN), G. Corcella (INFN-LNF Frascati), D. d'Enterria (CERN),
M. Dasgupta (U. Manchester), I. García (IFIC València), T. Gehrmann (U. Zürich),
O. Gituliar (U. Hamburg), K. Hamacher (B.U. Wuppertal), A.H. Hoang (U. Wien),
N.P. Hartland (NIKHEF, VU Amsterdam), A. Hornig (LANL Los Alamos),
S. Jadach (IFJ-PAN Krakow), T. Kaufmann (U. Tübingen), S. Kluth (T.U. München),
D.W. Kolodrubetz (MIT), A. Kusina (LPSC Grenoble), C. Lee (LANL Los Alamos),
G. Luisoni (CERN), V. Mateu (U. Salamanca), H. Matevosyan (U. Adelaide),
W. Metzger (Radboud U.), S.O. Moch (U. Hamburg), P.F. Monni (CERN),
B. Nachman (LBNL Berkeley), E.R. Nocera (U. Oxford), M. Perelló (IFIC València),
W. Placzek (Jagiellonian U., Kraków), S. Plätzer (IPPP Durham, U. Manchester),
R. Perez-Ramos (Paris), G. Rauco (U. Zurich), P. Richardson (CERN, IPPP-Durham),
F. Ringer (LANL Los Alamos), J. Rojo (NIKHEF, VU Amsterdam), Ph. Roloff (CERN),
Y. Sakaki (KAIST Daejeon), N. Sato (JLab Newport News), R. Simoniello (CERN),
T. Sjöstrand (Lund U.), P.Z. Skands (Monash U.), M. Skrzypek (INP Kraków),
G. Soyez (IPhT CEA-Saclay), I.W. Stewart (MIT), M. Stratmann (U. Tübingen),
J. Talbert (DESY, U. Oxford), S. Todorova (CNRS), S. Tokar (Comenius U.),
M. Vos (IFIC València), and A. Vossen (Indiana U.)

**Abstract**

This document collects the proceedings of the *"Parton Radiation and Fragmentation from LHC to FCC-ee"* workshop (http://indico.cern.ch/e/ee_jets16) held at CERN in Nov. 2016. The writeup reviews the latest theoretical and experimental developments on parton radiation and parton-hadron fragmentation studies –including analyses of LEP, B-factories, and LHC data– with a focus on the future perspectives reacheable in $e^+e^-$ measurements at the Future Circular Collider (FCC-ee), with multi-ab$^{-1}$ integrated luminosities yielding $10^{12}$ and $10^8$ jets from Z and W bosons decays as well as $10^5$ gluon jets from Higgs boson decays. The main topics discussed are: (i) parton radiation and parton-to-hadron fragmentation functions (splitting functions at NNLO, small-$z$ NNLL resummations, global FF fits including Monte Carlo (MC) and neural-network analyses of the latest Belle/BaBar high-precision data, parton shower MC generators), (ii) jet properties (quark-gluon discrimination, $e^+e^-$ event shapes and multi-jet rates at NNLO+N$^n$LL, jet broadening and angularities, jet substructure at small-radius, jet charge determination, $e^+e^-$ jet reconstruction algorithms), (iii) heavy-quark jets (dead cone effect, charm-bottom separation, gluon-to-$b\bar{b}$ splitting); and (iv) non-perturbative QCD phenomena (colour reconnection, baryon and strangeness production, Bose-Einstein and Fermi-Dirac final-state correlations, colour string dynamics: spin effects, helix hadronization).

### Speakers

D.P. Anderle (U. Manchester), F. Anulli (INFN Roma), V. Bertone (NIKHEF Amsterdam), C. Bierlich (Lund Univ.), G. Corcella (INFN-LNF Frascati), D. d'Enterria (CERN), M. Dasgupta (U. Manchester), K. Hamacher (Bergische U. Wuppertal), S. Jadach (IFJ-PAN Krakow), S. Kluth (T.U. München), W. Metzger (Radboud U.), V. Mateu (U. Salamanca), H. Matevosyan (U. Adelaide), S.O. Moch (U. Hamburg), P.F. Monni (CERN), B. Nachman (LBNL Berkeley), S. Plätzer (IPPP Durham, U. Manchester), R. Perez-Ramos (Paris), G. Rauco (U. Zurich), P. Richardson (CERN, IPPP-Durham), Y. Sakaki (KAIST Daejeon), N. Sato (JLab Newport News), M. Selvaggi (CERN), T. Sjöstrand (Lund U.), P.Z. Skands (Monash U.), G. Soyez (IPhT CEA-Saclay), J. Talbert (DESY), S. Todorova (CNRS), S. Tokar (Comenius U.), M. Vos (IFIC València), A. Vossen (U. Indiana)

### Additional Participants

N. Alipour Tehrani (CERN), S. Amoroso (CERN), A. Blondel (U. Genève), H. Brooks (IPPP/Durham U.), S. Carrazza (CERN), M. Dam (Niels Bohr Institute), N. Fischer (Monash U.), O. Fischer (U. Basel), J. Gao (Jiao Tong U., Shanghai), P. Gunnellini (DESY), I. Helenius (Tübingen U.), P. Janot (CERN), A. Jueid (Essaadi U.), D. Kar (U. Witwatersrand), M. Klute (MIT), E. Leogrande (CERN), T. A. Lesiak (INP-PAS, Krakow), H. Li (Monash U.), D. Liberati (CNR, Roma), A. Lifson (ETH Zurich), E. Locci (CEA-Saclay), L. Lönnblad (Lund U.), J.J. Lopez-Villarejo (Lausanne), G. Luisoni (CERN), S. B. Nielsen (U. Copenhagen, Niels Bohr Institute), V. Okorokov (MEPhI Moscow), F. Piccinini (INFN Pavia), W. Placzek (Jagiellonian U. Krakow) R. Rahmat (U. Iowa), P. Rebello-Teles (CBPF Rio de Janeiro), S. Richter (UC London), J. Rojo (NIKHEF, VU Amsterdam), M. Seidel (CERN), G. G. Voutsinas (CERN), B. Webber (Cambridge), L. Zawiejski (IFJ PAN Krakow)

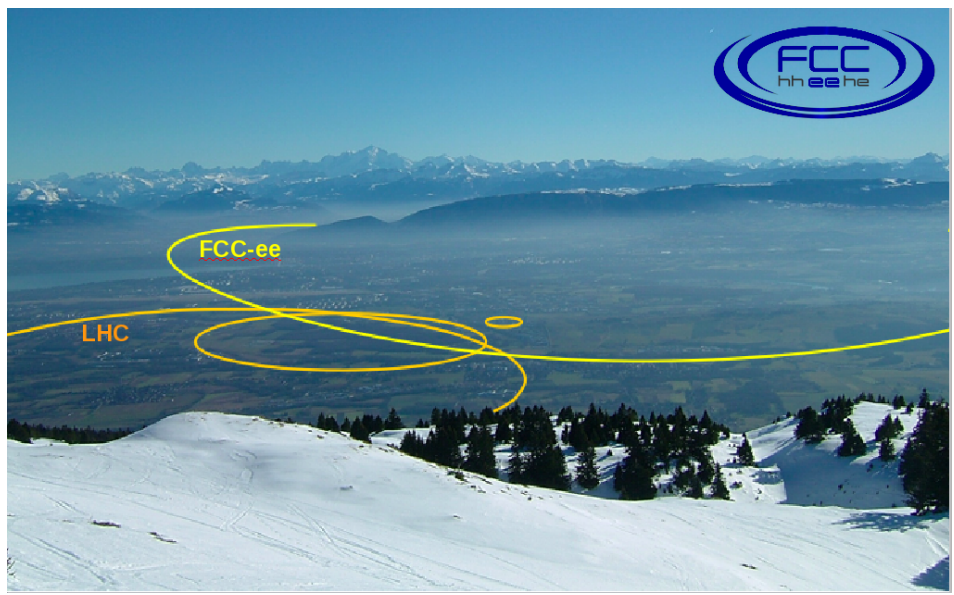

# 1 Introduction

The workshop *"Parton Radiation and Fragmentation from LHC to FCC-ee"* was held at CERN, Nov. 22–23, 2016, as part of the FCC-ee *QCD and $\gamma$-$\gamma$ physics* working group activities in the context of the preparation of the FCC-ee Conceptual Design Report in 2017. The meeting brought together experts from several different fields to explore the latest developments on our theoretical and experimental understanding of parton radiation and fragmentation, organized along four broad sessions:

1. **Parton-to-hadron fragmentation functions**, covering splitting functions at NNLO, small-$z$ NNLL resummations, global FF fits including Monte-Carlo and neural-network analyses of the latest Belle/BaBar high-precision measurements;

2. **Parton radiation and jet properties**, including talks on parton showers, quark-gluon discrimination, precision $e^+e^-$ event shapes and multi-jet rates, jet substructure and small-radius jets, jet charge determination; determination, and $e^+e^-$ jet reconstruction algorithms;

3. **Heavy-quark jets**, with talks on the dead-cone effect, charm-bottom separation, and gluon-to-$b\bar{b}$ splitting;

4. **Non-perturbative QCD phenomena**, with talks on colour reconnections, baryon and strangeness production, Bose-Einstein and Fermi-Dirac and final-state hadron correlations, and colour string dynamics: spin effects, helix hadronization.

About 65 physicists took part in the workshop, and 30 talks were presented. Slides as well as background reference materials are available on the conference website

<http://indico.cern.ch/e/ee_jets16>

These proceedings represent a collection of extended abstracts and references for the presentations, plus a summary of the most important results and future prospects in the field. Contents of these proceedings will be incorporated into the FCC-ee Conceptual Design Report under preparation.

CERN, January 2017

<div style="text-align:right">
Peter Skands<br>
David d'Enterria
</div>

# 2 Proceedings Contributions

————————

# Parton Fragmentation Functions

Anselm Vossen[1]

[1]Indiana University, USA

**Abstract:** This contribution gives an overview of the field of fragmentation functions. The emphasis is on recent experimental results on light quarks and gluons fragmenting into light hadrons. Some possibilities for the study of fragmentation functions at a future FCC-ee are discussed.

## Introduction

Perturbative QCD can be used to describe many high energy scattering processes. In most cases factorization theorems enable a separation of the respective cross-sections into parts dominated by short distances, which are perturbatively calculable and long distance parts which have to be measured experimentally [1].

If hadrons in the final state are identified, the non-perturbative functions describing the formation of these colorless bound final states are called fragmentation functions (FFs). For a more in depth overview of this field and a detailed list of available datasets and experimental results, see our recent review [2]. The study of fragmentation functions is complementary to the study of parton distribution functions (PDFs), which describe the initial state long distance behaviour of the collisions, i.e. the structure of the hadronic systems before the scattering. At variance to PDFs, FFs cannot be computed on the lattice, a challenge which is rooted in the difficulty to integrate over the spectators in the final state. Since FFs describe the formation of colorless bound states from colored partons they are conceptually a consequence of color confinement, one of the most intriguing problems in QCD.

In the following we will concentrate on leading twist (twist 2) FFs which have a probabilistic interpretation. For the exact field theoretic definition see again [2]. Here we will just give a working definition. The integrated fragmentation function $D_1^{h/q}(z)$ gives the probability that a quark $q$ fragments into hadron $h$ with $h$ carrying the fraction $z$ of the parent parton momentum. The subscript indicates here the leading twist and the superscript might be omitted if the meaning is otherwise clear. The definition is chosen such that the cross-section for semi-inclusive hadron production can be written using QCD factorization theorems as $\sigma(lp \to lhX) = \sum_q e_q^2 f_1^{q/p} \otimes D_1^{h/q}$ for Semi-Inclusive Deep-Inelastic Scattering (SIDIS), and $\sigma(pp \to hX) = \sum_{i,j,k} f_1^{i/p_a}(x_a) f_1^{j/p_b}(x_b) \otimes D_1^{h/k}$ for $pp$ and $\sigma(e^+e^- \to hX) = \sum e_q^2 D_1^{h/q}(z)$ for single-inclusive annihilation.

| H \ q | U | L | T |
|---|---|---|---|
| U | $D_1^{h/q}$ | | $H_1^{\perp h/q}$ |
| L | | $G_1^{h/q}$ | $H_{1L}^{\perp h/q}$ |
| T | $D_{1T}^{\perp h/q}$ | $G_{1T}^{h/q}$ | $H_1^{h/q}$  $H_{1T}^{\perp h/q}$ |

Table 1: Interpretation of FFs for quarks, see text for more details. The columns indicate the quark polarization — unpolarized (U), longitudinally polarized (L), transversely polarized (T). The rows indicate the hadron polarization.

Extending the integrated fragmentation function $D_1$ to different quark and hadron polarizations and allowing the FFs to depend on the transverse momentum $\vec{k}_T$ of the hadron with respect to the parent quark, we can classify fragmentation functions according to the so-called Amsterdam notation for FFs [3][4][5][6] as given in tbl. 1. Here we only introduce functions describing quark fragmentation into single hadrons and the integrated gluon FF. Quite a bit of theoretical and experimental work has been devoted recently to polarized gluon fragmentation functions and di-hadron fragmentation functions, both polarized and unpolarized. We will not discuss these much further in this write-up and instead refer to the literature. Note also that some of the FFs vanish if we integrate over $\vec{k}_T$. These are denoted with the superscripts $\perp$. Even though fragmentation functions, enter virtually all semi-inclusive cross-sections of hard-scattering processes, the precise knowledge of spin and $\vec{k}_T$ dependent FFs in the intermediate $z$ region plays an especially important role in hadronic physics, where the nuclear structure encoded in the PDFs has to be disentangled from fragmentation contributions. For some chiral-odd parton distribution functions, like the transversity $h_1(x)$, the only precise experimental data is sensitive to the combination with a chiral-odd FFs, like the transverse polarization dependent Collins FF $H_1^\perp$. In the following we will however concentrate on the quark polarization integrated FF $D_1$.

## Experimental Access to FFs

As alluded to in the introduction, FFs enter many hard scattering processes. Current extractions use Single Inclusive Annihilation (SIA), SIDIS and $pp$ data. The different configurations are complementary to each other, each having different advantages and disadvantages. The cleanest access is provided by SIA, where the factorized cross-section can be written as

$$\frac{1}{\sigma_{\text{tot}}}\frac{d\sigma^{e^+e^-\to hX}}{dz} = \frac{1}{\sum_q e_q^2}\left(2F_1^h(z,Q^2) + F_L^h(z,Q^2)\right) \quad (1)$$

and the structure function $F_1^h$ can be written in terms of FFs at NLO as

$$\sum_q e_q^2 \left(D_1^{h/q}(z,Q^2) + \frac{\alpha_S(Q^2)}{2\pi}\left(C_1^q \otimes D_1^{h/q} + C_1^g \otimes D_1^{h/g}\right)(z,Q^2)\right). \quad (2)$$

Here the superscript $q$ signifies quark related functions, $g$ gluonic functions. The $C$ are Wilson coefficients. Looking at the cross-sections, the advantages of SIA are the clean and direct access to the FFs. There is no contribution by the nucleon structure. Theory calculations are well under control. In this venue new NNLO calculations have been shown [7].

However, there are also obvious shortcomings. The separation of flavors is nontrivial, since the contribution of each FF differs only by the coupling constants. And the access to the gluon FFs comes only from scaling violations which require a large lever am in $Q^2$.

Some of these shortcomings can be mitigated somewhat. More flavor sensitivity can be reached by including data on the $Z^0$ pole as well as below, using polarized beams [8] and detecting two back-to-back identified hadrons in the final state. Other techniques, such as using three jet events to access the gluon FF at leading order, or using displaced vertices to identify charm and bottom production are ambiguous beyond leading order calculations. However, in current fits, the heavy quark tagged data is commonly used.

**SIA Data**

Experimental data on SIA exists over a wide range of energies, from $\sqrt{s} = 3$ GeV up to $\sqrt{s} = 209$ GeV [9][10]. For our knowledge of FFs the datasets taken by the B-factories on or near the $\Upsilon(4S)$ resonance from Belle [11][12] and BaBar experiments [13] as well as the data taken by LEP and SLC on the $Z^0$ resonance [14][15][16][17] are the most important. Many older datasets taken below the $\Upsilon(4S)$ resonance or in between the resonance and the $Z^0$ pole lack statistics and sometimes are not documented very well. Above the $Z$ pole, the $e^+e^-$ annihilation cross-section drops quite quickly, which will also mean that the $Z^0$ pole data collected by FCC-ee will play a dominant role compared to other planned center-of-mass energies. LEP collected over 200 pb$^{-1}$ on the $Z^0$ pole whereas Belle at KEKB collected about 1 ab$^{-1}$ on the $\Upsilon(4S)$ resonance (BaBar at PEP-II collected about half that). This corresponds to orders of magnitude more data than collected by other experiments, other than LEP, and allowed for the first time to measure the SIA cross-section for identified pion, kaons and protons at large $z > 0.5$. From both, the LEP experiments and the B-factories, we have precision data on $\pi$, $K$ and $p$ production. Belle recently [12] showed the first measurement of the cross-section for back-to-back production of identified hadrons.

**Fits using SIA Data**

The SIA data have been used to extract FFs in a number of fits. Here we note the HKNS fit [18] which was pioneering in the estimation of uncertainties of fragmentation functions. The SIA data used in this fit does not allow a flavor separation of the FFs beyond a distinction between favored and unfavored FFs and the gluon FF obtained from the evolution equation is only weakly constrained (the last publication does not use the large Belle dataset yet). During this workshop, a new fit by the NNPDF group showed a first extraction of FFs using a neural network approach but also restricted to SIA data [19]. It is encouraging that this is now available since it presumably gives uncertainties free of bias by the parametrization, e.g. in areas where there is no data.

**Access to Flavor information and Gluon FFs**

Data beyond SIA is needed to achieve flavor separation and access to the gluon FF. For flavor separation, in particular SIDIS data on identified hadron production is important. The main challenge in the use of the data is that many SIDIS experiments are at quite low $\sqrt{s}$. This is in particular true for the new JLab data which would offer unprecedented statistics. Recent efforts in theory are aimed at getting threshold, higher order and target mass effects under control such as the JAM collaboration which presented at this workshop [20].

In $pp$ collisions the gluon FF can be accessed at leading order. However, calculations are much more challenging since the cross-section includes the partonic structure of both colliding protons. Since the initial partonic kinematics in the process $pp \to hX$ are not known, integration over all $x$ and $z$ is necessary. This includes kinematic areas in which the PDFs might not be known so well and does not allow access to the $z$ dependence of the FFs. A new development to address this problem has been the formalism to extract fragmentation from hadrons-in-jets [21]. These calculations, currently available at NLO, allow the extraction of the $z$ dependence of the FFs.

The first fit to include SIDIS data along with the latest $e^+e^-$ data from Belle and BaBar as well as the latest $pp$ data from the LHC (but not yet hadron-in-jet measurements) is the DSS fit [23]. This fit uses a parametrized approach and also extracts uncertainties of FFs. As is HKNS, the DSS fit is done at NLO accuracy. Agreement with all datasets is good in general. However, to solve the disagreement of the previous, pre-LHC, version of the fit [22], a $p_T$ cut of 5 GeV on the $pp$ data was introduced. This removes the region in which the fit does not converge to a consistent description of PHENIX and ALICE data due to the ALICE data preferring a smaller contribution by the gluon fragmentation function. The disagreement of QCD calculations using a representative set of FFs with newly published LHC data was a surprise at the time [24] and is a reminder that even though a lot of progress has been made in the extraction FFs, they are still fits and the potential for surprises exists with new data. With the use of both, SIDIS and $pp$ data, the DSS fit extracts flavor separated FFs and a precise extraction of the gluon FF.

**Challenges at low and high $z$**

Similar to PDFs, the extraction of FFs at low and high $z$ is challenging for theory. At high $z$, roughly above 0.8, theoretical uncertainties rise due to threshold effects, whereas at low $z$ target mass corrections have to be applied and time like splitting functions for the FFs diverge. Target mass corrections are an issue in particular for data sets at lower $\sqrt{s}$ and as mentioned above, there are efforts to address this [20]. The divergence of the splitting functions in principle makes a resummation to all orders necessary. In contrast to PDFs, the divergence happens already at higher $z$. Therefore fits usually employ cuts of $z$ greater than about 0.1. This regime is of course of interest in particular at high $\sqrt{s}$, like at the LHC, because the majority of the particle production happens there. However, approximation schemes to the resummation have been known for quite some time, in particular the so-called Modified Leading Log Approximation (MLLA) and these were also discussed during this workshop [25]. For spin dependent FFs, this region is usually not that interesting, since the multiple splittings tend to average out any spin dependence. Modern fits, like DSS, do not always do a good job in describing the low $z$ region. However, as we have learned in this workshop [7] this is not necessarily caused by the insufficiency of the fixed order calculation including data below the usual cutoff in calculations at NNLO or even NLO leads to a good description even at low $z$ where one would describe the data with the MLLA.

**Transverse Momentum Dependence and Evolution**

Recently, significant experimental and theoretical efforts have been focused on the intrinsic transverse momentum dependence of the FFs. Here, in addition to the $z$ dependence, the dependence on $\vec{k}_T$ is considered, where $\vec{k}_T$ is the transverse momentum of the detected hadron with respect to the parent quark direction. One important motivation for the precision mapping of the transverse momentum dependence is the necessity to disentangle the intrinsic transverse momentum of partons in the nucleon from transverse momentum generated in the fragmentation in SIDIS experiments. But also beyond nuclear physics applications, it is interesting to explore the $\vec{k}_T$ dependence of FFs because the tools used to describe the $\vec{k}_T$ spectrum have certain universal aspects that apply to PDFs and FFs. Following one possible factorization scheme, the Collins-Soper-Sterman (CSS) formalism [1], one can decompose the FFs in a nonperturbative collinear part, $\vec{k}$ dependent perturbative and non-perturbative parts and a term that bridges the non-perturbative and the perturbative part. Since the non-perturbative parts are universal and spin independent, exploring Transverse Momentum Dependent (TMD) evolution effects in FFs would have an impact beyond the studies of FFs. Currently, data on the transverse momentum spectrum of $Z^0$ and Drell-Yan production has arguably the largest impact on TMD evolution studies due to the large lever arm in $\sqrt{s}$ [26]. Having $e^+e^-$ data covering a large range in $Q^2$ would add complimentary information from a process that is theoretically well under control. There are in principle two observables, which give access to the intrinsic transverse momentum in fragmentation. One is the $p_T$ imbalance of back-to-back hadrons, which is sensitive to the convolution of the transverse momenta. The other is measuring the transverse momentum relative to the thrust or jet axis. The later method has the caveat that identifying the quark axis with the thrust or jet axis is problematic beyond LO. However, given the efforts to measure TMDs in jets, e.g. in $pp$, it will be important to do both measurements and compare. At this moment, there are no measurements of the $\vec{k}_T$ dependence of unpolarized FFs in $e^+e^-$ available. There are publications of the $\vec{k}_T$ dependence of the Collins FF $H_1^\perp$ [27][28] (also see the next section). Since $H_1^\perp$ is a TMD, even the $\vec{k}_T$ integrated measurements need TMD evolution. Other results that are sensitive the transverse momentum dependence come from COMPASS and HERMES which measured the $p_T$ spectra in SIDIS [29][30].

**Spin Dependent Fragmentation**

We already mentioned spin dependent fragmentation functions several times in the previous sections. Here we will summarize some recent results that are sensitive to transverse quark spin dependent FFs and FFs where the produced hadron is polarized. The main motivation to study transverse spin dependent FFs is that they can give access to the chiral-odd transversity PDF $h_1$ [35] which cannot be accessed in inclusive measurements. Instead, $h_1$ can be measured by using a transverse spin dependent FF as a quark polarimeter. In unpolarized $e^+e^-$ annihilation one can exploit the fact that in $e^+e^- \to q\bar{q}$ production, the spins of the quarks are correlated and therefore spin dependent FFs can be measured in correlation measurements of back-to-back hadron pairs. The Collins FF $H_1^\perp$, which describes a correlation between the transverse polarization of the fragmenting quark and the transverse momentum of the produced unpolarized hadron, has recently been measured by Belle [31][32], Babar [27] and BES-III [28]. If two hadrons are detected in the final state, the correlation of the relative transverse momentum between the two hadrons and the quark polarization is described by the di-hadron fragmentation function $H_1^\sphericalangle$. Due to the additional degree of freedom provided by the other hadron, this effect survives an integration over the intrinsic transverse momentum in the fragmentation, so $H_1^\sphericalangle$ can be treated in a collinear framework. It has been extracted recently [36] from Belle measurements [33] and has been used for the first measurement sensitive to transversity in $pp$ [37]. Recently, Belle also showed the first measurement sensitive to $G_1^\perp$, which describes the azimuthal correlation of the relative momentum of an unpolarized hadron pair with the parent quark helicity [39]. While the previous examples are FFs, which are sensitive to the parent quark spin and where the produced hadron is spinless, one can also have FFs which describe correlations of the hadron polarization with the spin and/or momentum of the parent quark. Considering this, quite a large number of FFs can be constructed [38] but most remain unmeasured today. A notable exception is the polarizing FF $D_{1T}^{\perp \Lambda/q}$., which describes he transverse polarization of $\Lambda$ hyperons in the fragmentation of unpolarized quarks. This effect has been measured by Belle [39].

**QCD Vacuum effects on Fragmentation**

The FFs discussed previously describe correlations between microscopic quark properties and "macroscopic" properties in the final state. However, there is also a suggestion that fluctuations in the QCD vacuum can leave an imprint on the final state. In particular, coupling to sphalerons or instantons that mediate between QCD vacuum states with different winding numbers could lead to a measurable effects. Quite some time ago jet handedness correlations were suggested as a sensitive observable [40]. However, since these average over many events, they are not unambiguously connected to QCD vacuum fluctuations. More recently, event-by-event fluctuations have been proposed [41]. A precision measurement of this observable would need sufficient statistics of high multiplicity $e^+e^-$ annihilation which could be made available for the first time by the FCC-ee.

## Opportunities with future datasets from Belle II and FCC-ee

Before the turn-on of a FCC-ee, several other facilities will have collected large datasets that can be used for the precision study of FFs. Probably most relevant will be Belle II at SuperKEKB [42]. The successor of Belle will collect about 50 $ab^-1$ at the $\Upsilon(4S)$ resonance over a decade. This is a similar amount as FCC-ee aims for at the $Z^0$ resonance and would therefore be a good opportunity to study evolution effects in FFs and extract gluon FFs from scaling violations, even though they are not very strong in the relevant $Q^2$ region. The planned EIC will allow the study of FFs in SIDIS at much higher $\sqrt{s}$ than was achieved at previous experiments. For example, the plan for the proposed realization at BNL, the eRHIC [43] , is to have a staged approach with $\sqrt{s}$ starting at 25 GeV and reaching 140 GeV after some time. At these energies, higher order effects are heavily suppressed and the collider avoids nuclear effects that are present in fixed targets. In addition, fragmentation in jets can be studied using data from the current $pp$ facilities, RHIC and LHC. This will allow precision measurements of the gluon fragmentation functions. The advantage of an $e^+e^-$ machine is the degree of theoretical control. Particularly the existence of factorization proofs and availability of calculations at NNLO. Therefore, even though the most precise data on gluon FFs or transverse momentum in jet fragmentation might not come from the FCC-ee, the extraction of gluon FFs from FCC-ee and Belle II via evolution equations and the study of fragmentation in jets at FCC-ee would be a crucial input to our understanding of these FFs. Combined with the Belle II data or possibly with even lower energy data from BES III, the study of evolution, which is currently a topic of very active theoretic study in the nuclear physics community, would be very interesting at FCC-ee. The FCC-ee could be instrumental in studying FFs of heavy mesons that are non-trivial to reconstruct in $pp$ environments, for example heavy $\Lambda$ baryons. Together with Belle II data, the flavor structure of heavy mesons that can also be produced at Belle II could be studied as well. Since FCC-ee is accessing the same phase space as the LEP experiments but with much more precision, similar topics can be addressed in kinematic regions that were not accessible by LEP due to lack of statistics. Probably the most prominent example are heavy flavor FFs which we know mostly from LEP data. FCC-ee would help to measure those at mid to high $z$ and possibly access the $p_T$ dependence. In general, FCC-ee would allow unpolarized and polarized FFs to be studied at very high precision. It would add data at high $Q^2$ and high $z$. Due to the high $\sqrt{s}$, lower values of $z$ can also be accessed that remain out of reach at lower energies or come with stronger mass effects. Given the statistics, it is also expected that flavor separation using back-to-back hadrons or polarized beams, would be vastly improved compared to LEP. The exact impact of FCC-ee data on FF extraction can obviously only be evaluated doing a more quantitative study, such as using pseudo-data in a global fit. The FCC-ee is also is also expected to produce a significant amount of Higgs bosons (several 10k), which would give clean access to gluon or $b$-FFs. However, the collected statistics is still small in comparison with other ways to access this FF, but these measurements would certainly be complementary. Due to the large statistics and higher multiplicity, event-to-event fluctuations due to QCD vacuum effects in $e^+e^-$ could be studied for the first time at an FCC-ee.

I want to thank Emanuele Nocera, Ingazio Scimeni and Marco Stratmann for helpful discussions.

# Splitting Functions at NNLO

Oleksandr Gituliar[1] and **Sven-Olaf Moch**[1]

[1] *II. Institut für Theoretische Physik, Universität Hamburg,
Luruper Chaussee 149, D-22761 Hamburg, Germany*

**Abstract:** In this short talk we review the currently available higher-order corrections for splitting functions: corrections at next-to-next-to-leading (NNLO) and next-to-next-to-next-to-leading order (N$^3$LO) in massless QCD to space-like and time-like quantities, mixed QED/QCD corrections, and corrections beyond fixed-order perturbation theory through resummation. The discussion is organized in the context of semi-inclusive electron-positron annihilation and deep-inelastic scattering processes which allow for analyzing properties and relations between time-like and space-like splitting functions.

## Introduction

Higher-order QCD corrections are very important for the precision analyses of processes at high-energy colliders because they allow for an accurate determination of various parameters of QCD and the Standard Model, and for the study of properties of the recently discovered Higgs boson with great precision. This is even more essential in the case of high-luminosity colliders, like the FCC-ee, where the precision of experimental measurements will dominate over the uncertainty of the theoretical predictions.

In this overview we briefly discuss the theoretical uncertainties to processes with long-distance hadronic effects. In practice these effects are described by *parton distribution functions* (PDFs) for the initial state hadrons and by *fragmentation functions* (FFs) for the final state hadrons. So far, these universal functions cannot be calculated from first principles in QCD and have to be extracted from experimental data, usually taken at some fixed value for the relevant hard scale $Q$. Theoretical predictions at different scales, however, are related by the standard renormalization group equations (RGEs) governing the running of PDFs and FFs with the scale $Q$. The evolution kernels in these RGEs are given by the so-called *splitting functions*, which are calculable as a perturbative series in the relevant couplings, i.e., the strong coupling constant $\alpha_s$ in QCD or powers of $\alpha$ and $\alpha_s$ in the mixed QED/QCD case.

The knowledge of higher-order corrections to the splitting functions is crucial in order to minimize theoretical uncertainties to PDFs and FFs. This, in turn, increases the precision of phenomenological analyses in the Standard Model. To that end, we briefly overview the state of the currently available NNLO QCD corrections, i.e., $\mathcal{O}(\alpha_s^3)$, to the time-like splitting functions in QCD and discuss possible techniques for completing their calculation at NNLO and extensions to higher orders.

## Higher-Order Corrections

The RGE evolution of PDFs and FFs is governed by the splitting functions in the corresponding kinematics, i.e. *space-like* for PDFs and *time-like* for FFs. The first phenomenological analysis of the evolution effects at NNLO has been conducted in [1], where PDFs have been extracted from data on deep-inelastic scattering (DIS), although, at that time, by employing approximate information about NNLO splitting functions, derived in [2] on the basis of the calculations [4],[3],[5]. This uncertainty was eliminated several years later when the exact NNLO QCD corrections to the space-like splitting functions became completely available [6],[7], which are now generally used for PDF precision analysis, see, e.g. [8]. In addition, the NNLO corrections to the helicity-dependent splitting functions have recently been obtained [9]. Beyond the QCD corrections at NNLO, the large-$n_f$ contributions to the four-loop splitting functions in QCD have recently become available [11]. In addition, the mixed QED/QCD corrections to the splitting functions could also be taken into account. For the latter, the size of the expansion parameters in perturbation theory is of the same order, i.e., $\mathcal{O}(\alpha_s^3) \sim \mathcal{O}(\alpha_s \alpha)$, see [10] for a review.

On the other hand, global analyses of FFs, e.g., for protons and charged hadrons have long been available to NLO accuracy in QCD [12],[13],[14],[15] and the first analysis of the FFs at NNLO was performed only recently [16]. As in the case of the first NNLO analysis of PDFs this study was performed with the help of the NNLO time-like splitting functions [17],[18],[19] which for the time being are only known with a small uncertainty in the three-loop result for the quark-gluon splitting function $P_{qg}^{(2)T}$. This uncertainty is, however, numerically irrelevant for the phenomenological applications.

An attempt to calculate the exact NNLO corrections to the time-like splitting functions from first principles has been undertaken in [20] following the work of [21]. The basic idea is to calculate a semi-inclusive cross-section for the electron-positron annihilation process, i.e.,

$$e^+ + e^- \to \gamma^*(q) \to p(k_0) + \langle n \text{ partons}\rangle, \tag{1}$$

which can be parametrized as follows

$$\frac{1}{\sigma_{\text{tot}}} \frac{d^2\sigma}{dx\,d\cos\theta} = \frac{3}{8}(1+\cos^2\theta)\,\mathcal{F}_T(x,\epsilon) + \frac{3}{4}\sin^2\theta\,\mathcal{F}_L(x,\epsilon) + \frac{3}{4}\cos\theta\,\mathcal{F}_A(x,\epsilon), \tag{2}$$

where the scaling variable $x$ is defined as

$$x = \frac{2\,q\cdot k_0}{q^2}, \qquad\qquad q^2 = s > 0, \qquad\qquad 0 < x \leq 1\,. \tag{3}$$

Once $\mathcal{F}_T(x,\epsilon)$ in eq. (2) is known the universal structure of the bare quantity after renormalization given by the mass factorization in QCD can be employed to extract unknown contributions to the time-like splitting functions $P_{qq}^{(n)T}$ and $P_{gq}^{(n)T}$ at the $n$-th order. If, alternatively, one considers in eq. (1) the process with exchange of a scalar which couples to gluons, e.g., the Higgs boson in the effective theory, one obtains the corresponding splitting functions $P_{qg}^{(n)T}$ and $P_{gg}^{(n)T}$. The bottle-neck in this approach at NNLO consists of the computation of the phase space master integrals for the individual contributions with real emission partons [22].

It is possible, however, to pass over the direct calculation and to restore time-like splitting functions from the corresponding space-like expressions, which is actually what has been done in refs. [17],[18],[19]. To do that, one can use well-known relations between space- and time-like kinematics, i.e., the Drell-Yan-Levy relation for the analytic continuation in energy $q^2 \to -q^2$ and the Gribov-Lipatov relation in $x$-space [23][24]. The latter one states that $P_{ij}^{(n)T} = -xP_{ji}^{(n)S}(1/x)$, which is exact of leading order, but it is known that these relations are not directly applicable at higher orders, see for instance [25].

Despite of that it has been possible to restore $P_{ns}^{(2)T}$ for the flavor non-singlet [17] as well as $P_{qq}^{(2)T}$ and $P_{gg}^{(2)T}$ for the singlet diagonal [18] time-like splitting functions at NNLO. This has been achieved with the help of additional relations [26], based on a universal reciprocity-respecting evolution kernel, i.e., a function $P$ with the property $P(x) = -xP(1/x)$. Moreover, by using constraints from the momentum sum rule and the supersymmetric limit it was possible to restore the off-diagonal time-like splitting functions $P_{qg}^{(2)T}$ and $P_{gq}^{(2)T}$ in [19]. The result for the former is an approximation, though, as it carries the above mentioned uncertainty in terms proportional to $P_{qg}^{(2)T} \sim \pi^2 \beta_0 (C_A - C_F) P_{qg}^{(0)T}$ due to insufficient information for fixing all color coefficients. Here $\beta_0$ is the leading order coefficient of the beta-function and the color coefficients are $C_A = 3$, $C_F = 4/3$ in QCD.

Improvements beyond fixed order perturbation theory are based on all-order resummations of logarithmically enhanced contributions at the kinematic endpoints $x \to 0$ and $x \to 1$. For the phenomenological analysis of FFs the resummation of small-$x$ double logarithms is particularly important. These give huge contributions already at $x \sim 10^{-3}$ at fixed-order in perturbation theory and the resummation of these logarithms stabilizes and consolidates the theoretical predictions. This analysis has been recently discussed in [27],[28], where some representative analytical and numerical results has been presented, see also [29].

## Summary

In this short talk we have summarized the current status of the higher-order analytic corrections to the space- and time-like splitting functions in massless QCD. While great progress has been achieved in the last two decades in NNLO analyses of these quantities, there is still room for further improvements. The necessary advanced computations for phase space master integrals can be performed by applying contemporary tools and techniques for the calculation of Feynman diagrams.

# Towards a Neural Network determination of Pion Fragmentation Functions

**Valerio Bertone**[1,2], Stefano Carrazza[3], Emanuele R. Nocera[4],
Nathan P. Hartland[1,2], and Juan Rojo[1,2]

[1] Department of Physics and Astronomy, VU University Amsterdam,
De Boelelaan 1081, NL-1081, HV Amsterdam, The Netherlands,
[2] Nikhef, Science Park 105, NL-1098 XG Amsterdam, The Netherlands,
[3] Theoretical Physics Department, CERN, Geneva Switzerland,
[4] Rudolf Peierls Centre for Theoretical Physics, 1 Keble Road
University of Oxford, OX1 3NP, Oxford, United Kingdom.

**Abstract:** We present a first preliminary determination of a set of collinear fragmentation functions of charged pions based on the NNPDF methodology. This determination is based on a wide set of single-inclusive annihilation data, including the most recent and precise measurements from $B$-factory experiments. Our analysis is performed at leading, next-to-leading and next-to-next-to-leading order in quantum chromodynamics. We discuss the result of our fits, highlighting the quality of their description of the data and their stability upon inclusion of higher-order corrections.

## Introduction

In the framework of quantum chromodynamics (QCD), fragmentation functions (FFs) encode the information on how quarks and gluons are turned into hadrons [1]. They are an essential ingredient in the factorisation theorems which allow for a quantitative description of hard-scattering processes involving identified hadrons in the final state [2]. Because of their nonperturbative nature, FFs are typically determined from the data in a global QCD analyses combining results from a variety of processes. These include hadron production in electron-positron single-inclusive annihilation (SIA), in lepton-nucleon semi-inclusive deep-inelastic scattering (SIDIS), and in proton-proton ($pp$) collisions (see *e.g.* Ref. [3]).

In this contribution, we present some recent progress towards a first determination of FFs and their uncertainties based on the NNPDF methodology. Within this methodology, FFs are represented as a Monte Carlo sample, from which the central value and the uncertainty can be computed respectively as a mean and a standard deviation, and they are parametrised by means of neural networks with a very large number of parameters (for details see *e.g.* Ref. [4] and references therein). As compared to the approach used in all the determinations of FFs achieved so far, the NNPDF methodology aims at reducing potential biases related to the procedure used to extract FFs. The NNPDF framework has proven to be robust, and has been extensively used to extract unpolarised [4] and polarised [5] parton distribution functions (PDFs) of the proton.

It looks then sensible to employ the NNPDF methodology to determine also FFs. The aim of this contribution is to begin such a program. To start with, here we limit our analysis to the FFs of charged pions, $\pi^\pm$, based on SIA data only. A dedicated forthcoming publication [6] will discuss these results for the pion FFs in more detail, as well as present results for the FFs of other light hadrons, namely charged kaons, $K^\pm$, and protons/antiprotons, $p/\bar{p}$, which constitute the largest fraction in frequently measured yields of hadrons. The discussion is organised as follows. First, we describe the dataset considered. Then, we discuss the theoretical framework and the settings adopted in our analysis. Finally, we present our set of pion FFs.

## Experimental dataset

This determination of FFs is based on a comprehensive set of cross section data from electron-positron annihilation into charged pions. We include measurements from the experiments performed at CERN (ALEPH [7], DELPHI [8] and OPAL [9]), DESY (TASSO [10,11,12]), KEK (BELLE [13] and TOPAZ [14]), and SLAC (BABAR [15], HRS [16], TPC [17] and SLD [18]). On top of the inclusive measurements, we also include flavour-tagged SIA data from DELPHI [8], TPC [19] and SLD [18]. Specifically, we consider cross section measurements for the sum of light quarks ($u$, $d$, $s$) and for individual charm and bottom quarks ($c$, $b$).

The dataset included in this analysis is summarised in Tab. 1, where we specify the name of the experiments, their corresponding publication reference, the energy of the centre-of-mass system (c.m.s.) $\sqrt{s}$, the relative normalisation uncertainty (r.n.u.), and the number of data points included in the fit. The kinematic coverage of the dataset is displayed in Fig. 1.

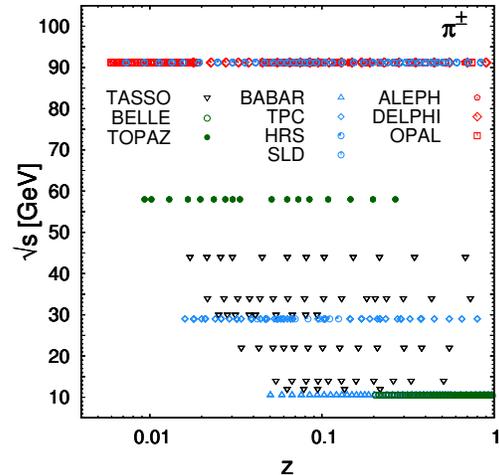

Figure 1: The kinematic coverage in the $(z, \sqrt{s})$ plane of the SIA data in Tab. 1. The data is from DESY (black), KEK (green), SLAC (blue) and CERN (red).

The bulk of the dataset comes from the CERN-LEP and the SLAC-SLC experiments at the scale of the $Z$-boson mass, $\sqrt{s} = M_Z$ (ALEPH, DELPHI, OPAL and SLD), and from the $B$-factory experiments at a significantly lower c.m.s energy, $\sqrt{s} \simeq 10.5$ GeV (BELLE and BABAR). All these experiments provide very precise data, with relative uncertainties at the few percent level, which account for about two thirds of the total dataset. The remaining data points settle at intermediate energy scales and are typically affected by larger uncertainties.

In this analysis, we retain only the data which falls in the range $[z_{\min}, z_{\max}]$, with $z_{\min} = 0.05$ for the experiments at $\sqrt{s} = M_Z$ and $z_{\min} = 0.1$ for all the other experiments, and $z_{\max} = 0.9$ for all the experiments. These cuts are meant to exclude kinematic regions where resummation effects, not taken into account in our fixed-order analysis, may become relevant. The number of data points before kinematic cuts is reported in parentheses in Tab. 1.

We gather all the information on statistical and systematic uncertainties, including their correlations, whenever available, and construct the covariance matrix for each experiment accordingly. Possible normalisation uncertainties, given as a percentage correction to the measured observable (see Tab. 1), are assumed to be fully correlated. Because of their multiplicative nature, which would lead to a systematically biased result [20], they are included through an iterative procedure (the so-called $t_0$ method [21]). As usual in the NNPDF framework, the covariance matrix is used to sample the probability distribution defined by the data, by generating a Monte Carlo pool of $N_{\rm rep} = 100$ pseudodata replicas according to a multi-Gaussian distribution. A different fit to each pseudodata replica is then performed as described in the next section.

| Exp. | Ref. | $\sqrt{s}$ [GeV] | r.n.u. [%] | $N_{\text{dat}}$ | $\chi^2/N_{\text{dat}}$ (LO) | $\chi^2/N_{\text{dat}}$ (NLO) | $\chi^2/N_{\text{dat}}$ (NNLO) |
|---|---|---|---|---|---|---|---|
| BELLE | [13] | 10.52 | 1.4 | 70 (78) | 0.54 | 0.13 | 0.12 |
| BABAR | [15] | 10.54 | 0.098 | 37 (45) | 1.04 | 1.28 | 1.37 |
| TASSO12 | [10] | 12.00 | 20 | 2 (5) | 0.71 | 0.88 | 0.84 |
| TASSO14 | [11] | 14.00 | 8.5 | 7 (11) | 1.54 | 1.60 | 1.68 |
| TASSO22 | [11] | 22.00 | 6.3 | 7 (13) | 1.28 | 1.65 | 1.62 |
| TASSO34 | [12] | 34.00 | 6.0 | 8 (16) | 1.09 | 1.08 | 0.99 |
| TASSO44 | [12] | 44.00 | 6.0 | 5 (12) | 1.96 | 2.00 | 1.85 |
| TPC (incl.) | [17] | 29.00 | — | 12 (25) | 0.79 | 1.02 | 1.13 |
| TPC ($uds$ tag) | [19] | 29.00 | — | 6 (15) | 0.70 | 0.66 | 0.62 |
| TPC ($c$ tag) | [19] | 29.00 | — | 6 (15) | 0.74 | 0.75 | 0.76 |
| TPC ($b$ tag) | [19] | 29.00 | — | 6 (15) | 1.59 | 1.58 | 1.57 |
| HRS | [16] | 29.00 | — | 2 (7) | 2.91 | 4.77 | 4.22 |
| TOPAZ | [14] | 58.00 | — | 4 (17) | 1.03 | 0.94 | 0.81 |
| ALEPH | [7] | 91.20 | 3.0 - 5.0 | 22 (39) | 0.78 | 0.64 | 0.68 |
| DELPHI (incl.) | [8] | 91.20 | — | 16 (23) | 2.63 | 2.62 | 2.59 |
| DELPHI ($uds$ tag) | [8] | 91.20 | — | 16 (23) | 1.99 | 2.00 | 1.93 |
| DELPHI ($b$ tag) | [8] | 91.20 | — | 16 (23) | 1.13 | 1.00 | 1.14 |
| OPAL | [9] | 91.20 | — | 22 (51) | 1.87 | 1.79 | 1.77 |
| SLD (incl.) | [18] | 91.20 | 1.0 | 29 (40) | 0.71 | 0.71 | 0.70 |
| SLD ($uds$ tag) | [18] | 91.20 | 1.0 | 29 (40) | 0.81 | 0.78 | 0.80 |
| SLD ($c$ tag) | [18] | 91.20 | 1.0 | 29 (40) | 0.61 | 0.65 | 0.65 |
| SLD ($b$ tag) | [18] | 91.20 | 1.0 | 29 (40) | 0.45 | 0.60 | 0.46 |
| Total | | | | 380 (602) | 0.995 | 0.963 | 0.958 |

Table 1: The dataset included in this analysis of charged pion FFs. The experiment, the publication reference, the c.m.s. energy $\sqrt{s}$, the relative normalisation uncertainty (r.n.u.), the number of data points after (before) kinematic cuts, and the $\chi^2$ per data point for the LO, NLO and NNLO analyses are displayed.

### Theoretical framework and analysis settings

The leading observable in our analysis is the SIA cross section involving the production of a charged pion $\pi^\pm$ in the final state. This is usually defined in terms of the *fragmentation* (structure) function $F_2^{\pi^\pm}$ as:

$$\frac{d\sigma^\pm}{dz}(z,Q^2) = \frac{4\pi\alpha^2(Q^2)}{Q^2} F_2^{\pi^\pm}(z,Q^2)\,, \tag{1}$$

where $z = E^{\pi^\pm}/E_b = 2E^{\pi^\pm}/\sqrt{s}$ is the energy of the observed pion, $E^{\pi^\pm}$, scaled by the energy of the beam, $E_b$, and $Q^2 > 0$ is equal to the c.m.s. energy squared, $s$. At leading twist, the factorised expression of the inclusive $F_2^{\pi^\pm}$ is given as a convolution between FFs and coefficient functions, by:

$$F_2^{\pi^\pm} = \langle e^2 \rangle \left[ D_\Sigma^{\pi^\pm} \otimes C_{2,q}^{\text{S}} + n_f D_g^{\pi^\pm} \otimes C_{2,g}^{\text{S}} + D_{\text{NS}}^{\pi^\pm} \otimes C_{2,q}^{\text{NS}} \right]\,, \tag{2}$$

where $n_f$ is the number of active flavours, $\langle e^2 \rangle = n_f^{-1} \sum_q^{n_f} \hat{e}_q$ (with $\hat{e}_q$ the effective electroweak charges, see *e.g.* Ref. [22] for their definition), $D_\Sigma^{\pi^\pm} = \sum_q^{n_f}(D_q^\pi + D_{\bar{q}}^\pi)$ is the singlet FF, $D_{\text{NS}}^{\pi^\pm} = \sum_q^{n_f}(\hat{e}_q^2/\langle e^2 \rangle - 1)(D_q + D_{\bar{q}})$ is a nonsinglet combination of FFs, $D_g^{\pi^\pm}$ is the gluon FF, and $C_{2,q}^{\text{S}}$, $C_{2,q}^{\text{NS}}$, and $C_{2,g}^{\text{S}}$ are the corresponding coefficient functions (the explicit dependence on the scales has been omitted for brevity). From Eq. (2), it is apparent that inclusive SIA data can constrain only three independent distributions, namely $D_\Sigma^{\pi^\pm}$, $D_g^{\pi^\pm}$, and $D_{\text{NS}}^{\pi^\pm}$. However, in the case of tagged data, the sums on $q$ inside Eq. (2) run only over tagged quarks. As a consequence, considering charm- and bottom-tagged data allows us to single out two more independent combinations of FFs.

In our analysis, we parametrise five independent FFs. On top of the singlet and the gluon FFs, $D_\Sigma^{\pi^\pm}$ and $D_g^{\pi^\pm}$, we choose the following nonsinglet combinations of FFs:

$$D_{T_3+\frac{1}{3}T_8}^{\pi^\pm} = \frac{2}{3}(2D_{u^+}^{\pi^\pm} - D_{d^+}^{\pi^\pm} - D_{s^+}^{\pi^\pm}), \quad D_{T_{15}}^{\pi^\pm} = \sum_{q=u,d,s} D_{q^+}^{\pi^\pm} - 3D_{c^+}^{\pi^\pm}, \quad D_{T_{24}}^{\pi^\pm} = \sum_{q=u,d,s,c} D_{q^+}^{\pi^\pm} - 4D_{b^+}^{\pi^\pm}, \quad (3)$$

where $D_{q^+} = D_q + D_{\bar{q}}$. The contribution of heavy quarks fragmenting into light hadrons is not well described by perturbative DGLAP evolution, and thus heavy-quark FFs need to be extracted from data. Each FF in our basis is parametrised as $D_i^\pi(z, Q_0) = \text{NN}_i(z) - \text{NN}_i(1)$, $i = g, \Sigma, T_3 + \frac{1}{3}T_8, T_{15}, T_{24}$, where $\text{NN}_i(z)$ are five independent neural networks (multi-layer feed-forward perceptrons) with 37 free parameters each. The subtraction of the term $\text{NN}_i(1)$ ensures that $D_i^\pi(z = 1, Q_0) = 0$, as it should.

The FFs are evolved from the initial parametrisation scale $Q_0$ to the scale of the data by solving time-like DGLAP equations. We use the zero-mass variable-flavour-number (ZM-VFN) scheme, with up to $n_f = 5$ active flavours, in which heavy-quark mass effects in the partonic cross sections are not taken into account. We choose $Q_0 = 5$ GeV, above the charm and bottom masses, but below the lowest value of $\sqrt{s}$ for which SIA data is available. This way, we avoid to deal with cross sections near the heavy-quark thresholds, which would instead be better described in a matched general-mass VFN scheme [24], especially in the presence of intrinsic heavy-quark components.

Our analysis is performed at leading, next-to-leading and next-to-next-to-leading order (LO, NLO and NNLO) accuracy in perturbative QCD. The computation of the cross sections and the evolution of the FFs is performed with the APFEL program [25], and has been extensively benchmarked in Ref. [26]. We use the value $\alpha_s(M_Z) = 0.118$ as a reference for the running of strong coupling at the mass of the Z boson, $M_Z = 91.1876$ GeV, and the values $m_c = 1.51$ GeV and $m_b = 4.92$ GeV for the charm and bottom masses. We also take into account running effects of the fine-structure constant $\alpha$ to LO, taking $\alpha(M_Z) = 1/127$ as a reference value.

Our FFs are fitted to the data by means of a Covariance-Matrix-Adaptation Evolution-Strategy (CMA-ES) learning algorithm [23], which ensures an optimal exploration of the parameter space and an efficient $\chi^2$ minimisation. In order to make sure that our fitting strategy provides a faithful representation of FFs and their uncertainties, we have validated it by means of *closure tests*. As discussed in detail in Ref. [4], closure tests are meant to quantify the robustness of the training methodology by fitting pseudodata generated using a given set of input FFs and checking whether the result of the fit is compatible with the input set. The successful outcome of our closure tests ensures that, in the region covered by the data included in the fit, procedural uncertainties (including those related to the parametrisation) are negligible, and that our extraction of FFs provides a faithful representation of the experimental uncertainties.

## Results

We now present the results of our FF fits to charged pion data. In Tab. 1, we report the values of the $\chi^2$ per data point for each experiment and for the whole dataset included in the fits corresponding to our LO, NLO, and NNLO analyses. We achieve a very good fit quality at all perturbative orders considered, with the global $\chi^2$ being close to one in all cases. The inclusion of higher-order corrections improves the global description of the data noticeably when going from LO to NLO, while only mildly when going from NLO to NNLO. In particular, the description of the BELLE measurements, which represent the most abundant and precise sample in our dataset, improves by a substantial amount. Simultaneously, the $\chi^2$ to the BABAR data deteriorates as more higher-order corrections are included in the fit. This points to a possible tension between BELLE and BABAR measurements, as also suggested in a previous dedicated analysis [27].

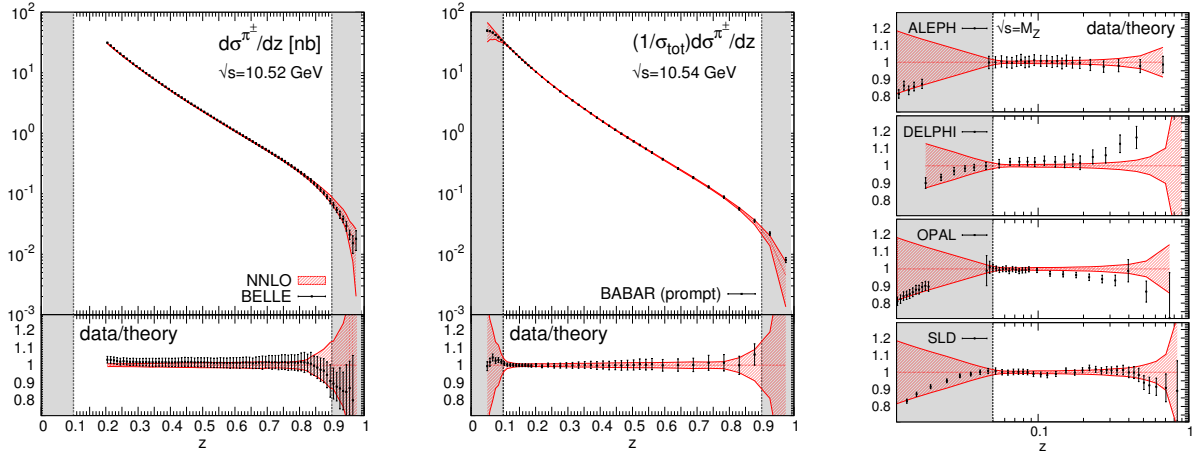

Figure 2: Data/theory comparison for BELLE (left), BABAR (center), and ALEPH, DELPHI, OPAL, and SLD (right) experiments. Predictions are obtained using our NNLO set of FFs. For BELLE and BABAR, we show both the original distributions (upper panel) and the data/theory ratio (lower panel); for the remaining experiments we show only the data/theory ratio. Shaded areas correspond to regions excluded by our cuts.

In Fig. 2 we compare predictions obtained with our NNLO fit with the bulk of our dataset, namely the low-energy measurements ($\sqrt{s} \simeq 10.5$ GeV) from BELLE (left plot) and BABAR (central plot), and the measurements at $\sqrt{s} = M_Z$ from ALEPH, DELPHI, OPAL, and SLD (right plot). We display both the original distributions (upper panels) and the data/theory ratios (lower panels) for BELLE and BABAR, while only the ratio plots for the other experiments. In all plots, shaded areas indicate the kinematic regions excluded by our cuts.

In general, our predictions provide a fairly good description of all datasets, indicating that (N)NLO QCD is able to bridge low- and high-energy data without significant tension. As expected, the agreement between data and predictions in the region allowed by our cuts is particularly good with the only exception of DELPHI, whose large-$z$ measurements tend to overshoot our predictions. This is reflected in the relatively poor $\chi^2$ reported in Tab. 1. Remarkably, most of the measurements falling in the regions excluded by our kinematic cuts are compatible with our predictions that, however, are affected by larger uncertainties there. This suggests that, especially at small values of $z$, NNLO QCD is able to catch most of the beyond-fixed-order effects that our cuts are meant to keep under control (see also Ref. [28]). Consequently, our cuts might be unnecessarily restrictive at NNLO.

Finally, we turn to show the FFs resulting from our fits. From left to right, the plots in Fig. 3 show the singlet, $D_\Sigma^{\pi^\pm}$, the gluon, $D_g^{\pi^\pm}$, the total charm, $D_{c^+}^{\pi^\pm}$, and the total bottom, $D_{b^+}^{\pi^\pm}$, FFs at $Q = 10$ GeV. The upper panel of each plot shows our FFs at LO, NLO, and NNLO, while the lower panel displays the ratio to the corresponding LO distributions. The bands represent the one-$\sigma$ uncertainty. These plots confirm our conclusions on the perturbative stability of our fits. In particular, we observe that in all cases the difference between LO and NLO is sizeable and the respective distributions are not compatible within uncertainty over most of the considered range in $z$. Conversely, the difference between NLO and NNLO is significantly smaller and the distributions are in much better agreement. We also note that the uncertainty band of the LO FFs is larger than that of the NLO and NNLO ones. A broadening of the uncertainties is indeed necessary to accommodate the absence of significant higher-order corrections. This effect, in conjunction with the deterioration of the $\chi^2$, emphasises the inadequacy of the LO approximation.

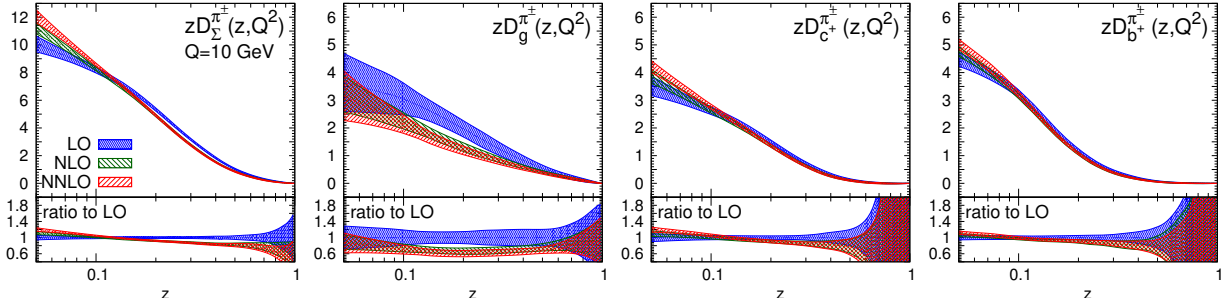

Figure 3: Comparison among our LO, NLO, and NNLO FFs at $Q = 10$ GeV. From left to right, the plots show the singlet $D_\Sigma^{\pi^\pm}$, the gluon $D_g^{\pi^\pm}$, the total charm $D_{c^+}^{\pi^\pm}$, and the total bottom $D_{b^+}^{\pi^\pm}$ FFs. The upper inset of each plot displays the FFs, while the lower inset displays their ratio to the corresponding LO FFs.

### Summary and outlook

We presented preliminary results of a determination of collinear FFs for charged pions based on the NNPDF methodology. The analysis was performed at LO, NLO, and NNLO in QCD and based SIA data only. We achieved a very good description of the data included in the fits and demonstrated perturbative convergence upon inclusion of higher-order corrections.

Our present results represent the first step towards a wider program. In the future, we plan to enlarge the fitted dataset by including hadron-production multiplicities in SIDIS and cross sections in $pp$ collisions. This will allow for a separation between favoured and unfavoured FFs and for a clearer investigation of the flavour dependence of the FFs, aspects not directly accessible from SIA data. Further theoretical sophistications might include the assessment of heavy-quark and resummation effects, which might be significant [24,28].

# First Monte Carlo determination of fragmentation functions from $e^+e^-$ annihilation into hadrons

Nobuo Sato[1]

[1]Jefferson Lab, Newport News, Virginia 23606, USA

**Abstract:** We report the results of the first Monte Carlo study of fragmentation functions extracted from from $e^+e^-$ annihilation data into pions and kaons. Using the iterative Monte Carlo method, the analysis largely eliminates the bias inherent in traditional analyses based on single fits and allows a rigorous determination of fragmentation function uncertainties.

### Introduction

High-energy collisions of electrons and positrons offer a valuable opportunity to study the formation of hadrons from quarks and gluons, providing a unique window on long-distance QCD dynamics. The collinear factorization framework [1] allows the single-inclusive annihilation (SIA) process $e^+e^- \to hX$, where the hadron $h$ is identified from the final state $X$, to be described in terms of a set of universal nonperturbative fragmentation functions (FFs) giving the probability of finding the hadron $h$ with a fraction $z$ of the parton's energy.

The need for accurate knowledge of FFs, especially for the production of kaons, has recently been highlighted [2] in the analysis of polarized semi-inclusive deep-inelastic scattering (SIDIS) asymmetries used to extract the strange quark polarization, $\Delta s$, in the nucleon. Inclusive DIS data alone give rise to a negative value of $\Delta s$ [4], while analysis of SIDIS data using the DSS [5] FFs suggests a positive $\Delta s$ at parton momentum fractions $x \sim 0.1 - 0.2$ [2,3]. Using the HKNS [6] FF parametrization instead, however, which gives somewhat smaller strange $\to$ kaon FFs than with the DSS fit [5], produces a negative $\Delta s$ consistent with the inclusive-only analysis. It is vital, therefore, to understand the differences between the FFs found in different analyses before definitive conclusions can be reached about sign and magnitude of $\Delta s$.

A feature common to all existing global FF analyses [5,6,7,8,9] is that they are obtained from single fits. To address the issues raised by the ambiguities in the strange FFs and the extraction of $\Delta s$, the Jefferson Lab Angular Momentum (JAM) Collaboration performed the first Monte Carlo analysis of FFs [10], using the Iterative Monte Carlo (IMC) methodology [4]. The IMC approach allows a full exploration of the parameter space when sampling initial priors for the fitting function, thereby eliminating bias introduced by fine-tuning specific parameters not well contrained by the data, and provides a robust statistical approach to determining FF uncertainties without the need for *ad hoc* tolerance criteria.

### Methodology

In choosing the appropriate functional form for the FFs, we note that SIA observables are sensitive only to the charge-even quark distributions $D^h_{q^+}(z, Q^2)$ and the gluon $D^h_g(z, Q^2)$, and decouple entirely from the charge-odd combinations $D^h_{q^-}(z, Q^2)$, where $q^\pm = q \pm \bar{q}$. SIA data can therefore only provide information on the $D^h_{q^+}$ and gluon distributions, and not on the separate $q$ and $\bar{q}$ FFs, which would require additional data, such as from SIDIS.

All previous global FF analyses have been based on single $\chi^2$ fits, in which it is not possible to determine *a priori* whether the results corresponds to parameter values from a $\chi^2$ stuck in a local minimum. Furthermore, some shape parameters that are difficult to constrain are typically fixed by hand, and since some of these are strongly correlated, this can significantly bias the results. The issues of multiple solutions and correlations are addressed in the IMC sampling of the parameter space [4,10], which allows exploration of all possible solutions.

The output of the IMC fitting procedure is an MC representation of the probability density for the parameters $\mathcal{P}(a|\text{data})$, which allows the expectation values and variances of the FFs to be estimated from

$$\text{E}[\mathcal{O}] = \int d^m a\, \mathcal{P}(\boldsymbol{a}|\text{data})\, \mathcal{O}(\boldsymbol{a}), \qquad (1)$$

$$\text{V}[\mathcal{O}] = \int d^m a\, \mathcal{P}(\boldsymbol{a}|\text{data})\, \big(\mathcal{O}(\boldsymbol{a}) - \text{E}[\mathcal{O}]\big)^2, \qquad (2)$$

where $\mathcal{O}$ are observables that depend on the FFs and $\boldsymbol{a}$ is the $m$-component vector representing the shape parameters of the FFs. Note that while the FF parametrization used in the IMC analysis [10] is not intrinsically more flexible than others, the MC representation is significantly more adaptable for describing FFs. Indeed, the resulting averaged central value of the FFs as a function of $z$ is a linear combination of many functional shapes, effectively increasing the flexibility of the parametrization.

**Data sets**

In the first IMC analysis of FFs from JAM [10], $\pi^\pm$ and $K^\pm$ cross sections from $e^+e^-$ SIA data were used from experiments at DESY (TASSO and ARGUS), SLAC (TPC, HRS and SLD), CERN (OPAL, ALEPH and DELPHI), and KEK (TOPAZ), as well as more recent data from Belle and BaBar at KEK and SLAC (see Ref. [10] for details). The total number of $\pi^\pm$ and $K^\pm$ data points is 459 and 391, respectively. The use of flavor-tagged data from the OPAL experiment in particular allowed the separation of hadron production from heavy and light quarks. To avoid inconsistencies between the theoretical formalism and the data, cuts were applied to exclude the small-$z$ region: $z > 0.1$ for pion data at energies below the $Z$-boson mass and $z > 0.05$ for data at $Q \approx M_Z$. For kaons, a cut of $z > 0.2$ was applied for the low-$Q$ kaon data sets from ARGUS and BaBar.

**Results and outlook**

To establish the stability of the IMC procedure, we examine the iterative convergence of the priors and the final posterior distributions by observing the variation of the volume $V = \prod_i \sqrt{W_i}$, where $W_i$ are the eigenvalues of the covariance matrix, with the number of iterations, as shown in Fig. 1. During the first $\sim 10$ iterations, the volume changes by some 9 orders of magnitude, indicating a very rapid variation of the prior distribution, but becomes stable after $\sim 30$ iterations. A final iteration is then performed with $10^4$ fits using the optimal MC priors sample.

For the $\pi$ data, good overall agreement with the data is obtained, with a total $\chi^2/N_{\text{dat}} \approx 1.31$. The Belle prompt pion data require an $\approx 10\%$ normalization, which may be related to the overall normalization correction from initial state radiation [13]. For the kaon cross sections, the overall agreement between theory and experiment is slightly better than for pions, mostly because of the relatively larger uncertainties on the $K$ data, with $\chi^2/N_{\text{dat}} \approx 1.01$.

The FFs from the IMC analysis are shown in Fig. 2 at $Q^2 = 1$ GeV$^2$ for light quarks and gluons and $Q^2 = m_q^2$ for the heavy flavors. Both the prior and posterior distributions are shown from selected iteration steps in the IMC chain, with the first and last rows representing the initial and final steps. After the initial iteration, the large spread in the prior FFs from the flat sampling of the parameter space is reduced significantly. We find stable distributions after $\approx 30$ iterations, consistent with the convergence of the volumes in Fig. 1.

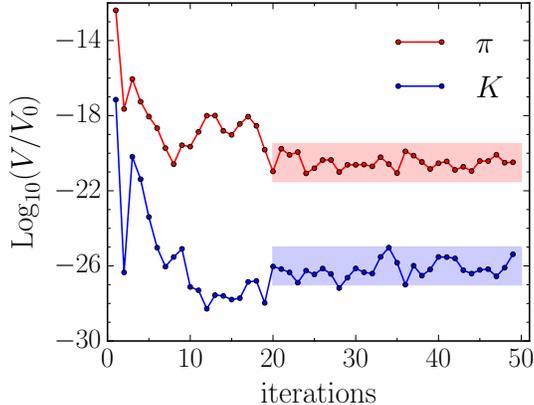

Figure 1: Normalized IMC volume vs. the number of iterations for $\pi$ (red lines) and $K$ (blue lines) mesons. The convergence of the volumes is indicated by the colored bands.

Generally the pion FFs are larger in magnitude than the kaon FFs, with the exception of the strange quark, where the $s^+$ to kaon FF $D^{K^+}_{s^+}$ is larger than that for the pion, $D^{\pi^+}_{s^+}$, over most of the $z$ range. For the fragmentation to kaons, one of the most conspicuous differences with pions is the large $D^{K^+}_{s^+}$, which is comparable to the $u^+$ and $d^+$ FFs to pions.

Compared with parametrizations from other global FF analyses [6,5,7], our fitted FFs are qualitatively similar for the most part, but have some important differences. For pions, our $u^+$ and $d^+$ FFs are $\sim 20\% - 30\%$ larger in magnitude at $z \lesssim 0.3$ compared with the DSS [5] and HKNS [6] results, while for kaons our $u^+$ FF is closer to HKNS and $s^+$ FF closer to DSS. On the other hand, the strange to kaon FF lies between the HKNS and DSS results at intermediate $z$ values, but coincides with the DSS at $z \gtrsim 0.5$. Interestingly, we do not observe the large excess of $s$ to $K$ fragmentation over $u$ to $K$ found in the DSS analysis, which has important phenomenological consequences for the extraction of the polarized strange quark PDF from semi-inclusive DIS data [2].

The kaon FFs, on the other hand, show greater deviation from the earlier results. Here, the favored $D^{K^+}_{s^+}$ function is similar in magnitude to that from the DSS parametrization [5] for $0.5 \lesssim z \lesssim 1$, but displays important differences at $z \lesssim 0.5$ that stem from the greater flexibility of the parametrization used in our analysis. We also find a larger magnitude of the $D^{K^+}_{u^+}$ FF at moderate to low $z$ values compared with the DSS fit in particular. In contrast, the gluon to kaon distribution, which peaks at very large $z$ values, $z \sim 0.85$, but with a very small magnitude, is consistent with the DSS result. The disparity between the fitted $D^{\pi^+}_g$ and $D^{K^+}_g$ functions is particularly striking. At energies on the order of the $Z$-boson mass, the evolved distributions are much more similar to those of the previous analyses, with the exception of the $D^{\pi^+}_g$ and $D^{\pi^+}_{s^+}$ FFs.

The partial separation of the FFs for the various quark flavors has been possible because of the tagged flavor data and the $Q^2$ dependence of SIA cross sections, from low $Q \sim 10$ GeV up to the $Z$-boson mass, selecting differently weighted combinations of FFs in the $\gamma$ and $Z$-exchange cross sections. To further decompose the quark and antiquark FFs, and better constrain the gluon fragmentation, additional information will be needed from SIDIS and meson production in $pp$ collisions. In addition, some tensions exist among high energy SIA data sets at overlapping kinematics, which will require new high precision measurements of SIA at future facilities like the FCCee. A combined analysis of polarized DIS and SIDIS data and SIA cross sections is currently in progress. Programs for generating the "JAM16FF" FFs obtained in this analysis can be downloaded from http://www.jlab.org/theory/jam.

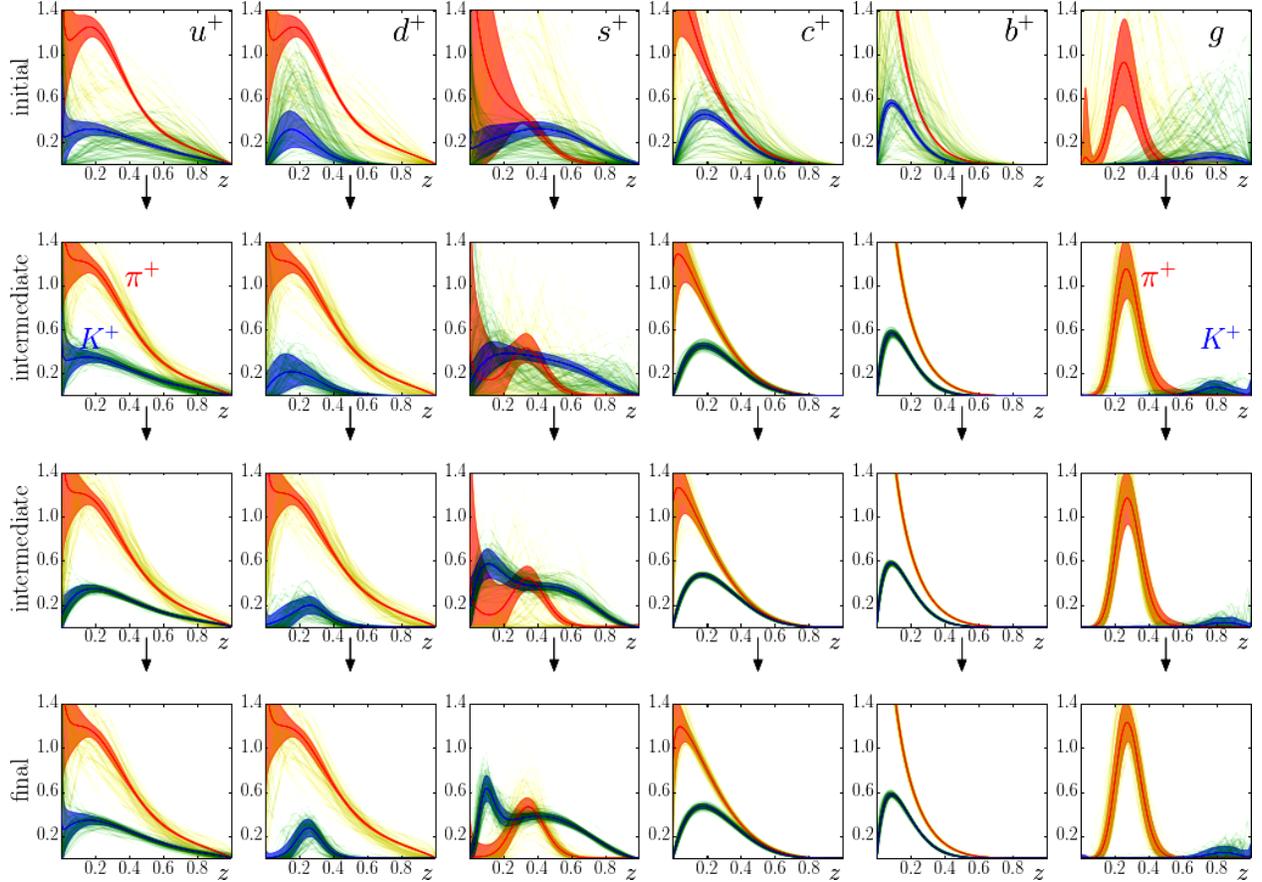

Figure 2: Iterative convergence of the $\pi^+$ (red) and $K^+$ (blue) FFs for various flavors. The first row is the initial flat priors (single yellow curves for $\pi^+$, green for $K^+$), and the corresponding posteriors (error bands). The second and third rows are intermediate IMC chain snapshots, with the last row showing the priors and posteriors of the final IMC iteration.

### Acknowledgments

I thank A. Accardi, J. J. Ethier, M. Hirai, S. Kumano and W. Melnitchouk for their collaboration on this analysis. This work was supported by the U.S. Department of Energy Contract No. DE-AC05-06OR23177, under which Jefferson Science Associates, LLC operates Jefferson Lab.

# Fragmentation Functions from BaBar

Fabio Anulli[1] (on behalf of the BABAR Collaboration)

[1] INFN, Sezione di Roma, I-00185 Roma, Italy

**Abstract:** The BABAR experiment had been taking data for the period 1999-2008 at the PEP-II $e^+e^-$ collider at SLAC. Data were recorded mostly at a center-of-mass (CM) energy of 10.58 GeV, corresponding to the peak of the $\Upsilon(4S)$ resonance, with about 10% of data 40 MeV below it, for an integrated luminosity of about 470 fb$^{-1}$. We present some of the most significant measurements of inclusive production cross sections of light and charmed hadrons, related to unpolarized fragmentation functions, as well as measurements of the spin-dependent Collins fragmentation functions.

## Introduction

Although BABAR [1] was designed and optimized for studying time-dependent $CP$ asymmetries in $B$-meson decays, the high luminosity and excellent detector performances allow also to investigate different aspects of strong interactions, in particular, measurements related to fragmentation functions (FFs) of light and heavy quarks. A FF quantifies the probability of producing a particular hadron $h$ in a jet initiated by a given parton (quark or gluon). In $e^+e^-$ annihilation the FFs are strictly connected to the hadron multiplicities, defined as:

$$F^h(z, Q^2) = \frac{1}{\sigma_{tot}} \frac{d\sigma(e^+e^- \to hX)}{dz}, \qquad (1)$$

where $Q^2 = s$, with $\sqrt{s}$ being the CM energy of the collision, $z \equiv 2E_h/\sqrt{s}$ is the fraction of the parton energy carried by the hadron, and $\sigma_{tot}$ is the total hadronic cross section. At the CM energies of a $B$-factory the process is mediated by a virtual photon, $e^+e^- \to \gamma^* \to q\bar{q}$ at leading order, while at higher energies also the $Z^0$ exchange diagram must be taken into account, as it modifies the total cross section and the flavor composition.

So far, BABAR has measured hadrons multiplicities for several light hadrons, namely $\pi^\pm$, $K^\pm$, $\eta$, and protons, as well as for the charmed baryons $\Lambda_c^\pm$, $\Xi_c^0$, and $\Omega_c^0$. Spin-induced correlations between particles in opposite jets, related to polarized fragmentation functions, have also been studied. A selection of these results is presented in the following sections.

## Inclusive production of light charged hadrons

The BABAR measurements of the inclusive production cross sections of the light hadrons $\pi^\pm$, $K^\pm$, $p/\bar{p}$ are based on a data sets of 0.91 fb$^{-1}$ [2]. In parallel, 3.6 fb$^{-1}$ of data recorded at the $\Upsilon(4S)$ resonance are also analyzed. The latter sample provides independent, stringent systematic checks, and the combined samples provide data-driven calibrations of tracking and particle identification performances. The total systematic uncertainty on the pion cross section is at the level of few percent in the full momentum range. It is dominated at low momenta by tracking efficiencies, and at high momenta by particle identification. The uncertainties on the kaon and proton cross sections have similar patterns, but are significantly larger. The results are presented including the decay products of $K_S^0$ and weakly decaying strange baryons (conventional cross section), or excluding them (prompt production). Figure 1a shows the differential cross sections for the prompt production of the three particles as a function of the scaled momentum $x_p = 2p^*/\sqrt{s}$ (black points), compared with the predictions of the three models JETSET, UCLA and HERWIG, which implement three different mechanisms for hadronization and for which the default parameter values have been used in the simulation. Both statistical and systematic errors are included. Note that the results are very precise and extend up to $x_p \sim 1$. All three models describe the bulk of the spectra qualitatively, but no model describes any spectrum in detail, with the largest deviations from data in the high momentum region. Large deviations are seen especially in the case of the proton cross section.

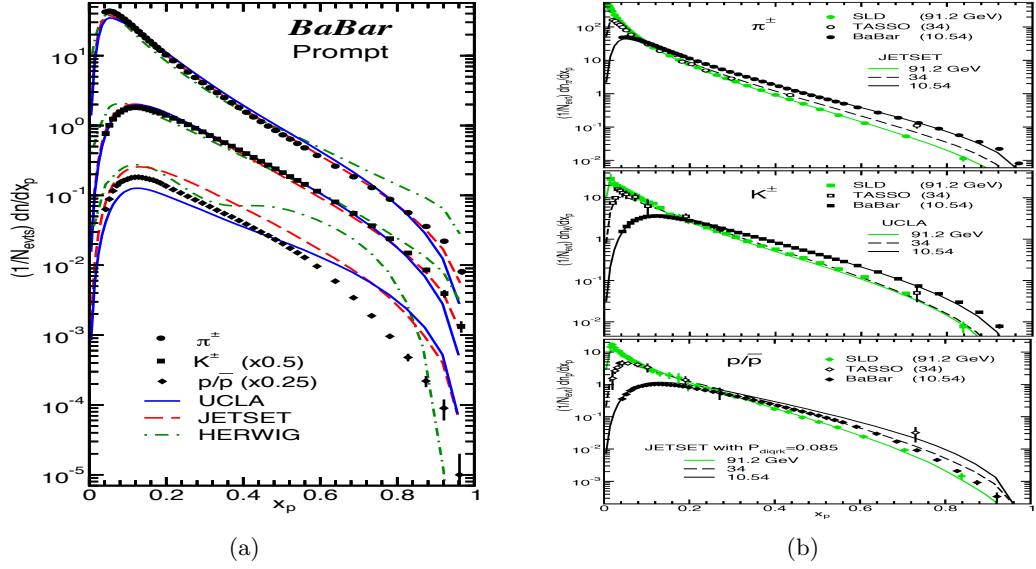

Figure 1: (a) Comparison of the prompt $\pi^\pm$, $K^\pm$ and $p/\bar{p}$ cross sections in $e^+e^- \to q\bar{q}$ events with the predictions of the UCLA, JETSET and HERWIG models. (b) Conventional $\pi^\pm$, $K^\pm$ and $p/\bar{p}$ cross sections measured at three different CM energies, compared with JETSET predictions.

BABAR data can be used together with the very accurate measurements at the $Z^0$ mass to test the scaling properties of hadronization. Scaling violation effects are expected at low $x_p$ due to the mass of hadrons, while at higher momentum a substantial scaling violation is expected because of the running of the strong coupling $\alpha_s$. As an example, Fig. 1b reports the differential cross sections measured at three different energies by the BABAR, TASSO [3] and SLD [4] experiments, and compares them with the predictions from JETSET. Strong scaling violation is observed for the pion data, correctly reproduced by model prediction at all energies for $x_p \gtrsim 0.1$, with only a few percent difference at very high momenta with BABAR data. Also kaon data are consistent with JETSET predictions, which indicates that the model handles correctly the different flavor content at CM energies of 10 ad 90 GeV. On the contrary, the proton data show scaling-violation effects at large $x_p$ smaller than model predictions (this is true also for UCLA and HERWIG, not shown here).

## Inclusive production of charmed hadrons

Heavy hadrons produced in $e^+e^-$ annihilations offer a tool for the study of heavy-quark jet fragmentation, in terms of both the relative production rates of hadrons with different quantum numbers and their associated spectra. The latter can be characterized in terms of a scaled momentum, defined in this case as $x_p = p^*/p^*_{max}$, where $p^*_{max} = \sqrt{s/4 - m^2}$ is the maximum momentum available to a particle of mass $m$ produced via $e^+e^- \to q\bar{q}$. BABAR studied the production of $\Lambda_c^\pm$ [5], $\Xi_c^0$ [6], and $\Omega_c^0$ [7] charmed baryons, containing zero, one and two strange valence quarks, respectively, in addition to the charm quark.

The $\Lambda_c$ study uses a sample of 9.5 fb$^{-1}$ of off-resonance data at $\sqrt{s} = 10.54$ GeV and 81 fb$^{-1}$ at the $\Upsilon(4S)$ peak, and is based on the reconstruction of the 3-body decay mode $\Lambda_c^+ \to pK^-\pi^+$. The invariant mass resolution of the reconstructed $\Lambda_c^+$ varies from 3.75 to 5.75 MeV with increasing $x_p$. Track efficiencies are evaluated from data in two-dimensional $(p, \theta)$ bins, and events are weighted by the efficiency matrix. The distribution of the efficiency-corrected invariant mass is then fitted in each $x_p$ bin to extract the signal yield.

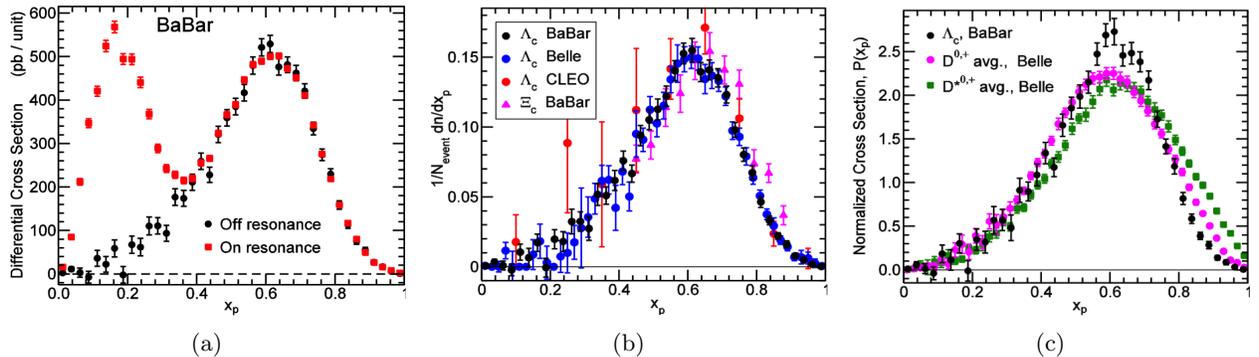

Figure 2: (a) Differential cross sections for $\Lambda_c^+ + \Lambda_c^-$ production in the off-(circles) and on-resonance (squares) data as functions of $x_p$. The errors are statistical only. (b) Differential $\Lambda_c$ production rate compared with previous measurements. The $\Xi_c^0$ rate is normalized to match the peak $\Lambda_c$ rate. (c) Comparison of $\Lambda_c$ data with charmed mesons rates from Belle [8].

The measured cross sections in the off- and on-resonance data sets are reported in Fig. 2a. There are two broad peaks in the on-resonance cross section, corresponding to the contributions from $\Upsilon(4S)$ decays at low $x_p$ and from $e^+e^- \to c\bar{c}$ events at high $x_p$. For $x_p > 0.47$, the kinematic limit for a $B$-meson decay including a $\Lambda_c^+$ and an antiproton, the two cross sections are consistent, indicating no visible contribution from $\Upsilon(4S)$ events. The cross section is obtained assuming a branching fraction $\mathcal{B}(\Lambda_c^+ \to pK^-\pi^+) = 5.0 \pm 1.3\%$.

The shape of the differential production rate is quite hard, as expected, peaking near $x_p = 0.6$; it is consistent with previous results and is measured more precisely, as shown in Fig. 2b, where the BABAR $\Lambda_c$ off-resonance data are compared with CLEO and Belle data, and with the analogous BABAR measurement of $\Xi_c^0$ inclusive production [6]. The peak of the $\Xi_c^0$ distribution, scaled to to the peak height of the $\Lambda_c$ distribution, is slightly shifted to an higher value of $x_p$.

Figure 2c compares the $\Lambda_c$ production rates normalized to unity area measured by BABAR with those obtained by Belle for the inclusive production of charmed $D$ and $D^*$ mesons [8]. Although qualitatively similar, the $D^{(*)}$ meson distributions show broader peaks than the $\Lambda_c$ distribution and differ greatly in the way they drop to zero at high $x_p$.

The measured $\Lambda_c^{\pm}$ differential cross section can be used to test several models of heavy-quark fragmentation, none of which, however, provides a complete description of the data (see [5] for details).

### Polarized fragmentation functions and Collins asymmetries

Transverse spin effects in fragmentation processes were first discussed by Collins, who introduced the chiral-odd polarized fragmentation function $H_{1,q}^{\perp h}(z, P_{h\perp})$ [9]. The probability that a transversely polarized quark ($q^{\uparrow}$), with momentum direction $\hat{\mathbf{k}}$ and spin $\mathbf{S}_q$, fragments into a spinless hadron $h$ with momentum $\mathbf{P}_h$ is defined in terms of the unpolarized $D_{1,q}^h$ and the Collins fragmentation functions:

$$D_{q^{\uparrow}}^h(z, P_{h\perp}^2) = D_{1,q}^h(z, P_{h\perp}^2) + H_{1,q}^{\perp h}(z, P_{h\perp}^2) \frac{(\hat{\mathbf{k}} \times \mathbf{P}_{h\perp}) \cdot \mathbf{S}_q}{z M_h}, \tag{2}$$

where $M_h$, $\mathbf{P}_{h\perp}$, and $z$ are the hadron mass, momentum transverse to $\hat{\mathbf{k}}$, and fractional energy, respectively, in the $e^+e^-$ CM energy. The term including $H_1^{\perp}$ introduces a modulation of the azimuthal angle distribution of the final-state hadrons around the direction of the fragmenting quark, called Collins asymmetry.

In $e^+e^- \to q\bar{q}$ events, the quantities $\hat{\mathbf{k}}$ and $\mathbf{S}_q$ of the two quarks are not experimentally accessible. However, the quarks must be produced back-to-back, with their spins aligned to each other and polarized along the $e^+$ or $e^-$ direction. This results in an azimuthal correlation between pairs of spinless hadrons $h_1$ and $h_2$ in the opposite jets originated by the $q-\bar{q}$ pair, reflecting the product of two Collins functions. The Collins asymmetries can be therefore studied through the process $e^+e^- \to q\bar{q} \to h_1 h_2 X$, where $X$ represents the remainder of the particles produced in the event.

Following the prescription given in Ref. [10], two different reference frames are used: RF12, where the azimutahl angles $\phi_1$ and $\phi_2$ of the two hadrons are defined with respect to a plane spanned by the thrust axis and the $e^+e^-$ axis, and RF0, where the the azimuthal angle $\phi_0$ of one hadron with respect to the plane made by the $e^+e^-$ axis and the momentum of the other hadron is defined.

The first measurements of the Collins effect in $e^+e^-$ annihilation experiments were performed by the Belle Collaboration [11], which studied the dependence of the asymmetry as a function of the pion fractional energies $z_1$ and $z_2$.

BABAR published two analyses on Collins asymmetries: the first one [12] reports the Collins asymmetries for charged pion pairs as a function of fractional energies and transverse momenta of the pions, while in the second analysis [13] a simultaneous extraction of the asymmetries as a function of $z_1$ and $z_2$ for $\pi\pi$, $K\pi$ and $KK$ pairs is performed.

The azimuthal distributions are strongly distorted by detector acceptances and possibly gluon radiation, which can hide the true asymmetry. To get rid of these effects which are independent of the hadrons electric charge, the selected candidate pairs are subdivided in two samples, formed by pairs of unlike-charge (U) and like-charge (L) pions, and the ratio of the two corresponding normalized yield is built. These so-called double ratios can be fitted with a function

$$F_i^{UL} = B_i^{UL} + A_i^{UL} \cos\phi_i, \tag{3}$$

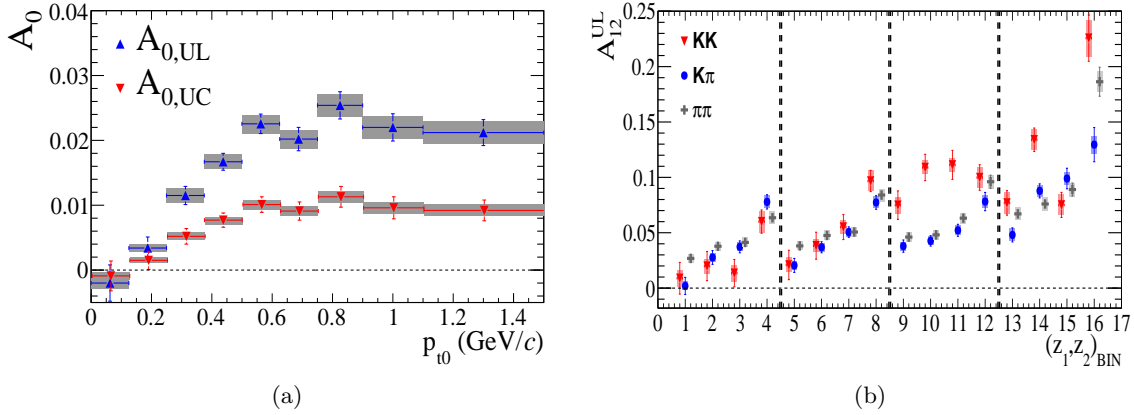

Figure 3: (a) Collins asymmetries for pions as a function of $p_{t0}$ in RF0. Statistic and systematic uncertainties are represented by the bars and the bands around the point, respectively. (b) Asymmetries in RF12 for $KK$, $K\pi$, and $\pi\pi$ pairs in 16 $(z_1, z_2)$ bins: in each interval between the dashed lines, $z_1$ varies in the following ranges: $[0.15, 0.2]$, $[0.2, 0.3]$, $[0.3, 0.5]$, and $[0.5, 0.9]$, while within each interval the points correspond to the analogous four bins in $z_2$.

where $\phi_i \equiv \phi_{12} = \phi_1 + \phi_2$ or $\phi_i \equiv 2\phi_0$ for RF12 and RF0, respectively. The fitted asymmetries $A_i$ are proportional to a particular combination of favored and disfavored Collins and unpolarized fragmentation functions. The $A_i$ need to be corrected by several experimental effects, such as resolution, particle misidentification, and background contamination (in particular by light mesons produced from weak decays in $e^+e^- \to c\bar{c}$ events). The results reported in Ref. [12] obtained for the asymmetries in the two-dimensional $(z_1, z_2)$ bins are generally consistent with those from Belle in the common explored range. The asymmetries are of the order of several percent, and clearly rise with increasing fractional energies. BABAR provided the only measurements of the asymmetries as a function of the pions transverse momenta, which can be used to study the $Q^2$ evolution of the Collins function. As an example, Fig. 3a shows the asymmetries $A_0^{UL}$ and $A_0^{UC}$ (where $UC$ stands for the ratio of unlike-charge over all charged pion pairs) as a function of $p_{t,0}$.

A slightly different event and track selection and a coarser binning is used for the second analysis, in which the asymmetries for $\pi\pi$, $K\pi$ and $KK$ pairs have been measured simultaneously [13]. The asymmetries measured as function of the fractional energies in RF12 for the $U/L$ sample are shown in Fig. 3b. These results provide the first information ever obtained in $e^+e^-$ annihilation on the kaon Collins function, which is sensitive to the strange quark.

The results by BABAR and Belle have been used in combination with data from Semi-Inclusive Deep-Inelastic Scattering experiments to extract simultaneously the Collins FFs and the *transversity* distribution function (see for example the recent works by Anselmino *et al.* [14]).

With the much larger data sets expected at Belle II, it will be possible to improve these studies by measuring the asymmetries in multi-dimensional bins of fractional energies, transverse momenta and polar angles, to access the fully differential cross sections.

# Fragmentation functions at NNLO and their small-z logarithmic corrections

**Daniele Paolo Anderle**[1], Tom Kaufmann[2], Felix Ringer[3], and Marco Stratmann[2]

[1] Lancaster-Manchester-Sheffield Consortium for Fundamental Physics, School of Physics & Astronomy, University of Manchester, Manchester M13 9PL, United
[2] Institute for Theoretical Physics, University of Tübingen, Auf der Morgenstelle 14, 72076 Tübingen, Germany
[3] Theoretical Division, MS B283, Los Alamos National Laboratory, Los Alamos, NM 87545, USA

**Abstract:** With the ever increasing amount of precise data available for hadron production processes, the perturbative QCD framework is being extended to explore effects and corrections that go beyond the next-to-leading order (NLO) accuracy. Fixed order calculations at next-to-next-to-leading order (NNLO) are becoming the new necessary standard required for precision predictions and, consequently, the analysis of the non-perturbative structure of the hadron has to align to this standard. Moreover, relevant effects specific to some kinematical regions, such as the small $z$- and large $z$-regions in Semi-Inclusive $e^+e^-$ Annihilation (SIA), can be investigated through the means of resummation techniques and can be also included in the analysis of final state parton distribution functions. In this talk we present a first analysis of parton-to-pion fragmentation functions at next-to-next-to-leading order based on single-inclusive pion production in electron-positron annihilation, together with its extension to the small $z$-region where an all order resummation of large logarithmic contributions has to be included to further extend the lower cuts on the fit's domain. Further measurements are shown to be necessary in order to extend high precision extractions of FF in the small $z$-region.

## Main Considerations

Fragmentation functions (FFs) $D_i^h(z, Q^2)$ are an integral part of the theoretical framework describing hard-scattering processes with an observed hadron in the final-state in perturbative QCD (pQCD) [1]. They parametrize in a process-independent way the non-perturbative transition of a parton with a particular flavor $i$ into a hadron of type $h$ and depend on the fraction $z$ of the parton's longitudinal momentum taken by the hadron and a large scale $Q$ inherent to the process under consideration [2]. The prime example is single-inclusive electron-positron annihilation (SIA), $e^-e^+ \to hX$, at some center-of-mass system (c.m.s.) energy $\sqrt{S} = Q$, where $X$ is some unidentified hadronic remnant.

Precise data on SIA [3],[4],[5],[6],[7],[8],[9], available at different $\sqrt{S}$, ranging from about $10\,\text{GeV}$ up to the mass $M_Z$ of the $Z$ boson, reveal important experimental information on FFs that is routinely used in theoretical extractions, i.e., fits of FFs [10],[11],[12],[13],[14],[15]. Processes other than SIA are required, however, to gather the information needed to fully disentangle all the different FFs $D_i^h$ for $i = u, \bar{u}, d, \bar{d}, \ldots$ quark and antiquark flavors and the gluon. Specifically, data on semi-inclusive deep-inelastic scattering (SIDIS), $e^\pm p \to hX$, and the single-inclusive, high transverse momentum ($p_T$) production of hadrons in proton-proton collisions, $pp \to hX$, are utilized, which turn extractions of FFs into global QCD analyses [10],[11],[12],[13]. Most recently, a proper theoretical framework in terms of FFs has been developed for a novel class of processes, where a hadron is observed inside a jet [16]. It is expected that corresponding data [17] will soon be included in global analyses, where they will provide additional constraints on, in particular, the gluon-to-hadron FF.

The ever increasing precision of all these probes sensitive to the hadronization of (anti-)quarks and gluons has to be matched by more and more refined theoretical calculations. One way of advancing QCD calculations is the computation of higher order corrections in the strong coupling $\alpha_s$. Here, next-to-leading order (NLO) results are available throughout for all ingredients needed for a global QCD analysis of FFs as outlined above. Specifically, they comprise the partonic hard scattering cross sections for inclusive hadron production in SIA [18],[19], SIDIS [18],[19],[20],[21], and $pp$ collisions [22] and the evolution kernels or time-like parton-to-parton splitting functions $P_{ij}^T$ [23],[24],[25],[26], which govern the scale $Q$ dependence of the FFs through a set of integro-differential evolution equations [27]. Such type of NLO global analyses of FFs represent the current state-of-the-art in this field. For instance, a recent extraction of parton-to-pion FFs $D_i^\pi$ at NLO accuracy can be found in Ref. [13]. A special role in this context plays SIA, where fits of FFs can be carried out already at the next-to-next-leading order (NNLO) level thanks to the available SIA coefficient functions [24],[26],[28],[29] and kernels $P_{ij}^T$ at NNLO [30]. This has not yet been achieved in the case of hadron production in SIDIS or in $pp$ collisions. A first determination of parton-to-pion FFs from SIA data at NNLO accuracy has been performed recently in [14]. Moreover, double-inclusive electron positron annihilation is among those processes involving FF that are not yet included in global analyses although theoretical calculations are available. In particular, expressions for cross sections differential in various kinematical variables are known at NLO accuracy (e.g. see [31],[32] and [33]) and could be used to achieve flavour discrimination in a $e^+e^-$ solely analysis. Since extending those calculations up to NNLO accuracy is an easier task than what would it be in the case of SIDIS or $pp$ collisions, it may be relevant in a near future to have an experimental counterpart of such observables.

Another important avenue for systematic improvements in the theoretical analysis of data sensitive to FFs, which was studied in [34], concerns large logarithms present in each fixed order of the perturbative series in $\alpha_s$ for both the evolution kernels $P_{ij}^T$ and the process-dependent hard scattering coefficient functions. More specifically, one has to deal with logarithms that in the limit of small momentum fractions $z$ can become large and, in this way, can spoil the convergence of the expansion in $\alpha_s$ even when the coupling is very small. Two additional powers of $\log^{2k}(z)$ can arise in each fixed order $\alpha_s^k$, which is numerically considerably more severe than in the space-like case relevant to deep-inelastic scattering (DIS) and the scale evolution of parton density functions (PDFs) and completely destabilizes the behavior of cross sections and FFs in the small-$z$ regime.

To mitigate the singular small-$z$ behavior imprinted by these logarithms, one needs to resum them to all orders in perturbation theory, a well-known procedure [35]. Knowledge of the fixed-order results up to N$^m$LO determines, in principle, the first $m+1$ "towers" of logarithms to all orders. Hence, thanks to the available NNLO results, small-$z$ resummations have been pushed up to the first three towers of logarithms for SIA and the time-like splitting functions $P_{ij}^T$ recently, which is termed the next-to-next-to-leading logarithmic (NNLL) approximation [36],[37]. Based on general considerations on the structure of all-order mass factorization, as proposed and utilized in Ref. [36],[37], in [34] the resummed coefficient functions for SIA and the evolution kernels $P_{ij}^T$ were re-derived and compared to the results available in the literature. Those expressions were also extended by restoring their dependence on the factorization and renormalization scales $\mu_F$ and $\mu_R$, respectively. We note that large logarithms also appear in the limit $z \to 1$. Their phenomenological implications have been addressed in the case of SIA in Ref. [38], [39], but not yet considered in the context of a FF analysis.

In [34] resummations were applied in the entire $z$ range, i.e., for the first time, FFs were extracted from SIA data with identified pions up to NNLO+NNLL accuracy, including a proper matching procedure. The phenomenological investigation was done by comparing the outcome of a series of fits to data both at fixed order accuracy and by including up to three towers of small-$z$ logarithms. The fits were done using a minimization of $\chi^2$-technique such as the one used in [10] and [11].

In terms of $\chi^2$ values the main results are presented in Tab. 1 whereas a visual representation of the same results is given in Fig. 1.

| accuracy | $\chi^2$ | $\chi^2/\text{dof}$ |
|---|---|---|
| LO | 1260.78 | 2.89 |
| NLO | 354.10 | 0.81 |
| NNLO | 330.08 | 0.76 |
| LO+LL | 405.54 | 0.93 |
| NLO+NNLL | 352.28 | 0.81 |
| NNLO+NNLL | 329.96 | 0.76 |

Table 1: The obtained $\chi^2$-values, and the $\chi^2/\text{dof}$ for the fits at fixed order and resummed accuracy as described in the text.

All fits are given with a central choice of scale $\mu_R = \mu_F = Q$. Results are given both for fits at fixed order (LO, NLO, and NNLO) accuracy and for selected corresponding fits obtained with small-$z$ resummations. Here, all cross sections are always matched to the fixed order results according to the procedures described in Sec. IIB and Sec. IIC of [34]. More specifically, the logarithmic order was chosen in such a way that resum logarithmic contributions which are not present in the fixed-order result are not resummed. For this reason,the LO calculation is only matched with the LL resummation as the only logarithmic contribution at LO is of LL accuracy. Using the same reasoning, NLO is matched with the NNLL resummed results. Finally, at NNLO accuracy five towers of small-$z$ logarithms are present. However, the most accurate resummed result currently available is at NNLL accuracy which includes the first three towers. Thus, NNLO is matched only with NNLL.

The main aspects of these fits can be read off directly from Tab. 1: a LO fit is not able to describe the experimental results adequately. The NLO fit already gives an acceptable result, which is further improved upon including NNLO corrections. Compared to the corresponding fixed-order results, the fits including also all-order resummations of small-$z$ logarithms exhibit, perhaps somewhat surprisingly, only a slightly better total $\chi^2$, except for the LO+LL fit, where resummation leads to a significant improvement in its quality. The small differences in $\chi^2$ between fits at NNLO and NNLO+NNLL accuracy are not significant. Hence, we must conclude that in the $z$-range covered by the experimental results, NNLO expressions already capture most of the relevant features to yield a satisfactory fit to the SIA data with identified pions. The same conclusions can be reached from Fig. 1, where the used inclusive pion multiplicity data in SIA are compared with the theoretical cross sections at different levels of fixed- and logarithmic-order obtained from the fits listed in Tab. 1. The vertical dotted lines in Fig. 1 indicate the lower cuts in $z$ applied for the data sets at different c.m.s. energies as discussed in caption and further on. The leftmost line (corresponding to $z_{\min} = 0.075$) is the cut used in the NNLO analysis in Ref. [14]. Both, the data and the calculated multiplicities are shown as a function of $\zeta \equiv -\log z$.

An important phenomenological question that arises in this context is how low in $z$ one can push the theoretical framework outlined in [34] before neglected kinematic hadron mass corrections become relevant. Hadron mass effects in SIA have been investigated to some extent in [41] but there is not a clear answer on how to systematically include them in a general process (see [42], and [43]), i.e.,

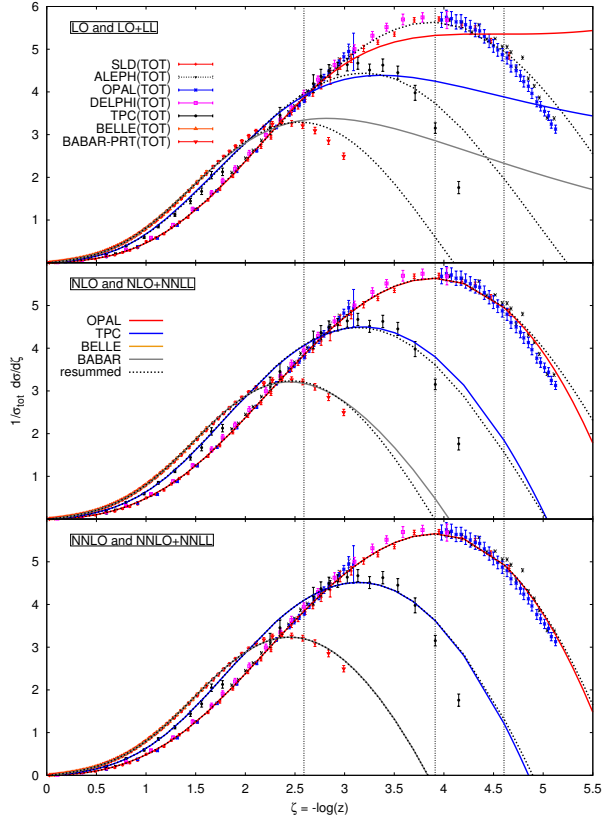

Figure 1: Pion multiplicity data [3],[4],[5],[6],[7],[8],[9] included in the analyses as a function of $\zeta = \log(1/z)$ compared to the results of various fits without (solid lines) and with (dotted lines) small-$z$ resummations. All curves refer to the central choice of scale $\mu = Q$. The top, middle, and lower panel shows the results at LO and LO+LL, NLO and NLO+NNLL, and NNLO and NNLO+NNLL accuracy, respectively. The vertical dotted lines illustrate, from left to right, the lower cuts $z_{\min} = 0.075$ adopted in [14], and $z_{\min} = 0.02$ and $0.01$ used in all our fits for the TPC data and otherwise, respectively.

ultimately in a global analysis of FFs. Therefore, one needs to determine a lower value of $z$, largely on kinematical considerations, below which fits of FFs make no sense. A straightforward, often used criterion to assess the relevance of hadron mass effects is to compare the scaling variable $z$, i.e. the hadron's energy fraction $z = 2\,E_h/Q$ in a c.m.s. frame, with the corresponding three-momentum fraction $x_p$ which is often used in experiments. Since they are related by $x_p = z - 2m_h^2/(zQ^2) + \mathcal{O}(1/Q^4)$ [19], i.e., they coincide in the massless limit, any deviation of the two variables gives a measure of potentially important power corrections. To determine the cut $z_{\min}$ for a given data set, it was demanded that $z$ and $x_p$ are numerically similar at a level of 10 to at most 15%. The BELLE data are limited to the range $z > 0.2$ [3], where $z$ and $x_p$ differ by less than 1%. BABAR data are available for $z \gtrsim 0.05$, which translates in a maximum difference of the two variables of about 14%. Concerning the TPC data, a lower cut had to be placed at $z_{\min} = 0.02$ to arrive at a converged fit, which corresponds to a difference of approximately 11% between $z$ and $x_p$

In general, it turns out, that in the range of $z$ where SIA data are available and where the framework can be applied, a fit at fixed, NNLO accuracy already captures most of the relevant small-$z$ behavior needed to arrive at a successful description of the data, and resummations add only very little in a fit.

In light of this observations, a better understanding of the interplay of resummations and other sources of potentially large corrections in the region of small momentum fractions is another important avenue of future studies for time-like processes. One if not the most important source of power corrections is the hadron mass, which is neglected in the factorized framework adopted for any analysis of fragmentation functions. At variance with the phenomenology of parton distributions functions, where one can access and theoretically describe the physics of very small momentum fractions, hadron mass corrections prevent that in the time-like case. In fact, they become an inevitable part and severely restrict the range of applicability of fragmentation functions and the theoretical tools such as resummations. In addition, resummations can and have been studied for large fractions of the hadron's momentum. With more and more precise data becoming available in this kinematical regime, it would be very valuable to incorporate also these type of large logarithms into the analysis framework for fragmentation functions at some point in the future.

# Soft parton-to-hadron FFs at NNLO*+NNLL

**David d'Enterria**[1] and Redamy Pérez-Ramos[1,2,3]

[1] CERN, PH Department, CH-1211 Geneva 23, Switzerland
[2] Sorbonne Universités, UPMC Univ Paris 06, UMR 7589, LPTHE, F-75005, Paris, France
[3] CNRS, UMR 7589, LPTHE, BP 126, 4 place Jussieu, F-75252 Paris Cedex 05, France

**Abstract:** The evolution of the parton-to-hadron fragmentation functions (FF) at low fractional hadron momenta $z$ is theoretically studied in a framework combining next-to-next-to-leading-order (NNLO) $\alpha_s$ corrections with next-to-modified-leading logarithmic resummations of soft and collinear parton radiation. The energy evolution of the moments of the low-$z$ FFs are thereby computed, and compared to the existing experimental $e^+e^-$ and DIS $e^{\pm}$p jet data. The impact of hadron-mass effects and higher-order corrections is presented. The data–theory comparison of the four FF moments (total hadron multiplicity, peak, width, and skewness) allow us to extract the QCD coupling $\alpha_s$ at approximate NNLO accuracy, $\alpha_s(m_Z^2) = 0.1205 \pm 0.0010^{+0.0022}_{-0.0000}$, in excellent numerical agreement with the current world average.

## Introduction

The distribution of hadrons in a jet is theoretically described by a fragmentation function (FF), $D_{i \to h}(z, Q)$, representing the probability that parton $i$ fragments into hadron $h$ carrying a fraction $z = p_{\text{hadron}}/p_{\text{parton}}$ of the parent parton momentum. Starting with a parton at a given energy $Q$, its evolution to another energy scale $Q'$ is driven by a branching process of parton radiation and splitting, resulting in a jet shower, which can be computed perturbatively using the DGLAP equations [1] at large $z \gtrsim 0.1$. The bulk of hadrons produced in the fragmentation of a jet come, however, from shower partons with low $z < 0.1$ values, a regime dominated by soft and collinear gluon bremsstrahlung that require proper resummation of their associated $\log(1/x)$ and $\ln\theta$-singularities. Indeed, the emission of successive gluons inside a jet follows a parton cascade where the emission angles decrease as the jet evolves towards the hadronisation stage ("angular ordering"). Due to colour coherence and interference of gluon radiation, the energy spectrum of the majority of intrajet partons adopts a typical "hump-backed plateau" (HBP), or distorted Gaussian (DG), shape as a function of the log of the inverse of $z$, $\xi = \ln(1/z)$. This final distribution is infrared-safe in the sense that it is directly imprinted, without modifications, into the final charged particles spectrum after parton hadronization.

Various perturbative resummation schemes have been developed to treat the soft and collinear singularities present in the shower evolution of a jet: (i) the Leading Logarithmic Approximation (LLA) resums single logs of the type $\left[\alpha_s \ln\left(k_\perp^2/\mu^2\right)\right]^n$ where $k_\perp$ is the transverse momentum of the emitted gluon with respect to the parent parton, (ii) the Double Logarithmic Approximation (DLA) resums soft-collinear and infrared gluons, $g \to gg$ and $q(\bar{q}) \to gq(\bar{q})$, for small values of $z$ and $\theta$ $[\alpha_s \ln(1/z) \ln\theta]^n \sim \mathcal{O}(1)$ [2,3], (iii) Single Logarithms (SL) [4,5] take into account the emission of hard collinear gluons ($\theta \to 0$), $[\alpha_s \ln\theta]^n \sim \mathcal{O}(\sqrt{\alpha_s})$, and (iv) the Modified Leading Logarithmic Approximation (MLLA), a SL correction to the DLA, resumming terms of order $[\alpha_s \ln(1/z) \ln\theta + \alpha_s \ln\theta]^n \sim [\mathcal{O}(1) + (\mathcal{O}(\sqrt{\alpha_s}))]$ [4]. We have developed a scheme to analytically compute the evolution of the HBP distribution of soft radiated partons in jets combining the next-to-MLLA approach [6] with fixed-order $\alpha_s$ corrections at an increasing level of accuracy: first at approximate next-to-leading-order, NLO* [7], followed by full-NLO [9], and then approximately next-to-NLO, NNLO* [10].

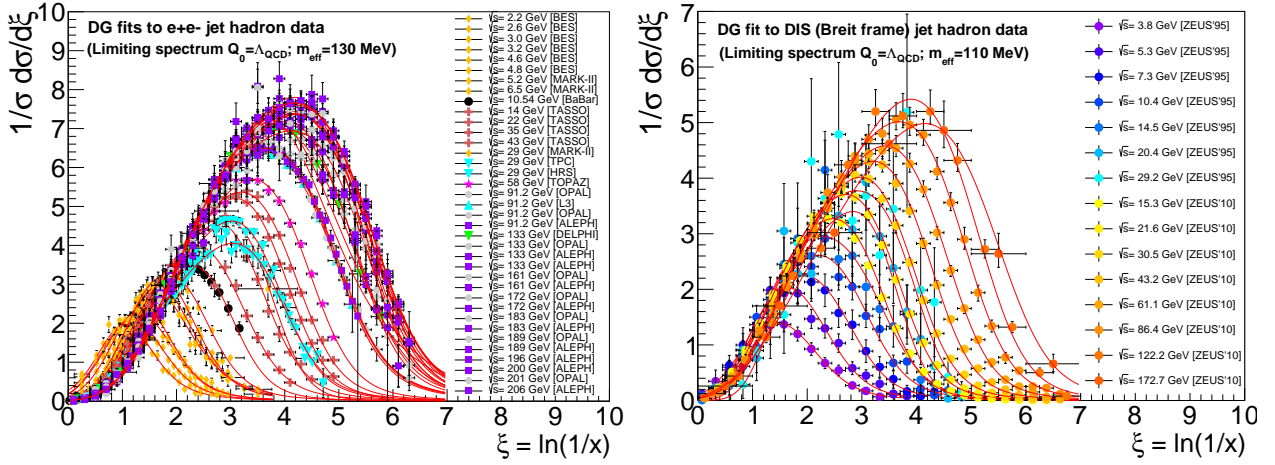

Figure 1: Charged-hadron spectra in jets as a function of $\xi = \ln(1/z)$ in $e^+e^-$ at $\sqrt{s} \approx$ 2–200 GeV (left), and $e^\pm, \nu$-p (Breit frame, scaled ×2 for the full hemisphere) at $\sqrt{s} \approx$ 4–180 GeV (right), individually fitted to Eq. (1) with the hadron mass corrections ($m_{\text{eff}} =$ 130,110 MeV) quoted.

We have used this framework to calculate the evolution of the jet FF, or rather its moments, as a function of the jet energy. The soft fragmentation function of a jet can be expressed, without any loss of generality, in terms of a distorted Gaussian (DG):

$$D(\xi, Y, \lambda) = \mathcal{N}/(\sigma\sqrt{2\pi}) \cdot e^{\left[\frac{1}{8}k - \frac{1}{2}s\delta - \frac{1}{4}(2+k)\delta^2 + \frac{1}{6}s\delta^3 + \frac{1}{24}k\delta^4\right]}, \text{ with } \delta = (\xi - \bar{\xi})/\sigma, \quad (1)$$

where $\mathcal{N}$ is the average hadron multiplicity inside a jet, and $\bar{\xi}$, $\sigma$, $s$, and $k$ are respectively the mean peak position, dispersion, skewness, and kurtosis of the distribution. Within our DGLAP+NMLLA framework, the evolution of the moments of the FF as a function of initial jet energy and shower energy cutoff depend only on $\Lambda_{\text{QCD}}$. By comparing the theoretical energy dependence of the FF moments to jet fragmentation measurements in $e^+e^-$ and deep-inelastic $e^\pm$p collisions, we are therefore able to extract a high-precision value of $\alpha_s$ at NNLO* accuracy.

### Energy evolution of the soft parton-to-hadron fragmentation functions

The system of equations for the FFs $D_{i \to h}(z, Q)$ can be written as an evolution Hamiltonian which mixes gluon and (anti)quark states expressed in terms of DGLAP splitting functions for the branchings $g \to gg$, $q(\bar{q}) \to gq(\bar{q})$ and $g \to q\bar{q}$, where $g$, $q$ and $\bar{q}$ label a gluon, a quark, and an anti-quark respectively. The set of DGLAP+NMLLA integro-differential equations for the FF evolution can be solved analytically (iteratively) by expressing the Mellin-transformed hadron distribution in terms of the anomalous dimension $\gamma$: $D \simeq C(\alpha_s(t)) \exp\left[\int^t \gamma(\alpha_s(t'))dt\right]$ for $t = \ln Q$, leading to a perturbative expansion in half powers of $\alpha_s$: $\gamma \sim \mathcal{O}(\alpha_s^{1/2}) + \mathcal{O}(\alpha_s) + \mathcal{O}(\alpha_s^{3/2}) + \mathcal{O}(\alpha_s^2) + \mathcal{O}(\alpha_s^{5/2}) + \cdots$. The full expansion including up to the complete set of $\mathcal{O}(\alpha_s^2) + \mathcal{O}(\alpha_s^{5/2})$ terms corresponds to theoretical results at NNLO+NNLL accuracy.

The anomalous dimension $\gamma$ allows one to calculate the moments of the DG through:

$$\mathcal{N} = K_0, \quad \bar{\xi} = K_1, \; \sigma = \sqrt{K_2}, \; s = \frac{K_3}{\sigma^3}, \; k = \frac{K_4}{\sigma^4}; \; K_{n\geq 0}(Y,\lambda) = \int_0^Y dy \left(-\frac{\partial}{\partial \omega}\right)^n \gamma_\omega \bigg|_{\omega=0}, \quad (2)$$

which are then inserted into Eq. (1). Corrections of $\gamma$ up to order $\alpha_s^{3/2}$ were computed in Refs. [7,8], followed by the full set of NLO $\mathcal{O}(\alpha_s^2)$ terms, including the two-loop splitting functions, in Ref. [9]. At NLO, the diagonalisation of the evolution Hamiltonian results in two eigenvalues $\gamma_{\pm\pm}$ in the $\mathcal{D}^\pm$ basis, where the relevant one for the calculation of the FF moments $\gamma_{++} \to \gamma_\omega^{\rm NLO+NNLL}$, reads:

$$\begin{aligned}\gamma_\omega^{\rm NLO+NNLL} &= \frac{1}{2}\omega(s-1) + \frac{\gamma_0^2}{4N_c}\left[-\frac{1}{2}a_1(1+s^{-1}) + \frac{\beta_0}{4}(1-s^{-2})\right] \\ &+ \frac{\gamma_0^4}{256N_c^2}(\omega s)^{-1}\left[4a_1^2(1-s^{-2}) + 8a_1\beta_0(1-s^{-3}) + \beta_0^2(1-s^{-2})(3+5s^{-2})\right. \\ &\left. - 64N_c\frac{\beta_1}{\beta_0}\ln 2(Y+\lambda)\right] \\ &+ \frac{1}{4}\gamma_0^2\omega\left[a_2(2+s^{-1}+s) + a_3(s-1) - a_4(1-s^{-1}) - a_5(1-s^{-3}) - a_6\right], \quad (3)\end{aligned}$$

where $\gamma_0^2 = \frac{4N_c\alpha_s}{2\pi} = \frac{4N_c}{\beta_0(Y+\lambda)}$ is the LL anomalous dimension, $s = \sqrt{1+\frac{4\gamma_0^2}{\omega^2}}$, $\beta_i$ the QCD $\beta$-function coefficients, $a_{1,2}$ and hard constants obtained in [7], and $a_{3,4,5,6}$ are new constants obtained from the full-resummed NNLL splitting functions [11]. In addition, terms from the NNLO $\alpha_s$ running expression and the full systematic expansion of the anomalous dimension from the NNLL small-$z$ resummed splitting functions have been added to the order $\mathcal{O}(\alpha_s^{5/2})$ [12]. In particular, the account of NNLO $\alpha_s$ provides corrections $\propto \beta_2$ which were not considered before, and which are needed for the extraction of accurate values of $\alpha_s$ from the data. Such terms should be added with those containing the small-$z$ resummed N$^3$LO splitting functions on equal footing and to the same order. (Since the splitting functions are only incorporated at NLO for the moment; our calculations can be considered of order NNLO$^*$+NNLL so far). Upon inverse-Mellin transformation, one obtains the energy evolution of the FF, and its associated moments, at NNLO$^*$+NNLL accuracy, as a function of $Y = \ln(E/\Lambda_{\rm QCD})$, for an initial parton energy $E$, down to a shower cut-off scale $\lambda = \ln(Q_0/\Lambda_{\rm QCD})$ for $N_f = 3, 4, 5$ quark flavors. The resulting formulae for the energy evolution of the moments depend on $\Lambda_{\rm QCD}$ as *single* free parameter. Relatively simpler expressions are obtained in the limiting-spectrum case (for $\lambda = 0$, i.e. evolving the FF down to $Q_0 = \Lambda_{\rm QCD}$) motivated by the "local parton hadron duality" hypothesis for infrared-safe observables which states that the HBP distribution of partons in jets is simply renormalized in the hadronization process without changing its shape. Thus, by fitting the experimental hadron jet data at various energies to Eq. (1), one can determine $\alpha_s$ from the corresponding energy-dependence of its FF moments.

Figure 2 shows the energy evolution of the zeroth (multiplicity), first (peak position, closely connected to the mean of the distribution), second (width), and third (skewness) moments of the FF, at four levels of accuracy (LO$^*$+LL, NLO$^*$+NLL, NLO+NNLL, and NNLO$^*$+NNLL). The hadron multiplicity increases exponentially with jet energy whereas the FF peak and width do it logarithmically, and the skewness features a slow power-law dropoff. The theoretical convergence of their evolutions are robust as proven by the small changes (10% max.) introduced by incorporating higher-order terms. On the other hand, the kurtosis (not shown), obtained from the fourth derivative of the anomalous dimension, features large non-convergent fluctuations in their jet-energy dependence from LO to NNLO$^*$.

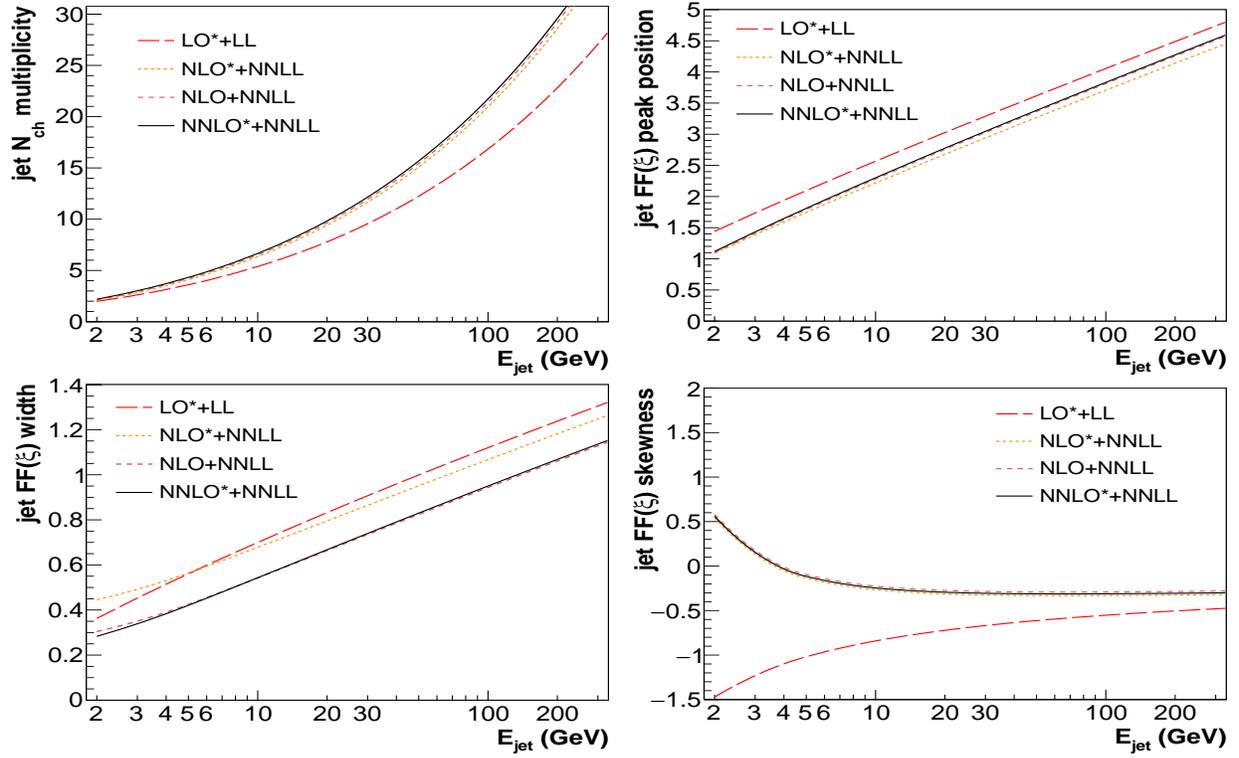

Figure 2: Comparison of theoretical predictions at increasing level of accuracy (LO$^*$ to NNLO$^*$) for the energy evolution of the jet charged-hadron multiplicity (top, left), FF peak position (top, right), FF width (bottom, left), and FF skewness (bottom, right).

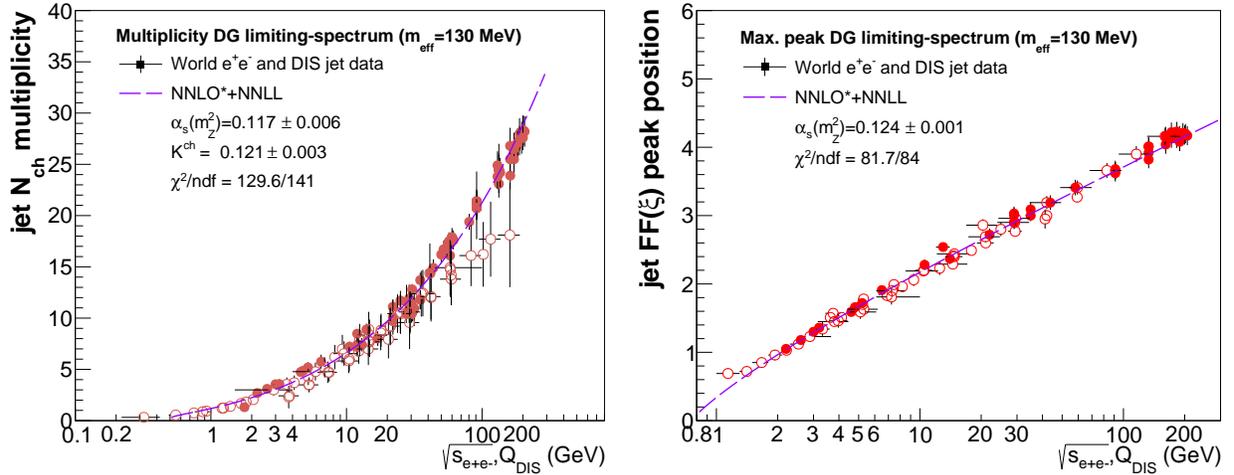

Figure 3: Energy evolution of the jet charged-hadron multiplicity (left) and FF peak position (right) in $e^+e^-$ and DIS jet data, fitted to the NNLO$^*$+NNLL predictions. The obtained $\mathcal{K}_{\mathrm{ch}}$ normalization constant, the individual NNLO$^*$ $\alpha_{\mathrm{s}}(\mathrm{m}_{\mathrm{Z}}^2)$ values, and $\chi^2/\mathrm{ndf}$ of the two fits, are quoted.

**Data-theory comparison and $\alpha_\text{s}$ extraction**

We have first fitted to Eq. (1) the existing experimental jet FFs measured in 34 data sets from $e^+e^-$ annihilation at $\sqrt{s} = 2.2$–206 GeV (amounting to 1 200 HBP data points, Fig. 1 left) as well as from 15 DIS data-sets of the ZEUS collaboration over $\sqrt{s} = 2.2$–206 GeV, measured in the so call "brick-wall" Breit frame where the incoming quark scatters off the photon and returns along the same axis, (Fig. 1, right). Finite hadron-mass effects in the DG fit are accounted for through a rescaling of the theoretical (massless) parton momenta with an effective mass $m_\text{eff} \approx m_\text{pion}$ as discussed in Refs. [7,8]. Also, the impact of particle decays on the extracted FF moments has been assessed comparing the BaBar data for prompt and inclusive hadrons [13], finding negligible effects for $\xi < 5$. To this first set of fitted FF moments, we have added the DG moments independently fitted by many other different $e^+e^-$ and DIS measurements and published in the literature, finally collecting a total of about 340 experimental FF moments.

The individual fits for the first two FF moments are shown in Fig. 3 as a function of the original parton energy (i.e. $\sqrt{s}/2$ in the case of $e^+e^-$, and invariant 4-momentum transfer $Q_\text{DIS}$ for DIS). The overall normalization ($\mathcal{K}_\text{ch}$) of the charged-hadron multiplicity of the jet, is an extra free parameter for this moment which, nonetheless, plays no role in the final $\Lambda_\text{QCD}$ extracted given that its value *just* depends on the *evolution* of the multiplicity, and not on its absolute value at any given energy. The NNLO*+NNLL limiting-spectrum ($\lambda = 0$) predictions for $N_f = 5$ active quark flavours[*], leaving $\Lambda_\text{QCD}$ as a free parameter, reproduce very well the data. The most "robust" FF moment for the determination of $\Lambda_\text{QCD}$ is the peak position $\xi_\text{max}$ which proves quite insensitive to most of the uncertainties associated with the extraction method (DG and energy evolution fits, finite-mass corrections) [8] as well as to higher-order corrections. The hadron multiplicities measured in DIS jets appear somewhat smaller (especially at high energy) than those measured in $e^+e^-$, due to limitations in the FF measurement only in half (current Breit) $e^\pm p$ hemisphere and/or in the determination of the relevant $Q$ scale [8].

The value of $\alpha_\text{s}(m_\text{Z}^2)$ obtained from the combined multiplicity+peak fit is $\alpha_\text{s}(m_\text{Z}^2) = 0.1205 \pm 0.0010$, where the error includes all uncertainties as discussed in Ref. [8]. A conservative theoretical scale uncertainty of $^{+0.0022}_{-0.0000}$ (obtained in [8] at NLO accuracy only) is added. This value is in perfect agreement with the current world-average, $\alpha_\text{s}(m_\text{Z}^2) = 0.1186 \pm 0.0012$, obtained at NNLO accuracy from the combination of 6 different sets of observables [14,15]. The precision of our result ($^{+2\%}_{-1\%}$) is clearly competitive with the other existing measurements, with a totally different set of experimental and theoretical uncertainties and, in particular, with an infrared-safe method that proves resilient against hadronization uncertainties. Upcoming full-NNLO corrections of the energy evolution of the FF moments will entail a reduction of our final $\alpha_\text{s}$ extraction and an eventual incorporation of the measurement into the world-average value. In the further future, the huge jet samples available at FCC-ee in the range of energies $\sqrt{s} = 90$–350 GeV (combined with detectors with advanced charged particle reconstruction down to momenta $p_\text{T} \approx 100$ MeV) will allow one to further reduce the experimental uncertainties, and extract $\alpha_\text{s}$ with permille precision.

---

[*]Few-% corrections are applied to deal with slightly different $N_f = 3, 4$ evolutions below charm, bottom thresholds.

# Small-Radius Jets and Substructure

Mrinal Dasgupta [1]

[1] School of Physics and Astronomy, University of Manchester

**Abstract:** We provide a review of our recent work on theoretical predictions for jets with a small radius, $R \ll 1$, which include the resummation of logarithms of $R$ to all orders in perturbation theory. Since the calculations we mention and the physics that we address are essentially independent of the hard process in which the observed jet is produced, we expect these results to be of interest for both hadron colliders such as the LHC as well as future $e^+e^-$ machines.

## Introduction

The study of jets and their shapes and substructure has been an integral part of collider phenomenology at particle colliders of the past such as LEP, HERA the Tevatron as well currently at the LHC [1]. Such studies have been crucial for example in establishing QCD as the theory of strong interactions. The definition of a jet requires an infrared and collinear safe algorithm as well as generally speaking involves a parameter $R$ interpreted as the jet radius, which characterises the angular size of the jet. While for testing perturbative QCD and extracting the strong coupling it is common to choose a value of $R \sim 1$, at hadron colliders in particular it is often desirable to choose a radius $R \ll 1$. One of the main reasons for this is to reduce the impact of contamination of jets from the underlying event as well as from pile-up. The accurate study of small $R$ jets involves the issue of all-orders resummation for terms that are logarithmically enhanced in $R$ as well as accounting for hadronisation corrections varying as $1/R$. These issues are also important for studies of jet substructure, which is probed via observables that involve small radii such as trimming and filtering. We describe below, the method and results for the all-orders resummation of $\ln R$ enhanced terms to leading (single) logarithmic accuracy for a variety of jet observables.

## Generating functionals and jet fragmentation functions

We address the issue of pure collinear enhancements to jet observables i.e. terms varying as $\alpha_s^n \ln^n R^2$. The origin of these terms, for *all* observables we consider here, are partonic cascades strongly ordered in emission angle i.e. for $n$ emissions from a hard initiating parton (quark or gluon) one has $\theta_1 \gg \theta_2 \gg \cdots \theta_n$.

In order to address the problem of small-$R$ resummation as generally as possible we first introduce quark and gluon generating functionals $Q[x,t]$ and $G[x,t]$ which encode the partonic configurations that are obtained when an initial quark or gluon with momentum fraction $x$ are probed on an angular scale corresponding to $t$ where we defined

$$t = \int_{R^2}^{1} \frac{\alpha_s(p_t\theta)}{2\pi} \frac{d\theta^2}{\theta^2} \sim \frac{\alpha_s}{2\pi} \ln \frac{1}{R^2}, \tag{1}$$

such that $t$ is a single-log evolution variable that takes account of the running of $\alpha_s$.

The evolution of an initial quark or gluon with respect to the angular scale $t$ may be represented via the DGLAP style evolution equations for the corresponding generating functionals

$$\frac{dQ(x,t)}{dt} = \int dz\, p_{qq}(z)\left[Q(zx,t)\,G((1-z)x,t) - Q(x,t)\right], \tag{2}$$

$$\frac{dG(x,t)}{dt} = \int dz\, p_{gg}(z)\left[G(zx,t)G((1-z)x,t) - G(x,t)\right] +$$
$$+ \int dz\, p_{qg}(z)\left[Q(zx,t)Q((1-z)x,t) - G(x,t)\right]. \tag{3}$$

The above coupled generating functional evolution equations may be solved using numerical (Monte Carlo) methods [2]. Alternatively it is also possible to obtain analytical or numerical results at any fixed-order in $t$ via Taylor expansion of the generating functionals. Results upto order $t^3$ for various quantities of interest can be found in Ref. [2]. Here we shall focus on all-orders resummed results.

The generating functionals can be used to derive resummed results for two types of jet fragmentation functions. We define $f_{j/i}^{\mathrm{incl}}(z,t)$ which gives the inclusive spectrum of microjets of flavour $j$ carrying momentum fraction $z$ of parton $i$. In several instances one is also interested in the hardest microjet that emerges from the fragmentation of an intial parton and here we define $f^{\mathrm{hardest}}(z,t)$ which gives the hardest microjet spectrum.

## Resummed results

We discuss below the resummed results that we obtained for the inclusive jet spectrum, jet substructure techniques such as filtering and for jet vetoes in Higgs productions.

### Inclusive jet spectrum

The inclusive jet spectrum is the archetypal jet observable for hadron colliders. One considers a $2 \to 2$ scattering with the fragmentation of outgoing partons to the measured jet. The result is then given by a convolution of the hard scattering spectrum with the inclusive jet fragmentation function:

$$\frac{d\sigma_{\mathrm{jet}}}{dp_t} \approx \frac{d\sigma_i}{dp_t} \int_0^1 dz\, z^{n-1} f_{\mathrm{jet}}^{\mathrm{incl}}(z,t) = \frac{d\sigma_i}{dp_t}\langle z^{n-1}\rangle. \tag{4}$$

where we assumed a power law fall off $p_t^{-n}$ for the partonic Born level spectrum. At the LHC typical $n$ values vary between 4 and 7 and higher at high $p_t$.

Figure 1 shows the results for $n=4$ for both initial quark and gluons. One notes 30 to 50 percent effects for $R=0.4$ and $R=0.2$ respectively for gluon jets at $p_t = 50$ GeV.

For detailed studies regarding the impact of small-$R$ resummation on the inclusive jet spectrum we refer the reader to Ref. [3].

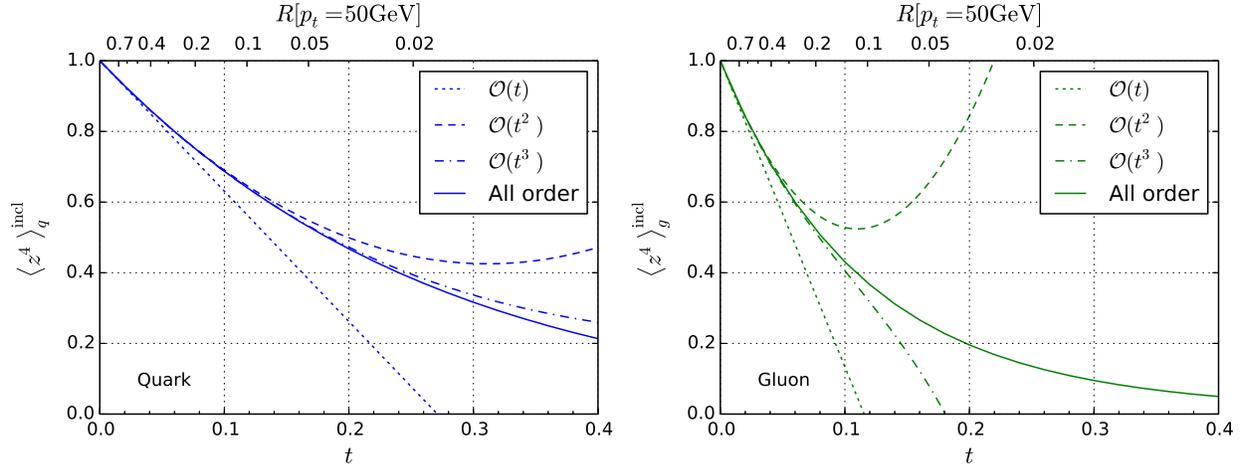

Figure 1: The result for $\langle z^4 \rangle^{\text{incl}}$ at all orders as a function of $t$ (lower axis), together with the first 3 orders of its expansion in $t$, shown for initiating quarks (left) and gluons (right). The upper axis gives the correspond $R$ values for a jet with $p_t$ of order 50 GeV. The factor $\langle z^4 \rangle^{\text{incl}}$, multiplied by a hard inclusive parton spectrum that goes as $p_t^{-5}$, gives the corresponding microjet spectrum.

**Filtering**

Filtering is a jet grooming tool that is used in substructure analyses and searches for boosted new particles [4]. Jet grooming is crucial at hadron colliders such as the LHC to reduce contamination of jets by the underlying event. In filtering, one takes a jet clustered with an initial radius $R_0$, reclusters its constituents on a smaller angular scale, $R_{\text{filt}} < R_0$, and then discards all but the $n$ hardest subjets. Whereas $t$ in the previous sections was defined as being $\frac{\alpha_s}{2\pi} \ln \frac{1}{R^2}$, plus higher orders from the running coupling, we now imagine taking a large-radius original jet, $R_0 = \mathcal{O}1$ and processing it with a small filtering radius, with $t$ defined in terms of the filtering radius, $t \simeq \frac{\alpha_s}{2\pi} \ln \frac{1}{R_{\text{filt}}^2}$, again plus higher orders from the running coupling. More generally, i.e. also for small $R_0$, $t \simeq \frac{\alpha_s}{2\pi} \ln \frac{R_0^2}{R_{\text{filt}}^2}$ plus higher orders, and the quantities we work out here will then relate the filtered jet to the original jet rather than to the original parton.

We define $f^{k\text{-hardest}}(z)$ to be the probability that the $k$-th hardest subjet carries a momentum fraction $z$ of the initial parton (or large-$R$ jet). We can then express the energy loss between the filtered jet and the initial parton as

$$\langle \Delta z \rangle^{\text{filt},n} = \left[ \sum_{k=1}^{n} \int dz\, z\, f^{k\text{-hardest}}(z) \right] - 1. \qquad (5)$$

Resummed results for $n=2$ and $n=3$ for the case of gluon jets are shown in Figure 2. For $n=3$ a filtered jet retains more of the initial parton's momentum (95 percent for $t=0.1$) than the hardest jet obtained after fragmentation of an initial gluon (75 percent for $t=0.1$). It is also evident that for $n=3$ there is a poor convergence of the fixed-order perturbative series signalling that here small R resummation is crucial for precision physics with filtered jets.

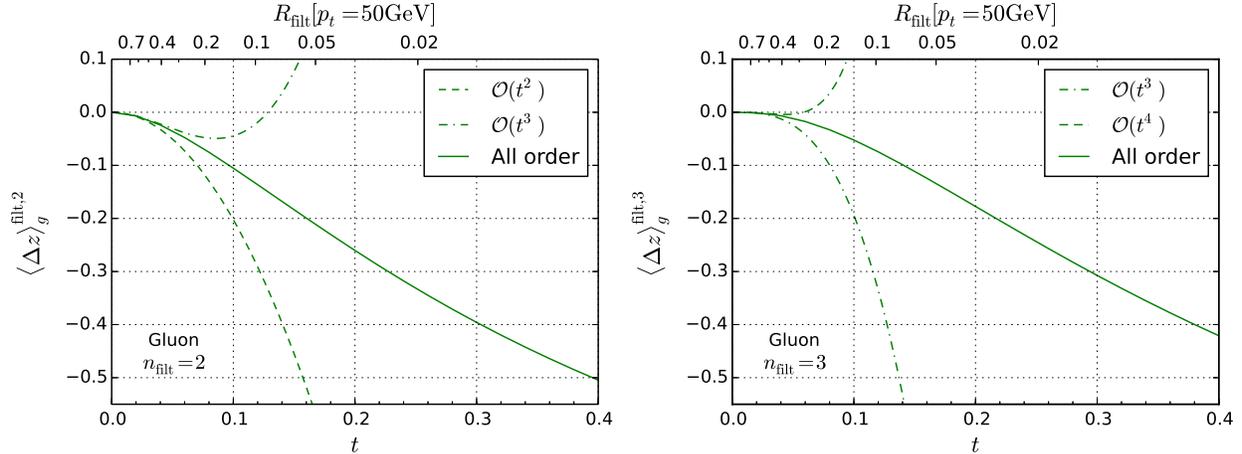

Figure 2: Average fractional jet energy loss $\Delta z$ after filtering with $n_{\text{filt}} = 2$ and $n_{\text{filt}} = 3$, as a function of $t$, for gluon-induced jets. Resummed results are represented as solid lines. The second, third and fourth orders in $t$ are represented as dashed, dash–dotted and dash–dash–dotted lines respectively.

**Jet vetoes in Higgs production**

Another area to which small $R$ resummation applies is the case of jet vetoes as used in Higgs production studies. Here there are multiple hard scales in the problem i.e. the Higgs mass $m_H$, the $p_t$ veto scale and the jet radius $R$. Resummation for the logarithms in $m_H/p_t$ have recently been carried out to next-to–next-to leading logarithmic accuracy [5],[6]. At the same time one may wish to consider the impact of small $R$ resummation for values of jet radius $R \sim 0.4$ and which are used in experimental analyses e.g. by the ATLAS collaboration.

The basic leading-logarithmic resummed result for the logarithms of $m_H/p_t$ may be written in terms of a Sudakov form factor which arises from a veto on primary partons emitted from the hard incoming partons:

$$P(\text{no primary-parton veto}) = \exp\left[-\int_{p_t}^{Q} \frac{dk_t}{k_t} \bar{\alpha}_s 2 \ln \frac{Q}{k_t}\right]. \quad (6)$$

Accounting for the fact that the veto is placed on microjets rather than primary partons one obtains an additional factor:

$$\mathcal{U} \equiv \frac{P(\text{no microjet veto})}{P(\text{no primary-parton veto})} = \exp\left[-2\bar{\alpha}_s(p_t) \ln \frac{Q}{p_t} \int_0^1 dz f^{\text{hardest}}(z, t(R, p_t)) \ln z.\right] \quad (7)$$

This $R$-dependent correction generates a series of terms $\alpha_s^{m+n}(Q) L^m \ln^n R^2$, while we have neglected terms suppressed by one or more powers of either $L = \ln Q/p_t$ or $\ln R^2$.

Eq. (7) shows that the key quantity for the small-$R$ part of the resummation is the first logarithmic moment of $f^{\text{hardest}}(z)$

$$\langle \ln z \rangle^{\text{hardest}} \equiv \int_0^1 dz\, f^{\text{hardest}}(z) \ln z\,. \quad (8)$$

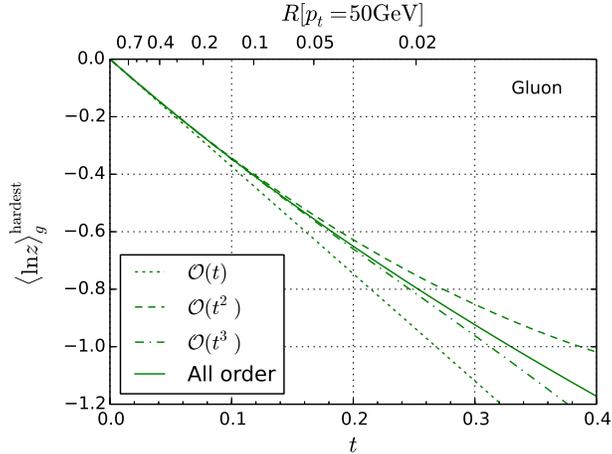

Figure 3: Average of the hardest microjet $\ln z$, as a function of t, shown for gluon-induced jets. The resummed results are represented as solid lines. The first three orders in $t$ are represented as dotted, dashed and dash-dotted lines respectively.

It is the logarithmic moment of the microjet spectrum from initial gluons that is relevant here, since the vetoed primary partons are all gluons. The resummed results along with the fixed-order results up to order $t^3$ are shown for gluon jets in Figure 3, where one observes, in contrast to the filtering case, a fairly stable perturbative expansion.

Detailed phenomenological studies for Higgs production with jet vetoes including the effects of small $R$ resummation may be found in Ref. [7].

## Summary and Outlook

In this article we have presented results for small-$R$ resummation for a number of jet observables of phenomenological interest. Examples discussed here are the inclusive jet spectrum, jet susbstructure observables and Higgs production with a jet veto. The resummation we have described addresses emissions in the collinear limit of a hard initial parton and as such is process independent. This in turn means that the results presented here can be adapted for use in both hadron collider as well as $e^+e^-$ jet observables. Two areas of future interest would be phenomenological studies at even smaller $R$ values than currently used (i.e. say $R = 0.1$) and on the theoretical side the extension of the leading logarithmic framework presented here to next-to–leading logarithmic accuracy. These developments will help to shed further light on collinear parton dynamics as well as help to isolate and better understand non-perturbative hadronisation corrections that become important at small $R$.

# Quark vs. Gluon Jets

Gregory Soyez[1]

[1] IPhT, CEA Saclay

**Abstract:** The ability to discriminate between quark and gluon jets has many applications in collider physics. In this contribution we briefly report on a work, initiated in the context of the 2015 Les Houches "Physics at TeV Colliders" workshop, which compares the quark/gluon tagging performance predicted by different Monte-Carlo generators. We discuss measurements at the LHC and at a FCC-ee that would further constrain quark/gluon tagging and Monte Carlo generators.

## Introduction

Designing an method to effectively separate quark- and gluon-initiated jets is a longstanding open question. (see e.g. [1] for a series of possible candidates). It is usually done via jet substructure observables like jet shapes which exploit differences in the radiation pattern of quarks and gluons. In general, we are interested in developing quark/gluon discrimination tools that go as far as possible beyond the naive $C_A$ v. $C_F$ Casimir scaling and are able to do so with limited and controlled theoretical uncertainties.

A key question in that respect is how well current (parton-shower) Monte-Carlo generators agree on their respective predictions for the quark-gluon discriminating power. In this contribution, we report on a study presented in Ref [2] where we show based on an idealised case that the results obtained for the quark/gluon discriminating power differ sizeably between Monte Carlo generators.

For more details, we refer directly to Section IV.5 of Ref. [2] and references therein. Most of the results presented below are taken from an extended version in preparation, Ref [3].

## Are quark and gluon jets well-defined?

Since quarks and gluons can branch into one another, are ill-defined concepts beyond the lowest order of the perturbative series, and are not directly observed in the final state of the collisions, the concept of a quark and a gluon jet might itself seem ill-defined at first sight.

Rather than trying to determine a truth definition of a quark or a gluon, our approach is to consider a more practical approach, tied to the hadronic final state. We therefore define *a phase space region (as defined by an unambiguous hadronic fiducial cross section measurement) that yields an enriched sample of quarks (as interpreted by some suitable, though fundamentally ambiguous, criterion)*. We note that one still needs to determine the criterion that corresponds to a successful quark enrichment and for that, we have to rely to some degree on a less well-defined notion of what a quark jet is.

In a way, we can see this as using "quark" and "gluon" as adjectives and not as nouns.

## Comparisons between different generators in an idealised study

We have systematically tested the performance of quark/gluon tagging predicted by different Monte Carlo generators in an idealised setup. We have considered $e^+e^- \to Z \to u\bar{u}$ as a source of quark jets and $\mu^+\mu^- \to H \to gg$ as a source of gluon jets.

As a discriminating variable, we have studied generalised angularities [4] for which (for $\kappa = 1$) there also exists analytic results at the NLL accuracy:

$$\lambda_\alpha^\kappa = \frac{1}{E_{\text{jet}}^\kappa R^\alpha} \sum_{i \in \text{jet}} E_i^\kappa \Delta R_i^\alpha. \tag{1}$$

The full study includes several working points in the $(\kappa, \beta)$ parameter space but we focus here on the IRC-safe "Les Houches Angularity" (LHA) $\lambda_{0.5}^1$.

To quantify the discriminating power, we use the following quantity

$$\Delta = \frac{1}{2} \int d\lambda \frac{(p_q(\lambda) - p_g(\lambda))^2}{p_q(\lambda) + p_g(\lambda)}, \tag{2}$$

built from the probability distributions $p_q$ and $p_g$ for the quark and gluon samples as a function of the LHA. In a way, $\Delta$ can be seen as a measure of the significance of the difference between the quark and gluon probabilities. It also has the advantage that the integrand can be plotted as a function of $\lambda$ to see where the discriminating power gets its larger contributions.

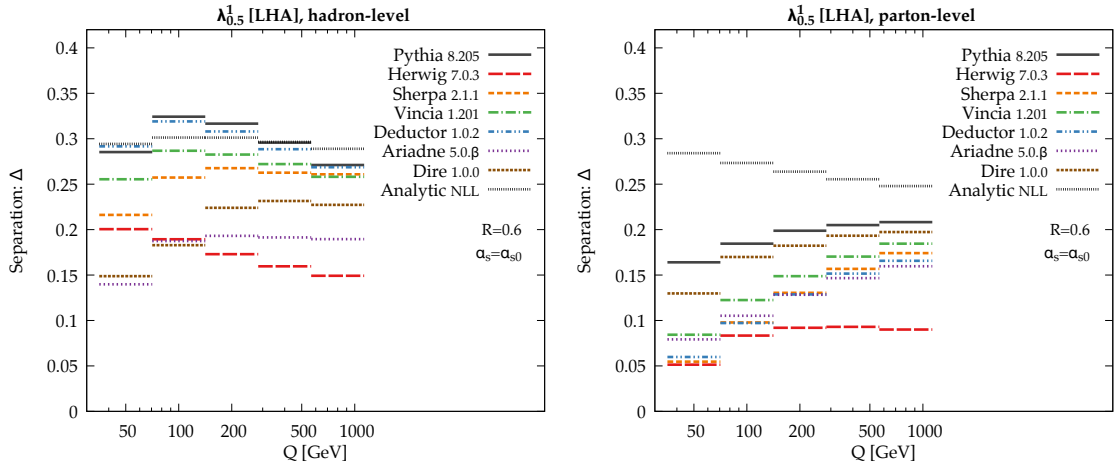

Figure 1: Dependence of the quark/gluon discriminating power ($\Delta$) as a function of the energy ($Q$) of the collision for different Monte Carlo generators. Left: hadron level; right: parton level.

The quark and gluon processes have been simulated using a series of Monte-Carlo generators. Currently: Pythia 8.205 [5], Herwig 7.0.3 [6], Sherpa 2.1.1 [7], Vincia 1.201 [8], Deductor 1.0.2 [9], Ariadne 5.0.$\beta$ [10] and Dire 1.0.0 [11]. We also include analytic results at NLL with a modelling of hadronisation effects. The probability distributions $p_q(\lambda)$ and $p_g(\lambda)$, as well as the discriminant $\Delta$ can be measured for different angularities, varying fundamental parameters like the energy $Q$ of the collision, the jet radius $R$ and the value used for the strong coupling constant at the $Z$ mass.

In order to pinpoint what ingredients in the generators drive the discriminating power, we have also varied a few chosen internal knobs in each of the generators. See Refs. [2] and [3] for details.

The code used for the analysis is developed in the Rivet 2.4 [12] framework with jet clustering and manipulation done using FastJet 3.1 [13]. It is publicly available in the `lh2015-qg` repository on GitHub.

Figure 1 shows an example of our findings: the dependence of $\Delta$ on the centre-of-mass energy of the collision. We observe rather large differences between the generators under consideration, both at parton and at hadron level, with Pythia predicting a large discriminating power and Herwig a much smaller one. We also see that non-perturbative effects have a large impact on $\Delta$. Differences are however already substantial at the perturbative level, i.e. in the parton shower. These differences can mostly be traced back to $p_g(\lambda)$ which is currently poorly constrained while $p_q(\lambda)$ is reasonably well constrained e.g. from LEP data. Large differences are also seen for other angularities (both IRC safe and unsafe) and quality measures and call for a better understanding and better constraints on both the perturbative shower and the non-perturbative corrections.

**Possible measurements at the LHC**

It is natural to wonder if one can perform dedicated measurements at the LHC to help constrain the large differences observed above. A simple option is to measure (generalised) angularity distributions[*] and the corresponding separation variables for dijet (gluon enriched) and $Z$+jet (quark enriched) events. Note that we want to report results directly for each processes without making any model-dependent effort to recover "quark" and "gluon" results. In particular, the separation $\Delta$ should be computed directly between the $Z$+jet and dijet distributions.

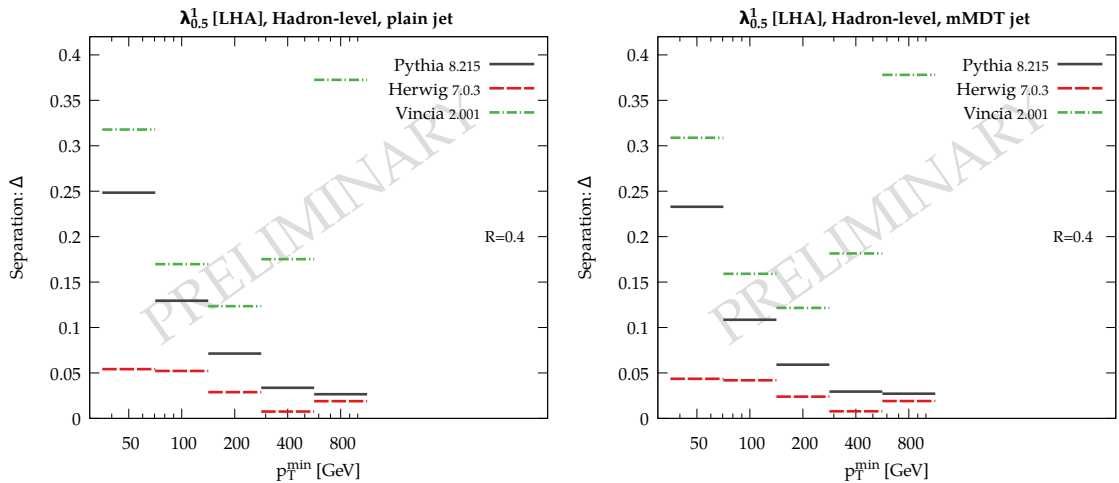

Figure 2: Separation $\Delta$ between the LHA measured on jets from $Z$+jet and dijet events as a function of the cut $p_T^{\min}$ on the jet transverse momentum. Left: angularities are computed using all the constituents of the jet; right: an mMDT procedure is applied before computing the angularities.

As for the $ee$ study-case presented above, there are several parameters that can be varied to further probe the kinematic dependence: the scale dependence can be probed by varying the cut on the jet $p_t$ and the angular dependence can be studied either by varying the jet radius, or by measuring the (generalised) angularities on a jet groomed with the (modified) mass-drop procedure (mMDT). This study — which can also be found on the `lh2015-qg` `GitHub` repository — is still in progress but preliminary results are presented in Figure 2, where we show the separation $\Delta$ as a function of the cut $p_T^{\min}$ on the jet transverse momentum. The same patterns as in the previous case are observed with Pythia and Vincia predicting much larger separations than Herwig. these differences remain after applying a mMDT procedure, suggesting that the differences are already present in the description of small-angle physics. Measuring these quantities at the LHC would definitely help to further constrain the Monte Carlo generators.

---
[*]now defined using the $p_t$ of the jet constituents instead of their energy

**Possible measurements at an FCC-ee**

With a lower hadronic activity, the environment of $e^+e^-$ collisions is far more conducive to precision measurements. Some additional information about quark/gluon discrimination and new constraints on parton-shower generators could possibly already be available from a re-analysis of LEP data with the tools described above, but a new circular collider at a higher energy and with higher statistics would definitely bring in valuable information in many respects.

Since LEP data is already extensively used in Monte Carlo tuning and provide mostly a quark-enriched sample of jets, one observes much smaller differences between generators for the distributions obtained from quark jets than for gluon jets. One should therefore target to build a clean gluon-enriched sample.

A first process of interest is be to look at 3-jet events with 2 $b$-tagged jets, where the third jet would provide a clean gluon enriched sample. This would largely benefit from the high luminosity and energy coverage expected at a FCC-ee.

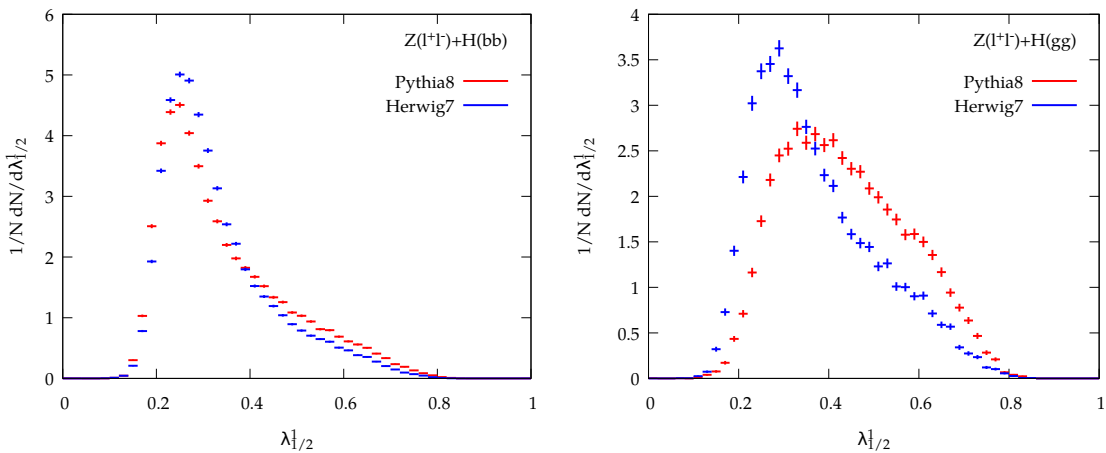

Figure 3: LHA distributions observed from Pythia8 and Herwig7 simulations for $e^+e^- \to Z(\ell^+\ell^-)H(b\bar{b})$ (left) and $e^+e^- \to Z(\ell^+\ell^-)H(gg)$ (right) events. We assume perfect $b$ tagging.

Another process of interest is associated Higgs production at $\sqrt{s} = 240$ GeV, where the Higgs can either decay to a $b\bar{b}$ pair or to gluons. Figure 3 shows the distributions obtained for the LHA after selecting a pair of leptons within 20 GeV of the $Z$ mass and requiring that the two jets are within 15 GeV of the Higgs mass. We clearly see a decent agreement between Pythia and Herwig for the $b\bar{b}$ sample, with much larger differences in the gluon-enriched sample.

The plot includes expected (ideal) statistical uncertainties for an integrated luminosity of 2.5 ab$^{-1}$, corresponding to 2.1 million $HZ$ events, including about 80000 events with the $Z$ decaying to a $e^+e^-$ or $\mu^+\mu^-$ pair and the Higgs boson decaying to a $b\bar{b}$ pair, and about 12000 events where the Higgs boson decays to a gluon pair instead.

We clearly see how such a measurement, possible at a FCC-ee, would bring crucial information for the development of Monte Carlo event generators. It would also help developing better quark/gluon taggers, a tool of broad application in collider physics.

# Gluon vs. quark fragmentation – from LEP to FCC-ee

Klaus Hamacher[1]

[1]Bergische U. Wuppertal, Germany

**Abstract:** Experimental results on gluon and quark fragmentation obtained by the LEP experiments are reviewed. The importance of colour coherence, transverse-momentum-scales and hadronisation corrections for jet measurements is emphasized. Precision results on multiplicities of three-jet events and the deduced multiplicity of a gluon-gluon colour neutral system are discussed. Identified particle results and colour octet neutralisation are addressed. Prospects for corresponding measurements at the FCC-ee are given.

## Introduction

At the time of LEP several topics of strong interaction physics were still incompletely understood. Many questions could be clarified in the very clean $e^+e^-$ environment with unprecedented statistical precision. Among those where basic predictions concerning the comparison of gluon and quark fragmentation. Such a comparison is attractive as the different colour structure of gluons and quarks should be immediately evident. In this talk I briefly review the basic experimental strategies. Then results for gluon and quark jets and the limitations of such measurements are discussed before coming to precision results on the multiplicity in three-jet events. Finally some results on identified particles and open topics on octet neutralisation are discussed before giving the outlook to FCC-ee.

## Basic Ideas and Experimental Strategy

In the hadronic final state of $e^+e^-$ annihilation, gluon and quark fragmentation can be compared in events with three jets, two of which originate from the initially produced $q\bar{q}$ pair, the third from a radiated hard gluon. The mutual assignment of partons to jets is done at tree-level, thereby limiting the quantitative comparison of gluon and quark jet properties to leading order precision. This basic limitation can only be overcome when studying more inclusive properties of three-jet events, like the event multiplicity or overall particle spectra. Still even then the identification of jets introduces some systematic dependence on the jet algorithm chosen. Moreover, the parton properties underlying a three-jet event need to be determined from the observed jets. This implies that hadronisation needs to be considered when determining the parton properties! This inevitably introduces bias and smearing which may strongly influence e.g. observed fragmentation functions.

Experimentally gluon jets are enriched using energy ordering and anti-tagged by identifying heavy hadron decays predominantly in b quark events. Gluon jets are then compared to a mixture of light quark and gluon jets in, again, anti-tagged light quark events or to events where the gluon is replaced by an isolated photon. The kinematics in these events is chosen similar to those containing the tagged gluons. The respective purities of gluon and quark jets are taken from simulation, pure gluon and quark distributions are inferred using matrix inversion.

Initially these techniques were applied to the so called Y-events [1,2,3,4], containing two jets with similar energy ($E \lesssim 25$ GeV) and angle with respect to the event axis. Also fully symmetric ("Mercedes"-star) events [5,6] and events where a gluon jet recoils with respect to two quark jets [7,8] were used, again comparing to properties of jets of similar energy. After recognising that the energy is an inappropriate scale, as it depends on the Lorentz-system [9], invariant transverse-momentum-scales have been introduced. The above technique was then expanded to general three-jet topologies. This allowed gluon to quark comparisons in a large kinematic range [10].

Table 1: Selected data on the hadron multiplicity ratio in gluon to quark jets as a function of jet energy.

| Exp. | | $E_{Jet}/\text{GeV}$ | $r_n$ |
|---|---|---|---|
| Cleo | [12] | 3.5 | $1.04 \pm 0.02 \pm 0.05$ |
| HRS | [13] | 9.7 | $1.29 \pm 0.2 \pm 0.2$ |
| Aleph | [14] | 24 | $1.249 \pm 0.084 \pm 0.022$ |
| Delphi | [5] | 24 | $1.241 \pm 0.015 \pm 0.025$ |
| Opal | [7] | 40 | $1.552 \pm 0.041 \pm 0.061$ |

### Initial Results on Jet Fragmentation

It is a basic QCD expectation that the hadron multiplicity is proportional to the colour charge of the radiating parton [11], i.e. the multiplicity ratio in gluon to quark jets should be $C_A/C_F = 9/4 = 2.25$. The experimentally seen ratio is far smaller (see Table 1), however, indicates a clear increase with energy [5].

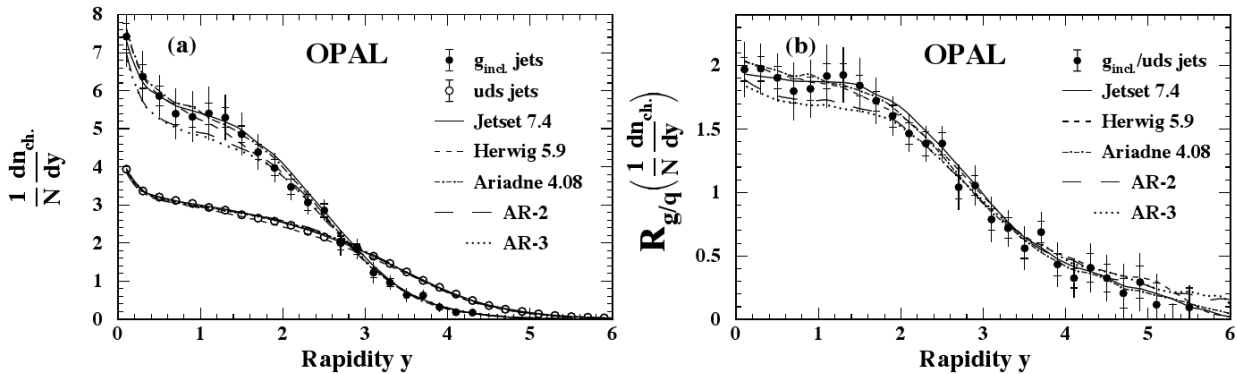

Figure 1: Rapidity-distribution for gluon & quark jets (left) and their ratio (right) [15].

The small multiplicity can be understood by inspecting the rapidity (or $y$) distribution of the produced particles with respect to the event axis (see Figure 1). Here the gluon distribution is taken from events where the gluon recoils with respect to two b-jets. The average gluon energy is $E_g = 40\text{GeV}$, the transverse momentum with respect to the quark-jets $\kappa_g \sim 37\text{GeV}$. The quark distribution stems from light quark events with $E_q = \kappa_q = 45.6\text{GeV}$ [7].

At small $y$, that is for particles produced first in time [17], the production rate is a factor $\lesssim 2$ higher for gluons compared to quarks. For $y \gtrsim 2$ the ratio falls off strongly and for $y > 3$ more particles are produced in quark compared to gluon jets. The latter to a large part is a consequence of energy conservation; there is not enough energy left in gluon jets to produce additional particles.

Moreover, particle production in quark jets may be eased as quarks are valence particles of hadrons. Overall this extra production diminishes the total multiplicity ratio. Still, even at small rapidity the expectation 2.25 is not met. This partly is due the difference in the relevant scales ($E_q, \kappa_g, \sim 20\%$) of the underlying partons, but also affected by coherent emission from the underlying $q\bar{q}g$-ensemble and the expected destructive interference (the so-called string effect, see next section).

In order to illustrate the evolution of the parton multiplicity the so-called subjet-rates $R_i$ and their derivatives have been compared for gluon and quark jets [16]. $R_1$ is the experimental equivalent to the Sudakov form factor $\Delta(y)$ (see e.g. [18]):

$$\Delta(y|y_0) = \exp\left\{-\int_{y_0}^{y} \Gamma_{p \to p'p''}(y')dy'\right\} \stackrel{!}{=} R_1(y) = \frac{N_1(y)}{N_0}$$
$$-\Gamma(y) = \tilde{D}_1(y) = \frac{1}{N_1(y)} \frac{\Delta N_1(y)}{\Delta y}$$
$$\Gamma_q(y, y_0) = \frac{C_F}{2\pi} \frac{\alpha_s}{y_0} \left(\ln \frac{y}{y_0} - \frac{3}{2}\right) \qquad \Gamma_g(y, y_0) = \frac{C_A}{2\pi} \frac{\alpha_s}{y_0} \left(\ln \frac{y}{y_0} - \frac{11}{6}\right)$$
(1)

$R_1(y)$ is the 1-jet rate, which is the probability, that at a given resolution, $y$, no splitting has happened in a jet, i.e. there is still only 1 "parton" present. $\Gamma_p(y)$ is the probability density for a parton $p = (q, g)$ to split at a given $y$, $\tilde{D}_1$ is the corresponding experimental observable. $D_1(y) = R_1(y) \cdot \tilde{D}_1(y)$ is the rate of parton splitting. This concept has been generalised for higher rank ($2 \to 3$, $3 \to 4 \ldots$) splittings [16,19].

A measurement of the splitting kernels (multiplied by $y$) is shown in Figure 2. The splitting probability ($\tilde{D}_1$) is markedly bigger for gluon compared to quark jets at high $y$. Deviations from the NLLA expectations $\Gamma_p$ (see Equation 1) due to hadronisation set in at higher $y$ for gluons compared to quarks. The ratio $r_1 = \tilde{D}_1^g/\tilde{D}_1^q$ for large y is initially close to $C_A/C_F$, but showing a hump-structure which is described by fragmentation models (see Figure 2). Most likely the hump is a consequence of hadronisation smearing. Such a structure would be expected, as the splitting kernels are strongly varying functions of $y$ below a maximal value of $y_0 \sim 1/3$.

Figure 3 shows the related jet and splitting rates and splitting probabilities for the first four splittings. The rates $R_i$ show a well known pattern from multi jet rates in $e^+e^-$ (in this case one usually starts from the initial 2 jets). The splitting rates $D_i$ are far higher at higher $y$ for gluons, but for each rank this initial lead is compensated at smaller $y$ in quark jets. The splitting probability $\tilde{D}_i$, which for the first rank $i = 1$ splitting is about two times higher for gluons, converges to similar values for the gluon and quark cases at higher rank splittings. This general pattern easily explains the observed small gluon to quark hadron multiplicity ratio.

Especially the gluon fragmentation function(s) $D_g^h(z)$ to a hadron $h$ is a quantity which was intensively studied at LEP in $e^+e^-$ three-jet events. Initial measurements were done for Y- and Mercedes-events [3,4,16] and later extended to more general topologies [10,20,21]. Also boosting to symmetric topologies was used in order to ease the assignment of hadrons to partons [22]. In view of the studies of the scale dependence of the gluon and quark fragmentation functions, the introduction of transverse-momentum-like scales was an important step [10,19,23]. Relevant scales are:

$$\text{quark:} \quad \kappa = E_q \sin \frac{\Theta_{qg}}{2} \qquad \text{gluon:} \quad p_\perp = \frac{1}{2}\sqrt{\frac{s_{qg}s_{\bar{q}g}}{s_{e^+e^-}}} \quad . \tag{2}$$

Here, $\Theta_{qg}$ is the angle between the gluon and the closest quark jet and $s$ is the Mandelstam variable. Essentially $p_\perp$ is the harmonic mean of the transverse momenta of the gluon with respect to quark or anti-quark, respectively. These scales, as well as the parton energy $E_p$ used in the denominator of the scaling variable $x_E \sim z = E_{had}/E_p$ need to be determined from the jet properties, therefore, they acquire experimental as well as hadronisation uncertainties.

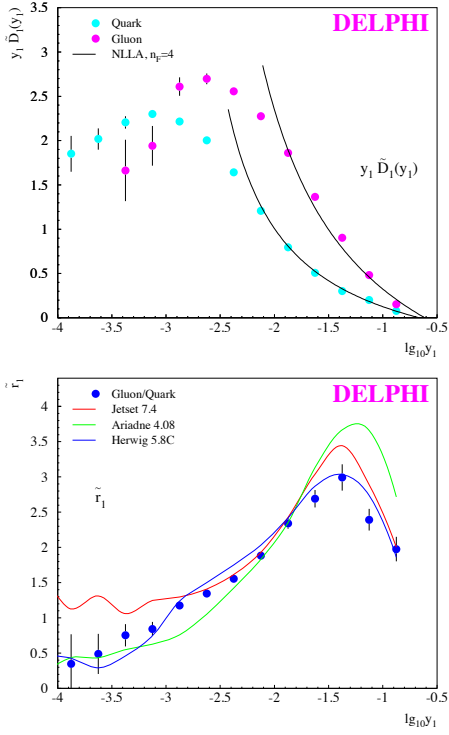
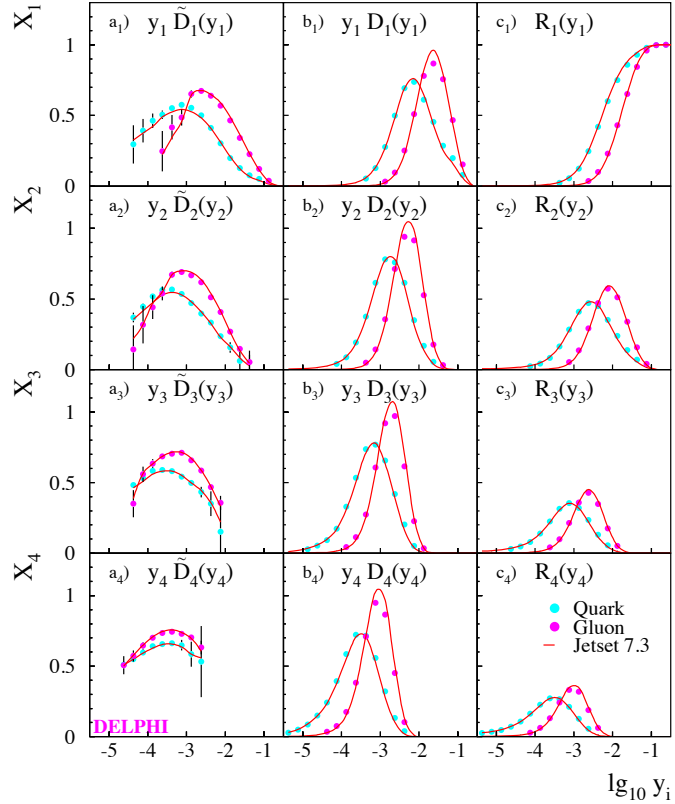

Figure 2: gluon & quark splitting kernels compared to the NLLA expectation (up) and their ratio compared to MC models (down) [16]. NLLA predicts $r_1 = 2.25$ ($n_f = 3$).

Figure 3: Gluon and quark subjet-rates ($R_i$), splitting-rates ($D_i$) and splitting probabilities/kernels $\tilde{D}_i$ compared to MC models [16].

The measured fragmentation function for quarks agrees well with the corresponding results from $e^+e^-$ at lower energies (see Figure 4 [21]) as well as with DGLAP fits. The corresponding gluon result is shown in Figure 5 and shows the expected stronger fall-off with $z \sim x_E$ as well as with the scale ($p_\perp$) compared to quarks. The ratio of the logarithmic slopes of the gluon and quark-fragmentation functions was the first measurement which quantitatively verified the colour factor ratio from a gluon measurement [10]: $C_A/C_F = 2.26 \pm 0.09 \pm 0.14$.

From a comparison of the data at high $x$ and the DGLAP fits it is obvious that the slope of the data exceeds that of the fits. This is a consequence of the irreducible smearing due to hadronisation and the corresponding uncertainty of the measured properties (momentum or energy $E_p$) of the partons in the event [24]. This uncertainty can be analytically estimated using the longitudinal phase space or tube model (see e.g. [18]) to be of the order $0.5\text{GeV}/E$. Applying this smearing to an analytic approximation of the data (taken from [10]) explains the observed deviation between the DGLAP fits (lower curves in Figure 6) and the data (represented by the upper curves). Due to the stronger fall-off of the fragmentation function with $z$ or $x$ for gluons the effect is here far more pronounced as for quarks. This basic problem inevitably appears whenever parton properties are measured from data. It is particularly big at small jet energy and for rapidly varying distributions

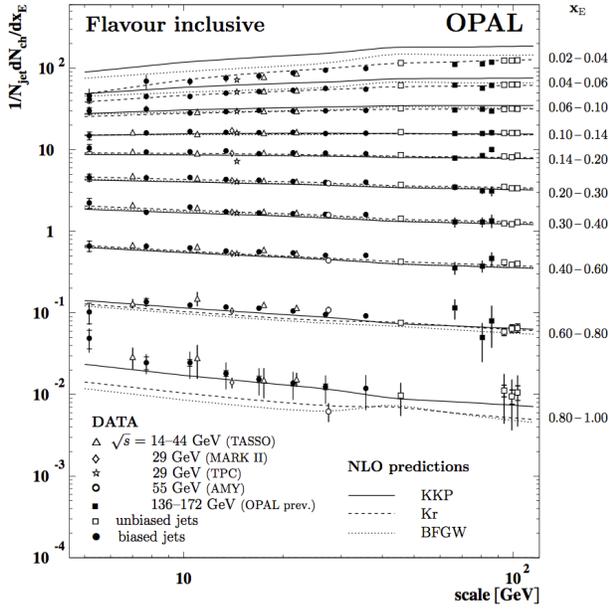
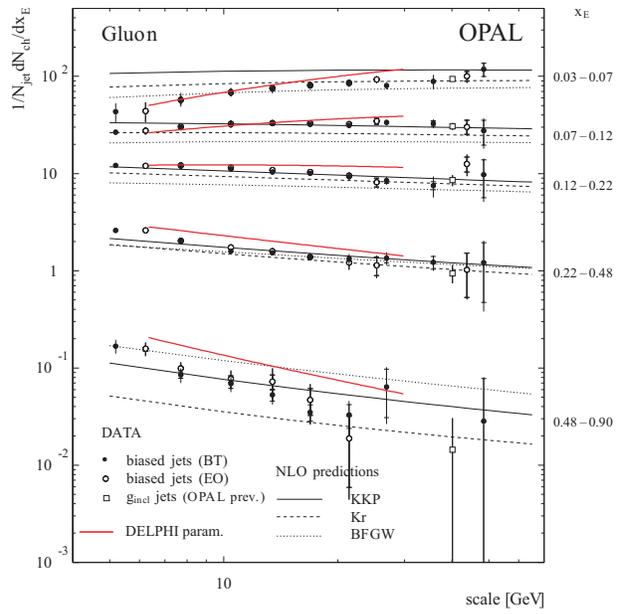

Figure 4: Quark fragmentation function as measured from three-jet events [21].

Figure 5: Gluon fragmentation function corresponding to Figure 4. The additional line represents a parameterisation of the data from [10].

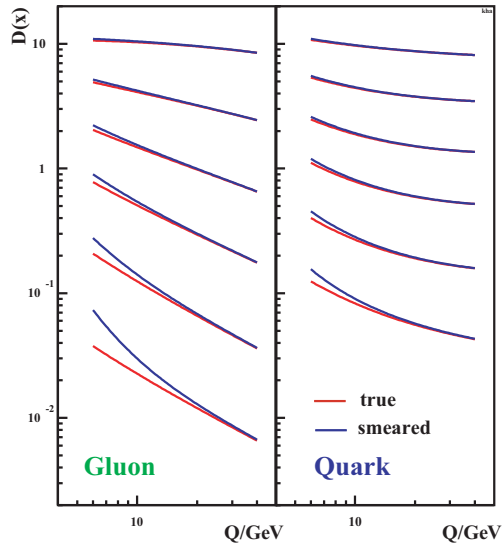

Figure 6: Effect of the hadronisation smearing of the parton properties on the fragmentation functions measured in three-jet events. The lower curves represent the true distributions, the upper curves the measured distributions.

*!
---
*Using error propagation the E-uncertainty is to be multiplied by the relative slope of the distribution.

## Multiplicity in Three-Jet Events

Coherent emission of hadrons from a $q\bar{q}g$-ensemble can be best observed using soft, low resolution, large wavelength hadrons emitted at large angles with respect to the underlying partonic system. A parameter free prediction for the ratio of particle rates produced perpendicular to the event plane of a three-jet event and the event axis in a $q\bar{q}$-event reads [25]:

$$\frac{N_\perp^{q\bar{q}g}}{N_\perp^{q\bar{q}}} = \frac{C_A}{C_F} \cdot r_t = \underbrace{\frac{C_A}{C_F}}_{\text{colour factor}} \cdot \frac{1}{4}\left[\widehat{q\,g} + \widehat{\bar{q}\,g} \underbrace{- \frac{1}{N_C^2}\widehat{q\,\bar{q}}}_{\text{destructive interf.}}\right] \quad , \tag{3}$$

with the number of colours $N_C = 3$ and (note the correspondence to Equation 2):

$$\widehat{i\,j} = 2\sin^2\frac{\Theta_{ij}}{2} \quad .$$

The destructive interference term represents the so called "string effect" [17] for this special kinematic situation. Note, that expression 3 is linear in the colour factor ratio and in the topological term $r_t$. For $r_t \sim 1$ the gluon and a quark jet are close by, the relevant colour charge is that of the initial quark. For large $r_t \to 2$ the gluon recoils with respect to the $q\bar{q}$ pair. The radiating colour charge then corresponds to two quark charges.

Experimentally expression 3 is studied [26,27] using two- and three-jet events selected using $k_t$-type jet algorithms with fixed $y_{cut}$. Events with more than three jets were discarded. Particle rates are measured in cones of $\sim 30°$ opening angle. The resulting ratio (see Figure 7) is insensitive with respect to variations of $y_{cut}$ and cone opening-angle.

The theoretical prediction given by Equation 3, including the destructive interference term, agrees well with the data for a large range of arbitrary and symmetric three-jet topologies. Exploiting the strictly linear $r_t$-dependence, the colour factor ratio can be extracted fitting the middle term of Equation 3 to the data (see Figure 8):

$$C_A/C_F = 2.211 \pm 0.014_{(\text{stat.})} \pm 0.045_{(\text{sys.})} \quad .$$

The systematic uncertainty enfolds a variation of the cone angle between $20°$ and $40°$ and a variation of $y_{cut}$ by a factor 2.

When measuring multiplicity the assignment of particles to jets imposes an obvious difficulty. Therefore, precision measurements rely on the overall multiplicity of three-jet events. A prediction for this multiplicity reads [17,28,29]:

$$N_{q\bar{q}g}(L_{q\bar{q}}, \kappa_{Lu}, \kappa_{Le}) = N_{q\bar{q}}(L_{q\bar{q}}, \kappa_{Lu}) + \frac{1}{2}N_{gg}(\kappa_{Le}) \tag{4}$$

$$L_{q\bar{q}} = \ln\left(\frac{s_{q\bar{q}}}{\Lambda^2}\right), \qquad \kappa_{Lu} = \ln\left(\frac{s_{qg}s_{\bar{q}g}}{s\Lambda^2}\right), \qquad \kappa_{Le} = \ln\left(\frac{s_{qg}s_{\bar{q}g}}{s_{q\bar{q}}\Lambda^2}\right)$$

This prediction takes coherence effects as well as the reduction of phase-space for gluon emissions from the quarks due to the emission of the leading gluon into account by the choice of scales. The sub-division of the multiplicity in a quark- ($N_{q\bar{q}}$) and a gluon part ($N_{gg}$) is motivated by the case when a photon replaces the gluon. An alternative choice assigning less multiplicity to the gluon is possible [28].

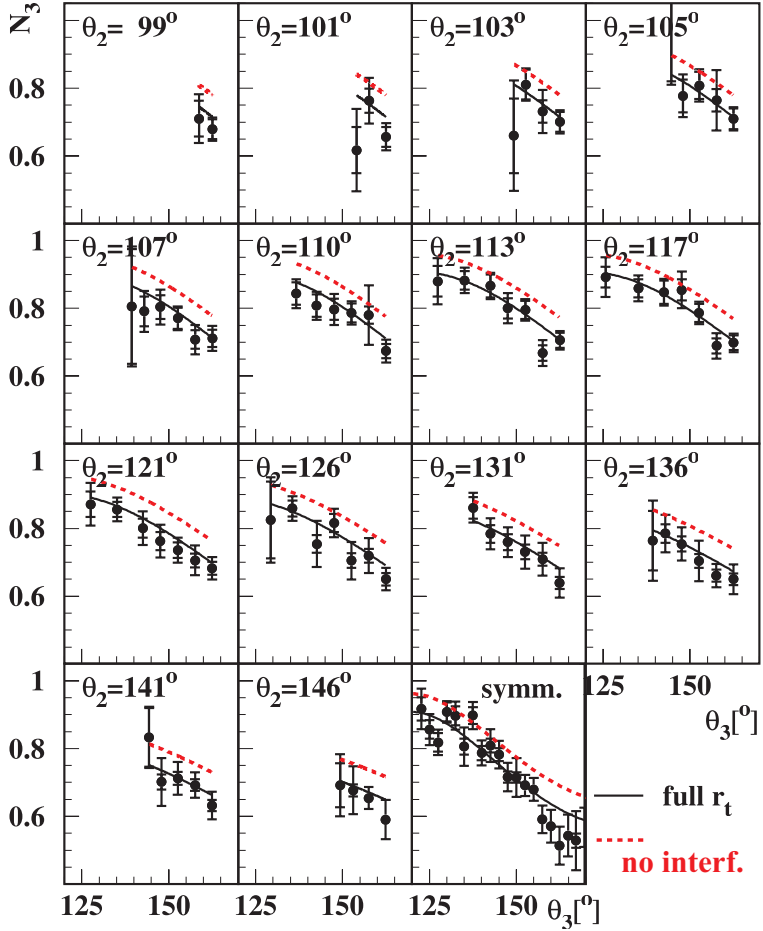
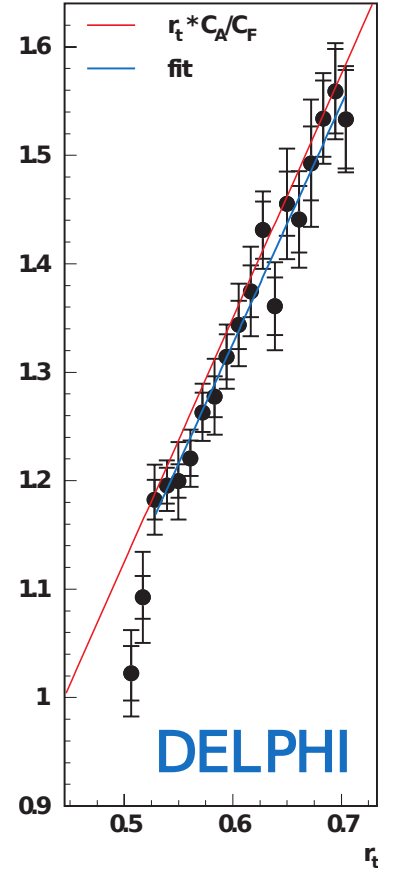

Figure 7: Ratio of particle production in 30° cones perpendicular to the event plane of three-jet events and the event axis of two-jet events [26]. The full line represents the parameter free prediction 3, for the dashed line destructive interference is omitted.

Figure 8: Data as in Figure 7, however, as function of $r_t$. The upper line is the expectation Equation 3, the lower line a fit to the data.

It has been observed experimentally [30] that the small gluon to quark multiplicity ratio is largely due to a non-perturbative offset (clearly evident from the almost equal gluon and quark multiplicity at small energy, see Table 1) and that it is more efficient to determine the colour factor ratio from the ratio of the energy slopes of the multiplicities. Asymptotically this ratio is identical to the multiplicity ratio because of de l'Hôpital's rule. In the dipole model the energy slope of the gluon and quark multiplicity is connected by the differential equation [28]:

$$\left.\frac{dN_{gg}(L')}{dL'}\right|_{L'=L+c_g-c_q} = \frac{C_A}{C_F}\left(1 - \frac{\alpha_0 c_r}{L}\right)\frac{d}{dL}N_{q\bar{q}}(L) \qquad (5)$$

$\alpha_0 c_r$ are known constants. $N_{q\bar{q}}(E)$ is known from $e^+e^-$ experiments. The solution of Equation 5 leaves a constant of integration free, which allows to accommodate non-perturbative differences

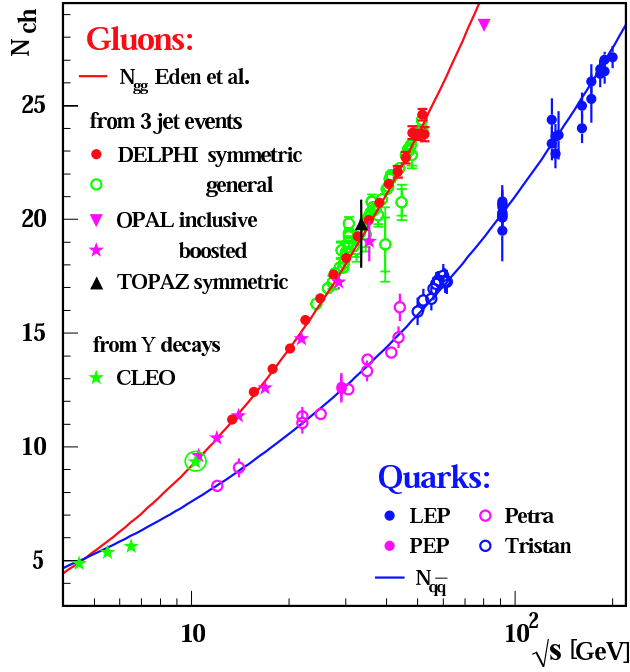

Figure 9: Hadron multiplicity in $e^+e^-$ annihilation and corresponding result for a colour singlett gluon pair [12] deduced from three-jet multiplicity measurements [6,7,22,31].

between the gluon and the quark multiplicity.

The three-jet multiplicity has been measured by several experiments [15,20,22,30,31] including related $p\bar{p}$-results [32]. Fitting the dipole prediction to symmetric and arbitrary three-jet topologies while leaving the offset floating, leads to a precise result for the colour factor ratio,

$$\frac{C_A}{C_F} = 2.261 \pm 0.014_{\text{stat.}} \pm 0.036_{\text{exp.}} \pm 0.052_{\text{theo.}} \pm 0.041_{\text{clus.}} \quad , \quad (6)$$

from the multiplicity slopes [31]. From a comparison of symmetric and arbitrary topologies it turned out that the alternative choice of scales (see [28] used in [15]) is unable to describe the full dataset.

Equation 4 can be solved for $N_{gg}$, the hadron multiplicity of a colour neutral gluon pair, shown in Figure 9. Note that the results obtained from three-jet events are well consistent with the only available direct $N_{gg}$ result so far from $\chi_b$-decays [12]. The gluon to quark colour factor ratio is immediately evident from the different energy slopes in this figure.

Essential outcome of the studies discussed so far is:
i) It is important to use the correct transverse-momentum-like scales which reflect the colour structure of the events. When inferring parton properties from jets, hadronisation effects should be considered.
ii) The influence of non-perturbative and finite-energy effects on gluon jets is stronger than for quark jets.
iii) Basic QCD properties can be more easily observed from dynamical variations of gluon and quark observables with the relevant scales.

## Identified Particles and Colour Octet Neutralisation

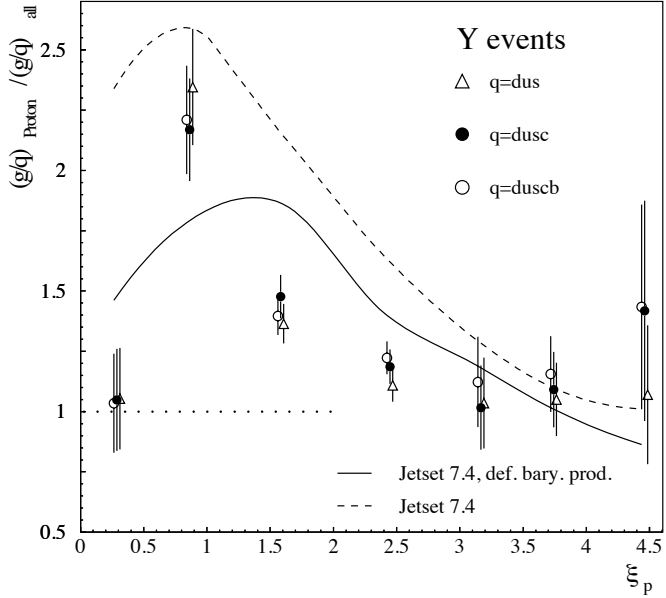
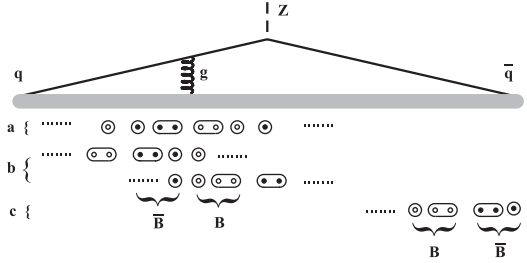

Figure 10: Chances for di-quark production in gluon (a,b) and quark jets (c) in the string model (up). Ratio of proton-production in gluon to quark jets normalised to the charged multiplicity ratio (left) [42].

Measurements of identified particle production have been performed at LEP in order to reveal possible differences in gluon and quark fragmentation. Hadronisation happens via an intermediate parton shower, which is by far dominated by gluons due to the structure of the splitting kernels. Differences are mainly expected for leading particles. For gluons there should be extra production of leading isoscalar particles [33,34,35,36] due to octet colour neutralisation (see Figure 11). Experimentally, no extra production of such states ($\Phi, \eta, \eta'$) could be observed so far at LEP [19,37,38,39]. There is however increased baryon production in gluon jets [38,40,41,42] as well as in $\Upsilon$-decays via gluons [43]. At LEP the extra production is focussed at small $\xi_p = -\ln x$, thus large $x$, see Figure 10. Here the double ratio of proton to hadron production is shown for gluon and quark jets in order to eliminate the general difference due to the colour factors. The excess at $\xi_p \sim 0.8$ can be understood in the string model [44]. Here an additional possibility for a splitting into a diquark anti-diquark pair exists (see Figure 10 case a), which via the Golden Rule leads to extra baryon production. This possibility is not present in cluster fragmentation.

In order to more generally search for isoscalar states, which are presumed to be heavy and for kinematic reasons must show a correspondingly hard fragmentation, it has been suggested to look for neutral leading systems with a rapidity gap [36]. The measurements [8,46,47] indeed indicate a small excess $\sim 2\%$ of leading neutral systems in gluon jets (see Figure 11). It has, however, so far not been possible to clarify this excess in detail. From the mass spectra there is an indication for an excess at masses $\lesssim 2$GeV.

In [48] deviations of the overall momentum spectra for gluons have been reported, while the quark spectra are perfectly described by the tuned MC models. The deviation starts in the model to data ratio at $x \sim 0.2$ and increases to about $-30\%$ at very high $x$. As the comparison has been done with fully simulated events, no hadronisation smearing enters. Overall the size of the effect amounts to about the same size as that observed for leading neutral systems.

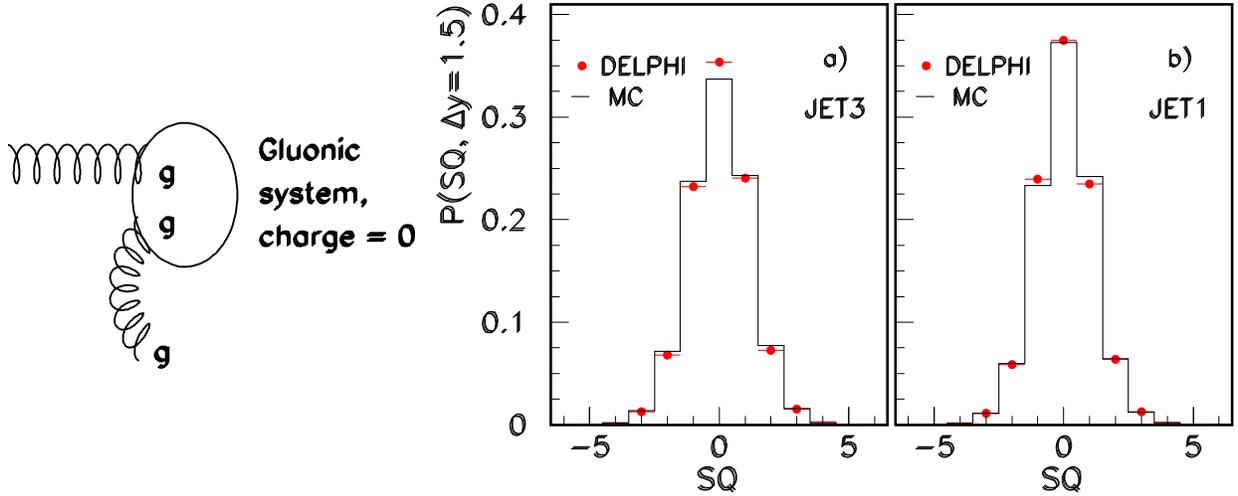

Figure 11: Octet neutralisation (left). Fraction of leading systems with charge SQ in gluon (a) and quark (b) jets. Lines are a prediction of Ariadne [45] .

**Outlook to FCC-ee**

FCC-ee will provide an enormous number of events exceeding the total statistics at LEP between 40GeV(accessible by radiative events) and 240GeV, even at 350GeV almost the LEP statistics is reached. About $10^{10}$ tagged three-jet events shall be provided at the Z. For gluon to quark comparisons this should allow to study "any" dynamical dependence (fragmentation functions, splitting kernels ... ) with a negligible statistical uncertainty. Systematics can be largely mitigated by unfolding and controlled using the energy dependence. The topology dependence can be explicitly compared to the energy dependence.

Deep understanding of gluon and quark jets (like for the differing width) will influence many other important measurements, e.g. $h \to gg$, which itself is a testing bed for $gg$ colour singlet fragmentation.

The high statistic will allow for studies of rare or difficult to measure processes. Leading particles in gluon and quark jets can be compared in order to search for octet fragmentation in jets, isoscalars and glueballs. Detailed studies and comparisons of mass plots will become possible and allow to search for meson as well as baryon resonances $(\Delta, \Lambda(1520), \dots)$. Additional questions will, in the meanwhile, be raised by low energy and lattice results.

For an FCC-ee experiment a high resolution electromagnetic calorimetry as well as very good particle identification and tracking are desirable. These should allow optimal determination of the jets as well as measurements of identified (neutral) particles and resonances.

# Distinguishing quark and gluon jets at the LHC

Giorgia Rauco[1] (on behalf of the ATLAS and CMS Collaborations)

[1] Universität Zürich, Zürich, Switzerland

**Abstract:** Studies focused on discriminating between jets originating from gluons and quarks at the LHC are presented. The results here discussed are obtained with proton collisions collected by the ATLAS experiment [1] at $\sqrt{s}=8$ TeV and by the CMS experiment [2] at $\sqrt{s}=13$ TeV.

## Introduction

Partons emitted from hard scattering process at the LHC form, due to QCD confinement and hadronization process, hadronic jets, which can be revealed with tracking and calorimeter systems. As known from theoretical principles and from experimental measurements, reconstructed jets show different properties depending on flavor of original parton. In general, due to the large color factor of gluons, gluon-initiated jets have higher particle multiplicity, a softer fragmentation function, and are less collimated than quark-initiated jets. These differences can be exploited to tag jets, and such a capability plays a fundamental role in several physics analyses. It results in an increased ability to discriminate full-hadronic final searches - composed mainly by quark-originated jets, from QCD background - where the gluon component is predominant.

## ATLAS studies

The ATLAS Collaboration, within a method based on data-driven template extraction of light-quark and gluon jet properties, tested a variety of discriminants and derived the systematic uncertainties on their performances [3]. Additionally, a precision measurement of the jet constituents multiplicity has been carried on [4]. Both approaches are presented in the following.

### Discriminant variables and data-driven templates

Templates of light-quark and gluon jet properties are derived from data, exploiting Z/γ + jets (quark-enriched) and dijets (gluon-enriched) events. Assuming the samples to be independent, the shape of an inclusive jet distribution is computed as the a linear combination of the pure light quark shape, weighted by the amount of light-flavor quarks, and the pure gluon shape, weighted by the amount of gluons, plus contamination from heavy flavor partons, weighted by the amount of contamination.

As an example, Fig. 1 shows the extracted templates for the calorimetric and tracker based jet width ($w_{calo}$ and $w_{trk}$), comparing the two parton flavors and the data with two parton shower generators. A good agreement between the data and the hadronizers is found for light-flavor quarks, while data are falling between the two parton showers in the case of gluon-like jets.

In addition to the samples used for the template extraction, other events have been selected to validate the extrapolated distributions. By selecting particular regions of the phase space, γ+jets and trijets events have been purified, becoming highly quark-like in the former sample and gluon-like in the latter sample.

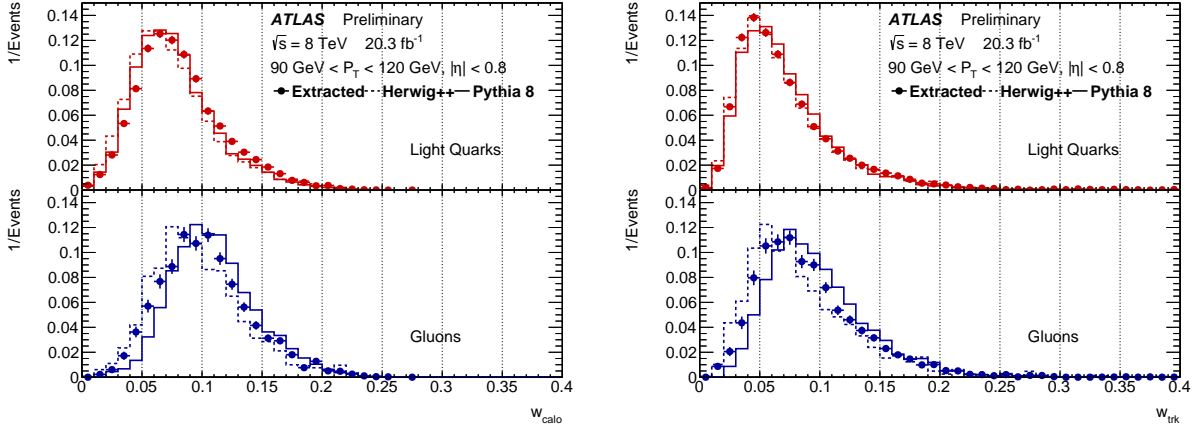

Figure 1: Extracted templates for $w_{calo}$ (left) and $w_{trk}$ (right) comparing data (points), PYTHIA (solid line) and HERWIG (dashed line) [3].

The comparison for the mean values of $w_{calo}$ (left) and $n_{trk}$ (right) along the jet $p_T$ spectrum between the extracted and validated templates are shown in Fig. 2, resulting in a good agreement for quarks and in a 15% disagreement in gluon distributions.

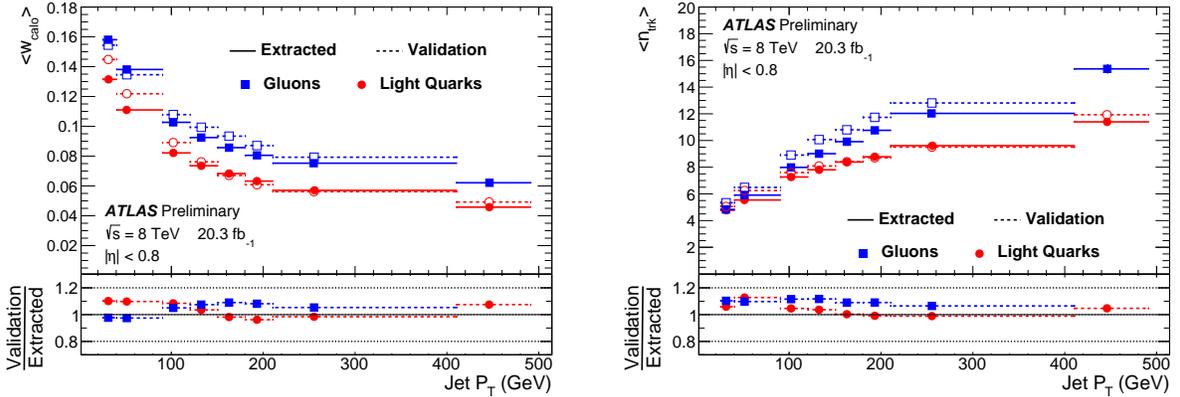

Figure 2: Comparison between the mean values of $w_{calo}$ and $n_{trk}$ as a function of $p_T$ [3].

### Measurement of the charged constituents of the jet

An additional study to enhance the discrimination between quark-like and gluon-like jets is the precision measurement of the jet constituents multiplicity, performed by applying unfolding techniques to remove distortions due to detector effects and by comparing several particle-level models. Figure 3 shows the jet $p_T$ dependence of the average charged-particle multiplicity for quark- and gluon-initiated jets, extracted with the gluon fractions from PYTHIA, along with the N3LO pQCD prediction. As expected, for both the quark-initiated jets and gluon-initiated jets the average multiplicity increases with jet $p_T$, but the increase is faster for gluon-initiated jets, for which the multiplicity is also higher.

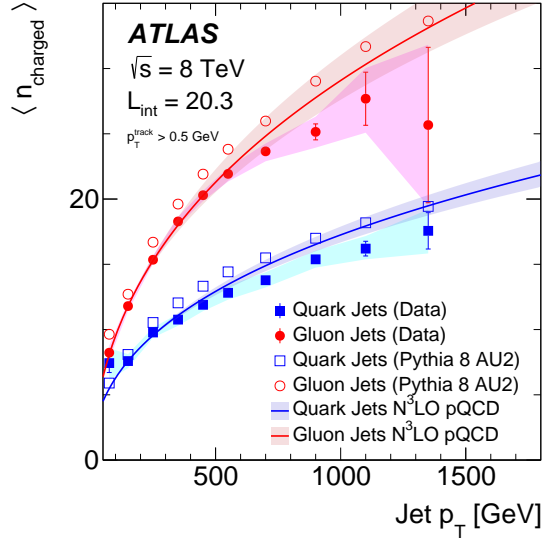

Figure 3: The average charged-particle multiplicity for quark- and gluon-initiated jets depending on the jet $p_T$ spectrum [4].

## CMS studies

The CMS Collaboration optimized observables based on performances established with Monte Carlo simulated QCD events. The CMS Collaboration optimized a likelihood-based discriminator based on performances established with Monte Carlo simulated QCD events and validated it using the data. A data-driven method is employed to derive corrections to account for observed differences with the data. Effect of using different parton shower models on the performance of the discriminator is also compared[5][6].

### Construction of a quark-gluon likelihood discriminant and its validation on data

Following the theoretical differences between quark- and gluon-like jets as explained in the Introduction, three observables are investigated: (i) the jet constituents multiplicity; (ii) the jet minor axis of the ellipse cone projected on the $\eta - \phi$ plane; (iii) the jet fragmentation distribution $p_T D$, defined as $\sqrt{\sum_i p_{T,i}^2}/\sum_i p_{T,i}$. The probability density functions of these variables, extracted in QCD events (showered with PYTHIA) where the jets have been tagged as gluon or quark, are used to build a likelihood product, which will have an output value in the [0,1] range, expressing the probability for a jet to be a quark-like jet. The performances of the quark-gluon likelihood (QGL) discriminator are checked in QCD simulated events by using the so-called ROC curves and are shown in Fig. 4.

The training observables and the taggers have been validated using 13 TeV collisions data in two control regions: Z+jets events, which are quark enriched and dijets events, gluon-rich. For both control regions, the full 2015 dataset has been used, corresponding to an integrated luminosity of 2.6 fb$^{-1}$ for the Z+jets events and of 23 nb$^{-1}$ for the dijets ones. Distributions of these variables observed in data are reasonably well described by the MC simulation, as shown in Fig. 5 [6].

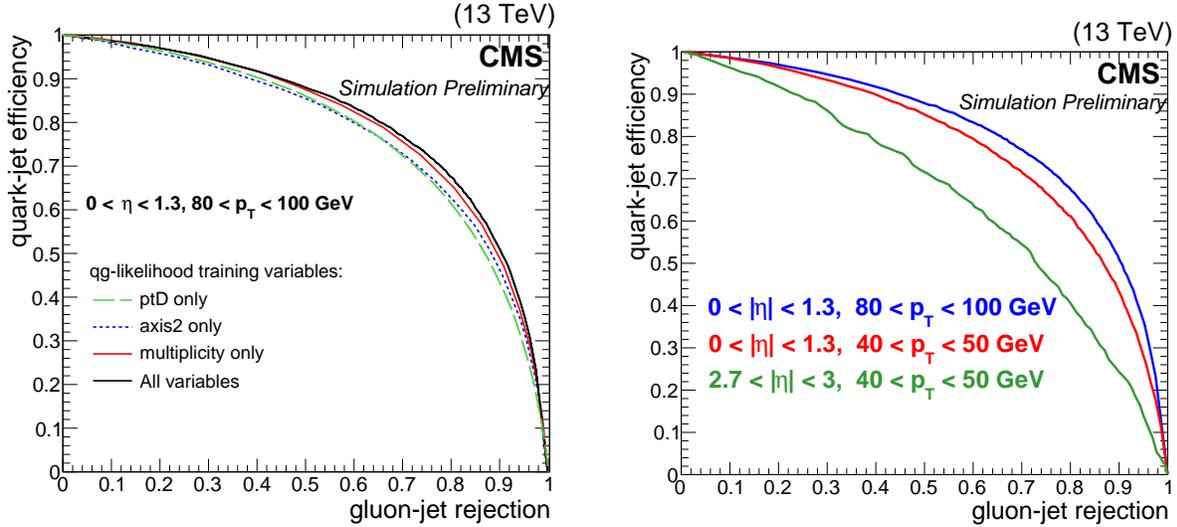

Figure 4: Quark jet tagging efficiency as a function of the gluon jet rejection: (left) individual variable discrimination compared to the full QGL (right) QGL performance in different kinematic regions [6].

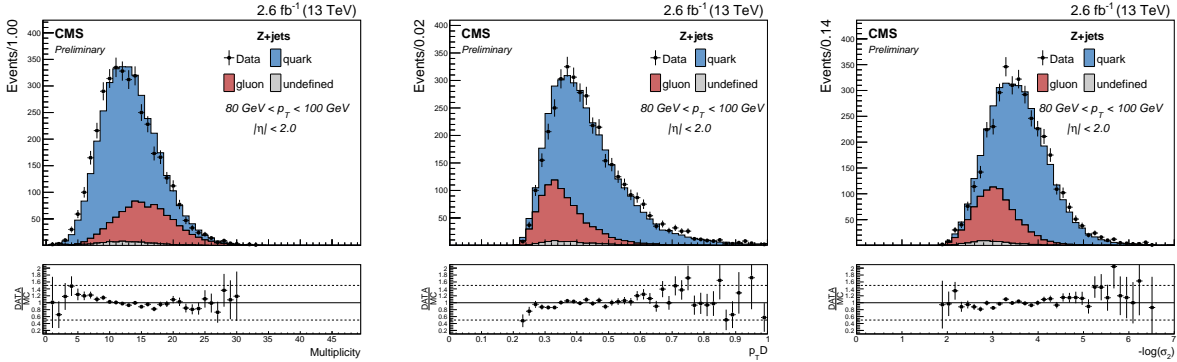

Figure 5: Comparisons of the training observables in Z+jets events in data and simulation [6].

**Systematic uncertainties and tagging efficiencies**

To estimate shape uncertainties on the QGL, a reweighting based method, taking into account the discriminator shape variations observed in the validation of the simulated samples, is pursued.

The chosen approach is the reshaping of the Monte Carlo distribution of both parton flavors components, with weights distributions binned in the likelihood output, by constraining them using the yields observed in the data. Both control regions used in the validation, Z+jets and dijets, have been simultaneously exploited and the method resulted in a relevant improvement in the data/MC agreement on the QGL distribution, as shown in Fig.6.

Validation on data and reshaping have been performed on HERWIG generated samples too and a comparison between the two hadronizers in the efficiencies to select gluon- and quark-jets with using a fixed cut on the likelihood output are shown in Fig. 7, before and after the application of the data-driven reweighting procedure. The efficiencies obtained with the data-driven corrected performances are very similar (within the percent level) for both parton flavors, comparing the two parton showers.

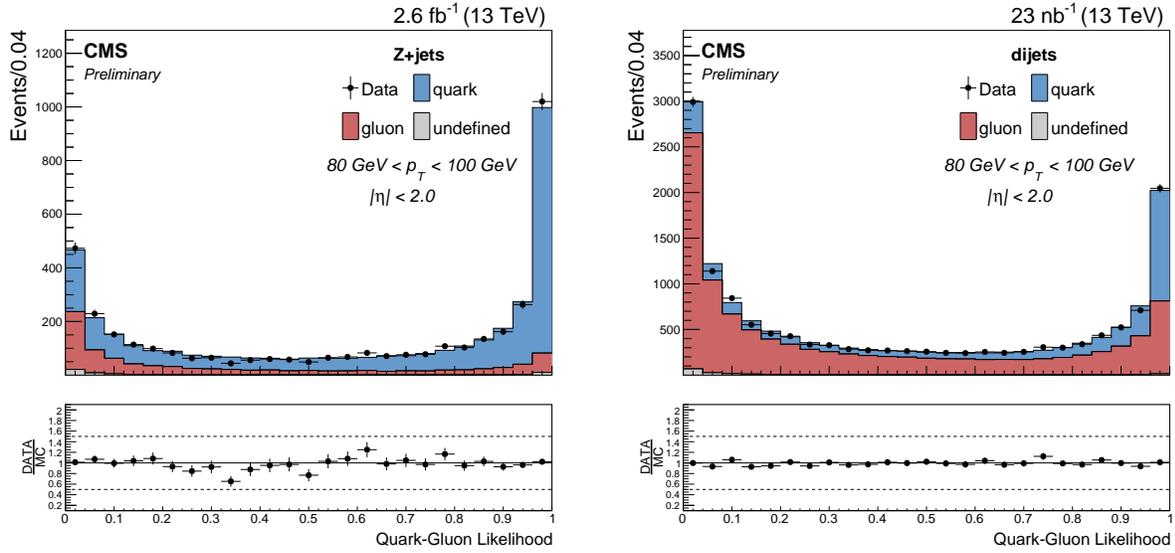

Figure 6: Data-MC comparisons on the tagger on Z+jets events (left) and dijets events (right) after the reshaping of the MC distributions [6].

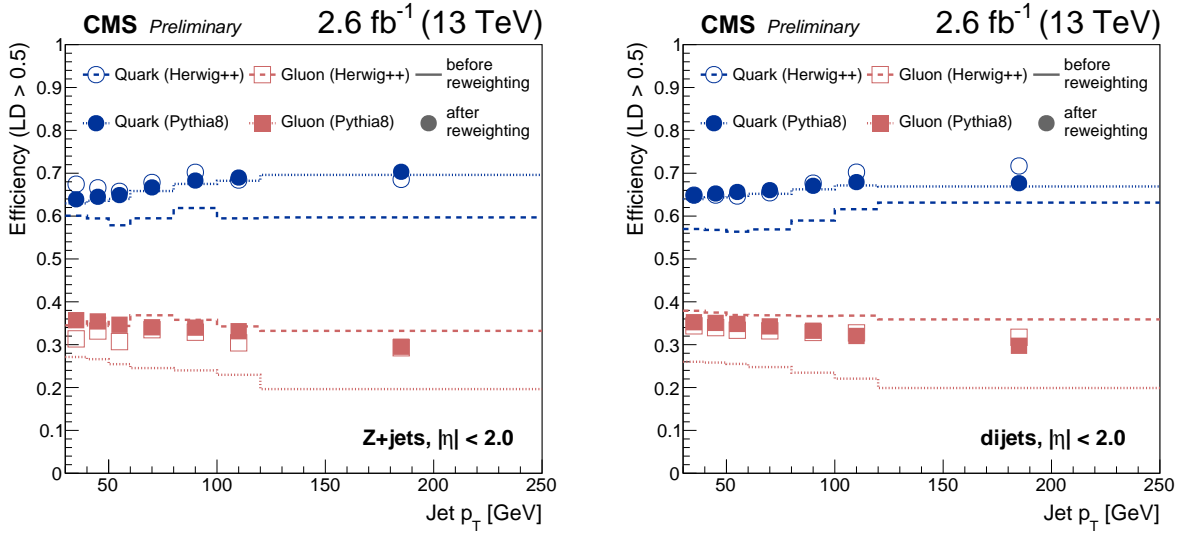

Figure 7: HERWIG and PYTHIA selection efficiencies by applying a fixed cut on the likelihood output LD>0.5. Efficiencies are evaluated in dijet events (left) or Z+jet events (right), as a function of the jet $p_T$ with or without the data-driven shape reweighting of the outputs [6].

**Conclusions**

The efforts made by the ATLAS and CMS Collaboration to experimentally distinguish between quark-like and gluon-like jets have been presented. The former extracted data-driven templates for several discriminating variables and performed also an independent precision measurement of the jet constituents multiplicity. The latter built a likelihood-based discriminant, validated it on data and compared its performances on two hadronizers.

# Jet charge determination at the LHC

Stano Tokar[1]

*On behalf of the ATLAS and CMS Collaborations*

[1] Comenius Univ. Bratislava, Slovakia

**Abstract:** Knowing the charge of the parton initiating a jet could be very useful both for testing different aspects of the Standard model and for searching signals of a beyond-the-standard-model physics. A weighted sum of the charges of jet constituents can be used at the LHC experiments to distinguish among jets from partons with different charges. A few applications of the jet charge variable are presented here. The jet charge was used to distinguish jets initiated by $b$ quarks from those initiated by $\bar{b}$ quarks, for distinguishing between boosted hadronically decaying $W^+$ and $W^-$, and for distinguishing jets initiated by quarks from those initiated by gluons.

## Introduction

The LHC experiments, ATLAS [1] and CMS [2], are aimed at precision measurements on deep inelastic scattering processes. To test different aspects of the strong interaction requires to reconstruct identity of the produced final state partons. In the case of quarks and gluons their identity is diluted by hadronization. Using jet charge as an observable sensitive to the electric charge of quarks defined as the momentum weighted charge sum constructed from charged-particle tracks in a jet, was suggested in Ref. [3]. Since then the jet charge was investigated from the theoretical as well as from the experimental point of view. For the theoretical studies see Ref. [4] and references inside. The first experimental use of jet charge was in deep inelastic scattering studies (neutrino-proton scattering) [6], [7] providing evidence of quarks in nucleons. In the LEP era, the jet charge variable was employed for tagging the charge of $b$-quark jets which was used for determination of asymmetry in the production of $b$ quark pairs ([8], [9]) and for neutral $B$ meson oscillation studies, see Refs. [10] and [11]. Later the jet charge technique was applied for tagging of $b$-quark-jet type within determination of top-quark charge ([12], [13], [14]), for boosted $W$ decaying hadronically – to distinguish them from quark and gluon jets ([15], [16], [17]) as well as for distinguishing jets from quarks and gluons ([18], [19], [20]).

There are a few approaches used for calculation of jet charge:

$$Q_{\rm J}^{(1)} = \frac{1}{p_{\rm T,J}^\kappa} \sum_{h \in {\rm Jet}} q_h \times (p_{{\rm T},h})^\kappa, \ \ Q_{\rm J}^{(2)} = \frac{\sum_{h \in {\rm Jet}} q_h |\vec{j} \cdot \vec{p}_h|^\kappa}{\sum_{h \in {\rm Jet}} |\vec{j} \cdot \vec{p}_h|^\kappa}, \ \ Q_{\rm J}^{(3)} = \sum_{h \in {\rm Jet}} z_h^\kappa q_h, \ z_h = \frac{E_h}{E_{\rm J}} \quad (1)$$

where $q_h$, $p_{{\rm T},h}$, $E_h$ and $\vec{p}_h$ are the hadron ($h$) track charge, transverse momentum, energy and momentum, respectively, $\kappa$ is an exponent (a free parameter), $E_{\rm J}$ is the jet energy, and $\vec{j}$ is the jet direction unit vector.

## Theoretical approach

Calculation of jet charge is challenging as it is not an infrared-safe quantity [4]. Difficulties arise due to fact that the jet charge is sensitive to hadronization and also knowledge of the fragmentation functions is needed. In theoretical framework the jet charge is calculated within a soft-collinear effective theory (SCET) – see Ref. [21]. Using the SCET approach the average jet charge is

$$\langle Q_\kappa^i \rangle = \int dz z^\kappa \sum_h Q_h \frac{1}{\sigma_{\text{jet}}} \frac{d\sigma_{h \in \text{jet}}}{dz} = \frac{1}{16\pi^3} \frac{\tilde{J}_{ii}(E, R, \kappa, \mu)}{J_i(E, R, \mu)} \sum_h Q_h \tilde{D}_i^h(\kappa, \mu) \quad (2)$$

where $z = E_h/E_{\text{jet}} \approx p_T^h/p_t^{\text{jet}}$, $\tilde{J}_{ii}$ are coefficients depending on jet definition and flavor $i$, $J_i$ is the jet function depending on jet energy ($E$), and jet cone ($R$), $\tilde{D}_i^h(\kappa, \mu) = \int_0^1 dx x^\nu D_i^h(x, \mu)$ is the Mellin moment of the fragmentation function $D_i^h$, and $\mu$ is the factorization scale.

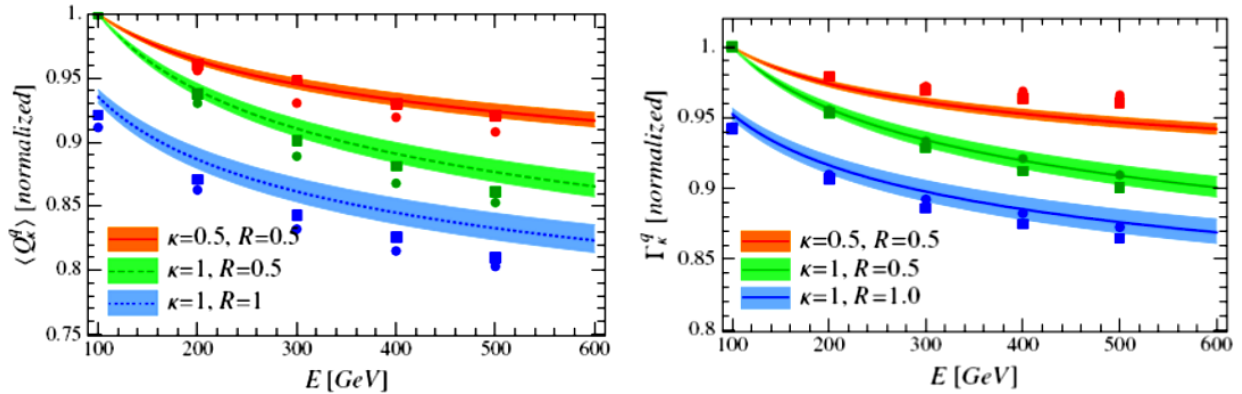

Figure 1: Comparison of theory prediction (bands) for the average (left) and width (right) of the jet charge distribution to PYTHIA 8 (squares and circles for $d$ and $u$ quarks) for $e^+e^-$ collisions – see text.

Within the SCEP approach, also the jet charge width $(\Gamma_\kappa^i)^2 = \langle (Q_\kappa^i)^2 \rangle - \langle Q_\kappa^i \rangle^2$ can be calculated. The calculation is similar to that of the average jet charge but is a bit more complicated as the correlation between hadrons should be taken into account [4]. The theoretical prediction for the average and width of the jet charge is shown in Fig. 1 as a function of jet energy ($E$) for the exponent $\kappa = 0.5$ and 1 and the shower size parameter: $R = 0.5$ and 1.0. The theoretical uncertainty is found from varying the factorization scale by a factor of 2. The distribution are normalized to 1 at $E = 100$ GeV and $R = 0.5$ and compared to PYTHIA 8 [22] predictions – square and circles represent $d$ and $u$ quarks, respectively.

### Determination of top-quark charge

The jet-charge technique was used by the ATLAS experiment for determination of the top-quark charge [14], where it was used to distinguish between two hypotheses: the SM hypothesis, which assumes the top quark with the electric charge 2/3 and the decay $t \to W^+ b$, and the exotic hypothesis based on an exotic quark with the charge -4/3 which assumes the decay $t_X \to W^- b$. Solution of the problem requires to distinguish between $b$ jets from $b$ quark and $\bar{b}$ quark. ATLAS used a data sample of 2.05 fb$^{-1}$ collected in proton-proton ($pp$) collisions at a center-of-mass energy $\sqrt{s} = 7$ TeV. For the analysis, the $t\bar{t}$ events of the lepton+jets channel, $t\bar{t} \to WWb\bar{b} \to (\ell\nu_\ell)(j_1 j_2)\bar{b}$ with two $b$ tags, were used. To determine the top-quark charge one needs to know the charges of the top quark decay products, $W$ boson and $b$ quark, and also a correct $W$–$b$-jet pairing is needed to provide they come from the same decaying object. The charge of $W$ can be found through its leptonic decay, $W^\pm \to \ell^\pm \nu_\ell(\bar{\nu}_\ell)$

– sign of lepton, $\ell$, is the same as that of $W$. As to the $b$-quark charge, a correlation between $b$-jet charge, calculated using the second term in Eq. 1, and the charge of the initiating quark, was used. Within the SM a correct pairing requires $\ell^+$ to be associated with $b$ quark ($Q_b = -1/3$), if they come from the same top quark, while in the exotic case the pairing of $\ell^-$ with $b$ should occur. For the lepton–$b$-jet pairing the following condition based on invariant mass of lepton and $b$-jet, $m(\ell, b)$, was used: $m(\ell, b_{\text{jet}}^{(1)}) < m_{\text{cr}}$ and $m(\ell, b_{\text{jet}}^{(2)}) > m_{\text{cr}}$. The threshold $m_{\text{cr}} = 155$ GeV was found by optimization. As a sensitive variable, to decide between these two hypotheses, a combined charge $Q_{\text{comb}} = Q_\ell \times Q_{\text{bjet}}$ was employed. If the average $Q_{\text{comb}}$ is less than zero, then the SM hypothesis is valid and if it is bigger than zero, the exotic hypothesis occurs. Distribution of $Q_{\text{comb}}$ reconstructed from the data and compared to MC expectations for the SM and exotic hypotheses is shown in Fig. 2 for the muon+jets channel. A similar distribution also for electron+jets channel is shown in Ref. [14]. As is seen from this figure, the data are in excellent agreement with the SM expectation. Statistical treatment based on pseudoexperiments and taking into account all uncertainties (theoretical and experimental) revealed that the exotic hypothesis was excluded at a confidence level better than $8\sigma$ [14].

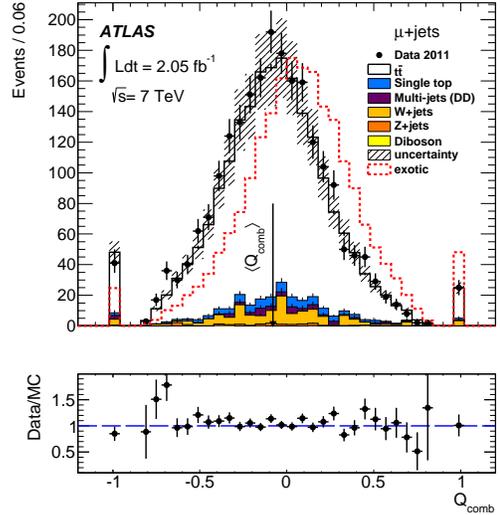

Figure 2: Distribution of $Q_{\text{comb}}$ reconstructed from data is compared to MC expectations for the SM and exotic hypotheses.

### Boosted $W$ boson and jet charge

The CMS collaboration employed the jet charge along with five other variables for the $W$ boson identification in the boosted regime [23]. The analysis was carried out at $\sqrt{s} = 8$ TeV using data sample of 19.7 fb$^{-1}$. The boosted $W$ boson was studied in topologies of $t\bar{t}$ ($\ell$+jets), $W$+jets and dijet events. In the $\ell$+jets $t\bar{t}$ topology events contain in final state two $b$ quarks and two $W$ bosons, one of which decays leptonically and the other one hadronically. The $W$ boosted topology was selected requiring the $W$ jet mass, $m_{\text{jet}}$, and its $p_T$ to fulfill: 60 GeV $< m_{\text{jet}} <$ 100 GeV and 400 GeV $< p_T <$ 600 GeV. The jet charge was calculated using the first term of Eq. 1. The same criteria were applied for boosted jets also in other topologies. Fig. 3 (left) shows the expected jet charge distributions of boosted $W^+$ and $W^-$ bosons with pileup and detector simulation (histograms), and without pileup and detector simulation (dashed thick lines). The color lines correspond to the boosted $W$ bosons from $t\bar{t}$ events while the black lines represent $W$+jets events. Fig. 3 (right) compares the $t\bar{t}$ boosted $W^+$ and $W^-$ jet charge distributions reconstructed from data with the simulated ones (POWHEG [24] with PYTHIA 6) representing a sum of signal and background. Good agreement between the data and simulated jet charge distributions can be stated. In addition, the $W^+$ and $W^-$ jets distributions for the $t\bar{t}$ data can be separated with $\geq 5\sigma$ [23].

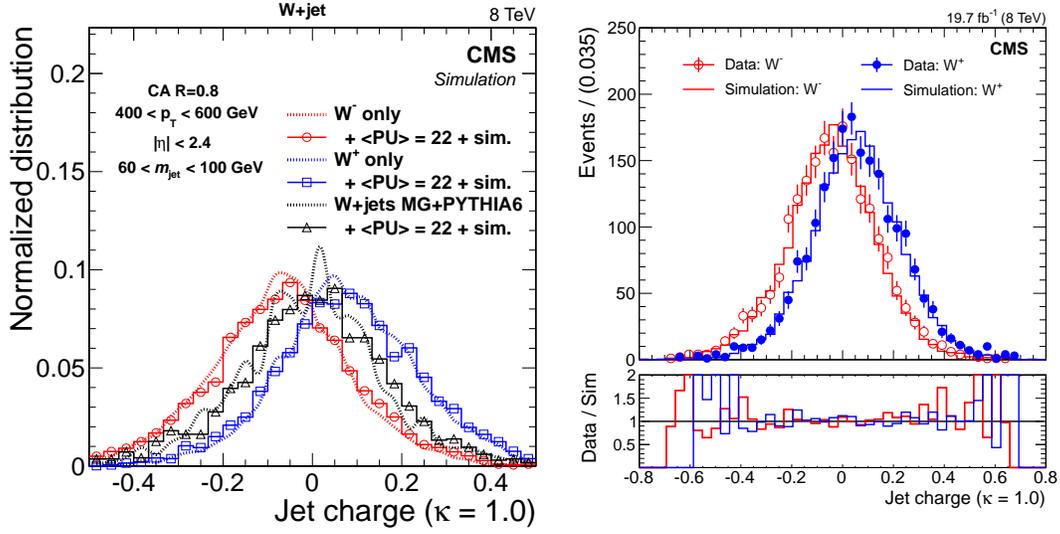

Figure 3: Jet charge boosted $W^+$ and $W^-$ with and without pileup compared with MADGRAPH/PYTHIA W+jets (left) and the jet charge for $W^+$ and $W^-$ reconstructed from data and compared with MC (right).

### Jet charge in dijet events

A measurement of jet charge in dijet events was carried out by ATLAS in $pp$ collisions at $\sqrt{s} = 8$ TeV using a sample of 20.3 fb$^{-1}$ [25]. Events were selected by a single jet trigger with jet $p_T$ threshold from 25 to 360 GeV. The jet charge was calculated using the first term of Eq. 1. Within the analysis an unfolding of jet charge distribution to particle level as a function of jet $p_T$ was performed. The systematic uncertainty was estimated to be a few percents.

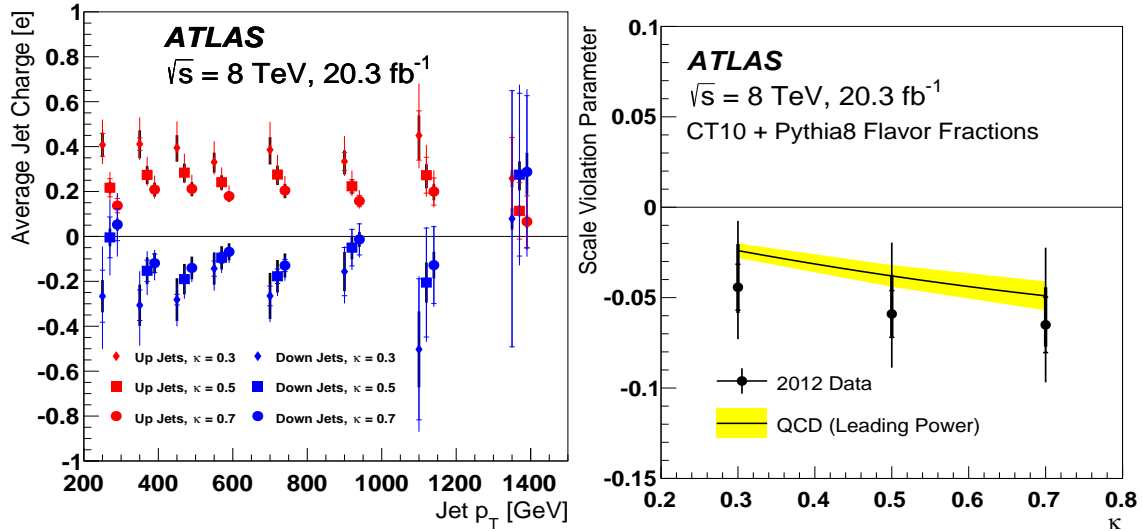

Figure 4: The extracted average $u$ and $d$ quark jet charges in bins of jet $p_T$ for $\kappa$=0.3, 0.5, and 0.7 (left) and the extracted scale violation parameter $c_\kappa$ from the data compared to theoretical calculations [5]. The error bars include statistical, experimental systematic, and PDF uncertainties added in quadrature (right).

Fig. 4 (left) shows the extracted average $u$- and $d$-quark-jet charges as a function of jet $p_T$ for $\kappa=0.3$, 0.5, and 0.7. Theory predicts that the energy dependence of jet charge moments is calculable perturbatively – see e.g. Refs. [4] and [5]. At the leading power of $\alpha_S$, the $p_T$ dependence of the average jet charge reads

$$\langle Q_J \rangle = \bar{Q}\left(1 + c_\kappa \ln\left(p_T/\bar{p}_T\right)\right) + O\left(c_\kappa^2\right), \ c_\kappa \approx -0.38 \pm 0.006 \ \kappa = 0.5 \qquad (3)$$

where $\bar{Q} = \langle Q_J \rangle (\bar{p}_T)$ for some fixed $\bar{p}_T$, $c_\kappa$ is the scaling violation parameter [25]. The measured values of $c_\kappa$ are, within uncertainties, in agreement with the theoretical prediction, as can be seen from Fig. 4 (right).

Jet charge in dijet events was also investigated by CMS. The analysis was carried out with a data sample of 19.7 fb$^{-1}$ collected in $pp$ collisions at $\sqrt{s} = 8$ TeV [26]. A measurement of three different charge observables of the dijet leading jet was performed. The first two observables $Q$ and $Q_L$ are identical with the first two terms of Eq. 1, while the third observable, $Q_T$, is similar to $Q_L$ but track perpendicular momentum to the jet axis is taken instead of the parallel momentum component. The measured jet charge distribution is unfolded from detector to particle level. Fig. 5 shows a comparison of unfolded leading jet charge distributions for the charges $Q$, $Q_L$ and $Q_T$ with PYTHIA 6, HERWIG++ generators. From the comparison it follows that the measured jet charge distributions are in good agreement with expectations.

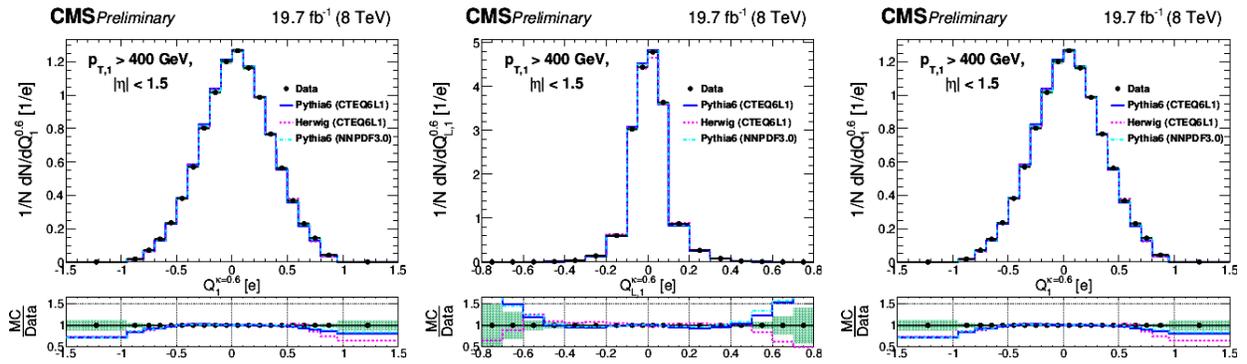

Figure 5: Comparison of unfolded leading jet charge distributions for the charges $Q$ (left), $Q_L$ (middle) and $Q_T$ (right) with PYTHIA 6, HERWIG++ [27] generators, the used $\kappa = 0.6$.

## Conclusion

The ATLAS and CMS experiments have shown that in $pp$ collisions the jet charge variable can be effectively used to distinguish jets initiated by partons of different electric charges. The jet charge, especially when combined with other variables within multivariate techniques, can be used in: studies of asymmetries in $q\bar{q}$ production - to distinguish quarks from antiquarks, studies with $W$ bosons decaying hadronically and many other studies where flavour of jets should be determined. A good perspective of using jet charge is in boosted approaches, especially at 13 – 14 TeV collisions, to distinguish heavy charged and neutral vector bosons.

# Application of quark-gluon jet discrimination and its uncertainty

Yasuhito Sakaki[1]

[1]KAIST, Daejeon, South Korea

**Abstract:** Studies on quark- and gluon-initiated jet discrimination have evolved significantly in recent years. We study the impact of the discrimination technique in the search for strongly interacting supersymmetric particles at the LHC. Taking the example of gluino pair production, considerable improvement is observed in the LHC search reach on including the jet substructure observables to the standard kinematic variables within a multivariate analysis. We also examine evolution variable dependence of the jet substructure by developing a parton shower generator that interpolates between different evolution variables using a parameter $\alpha$ for making clear a reason of Monte Carlo event generator dependence on QCD jet substructure. Jet shape variables and associated jet rates for quark and gluon jets are used to demonstrate the $\alpha$-dependence of the jet substructure.

## Introduction

The discrimination of quark-initiated jets from gluon-initiated ones is an important subject involving jet substructure, and has a lot of potential in improving the search for new physics. Theoretical estimates for the performance of such tagging algorithms are primarily carried out with the help of Monte Carlo (MC) simulation tools, such as, `Pythia`, `Herwig` and `Sherpa`. Even though qualitative features are in agreement, differences in the predictions of the different MC's have been noted as far as quantitative estimates of the quark-gluon tagger performance is concerned.

In Sec. 2, we describe the quark-gluon separation variables used to define a multivariate discriminant, our Monte Carlo simulation of the signal and background processes as well as the kinematic selection of the signal region. We obtain the signal and background likelihood distributions using the multivariate analysis procedure for combining the information from both kinematics and jet substructure. The distributions are used to estimate the expected LHC search reach using different methods in the gluino-neutralino mass plane. In Sec. 3, we consider Monte Carlo dependences on QCD jet substructure, and especially focus on those arising from the difference of evolution variable in the parton shower formalism. A new evolution variable parametrized by a parameter $\alpha$ is introduced, and we implement the variable into a parton shower. Using the shower generator, we show $\alpha$-dependence of one important jet shape observable $C_1^{(\beta)}$ and associated jet rate observables. We summarize our findings in Sec. 4.

## Quark-gluon discrimination in the search for gluino pair production

The goal in this section is to evaluate the expected improvement in the search for gluino pair production at the LHC by including the quark-gluon tagging observables to the standard supersymmetry search strategy in the multijet and missing transverse momentum channel. After including initial and final state parton shower effects to leading order matrix elements, it is estimated that while the third and fourth highest transverse momentum jets in gluino-pair events are expected to be quark-initiated, in the dominant $V$+jets ($V = Z, W$) backgrounds, they are more likely to be gluon-initiated. This leads to a considerable improvement in the signal to background ratio, when jet substructure based observables are utilized.

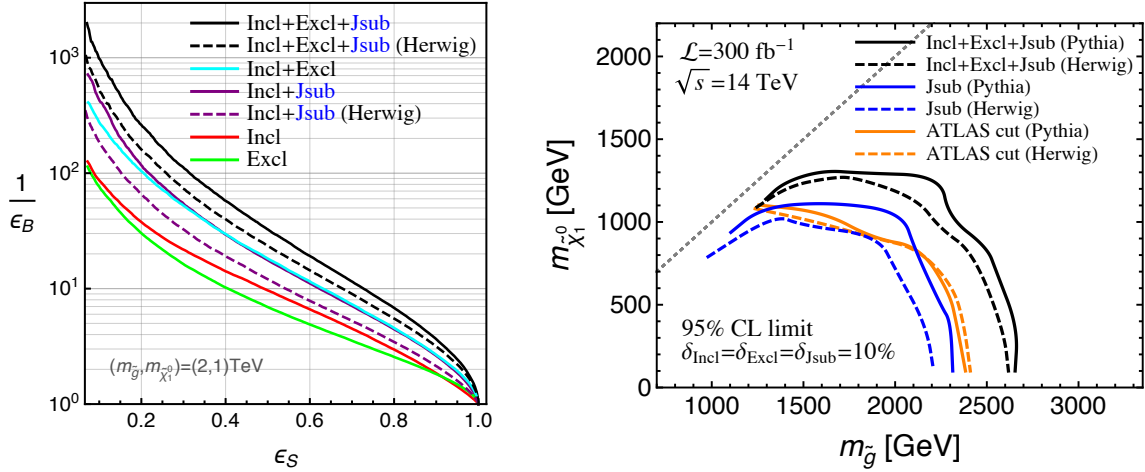

Figure 1: In the left panel, the red, green and cyan curves show the ROC curves for the MVA analyses. The right panel shows that exclusion contours for each MVA analyses.

Based on the difference in splitting probabilities in a parton-shower picture, different possible variables have been proposed for quark-gluon discrimination, which essentially rely on the fact that a gluon produced with similar kinematics leads to a larger multiplicity of soft emissions compared to a quark, and a gluon-initiated jet is wider than a quark-initiated one. These differences follow from the higher colour charge-squared of the gluon, $C_A = 3$, versus $C_F = 4/3$ for a quark. As demonstrated in previous studies, based on both perturbative methods as well as MC simulations, it is found that the following variables lead to a better quark-gluon separation:

1. Energy-energy-correlation (EEC) angularity [1] defined in terms of the charged track momenta, $C_1^{(\beta)} = \sum_{i,j \in Tracks} p_{T,i} p_{T,j} \Delta R_{ij}^\beta / p_{T,J}^2$, where $p_{T,i}$ and $p_{T,j}$ are the energies of the particles labeled by $i$ and $j$ in the jet, $p_{T,J}$ is the jet energy, and $\Delta R_{ij}$ is the angle between $i$ and $j$. The sum runs over all distinct track pairs of particles in the jet.

2. Jet mass ($m_J$) scaled by its transverse momentum $m_J/p_{T,J}$.

3. The number of charged tracks inside the jet cone ($n_{ch}$), with each charged track having $p_T > 1\text{GeV}$, where $p_T$ denotes its transverse momentum.

Using the variables as inputs, we first build a function which returns a single discriminant using a multivariate analysis. We write the discriminant as $B$. The multivariate analysis has been carried out by employing a Boosted Decision Tree (BDT) algorithm with the help of the TMVA-Toolkit in the ROOT framework. The training of the BDT classifier has been performed using the $Z + q$ and $Z + g$ processes at the Born level.

As a final ingredient to our analysis, we perform a further MVA study with ten input variables containing: "Inclusive" variables $\{M_{\text{eff}}, H_T\}$, "Exclusive" variables $\{p_{T,1}, p_{T,2}, p_{T,3}, p_{T,4}\}$, and Jet substructure variables $\{B_1, B_2, B_3, B_4\}$, where $M_{\text{eff}}$ and $H_T$ are the effective mass and scalar sum of jet transverse momenta, and $p_{T,i}$ and $B_i$ are the transverse momentum and the quark-gluon discriminant for $i$-th jet. This defines a signal and background likelihood with all the kinematic and jet substructure information of the event. The BDT score cut is chosen to maximize the exclusion (or discovery) significance for a given model point.

In Fig. 1, the left panel shows ROC curves for the MVA analyses based on inclusive variables (Incl), exclusive variables (Excl), and/or jet substructure variables (Jsub). These sets carry independent information, and therefore the background rejection using the combined set increases compared to the ones using the individual sub-sets. By the fact, the $\epsilon_S$ and $1/\epsilon_B$ values on the cyan curve (Incl+Excl) is roughly given by the product of their corresponding values on the red (Incl) and green (Excl) curves. We can find that the background regection clearly improves by using the jet substructure variables. The right panel shows that exclusion contours for each MVA analyses. We can see that MC dependence on the results are clearly appear by using jet substructure variables. We will consider the source of the MC dependence on QCD jet substructure in the next section.

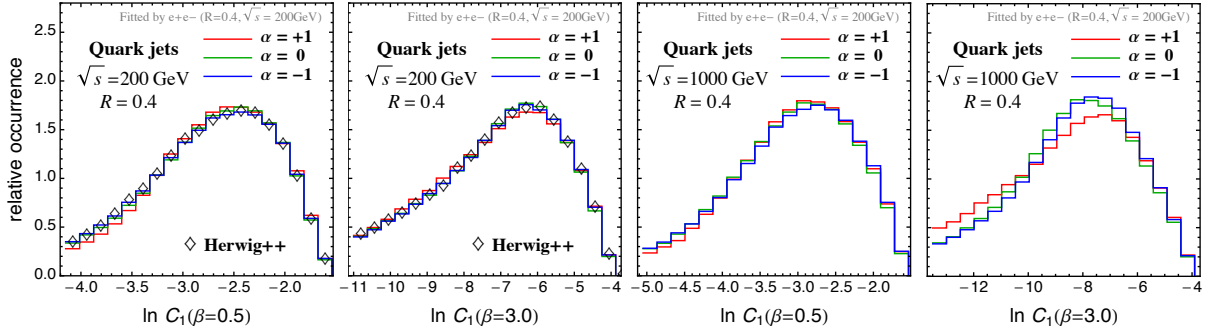

Figure 2: The first and second figures from the left show the fitted results for $C_1^{(0.5)}$ and $C_1^{(3)}$ distributions. The third and fourth figures from the left show the same distributions with a higher value of the center of mass energy, $\sqrt{s} = 1000$ GeV.

### Evolution variable dependence of QCD jet substructure

Although it might be difficult to pinpoint the reason for such differences in the jet substructure observables predicted by different generators, understanding the difference between the central components of the MC's can be useful in developing more precise simulation tools. The substructure of a quark or a gluon jet is governed by the pattern of QCD radiation, which is controlled by the parton shower algorithm. One of the core variables of a parton shower is the evolution variable, different choices for which are made in different MC's. In this section, our aim is to understand the effect of modifying the evolution variable and access its impact on jet substructure observables. We also ask the question whether certain choice of evolution variables can better reproduce the data on quark-gluon tagging observables.

With this goal in mind, we simulate jet substructure related observables with the following generalized evolution variable:

$$Q_\alpha^2 = [4z(1-z)]^\alpha q^2, \tag{1}$$

where, $\alpha$ is treated as a free parameter. For final state radiation, the above variable with $\alpha = 1$ and $-1$ correspond to the evolution variables employed in `Pythia8` and `Herwig++` respectively. Even though there are various recent parton shower formalisms, we implement the evolution variable into a traditional formalism based on refs. [4,5], which is used in `Herwig++`.

The parton shower depend on parameters $\alpha_S(m_Z)$, $m_{\rm qg}$, and $Q_{\rm cut}$, where $\alpha_S(m_Z)$ is the strong coupling constant at the scale is $Z$ boson's mass, $m_{\rm qg}$ is a effective mass of light quarks and gluons for avoiding the soft divergence, and $Q_{\rm cut}$ is a stopping scale. We utilize the $e^+e^- \to q\bar{q}$ events generated by `Herwig++` with hadronization switched off as our data. The $C_1^{(0.5)}$, $C_1^{(2)}$ and $C_1^{(3)}$ distributions have been used to tune the above parameters. The first emission in the jet has a significant effect on the jet shape, which can be parametrized by the momentum fraction $z$ and the angle $\theta$. Therefore, two independent $C_1^{(\beta)}$ distributions contain the necessary information about the jet shapes. Here, we use three variables in order to further examine the $\beta$ dependence of the QCD jet substructure.

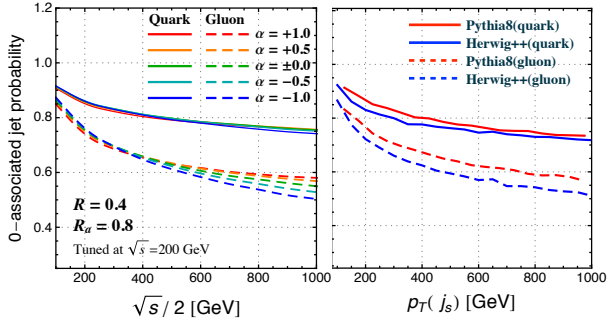
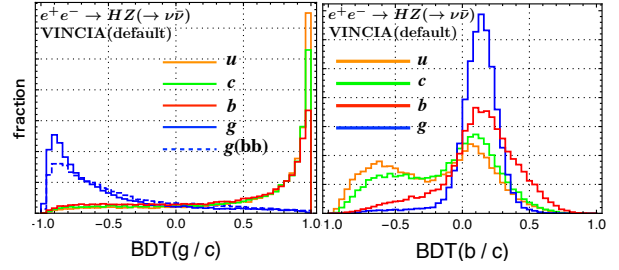

Figure 3: The left panel shows the associated jet rate for evolution variables parametrized by various $\alpha$. The right panel shows the rate with `Pythia8` and `Herwig++`.

Figure 4: The left (right) panel shows distributions of BDT score using $H \to gg$ ($b\bar{b}$) and $H \to c\bar{c}$ as training samples on $e^+e^- \to H + Z(\to \nu\bar{\nu})$.

In Fig. 2, the first and second figures from the left show the fitted results for $C_1^{(0.5)}$ and $C_1^{(3)}$ distributions, and hence the distributions are in good agreement with `Herwig++` predictions. The third and fourth figures from the left show the same distributions with a higher value of the center of mass energy, $\sqrt{s} = 1000$ GeV. As we can see, the $\alpha$-dependence of the shapes is found to be higher for higher energy jets. The larger the parameter $\beta$ in $C_1^{(\beta)}$ is, the larger the differences become. This implies that the wideness of the emissions, especially for the hardest emission in the jets, is different for each $\alpha$. This is because, the larger $\beta$ is, the larger the contribution to $C_1^{(\beta)}$ from the emission angle of the hardest emission becomes, which is understood from the following approximated formula, $C_1(\beta) \sim \ln z + \beta \ln \theta$, where $z$ and $\theta$ are the energy fraction and emission angle for the hardest emission in a jet.

Associated jet rates defined in Ref. [6] directly reveal the wideness of the emissions in jets. Associated jets are jets nearby a hard jet, and are defined by two parameters, $R_a$ and $E_a$. Here, $R_a$ is the maximum allowed angle between the momentum directions of the hard jet and the associated jet, and $E_a$ is the minimum energy of the associated jets. We set the value to $E_a = 20$ GeV in this study.

A high probability for having no associated jet implies that the probability of wide emissions occurring around the hard jet is low. Such probabilities have been obtained by using `Pythia8`, `Pythia6`, and `Herwig++`, and it has been found that the no associated jet probability predicted by `Pythia` is higher than the one obtained with `Herwig++` [6] as shown in the right panel of Fig. 3. The no associated jet probabilities calculated with $Q_1$, $Q_{0.5}$, $Q_0$, $Q_{-0.5}$ and $Q_{-1}$ are shown in the left panel of Fig. 3. We can see that no associated jet probabilities are similar for each $\alpha$ at the low energy range. This is expected as the parameters have been tuned at $\sqrt{s} = 200$ GeV. The $\alpha$ dependence is enhanced at the high energy range. The larger $\alpha$ is, the larger the no associated jet probabilities become. Therefore, an angular ordered shower ($\alpha = -1$) predicts wider jets, while a $p_\perp$ ordered shower ($\alpha = 1$) predicts narrower jets. This result is qualitatively in agreement with the missing phase space of the $p_\perp$ ordered shower [7]. The wideness of the emissions in the jets are thus tunable by changing the parameter $\alpha$ in the evolution variable continuously.

## Conclusion and discussion

In this paper, we have shown that a performance of background rejection clearly improve by using the quark-gluon discrimination technique, and reachable mass bounds for gluino and neutralino also increase in Fig. 1. We can see Monte Carlo generator dependences, which are stemming from that QCD jet substructure describing each generators are different.

To reveal a reason for the Monte Carlo dependences, we focused on a evolution variable dependence on QCD jet substructure. For the purpose, we have defined a generalized evolution variable in Eq. 1 which is specified by a parameter $\alpha$, and implemented the variable in to a parton shower. We have found that wideness of the soft emissions, especially the first ones in a jet are different for each $\alpha$. We can examine this directly by studying the associated jet probability. A high probability for having no associated jet simply means that the probability of wide emissions occurring around a hard jet is low. We have found that the larger $\alpha$ is, the larger the no associated jet probability becomes. This gives us a qualitative understanding of the generator dependence of associated jet rates, especially between `Pythia8` and `Herwig++`. Our results open up the possibility that we might be able to reproduce the wideness of jets observed in real data by varying the value of $\alpha$ in the evolution variable continuously.

For future $e^+e^-$ machine, the quark gluon discrimination should be useful. As a possible example, we would be able to apply the technique to the measurement of Higgs branching ration for the purpose of discrimination between $H \to b\bar{b}, c\bar{c}$ and $gg$. The left (right) panel in Fig. 4 shows distributions of BDT score using $H \to gg$ ($b\bar{b}$) and $H \to c\bar{c}$ as training samples on $e^+e^- \to H + Z(\to \nu\bar{\nu})$. The score is evaluated by $C_1^{(0.5)}$ and $C_1^{(3)}$ for the first and second highest energy jets. We can see that clear difference between the Higgs gluon decay and the quark decays in the left figure. The blue dashed curves shows a result for the $H \to gg$ sample containing $g \to b\bar{b}$ emissions at least onetime. The right figure shows that we can observe the difference between $H \to b\bar{b}$ and $H \to c\bar{c}$ using jet substructure variables. For a successful discrimination, we need to validate gluon jet distributions discribed by MC's with data. We would be able to obtain the outstanding quality of the gluon data from $e^+e^- \to q\bar{q}g$ or $b\bar{b}g$ at FCC-ee as stated in the Klaus's and Gregory's talks.

# Angularities from LEP to FCC-ee

Guido Bell[1], Andrew Hornig[2], Christopher Lee[2] and **Jim Talbert** [3,4]

[1] Theoretische Physik 1, Naturwissenschaftlich-Technische Fakultät, Universität Siegen, Walter-Flex-Strasse 3, 57068 Siegen, Germany
[2] Theoretical Division, Group T-2, M2 B283, Los Alamos National Laboratory, P.O. Box 1663, Los Alamos, NM 87545, USA
[3] Theory Group, Deutsches Elektronen-Synchrotron (DESY), D-22607 Hamburg, Germany
[4] Rudolf Peierls Centre for Theoretical Physics, University of Oxford, 1 Keble Road, OX1 3NP, Oxford, UK

**Abstract:** We present a preliminary NNLL' resummation of the event shape *angularities* and compare it to LEP data at $Q = 91.2$ GeV. Our calculation permits a future precision determination of the strong coupling $\alpha_s(m_Z)$ from a fit to the experimental distributions. As the angularities are sensitive to the same non-perturbative parameter $\mathcal{A}$ that shifts the thrust distribution, our analysis may help to lift current degeneracies in the two-dimensional $\alpha_s(m_Z) - \mathcal{A}$ fits.

## Introduction

Event-shape variables [1] characterize the geometric properties of a final-state distribution (e.g. dijet, three-jet-like, spherical, etc.) in collider processes. They are generally global observables that do not reject any final-state hadrons. Event shapes can be studied at hadron or $e^+e^-$ colliders, though we focus here on the latter where a wealth of experimental data already exists from the Large Electron-Positron Collider (LEP) and where an $e^+e^-$ Future-Circular-Collider (FCC-ee) could help to alleviate tensions in a number of $\alpha_s$–determinations that are based on different theoretical methods (cf. contributions from *V. Mateu* and *P. Monni*).

We focus on a class of event shapes generically defined as

$$e(X) = \frac{1}{Q} \sum_{i \in X} |\mathbf{p}_\perp^i| \, f_e(\eta_i), \qquad (1)$$

where $\eta_i$ is the rapidity of the i'th final-state particle with respect to the thrust axis and $\mathbf{p}_\perp^i$ its transverse momentum. The function $f_e(\eta)$ determines the specific observable. For example, for the two well-known event shapes *thrust* $T \equiv 1 - \tau$ [2] and (total) *jet broadening* $B_T$ [3], one has $f_\tau(\eta) = e^{-|\eta|}$ and $f_{B_T}(\eta) = 1$, respectively.

Both thrust and broadening can be generalized into a class of observables known as *angularities* [4,5],

$$f_{\tau_a}(\eta) = e^{-|\eta|(1-a)} \qquad \longleftrightarrow \qquad \tau_a(X) = \frac{1}{Q} \sum_{i \in X} E_i \, |\sin\theta_i|^a \, (1 - |\cos\theta_i|)^{1-a}, \qquad (2)$$

where $E_i$ is the energy and $\theta_i$ the angle of the i'th particle with respect to the thrust axis. The angularities thus depend on a continuous parameter $a$, which fulfils $-\infty < a < 2$ due to infrared (IR) safety. For $a = 0$, the angularity reduces to thrust, $\tau_0 = \tau$, and for $a = 1$, it reduces to broadening $\tau_1 = B_T$.

Like other QCD observables that depend on widely separated energy scales, event shapes are affected by logarithmic enhancements to the perturbative QCD (pQCD) expansion, which must be *resummed* to all orders. Many analyses have been performed to this end, both with standard pQCD and, more recently, also with Soft-Collinear Effective Theory (SCET) [6,7,8,9]. SCET formally separates the relevant scales present in collider processes, and it provides an elegant means to establish factorization theorems. For example, the angularity distribution factorizes in the dijet limit $\tau_a \to 0$ into a *hard* function $H(\mu, \mu_H)$ that encodes the matching of SCET to QCD, two *jet* functions $J(\mu, \mu_J)$ describing the evolution of the coloured partons into collimated jets, and a *soft* function $S(\mu, \mu_S)$ describing low-energy, wide-angle background radiation, all of which live at an associated scale $\mu_H \gg \mu_J \gg \mu_S$ [4,10]. The dependence of $H$, $J$, and $S$ on the factorization scale $\mu$ is controlled by renormalization group (RG) equations, which can be used to resum large logarithms present in each function. Indeed, many of the most precise event-shape resummations have been achieved with SCET techniques, with thrust and broadening currently resummed to $N^3LL$ [11,12] and NNLL [13,14] accuracy, respectively[*].

An immediate goal of this note is to use methods from SCET to predict angularity distributions to NNLL' accuracy [15], thereby realizing an improvement on a prior NLL' resummation [10,16]. Our calculation is based on a recent two-loop calculation of the angularity soft function [17], which we use to extract the missing NNLL' ingredients. We are also motivated by the presence of L3 Collaboration data [18], which measured the angularity distributions at 8 different values of $a \in \{-1, -0.75, -0.5, -0.25, 0, 0.25, 0.5, 0.75\}$ at both $Q = 91.2$ GeV and $Q = 197.0$ GeV. This will allow for a future extraction of $\alpha_s(m_Z)$ and the non-perturbative (NP) shift parameter $\mathcal{A}$ as discussed below.

### Sensitivity to Non-Perturbative Effects

As with any hadronic observable, event shapes are sensitive to low-energy QCD radiation. The importance of these NP effects depends on the domain of $\tau_a$ considered. For angularities with $a < 1$ in the (near-)tail region, power corrections from the collinear sector are suppressed with respect to those from the soft sector[†] [19,20]. The NP effects can then be parameterized into a *shape function* that is convolved with the perturbative distribution [21]. In the tail region, it can be shown rigorously via an operator-product-expansion (OPE) that the dominant NP effect results in a shift of the perturbative distribution [19][‡]

$$\frac{d\sigma}{d\tau_a}(\tau_a) \xrightarrow[\text{NP}]{} \frac{d\sigma}{d\tau_a}\left(\tau_a - c_{\tau_a}\frac{\mathcal{A}}{Q}\right). \tag{3}$$

Here $\mathcal{A}$ is a *universal* NP parameter that is defined as a vacuum matrix element of soft Wilson lines and a transverse energy-flow operator (for details, see [19]), while $c_{\tau_a}$ is an exactly calculable

---

[*]For a thorough elaboration of the logarithmic enhancements captured in a $N^k LL$ ($k \in \{0, 1, ...\}$) resummation and the subtle differences between primed and unprimed accuracies, see [16].

[†]The endpoint $a = 1$ corresponds to the onset of $SCET_{II}$ physics. We will not discuss the subtle differences between $SCET_I$ and $SCET_{II}$ observables, though thrust and angularities are examples of the former (for $a < 1$). At the broadening limit, the angularity reduces to a $SCET_{II}$ observable and therefore predictions based on a $SCET_I$ factorization theorem should become progressively worse as $a \to 1$. We observe this effect.

[‡] In the peak region, the OPE does not apply and a full shape function is required to capture the non-perturbative effects. Furthermore, the result in (3) is not only leading-order in the OPE, it is also subject to other corrections like finite hadron masses and perturbative renormalization effects on the quantity $\mathcal{A}$, as described in [22].

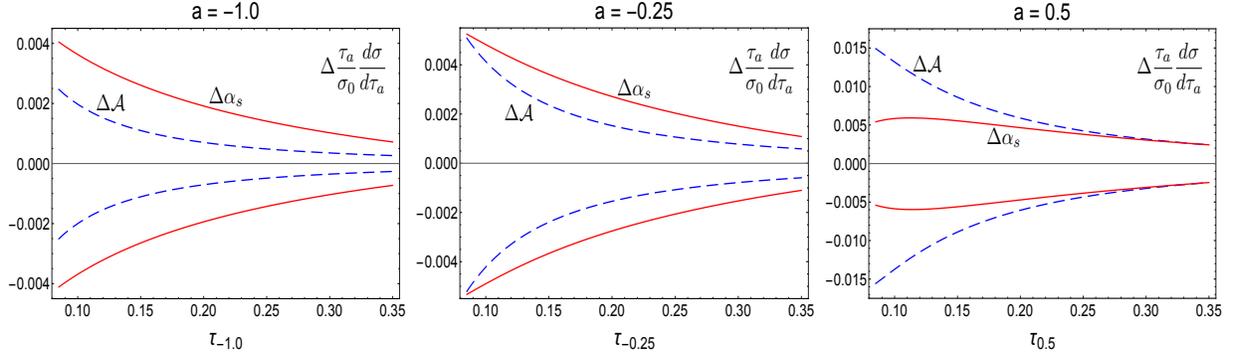

Figure 1: Difference distributions between central curves and curves evaluated with single variations of either $\mathcal{A}$ (dashed, blue) or $\alpha_s(m_Z)$ (solid, red) at three values of $a \in \{-1, -0.25, 0.5\}$. $Q = 91.2$ GeV in all three plots.

*observable-dependent* coefficient. For the angularities, it is given by[§]

$$c_{\tau_a} = \int_{-\infty}^{\infty} d\eta \, f_{\tau_a}(\eta) = \frac{2}{1-a}. \tag{4}$$

Hence, in any attempt to extract a value of the strong coupling by comparing data to theoretical predictions, one is simultaneously sensitive to $\alpha_s(m_Z)$ and $\mathcal{A}$. Indeed, the most precise extractions employing analytic treatments of NP effects [12,23] report values in an $\alpha_s(m_Z) - \mathcal{A}$ plane (cf. contribution from *V.Mateu*). Furthermore, the extracted values of $\alpha_s(m_Z)$ from these analyses are consistently (and often dramatically) lower than the world average, which is currently dominated by lattice-QCD calculations (cf. $0.1123 \pm 0.0015$ [23] to the world average $0.1181 \pm 0.0011$ [24]). It can be shown that the event-shape extractions are driven to small values precisely due to NP effects, and so any elucidation of these discrepancies requires a disentangling of perturbative and non-perturbative contributions.

Our proposal is to perform a future extraction of both $\alpha_s(m_Z)$ and $\mathcal{A}$ along the lines of previous SCET treatments, but at multiple values of the angularities $a$. The critical point is that the leading NP shift in (3) is $a$-dependent. Therefore, an extraction at a single centre-of-mass energy $Q$, but different values of $a$, will have a discriminating sensitivity to $\mathcal{A}$ and $\alpha_s(m_Z)$ in a similar way as varying $Q$. For example, angularities for $-2 \leq a \lesssim 0.5$ exhibit a factor of six variance in the overall NP shift. This sensitivity is essentially equivalent to measurements made between $Q = 35$ GeV and $Q = 207$ GeV, as analyzed for thrust e.g. in [12]. In Figure 1 we show the difference $(d\sigma/d\tau_a)_{\text{central}} - d\sigma/d\tau_a$ over the range $0.085 \leq \tau_a \leq 0.35$ for $a \in \{-1, -0.25, 0.5\}$, where $(d\sigma/d\tau_a)_{\text{central}}$ is an (unmatched) NNLL$'$ resummed distribution evaluated at $\alpha_s(m_Z) = 0.1161$ and $\mathcal{A} = 0.283$ GeV. For $(d\sigma/d\tau_a)$ we have varied $2\mathcal{A}$ by $\pm 0.1$ GeV and $\alpha_s(m_Z)$ by $\pm 0.001$, corresponding to the blue and red curves, respectively. These plots are analogous to Figure 10 in [12], where the same variations were made but at different values of $Q$, rather than $a$. Indeed, we find that varying $a$ ($Q$) down (up) from high (low) values leads to an enhanced sensitivity of the distributions to the relative effects of $\mathcal{A}$ and $\alpha_s(m_Z)$ variation. We are therefore optimistic that the $a$-dependence of the angularities can help to lift the degeneracies between $\alpha_s(m_Z)$ and $\mathcal{A}$ in the two-parameter fits.

---

[§]The expression for $c_{\tau_a}$ diverges in the limit $a \to 1$, where the SCET$_{\text{I}}$ factorization theorem we use breaks down. A careful analysis revealed that the NP effects to the broadening distributions are enhanced by a rapidity logarithm, $c_{B_T} = \ln Q/B_T$ [20].

## Angularities at NNLL′ Accuracy

The resummed cumulative distribution in $\tau_a$, $\sigma_c(\tau_a) = (1/\sigma_0) \int_0^{\tau_a} d\tau_a'(d\sigma/d\tau_a')$, will ultimately be given by [10,16]

$$\sigma_c(\tau_a) = e^{K(\mu,\mu_H,\mu_J,\mu_S)} \left(\frac{\mu_H}{Q}\right)^{\omega_H(\mu,\mu_H)} \left(\frac{\mu_J^{2-a}}{Q^{2-a}\tau_a}\right)^{2\omega_J(\mu,\mu_J)} \left(\frac{\mu_S}{Q\tau_a}\right)^{\omega_S(\mu,\mu_S)}$$
$$\times H(Q^2,\mu_H) \, \widetilde{J}\Big(\partial_\Omega + \ln\frac{\mu_J^{2-a}}{Q^{2-a}\tau_a}, \mu_J\Big)^2 \, \widetilde{S}\Big(\partial_\Omega + \ln\frac{\mu_S}{Q\tau_a}, \mu_S\Big) \frac{e^{\gamma_E \Omega}}{\Gamma(1-\Omega)} \,,$$

where $\sigma_0$ is the Born cross-section summed over massless quark flavours $f = \{u,d,s,c,b\}$, $H$ is the hard function, $\tilde{J}$ and $\tilde{S}$ are the Laplace-space jet and soft functions, and $K$, $\Omega$ and $\omega_{H,J,S}$ are evolution kernels that depend on the anomalous dimensions of the functions $H$, $\tilde{J}$ and $\tilde{S}$. The anomalous dimensions and the fixed-order functions have expansions in $\alpha_s$, such that resummations of higher logarithmic accuracy require increasingly higher-order terms.

To achieve NNLL′ accuracy, one needs all of the ingredients from Table 5 of [16]. In particular, the two-loop jet and soft anomalous dimensions and the respective finite (non-logarithmic) terms were not previously known. Calculating the soft variants has now been achieved in [17] via a generic algorithm for the numerical evaluation of two-loop dijet soft functions. The remaining two-loop jet anomalous dimension can then be calculated using RG consistency relations, and the finite term in the two-loop expansion of $\tilde{J}$ can be extracted via a comparison with a fixed-order code, for which we use the EVENT2 generator [25] (details will be given in [15]).

## Matching

SCET is an effective theory of QCD that predicts the singular terms in the cross section as $\tau_a \to 0$ and resums them to all orders. To obtain a reliable description in the large $\tau_a$ domain, one then needs to match the resummed distribution to the fixed-order QCD result. To this end, we utilize EVENT2 to generate the differential distribution up to $\mathcal{O}(\alpha_s^2)$.

Furthermore, we have designed *profile functions* [12] that smoothly interpolate between the peak region (where $\mu_H \gg \mu_J \gg \mu_S \sim \Lambda_{QCD}$), the tail region (where $\mu_H \gg \mu_J \gg \mu_S \gg \Lambda_{QCD}$) and the far-tail region (where $\mu_H \sim \mu_J \sim \mu_S \gg \Lambda_{QCD}$). In the peak region the soft scale is very nearly NP although, as we do not employ a model shape function in this analysis, we will not show predictions in this region anyway. On the other hand, the scales are well separated in the tail region, which is the region where resummation is most important. Finally, our predictions should match onto fixed-order perturbation theory in the far-tail region. Resummations should therefore be switched off, and the scales should merge at $\mu_{H,J,S} = Q$. While we do not show the explicit functional form of our profile scales, they are similar to those in [26]. The final theory errors presented below reflect independent variations of the hard, jet and soft scales added in quadrature.

## Results

Some benchmark *preliminary* results for NNLL′ resummed and $\mathcal{O}(\alpha_s^2)$ matched distributions are shown in Figure 2. Our results are for $a \in \{-0.5, 0.5\}$ and $Q = 91.2$ GeV, and we have set $\alpha_s(M_Z) = 0.1161$ and $\mathcal{A}_{\text{NNLL}'} = 0.283$ GeV as in [12]. The plots show the curves without (blue) and with the NP shift (green), and they also display the data points from [18]. We focus here on the central (or tail and far-tail) $\tau_a$ domain where the effect of resummation is most relevant. Plots including the peak region will be left for future studies. For $a = -0.5$ the difference between the perturbative (blue) and the NP shifted curve (green) is too small for one to be clearly preferred by the experimental data. For $a = 0.5$, on the other hand, the NP effect is sizeable and, indeed, necessary to accurately describe the data. This is a clear visual confirmation of the leading-order prediction in (3). Note that the error bars in Figure 2 *do not* include any error estimate coming from the EVENT2 extraction of the two-loop jet constant nor from matching to QCD. This will be addressed in [15].

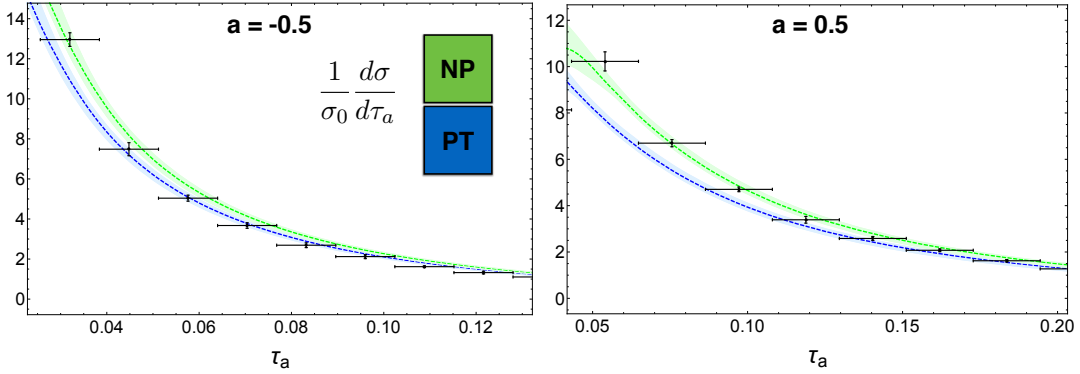

Figure 2: Preliminary NNLL' resummed and $\mathcal{O}(\alpha_s^2)$ matched angularity distributions at two values of the parameter $a \in \{-0.5, 0.5\}$. The blue (PT) curves represent the purely perturbative result, whereas the green (NP) curves includes the NP shift according to (3). $Q = 91.2$ GeV in both plots.

### Moving from LEP to FCC-ee

We argued that an $\alpha_s$–extraction using angularities could potentially alleviate the current degeneracies in the $\alpha_s(m_Z) - \mathcal{A}$ plane, due to the dependence of the leading NP shift on $a$. Of course, one also notes from (3) that the power correction is sensitive to the centre-of-mass energy $Q$. Therefore an FCC-ee operating at different energies could be an even greater probe in disentangling hadronization effects in $e^+e^-$ event-shape distributions. In Figure 3 we have demonstrated the minimization of NP effects as $Q$ increases from $91.2 \rightarrow 400$ GeV. Not only does one notice that the distributions are larger and more peaked in the low-$\tau_a$ region, one observes that the correction moves from a 9% effect at $Q = 91.2$ GeV to a 2% effect at $Q = 400$ GeV (calculated at $\tau_{0.25} = 0.15$). It is clear that the combined dependence of NP effects on $a$ and $Q$ could be significant. Regardless, precision resummations as presented in this note represent critical first steps in pursuing these goals.

*Acknowledgements:* J.T. acknowledges the hospitality and support of LANL, where this work began. The work of C.L. was supported in part by the U.S. Department of Energy, Office of Science, Office of Nuclear Physics under Contract DE-AC52-06NA25396 and an Early Career Research Award. Preprint numbers: LA-UR-16-29580, SI-HEP-2016-28.

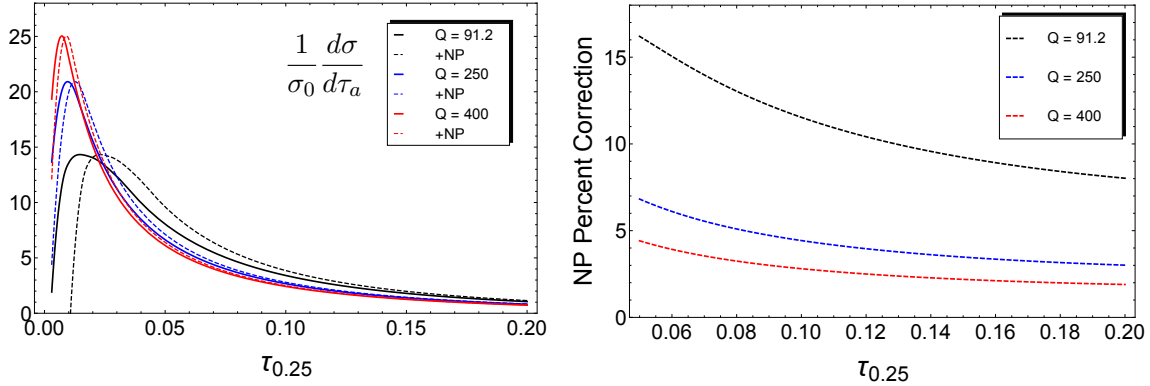

Figure 3: *Left:* Correction to (unmatched) differential angularity distributions from the leading NP power correction at various values of $Q$ (for $a = 0.25$). *Right:* Percent correction from the same effect.

# References

[1] M. Dasgupta and G. P. Salam, J. Phys. G **30** (2004) R143 [hep-ph/0312283].

[2] E. Farhi, Phys. Rev. Lett. **39** (1977) 1587.

[3] P. E. L. Rakow and B. R. Webber, Nucl. Phys. B **191** (1981) 63.

[4] C. F. Berger, T. Kucs and G. F. Sterman, Phys. Rev. D **68** (2003) 014012 [hep-ph/0303051].

[5] C. F. Berger and G. F. Sterman, JHEP **0309** (2003) 058 [hep-ph/0307394].

[6] C. W. Bauer, S. Fleming and M. E. Luke, Phys. Rev. D **63** (2000) 014006 [hep-ph/0005275].

[7] C. W. Bauer, S. Fleming, D. Pirjol and I. W. Stewart, Phys. Rev. D **63** (2001) 114020 [hep-ph/0011336].

[8] C. W. Bauer, D. Pirjol and I. W. Stewart, Phys. Rev. D **65** (2002) 054022 [hep-ph/0109045].

[9] M. Beneke, A. P. Chapovsky, M. Diehl and T. Feldmann, Nucl. Phys. B **643** (2002) 431 [hep-ph/0206152].

[10] A. Hornig, C. Lee and G. Ovanesyan, JHEP **0905** (2009) 122 [arXiv:0901.3780 [hep-ph]].

[11] T. Becher and M. D. Schwartz, JHEP **0807** (2008) 034 arXiv:0803.0342 [hep-ph].

[12] R. Abbate, M. Fickinger, A. H. Hoang, V. Mateu and I. W. Stewart, Phys. Rev. D **83** (2011) 074021 [arXiv:1006.3080 [hep-ph]].

[13] T. Becher and G. Bell, JHEP **1211** (2012) 126 [arXiv:1210.0580 [hep-ph]].

[14] A. Banfi, H. McAslan, P. F. Monni and G. Zanderighi, JHEP **1505** (2015) 102 [arXiv:1412.2126 [hep-ph]].

[15] G. Bell, A. Hornig, C. Lee and J. Talbert, in preparation.

[16] L. G. Almeida, S. D. Ellis, C. Lee, G. Sterman, I. Sung and J. R. Walsh, JHEP **1404** (2014) 174 [arXiv:1401.4460 [hep-ph]].

# Resummation of jet rates and event-shape distributions in $e^+e^-$

Thomas Gehrmann[1], Gionata Luisoni[2] and **Pier Francesco Monni**[2]

[1] Physik-Institut, Universität Zürich, CH-8057 Zürich, Switzerland
[2] CERN, Theoretical Physics Department, CH-1211 Geneva 23, Switzerland

## Introduction

Two-jet global event-shape variables in $e^+e^-$ annihilation are among the most studied QCD observables. They were originally designed to test the non-abelian nature and the dynamics of strong interactions and, owing to their sensitivity to the pattern of QCD radiation, they have been also widely used to perform extractions of $\alpha_s$, and to develop and test non-perturbative hadronisation models (see e.g. ref. [1] and references therein). Motivated by the latter phenomenological applications, the study of event shapes also led to important advances in fixed-order predictions and in the understanding of all-order properties of QCD radiation. Fixed order predictions for observables involving up to three jets in $e^+e^-$ collisions are known up to next-to-next-to-leading order (NNLO), and have been obtained both within the antenna subtraction method [2],[3], and more recently with the `CoLoRFulNNLO` method [4]. Electroweak corrections up to next-to-leading-order (NLO) have been also obtained in ref. [5].

While fixed-order calculations provide a good approximation of the hard radiation, which contributes to the region where event shapes have rather large values, resummed calculations are required where the bulk of data lies, i.e. in the region dominated by multiple soft-collinear emissions. In order to have a reliable theory prediction across the event-shape spectrum, it is fundamental to match the fixed-order prediction to the all-order computation that accounts for the logarithmic effects of soft and/or collinear radiation. In this talk we briefly review the current status of the predictions for event-shape observables, and their implications on the determination of the strong coupling constant. We also discuss the perspectives for the achievable theory precision for event shape distributions at a future FCC-ee machine.

## Determinations of the strong coupling from event shape distributions

Owing to the available data covering a broad spectrum of energies, and to the precision of the state-of-the-art theory predictions, event-shape observables are a customary tool used to fit the strong coupling constant. During the LEP era a set of 6 different event shapes was commonly used to perform fits of $\alpha_s$. This set involves thrust ($T$), C-parameter ($C$), heavy jet mass ($\rho$), total- and wide jet broadening ($B_{\rm T}$, $B_{\rm W}$) and two-to-three jet resolution parameter ($y_{23}^{\rm D}$) in the Durham jet-algorithm. Thanks to the precise measurements from LEP and PETRA for the six event shapes, several extractions of $\alpha_s$ were carried out using either predictions at NNLO or rather matched to next-to-leading logarithmic (NLL) calculations, which were subsequently corrected for hadronisation effects by means of Monte Carlo (MC) hadronisation generators. The results of these fits are reported in the upper block Figure 1. Other predictions only use data for a single event shape, as listed in the second block of Figure 1, and quantify hadronisation corrections in the context of a specific analytic model of non-perturbative dynamics. Power corrections are usually implemented in practice by computing the scaling of the leading power correction and then fitting its coefficient $\Omega$ simultaneously with $\alpha_s$. While different models for non-perturbative corrections produce the same leading scaling, the precise physical meaning of the coefficient depends on the model at hand. These are labelled with an additional '1/Q' in Figure 1, and make use of either thrust or C-parameter data from several experiments, spanning a center-of-mass energy range between 14 GeV and 206 GeV. The third block of $\alpha_s$ determinations in Figure 1 fits uses jets rates data [33],[7],[8] whereas the last two determinations are based on event-shape moments and uses again analytical non-perturbative corrections [9],[10].

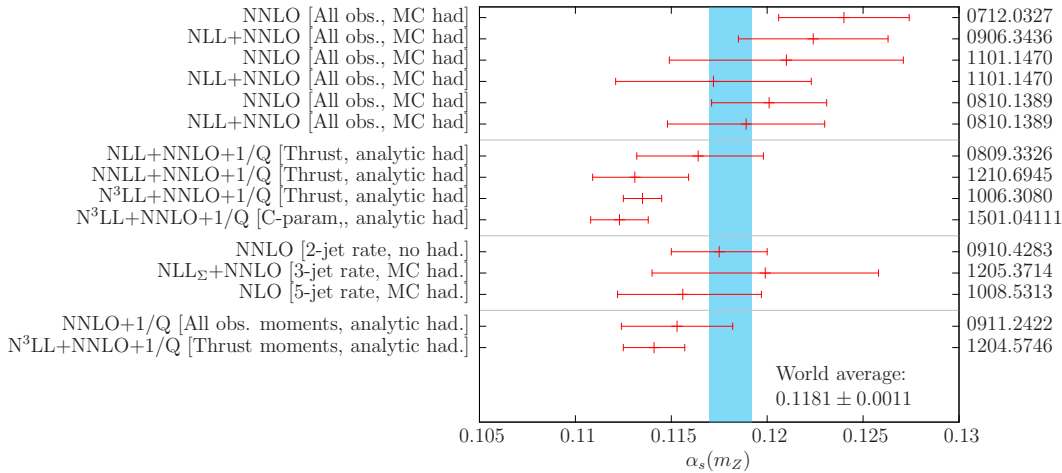

Figure 1: Summary of the recent determinations of $\alpha_s$ using NNLO predictions for event shape observables and jet rates.

It appears immediately clear that fits obtained using MC hadronisation corrections predict higher values of $\alpha_s$ compared to those based on analytical non perturbative corrections. The latter are in turn in tension with lattice calculations and some recent LHC measurement [11] that predict values closer to the world average. The origin of the higher $\alpha_s$ values in fits performed using MC hadronisation in Figure 1 is in part due to the discrepancy between the perturbative predictions used in the fits and the leading-order (LO) results of the MC generators used to tune the hadronisation models, that often results in an overestimate of non-perturbative corrections. A detailed observable-specific comparison of these differences at LEP I was performed in [12].

Moreover, the simultaneous fits of $\alpha_s$ and $\Omega$ often suffer from a very strong correlation between the two parameters, which implies that the resulting perturbative strong coupling value is very much affected by the precision with which the non-perturbative corrections are known. This is clearly an important limitation for this kind of extractions of $\alpha_s$. To improve on this, a good separation between perturbative and non-perturbative effects must be ensured either theoretically or experimentally.

To better disentangle the two effects, different viable solutions are possible. The use of precise data over a broad range of energies would in principle help resolve the degeneracy. However, the available low energy data has sizeable uncertainties and therefore it comes in with a small weight in the fitting procedure. In this regard, FCC-ee measurements would bring high-statistics data at higher centre-of-mass energies that would allow one to decrease the correlation between $\alpha_s$ and $\Omega$. Moreover, owing to the $\sim 1/Q$ scaling of the non-perturbative corrections, higher-energy measurements are less affected by hadronisation corrections, therefore the corresponding uncertainty in the strong coupling determination will be much reduced. However, the energies achievable at a FCC-ee collider may not be high enough for the hadronisation corrections to become unimportant. We will comment briefly on this point later on.

Among the measured observables, one could select those with lower sensitivity to hadronisation corrections, such as jet rates. To explore more this strategy, it is important to devise new observables which are more resilient to soft physics. One possible way is to use recently developed jet substructure techniques [13],[14] to groom the soft physics away from event-shape distributions.

A simultanous fit involving multiple differential distributions for different observables with different sensitivity to soft radiation would help disentangle perturbative and non-perturbative physics. Therefore, it is important to perform a global fit using analytical non-perturbative corrections for the entire set of 6 event shapes listed above computed at the same perturbative order. Until very recently this was only possible at NLL+NNLO [15],[12],[16] since only the thrust, the heavy-jet mass, and the $C$ parameter were known to a higher accuracy [17],[18],[19],[20]. Indeed, only fits using thrust and C-parameter data have been performed by means of NN(N)LL+NNLO theory predictions that include hadronisation corrections [21],[22],[23]. These two event shapes share very similar perturbative and non-perturbative features, resulting in similar results for the strong coupling. More recently, techniques that allow one to obtain NNLL+NNLO predictions for all of the global 2-jet event shapes have been developed. This progress makes it feasible to perform a global fit for different event shape observables using NNLL+NNLO predictions throughout.

## State of the art of theory predictions

### Resummation of event shapes at NNLL

As already mentioned in the introduction, the known NNLO fixed-order calculations for event-shape distributions are to be matched to a resummed calculation in order to have a robust theory control in the region where the radiation becomes soft or collinear. Until recently, next-to-leading logarithmic (NLL) resummations were the state of the art for 2-jet global event-shapes. At this order one computes all terms of order $\mathcal{O}(\alpha_s^n L^n)$ for the logarithm of integrated distribution. In the region of the spectrum where Sudakov logarithms become large $L \sim 1/\alpha_s$, NLL predictions are accurate at leading-order (LO). Moreover, the residual scale uncertainties of the latter predictions is still sizeable in most cases, therefore higher-order resummations become necessary to match the accuracy that fixed-order results achieve in the region where the radiation is hard.

In the past decade, considerable progress has been made towards an understanding of the radiation dynamics beyond NLL accuracy, and a number of computational techniques have been developed throughout the years. These resummations have been so far mainly obtained through observable-dependent factorisation theorems which lead to a full decomposition of the cross section in the infrared and collinear limit in terms of different kinematical subprocesses (i.e. soft, collinear, hard) which are then resummed individually through renormalisation-group evolution equations. Despite being systematically extendable to all logarithmic orders, this approach is strictly observable-dependent and relies fully on the fact that the observable can be factorised explicitly in some conjugate space. In particular, full next-to-next-to-leading logarithmic NNLL ($N^3LL$ corrections, with the exception of the four-loops cusp anomalous dimension, are known for $1 - T$, $\rho_H$ and $C$) predictions are available for a number of event shapes at lepton colliders like thrust $1 - T$ [17],[19], heavy jet mass $\rho_H$ [18], jet broadenings $B_T$, $B_W$ [24], $C$-parameter [20], and energy-energy-correlation [25].

To go beyond the above list of simple observables one must deal with the factorisation properties of more complicated observables. While strictly speaking a factorisation of the singular modes (soft and/or collinear) from the hard modes must occur for an observable to be resummable at all orders, in the presence of an arbitrary amount of radiation emissions from different singular regions of the phase space can be related in a highly non-trivial way. In this case, soft and collinear modes are entangled and deriving an analytic factorisation theorem can become very cumbersome if not impossible.

The idea of moving beyond the limitation of factorisation theorems led to the development of a general and systematic method to compute logarithmic corrections at all orders.

The resummation method formulated in refs. [26],[27],[28] relies on a property known as recursive infrared and collinear (rIRC) safety [26], and does not require any all-order factorization other than the one for the squared amplitudes in the soft and collinear limits. rIRC safety allows one to single out in the observable the subset of emissions which are closest to the QCD singularities and to perform an analytic subtraction of the IRC divergences at all orders against the corresponding virtual corrections. This operation results in a Sudakov radiator that expresses the no-emission probability above the scale of the hardest real emission. The properties of rIRC safety also ensure that the resolved real corrections at all perturbative orders can be expressed in terms of four-dimensional phase space integrals to all orders, and can be efficiently implemented using Monte Carlo techniques. This makes it possible to handle any complex continuously global rIRC safe observable, without any additional requirement on factorisability of the observable. In the context of $e^+e^-$ collisions, an automated implementation of the method at NNLL has been designed in the program ARES, that has been used to obtain the first NNLL+NNLO predictions for observables for which a factorisation theorem is not available such as the Thrust Major [27] and the 3-jet resolution parameter in different $e^+e^-$ jet clustering algorithms [28].

**Theory uncertainties at FCC-ee energies**

The higher centre-of-mass energy and luminosity achievable at the FCC-ee makes it natural to think of event-shape distributions as a tool for precision physics. Their sensitivity to radiation and the clean environment of the $e^+e^-$ collisions make them a powerful tool for a number of precision measurements such as the extraction QCD strong coupling constant or of the Higgs-boson light Yukawa couplings [29]. It is therefore important to analyse what the achievable theory precision is for such observables.

The residual QCD scale uncertainty in event-shape distributions at NNLL+NNLO is currently below 5% for LEP energies [27]. This will be further reduced at FCC-ee energies due to the milder running of the coupling at those scales, leading to a perturbative uncertainty of a few percent. An example for the $y_{23}$ three-jet resolution parameter in the Durham and Cambridge algorithms is shown in Figure 2.

While weak corrections are still negligible at FCC-ee energies, QED corrections also have a few-percent impact on the distributions, mainly due to the effect of initial-state photon radiation. Furthermore, residual effects of the radiative return can be still present at $\sqrt{s} = 240\,\text{GeV}$ for specific observables [5].

Although the impact of hadronisation is suppressed by the inverse of the centre-of-mass energy, non-perturbative effects are in general non-negligible for event shapes at $\sqrt{s} = 240\,\text{GeV}$, and can reach the several percent level. Therefore, non-perturbative still constitute the dominant uncertainty for sevelar event-shape distributions at FCC-ee energies.

Unless a better understanding of the non-perturbative dynamics will be attained by the time of the FCC-ee, the only way to perform percent-level precision physics with event-shape distributions with such a machine will be through observables which are less affected by hadronisation. An example of observables of this type are the differential jet rates obtained with jet algorithms that are resilient to soft physics [31],[32], which feature a faster reduction of hadronisation corrections with increasing centre-of-mass energy compared to other event shapes and $e^+e^-$ clustering algorithms (see e.g. [34]).

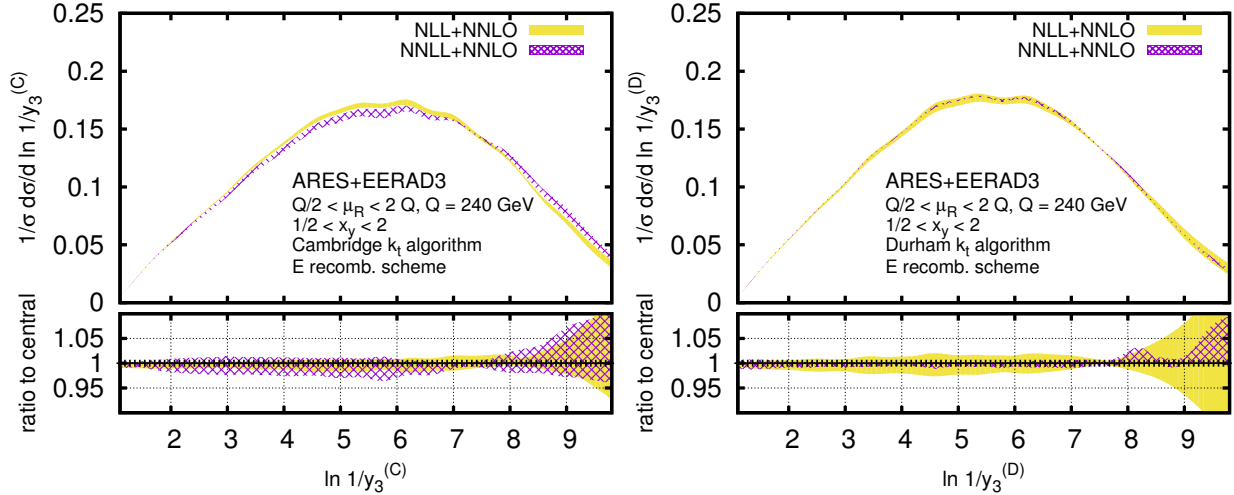

Figure 2: Distributions for the three-jet resolution parameter in the Durham (right) and Cambridge (left) algorithm at NNLL+NNLO for $\sqrt{s} = 240\,\text{GeV}$. The predictions have been obtained by combining the resummed calculation of ref. [28] with the NNLO obtained with the program `EERAD3` [30].

# C Parameter at N³LL

André H. Hoang[1,2], Daniel W. Kolodrubetz[3], **Vicent Mateu**[4], and Iain W. Stewart[3]

[1] University of Vienna, Faculty of Physics, Boltzmanngasse 5, A-1090 Wien, Austria
[2] Erwin Schrödinger International Institute for Mathematical Physics, University of Vienna, Boltzmanngasse 9, A-1090 Vienna, Austria
[3] Center for Theoretical Physics, Massachusetts Institute of Technology, Cambridge, MA 02139, USA
4 Departamento de Física Fundamental e IUFFyM, Universidad de Salamanca, E-37008 Salamanca, Spain

**Abstract:** The C-parameter event-shape distribution for $e^+e^-$ annihilation into hadrons is computed in the framework of SCET including input from fixed-order perturbation theory. We calculate all missing ingredients for achieving N³LL resummation accuracy in the cross section, which is then matched onto $\mathcal{O}(\alpha_s^3)$ fixed-order results. Hadronization power corrections are incorporated as a convolution with a nonperturbative shape function. Wide-angle soft radiation effects introduce an $\mathcal{O}(\Lambda_{\rm QCD})$ renormalon ambiguity in the cross section, which we cure by switching to the Rgap short-distance scheme. We also include hadron mass effects, but find their effect is rather small. Performing fits to the tail of the C-parameter distribution for many center of mass energies we find that the strong coupling constant is $\alpha_s(m_Z) = 0.1123 \pm 0.0015$, with $\chi^2/{\rm dof} = 0.99$.

## Introduction

The LEP $e^+e^-$ collider, previously located at CERN, has delivered an enormous amount of highly accurate experimental data, which can be used to explore the theory of strong interactions in its high-energy regime. To study Quantum Chromodynamics (or QCD) at high energies one needs to deal with jets: highly boosted and collimated bunches of particles that can be seen as the remnants of the underlying partons created at very short distances. One appealing strategy for describing jet dynamics is through event shapes, infrared- and collinear-safe observables which are constructed from the energy and momenta of all the produced hadrons (in this sense event shapes are global quantities). They are designed to measure geometrical properties of the distribution of particles, and in particular they quantify how "jetty" the final state is. Additionally, being global observables, it is possible to compute high-order perturbative corrections, carry out log resummation to higher order, show factorization and exponentiation properties, and use factorization to control power corrections.

One of the main uses of event-shape distributions is the determination of the strong coupling constant $\alpha_s$. The advantage of event shapes over other inclusive observables is that they are essentially proportional to $\alpha_s$, rather than probing $\alpha_s$ via corrections to a leading term (as is the case, for example, of DIS or the total hadronic cross section). Thus, event shapes are very sensitive to the strong coupling constant. On the other hand, event shapes are afflicted by nonperturbative power corrections and by large double Sudakov logarithms, which necessitate resummation.

Here we study the C-parameter which can be written as [1],[2]:

$$C = \frac{3}{2} \frac{\sum_{i,j} |\vec{p}_i||\vec{p}_j| \sin^2 \theta_{ij}}{\left(\sum_i |\vec{p}_i|\right)^2} \,. \tag{1}$$

It is interesting to compare C-parameter with thrust [3],

$$\tau = 1 - T = \min_{\vec{n}} \left( 1 - \frac{\sum_i |\vec{n} \cdot \vec{p}_i|}{\sum_j |\vec{p}_j|} \right), \quad (2)$$

where $\vec{n}$ is referred to as the thrust axis. The three main differences between $C$ and $\tau$ are: a) $C$ does not require a minimization procedure for its computation, whereas $\tau$ does (namely finding the thrust axis, event by event); b) $C$ is defined through a double sum, whereas $\tau$ sums only particle by particle; c) the fixed-order prediction of C-parameter develops an integrable singularity at $C_{\text{shoulder}} = 0.75$, whereas thrust is always smooth. By shoulder we refer to the fact that the partonic cross section attains an integrable singularity [4] at $C_{\text{shoulder}}$, and only non-planar configurations contribute for $C > C_{\text{shoulder}}$.[*] There are also a number of similarities between $C$ and $\tau$, and perhaps the most remarkable one is that in the dijet limit ($C, \tau \ll 1$) and up to and including NLL resummation, both partonic cross sections are related in a simple way, which can be schematically expressed as follows: $\tau_{\text{NLL}} = C_{\text{NLL}}/6$ [5]. Some other similarities will be highlighted later.

Previous analyses of the thrust distribution using SCET at N$^3$LL and analytic power corrections have found rather small (albeit precise) values of $\alpha_s$ [6],[7][†]. Two motivations for carrying out this new analysis are providing an additional determination of $\alpha_s$ and studying the universality of the leading power correction between thrust and C-parameter. In this proceedings we summarize work done in Refs. [16],[10].

## Theoretical developments

Until a few years ago, theoretical uncertainties were larger than the corresponding experimental ones and hadronic power corrections were not understood from ab-initio QCD considerations. The situation on the theory side has dramatically changed with the advent of Soft-Collinear Effective Theory (SCET) [11], [12], [13], [14]. This effective field theory separates the relevant physics occurring at the various scales which play a role when jets are being produced: hard scale $\mu_H$, of the order of the center of mass energy $Q$ (describes the production of partons at very short distances), jet scale $\mu_J \sim Q\sqrt{C/6}$ (describes the formation and evolution of jets at intermediate energies), and the soft scale $\mu_S \sim QC/6$ (describes wide angle soft radiation at longer distances). All three scales are widely separated for $C \ll 1$, and there is one function associated to each one of them: the hard coefficient $H_Q$ (the modulus square of the QCD to SCET matching coefficient), which is common to all event-shape factorization theorems; the Jet function $J_\tau$ (built up with collinear Wilson lines), which is common for thrust, C-parameter [16] and Heavy Jet Mass ($\rho$) [15]; and the Soft function $S_C$ (defined through soft Wilson lines), which in general depends on the specific form of the event shape. Whereas the former two are perturbative ($\mu_H, \mu_J \gg \Lambda_{\text{QCD}}$), permitting the calculation of the hard and jet functions in powers of $\alpha_s$, the soft function also has nonperturbative corrections that need to be accounted for ($\mu_S \gtrsim \Lambda_{\text{QCD}}$). Renormalization evolution among the three scales sums up large logarithms to all orders in perturbation theory. It turns out that the anomalous dimensions for the soft function for $C$ and $\tau$ are identical, see [16].

---

[*]In Ref. [4] it is shown that soft gluon resummation at $C_{\text{shoulder}}$ makes up for a smooth distribution at LL order.

[†]Other lower-order resummation event-shape analyses have also found small (although less accurate) values of $\alpha_s$ [8], [9].

The soft function can be further factorized into a partonic soft function $\hat{S}_{\widetilde{C}}$, calculable in perturbation theory, and a nonperturbative shape function $F_C$, which has to be obtained from fits to data [‡]. The treatment of hadronic power corrections greatly simplifies in the tail of the distribution, defined by $QC/6 \gg \Lambda_{\rm QCD}$, where the shape function can be expanded in an OPE. The leading power correction is parametrized by $\Omega_1^C$, the first moment of the shape function. Interestingly, if one ignores hadron mass effects [20],[21], this matrix element can be related to the corresponding one in thrust in a trivial manner: $\Omega_1^\tau/2 = \Omega_1^C/(3\pi) \equiv \Omega_1$ [22]. The main effect of this leading power correction is a shift of the cross section, $d\hat{\sigma}(C) \to d\hat{\sigma}(C - \Omega_1^C/Q)$. When presenting the results of our fits, we will employ the power correction parameter $\Omega_1$ to ease comparison.

The leading SCET factorization for the partonic C-parameter distribution can be written as [23],[10]:

$$\frac{1}{\sigma_0}\frac{d\hat{\sigma}_s}{dC} = \frac{Q}{6} H_Q(Q,\mu) \int ds\, J_\tau(s,\mu)\, \hat{S}_{\widetilde{C}}\left(\frac{QC}{6} - \frac{s}{Q}, \mu\right). \qquad (3)$$

It describes the most singular (and numerically dominating) partonic contributions in the dijet limit. The resummation of large logarithms is achieved by evolving the functions $H_Q$, $J_\tau$, and $\hat{S}_{\widetilde{C}}$ from their respective natural scales $\mu_H$, $\mu_J$ and $\mu_S$, where logs are small, to a common scale $\mu$ (which without loss of generality can be chosen to be, for instance, $\mu_J$). In Eq. (3) $\hat{S}_{\widetilde{C}}$ is also in the $\overline{\rm MS}$ scheme, and suffers from an $\mathcal{O}(\Lambda_{\rm QCD})$ renormalon. We can switch to the renormalon-free Rgap scheme [24] by performing subtractions on the partonic soft function (through an exponential of a derivative operator) and simultaneously allow the same terms to change $\Omega_1^C$ from the $\overline{\rm MS}$ scheme to the Rgap scheme. Our strong coupling $\alpha_s$ will always be in the $\overline{\rm MS}$ scheme. Adding these subtractions plus the renormalization group evolution kernels gives:

$$\begin{aligned}\frac{1}{\sigma_0}\frac{d\hat{\sigma}_s}{dC} &= \frac{Q}{6} H_Q(Q,\mu_H)\, U_H(Q,\mu_H,\mu) \int ds\, ds'\, dk\, J_\tau(s,\mu_J)\, U_J^\tau(s-s',\mu,\mu_J) \\ &\times U_S^\tau(k,\mu,\mu_S)\, e^{-3\pi\frac{\delta(R,\mu_S)}{Q}\frac{\partial}{\partial C}}\, \hat{S}_{\widetilde{C}}\left(\frac{QC - 3\pi\overline{\Delta}(R,\mu_S)}{6} - \frac{s}{Q} - k, \mu_S\right),\end{aligned} \qquad (4)$$

where $\delta(R,\mu_S)$ is a series in powers of $\alpha_s(\mu_S)$ that can be computed directly from the thrust partonic soft function in Fourier space. For the renormalon to be properly canceled by the subtractions, it is crucial that the exponential and the partonic soft function are consistently expanded out to a given order in $\alpha_s(\mu_S)$. The subtractions introduce a scale $R$, which is close to the soft scale $\mu_S$ and can be used to sum up large logs in the subtraction series through the finite shift parameter $\overline{\Delta}(R,\mu_S)$. The dependence on $R$ formally cancels between $\delta(R,\mu_S)$ and $\overline{\Delta}(R,\mu_S)$ order by order in perturbation theory, but the $R$ parameter is crucial to eliminate the $\Lambda_{\rm QCD}$ renormalon. Nonperturbative corrections are incorporated though a convolution with the shape function $F_C(p)$ whose first moment is $\Omega_1^C$. The hadron level prediction for the distribution is

$$\frac{1}{\sigma_0}\frac{d\sigma}{dC} = \int dp\, \frac{1}{\sigma_0}\frac{d\hat{\sigma}}{dC}\left(C - \frac{p}{Q}\right) F_C(p), \qquad \frac{d\hat{\sigma}}{dC} = \frac{d\hat{\sigma}_s}{dC} + \frac{d\hat{\sigma}_{\rm ns}}{dC}, \qquad (5)$$

and also includes the nonsingular contributions, $d\hat{\sigma}_{\rm ns}/dC$, which in the dijet limit contains all terms which are kinematically suppressed by additional powers of $C$.

---

[‡]Power corrections for the C-parameter distributions have been studied in other frameworks, see e.g. Refs. [17],[18],[19].

For our analysis we include perturbative corrections to the matrix elements $H_Q$, $J_\tau$ and $\hat{S}_{\widetilde{C}}$ to $\mathcal{O}(\alpha_s^3)$. For $H_Q$ they are known analytically, whereas for $J_\tau$ and $\hat{S}_{\widetilde{C}}$ only the logarithmic terms at $\mathcal{O}(\alpha_s^3)$ are known (since the anomalous dimensions are known at three loops). These non-logarithmic terms are added as unknown coefficients that are varied when we estimate the theory uncertainties. At $\mathcal{O}(\alpha_s^2)$ the jet function can be directly taken from Ref. [25]. The soft function needs to be computed to $\mathcal{O}(\alpha_s^2)$ [10], which can be done analytically at $\mathcal{O}(\alpha_s)$ and for the logarithmic corrections at $\mathcal{O}(\alpha_s^2)$. For the non-logarithmic $\mathcal{O}(\alpha_s^2)$ terms our evaluation uses numerical output of the parton level MC EVENT-2 [26],[27].

Through RGE evolution we achieve N$^3$LL resummation of the logarithmic terms. The anomalous dimensions required for solving the running equations can be taken directly from thrust. The only missing term is the four-loop cusp anomalous dimension, which is estimated using Padé approximants but is nevertheless varied in a wide range when estimating perturbative uncertainties. Its effect is in any case negligibly small. The required components for a given resummation order are specified in Table 2. We introduce a primed counting, which is defined as the regular (unprimed) one, but with the matrix elements being included to one order higher. For consistency, the renormalon subtraction series are including to one order higher as well. The primed counting achieves a better description of data and allows the correct summation of logs at the level of the distribution (for an extended discussion of this the reader is referred to [28]).

We include nonsingular terms at the same order as the functions $H_Q$, $J_\tau$, $\hat{S}_{\widetilde{C}}$. These can be obtained by subtracting the fixed-order singular cross section as described by the SCET factorization theorem from the full QCD partonic distribution. The latter can be computed analytically at $\mathcal{O}(\alpha_s)$, and determined numerically at $\mathcal{O}(\alpha_s^2)$ and $\mathcal{O}(\alpha_s^3)$ from the parton-level MC programs EVENT2 and EERAD3 [29],[30], respectively. For the $\mathcal{O}(\alpha_s^2)$ and $\mathcal{O}(\alpha_s^3)$ nonsingular contributions our numerical procedure entails uncertainties which are accounted in the estimate of the theoretical uncertainty.

It is sometimes customary to write the most singular terms of an event-shape cumulant cross section in the following exponentiated form:

$$\hat{\Sigma}(C) = \frac{1}{\hat{\sigma}} \int_0^C dC' \frac{d\hat{\sigma}}{dC'} = \left(1 + \sum_{m=1}^\infty B_m \left(\frac{\alpha_s(Q)}{2\pi}\right)^m\right) \\ \times \exp\left(\sum_{i=1}^\infty \sum_{j=1}^{i+1} G_{ij} \left(\frac{\alpha_s(Q)}{2\pi}\right)^i \ln^j\left(\frac{6}{C}\right)\right). \quad (6)$$

From the result for the factorization theorem in Eq. (3) one can determine the $G_{ij}$ and $B_i$ coefficients as shown in Table 1, see [16].

| Resummation Order | Calculable $G_{ij}$'s and $B_i$'s |
|---|---|
| LL | $G_{i,\,i+1}$ |
| NLL$'$ | $G_{i,\,i}$ and $B_1$ |
| N$^2$LL$'$ | $G_{i,i-1}$ and $B_2$ |
| N$^3$LL$'$ | $G_{i,i-2}$ and $B_3$ |

Table 1: Hierarchy of $G_{ij}$'s at each given order of resummation.

|        | cusp | non-cusp | matching | $\beta$ | ns | $\delta$ |
|--------|------|----------|----------|---------|-----|----------|
| LL     | 1    | -        | tree     | 1       | -   | -        |
| NLL    | 2    | 1        | tree     | 2       | -   | -        |
| N$^2$LL | 3   | 2        | 1        | 3       | 1   | 1        |
| N$^3$LL | 4$^p$ | 3      | 2        | 4       | 2   | 2        |
| NLL$'$ | 2    | 1        | 1        | 2       | 1   | 1        |
| N$^2$LL$'$ | 3 | 2      | 2        | 3       | 2   | 2        |
| N$^3$LL$'$ | 4$^p$ | 3  | 3        | 4       | 3   | 3        |

Table 2: Loop corrections for primed and unprimed orders. For the anomalous dimensions of $\overline{\Delta}(R, \mu_S)$ one uses the same orders as for other non-cusp anomalous dimensions. The superscript "p" indicates that a Padé approximation is being used.

### Setting the Renormalization Scales

The C-parameter can be divided into three distinct regions, in which the renormalization scales must satisfy different constraints

$$
\begin{aligned}
&1)\ \text{nonperturbative:}\ C \lesssim 3\pi\, \Lambda_{\text{QCD}} \\
&\quad \mu_H \sim Q,\ \mu_J \sim \sqrt{\Lambda_{\text{QCD}} Q},\ \mu_S \sim R \sim \Lambda_{\text{QCD}}\,, \\
&2)\ \text{resummation:}\ 3\pi\, \Lambda_{\text{QCD}} \ll C < 0.75 \\
&\quad \mu_H \sim Q,\ \mu_J \sim Q\sqrt{\frac{C}{6}},\ \mu_S \sim R \sim \frac{QC}{6} \gg \Lambda_{\text{QCD}}\,, \\
&3)\ \text{fixed-order:}\ C > 0.75 \\
&\quad \mu_H = \mu_J = \mu_S = R \sim Q \gg \Lambda_{\text{QCD}}\,.
\end{aligned}
\quad (7)
$$

These three regions are sometimes referred to as the peak, tail and far-tail regions, respectively. In order to satisfy these requirements we need to use renormalization scales that depend on the value of $C$, called *profile functions*. The constraints in Eq. (7) do not fully specify the profile functions, but this ambiguity cancels order-by-order in perturbation theory. This allows variations of the profiles to be used to estimate perturbative uncertainties. The specific form of the profile functions and the variation of their parameters are given in Ref. [10], and illustrated in Fig. 1.

In Fig. 2 we show our C-parameter cross section for $Q = m_Z$, together with experimental data. This figure is produced with our best theoretical prediction and uses our central values for $\alpha_s(m_Z)$ and $\Omega_1$ presented in Sec. 2. The center blue line corresponds to the prediction for our central profiles, whereas the light blue band shows the perturbative uncertainty.

### Fit results

Fitting for $\Omega_1$ together with $\alpha_s(m_Z)$ accounts for hadronization effects in a model-independent way. In order to determine $\alpha_s(m_Z)$ and $\Omega_1$ in the same fit, one needs to perform a global analysis that includes data at many center of mass energies $Q$. For each $Q$ the differential distribution has a noticeable degeneracy between the two fit parameters, and the use of data from the different $Q$ values breaks the degeneracy. Hence LEP and SLAC data are employed together with data from lower energy experiments such as TRISTAN and PETRA. For our analysis we use all available experimental data with energies between 35 GeV and 207 GeV in the tail region. To estimate theoretical uncertainties we perform 500 fits at each order in the resummation, NLL', N$^2$LL', and N$^3$LL', with theory parameters randomly chosen for each fit. These parameters specify: the profile functions, unknown perturbative coefficients, or statistical uncertainties on the numerical determination of the non-singular contributions. The result of these many fits are shown graphically as dots in Fig. 3. We show two projections: $\alpha_s$ vs $2\Omega_1$ in panel (a), and $\alpha_s$ vs $\chi_{\min}$/dof in panel (b). As the resummation order increases the perturbative uncertainty decreases as expected, and the $\chi^2$/dof also decreases significantly. The corresponding numerical results and uncertainties are shown in Table 3 [10].

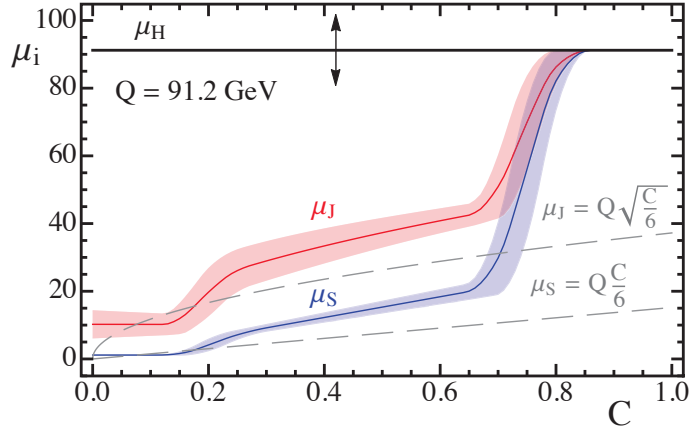

figure 1: Profile functions for the renormalization scales $\mu_H$, $\mu_J(C)$, and $\mu_S(C)$, when using the default profile parameters (center thick line) and when varying them (light band). Fully canonical profiles are shown in gray.

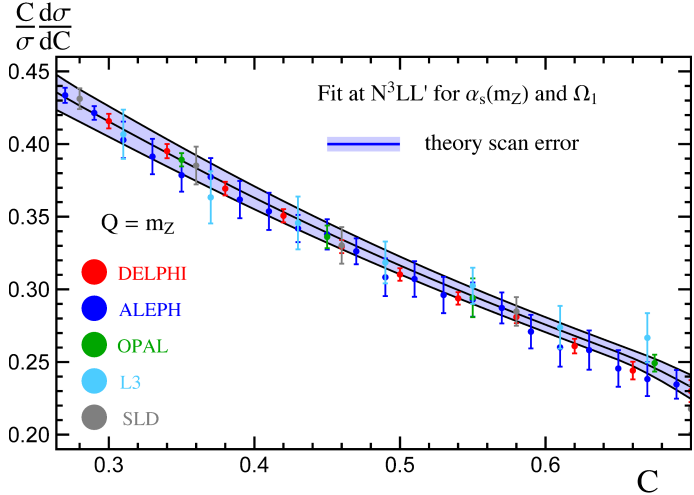

Figure 2: Theoretical prediction for the C-parameter distribution at N$^3$LL' order for $Q = m_Z$, using the best fit values for $\alpha_s(m_Z)$ and $\Omega_1$. The blue band corresponds to the theory uncertainty as described in the text. Experimental data for various experiments are also shown.

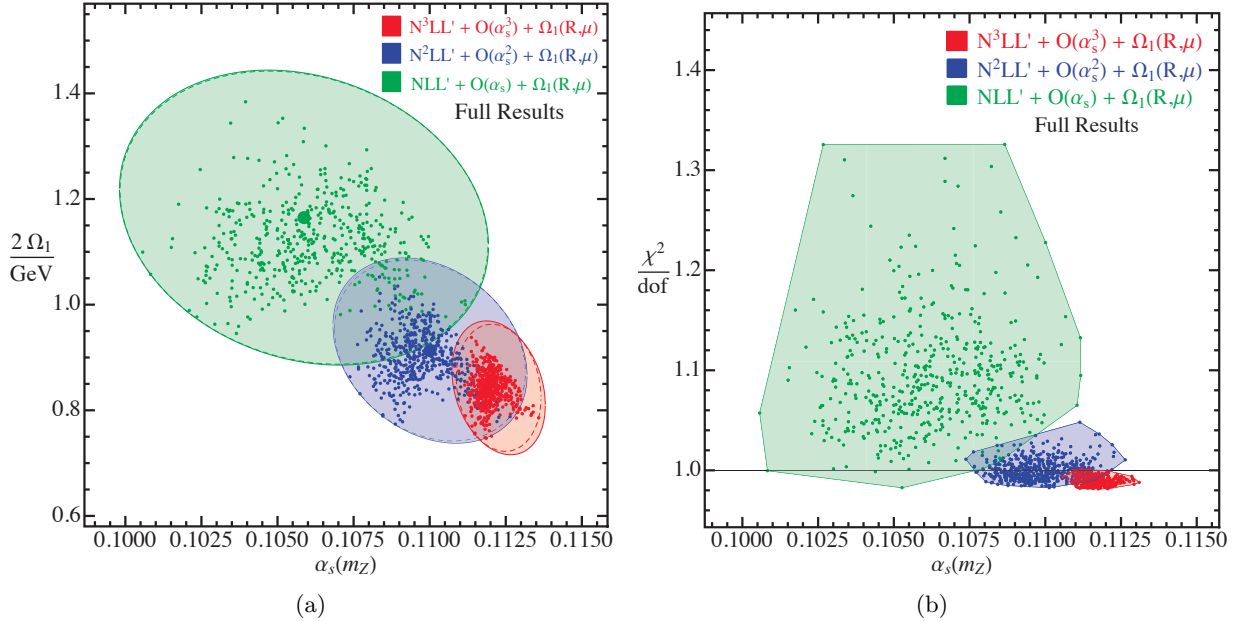

Figure 3: The left panel (a) shows the distribution of best fit points in the $\alpha_s(m_Z)$-$2\Omega_1$ plane for fits performed with our best theoretical predictions: resummation of large logs and power corrections defined in the Rgap scheme with renormalon subtractions. The dashed lines corresponds to an ellipse fit to the contour of the best-fit points to determine the theoretical uncertainty. The total (experimental + theoretical) 39% CL standard error ellipses are displayed (solid lines), which correspond to 1-$\sigma$ (68% CL) for either one-dimensional projection. The big points represent the central values in the random scan for $\alpha_s(m_Z)$ and $2\Omega_1$. The right panel (b) shows the distribution of best fit points in the $\alpha_s(m_Z)$-$\chi^2$/dof plane, corresponding to the points given in panel (a).

| order | $\alpha_s(m_Z)$ (with $\overline{\Omega}_1$) | $\alpha_s(m_Z)$ (with $\Omega_1(R_\Delta, \mu_\Delta)$) |
|---|---|---|
| NLL' | 0.1071(60)(05) | 0.1059(62)(05) |
| NNLL' | 0.1102(32)(09) | 0.1100(33)(06) |
| N³LL' | 0.1117(16)(06) | **0.1123(14)(06)** |

Table 3: Best fit values for $\alpha_s(m_Z)$ at various orders with theory uncertainties from the theory scan (first value in brackets), and experimental and hadronic error added in quadrature (second value in brackets). Our final result at N³LL' is shown in bold face.

In Fig. 4 we show determinations of $\alpha_s(m_Z)$ from fits to the C-parameter distribution with different levels of theoretical sophistication. From left to right they are: fixed order with $\mathcal{O}(\alpha_s^3)$ (large logs not yet summed up), N³LL' resummation (no $\Omega_1$ in the fit), with power corrections included (not yet removing the renormalon), including Rgap scheme (not yet accounting for hadron masses), and with hadron mass effects. All error bars are perturbative, so the error bars of the first two determinations to the left of the vertical dashed line do not account for the neglect of power corrections. Including the nonperturbative power corrections by fitting $\Omega_1$ has the greatest effect on $\alpha_s(m_Z)$.

Hadron mass effects give negligible contributions within current uncertainties.

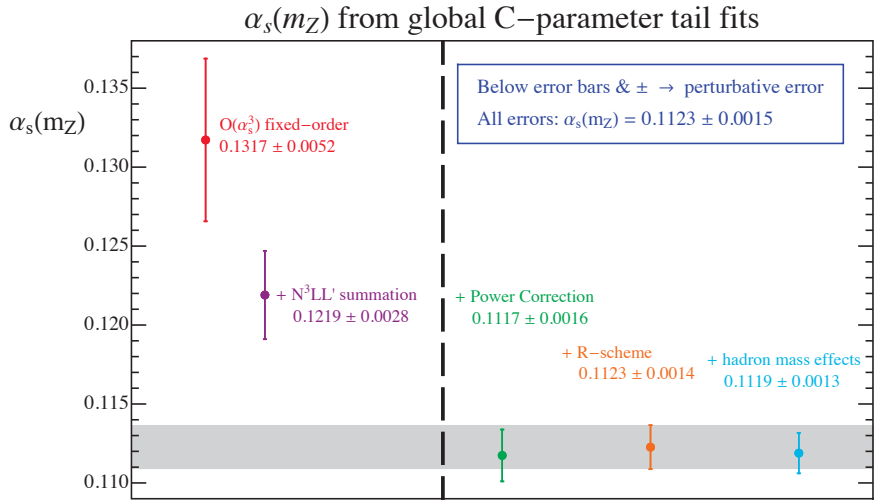

Figure 4: Impact of the different components of our theoretical setup on the determination of $\alpha_s(m_Z)$.

## Conclusions

We have presented an accurate determination of $\alpha_s(m_Z)$ from fits to the C-parameter distribution. The key points to our precise theoretical prediction are: a) higher order resummation accuracy (N³LL), achieved through the SCET factorization theorem, b) inclusion of $\mathcal{O}(\alpha_s^3)$ matrix elements and fixed-order kinematic power corrections, c) field-theoretical treatment of nonperturbative power corrections, and d) switching to a short-distance Rgap scheme, in which the sensitivity to infrared physics is reduced. We have not discussed hadron mass effects, as their effect is quite small. A thorough discussion can be found in [16].

Our final results from the global analysis reads [10]

$$\alpha_s(m_Z) = 0.1123 \pm (0.0002)_{\text{exp}} \pm (0.0007)_{\text{hadr}} \pm (0.0014)_{\text{pert}},$$
$$= 0.1123 \pm (0.0015)_{\text{tot}}$$

We conclude by presenting a comparison of our $C$-parameter fit with the determinations of $\alpha_s$ and $\Omega_1$ from our previous thrust analysis [6] of the thrust distribution, see Fig. 5. The figure shows that the determination of the strong coupling constant for both event shapes is compatible. Moreover there is universality in the result for the leading power correction $\Omega_1 = \Omega_1^C/(3\pi) = \Omega_1^\tau/2$ determined in both analyses. The two independent determinations agree within their 1-$\sigma$ uncertainties, where the precision of the extractions is greater than that achieved in the past. Note that without including the respective prefactors ($3\pi$ and 2) (shown in green) the values of $\Omega_1^\tau = 0.329\,\text{GeV}$ and $\Omega_1^C = 1.98\,\text{GeV}$ numerically differ by $\approx 4.5$-$\sigma$, so the agreement nicely demonstrates the consistency of our theoretical predictions. A detailed comparison with other $\alpha_s(m_Z)$ determinations can be found in Ref. [10].

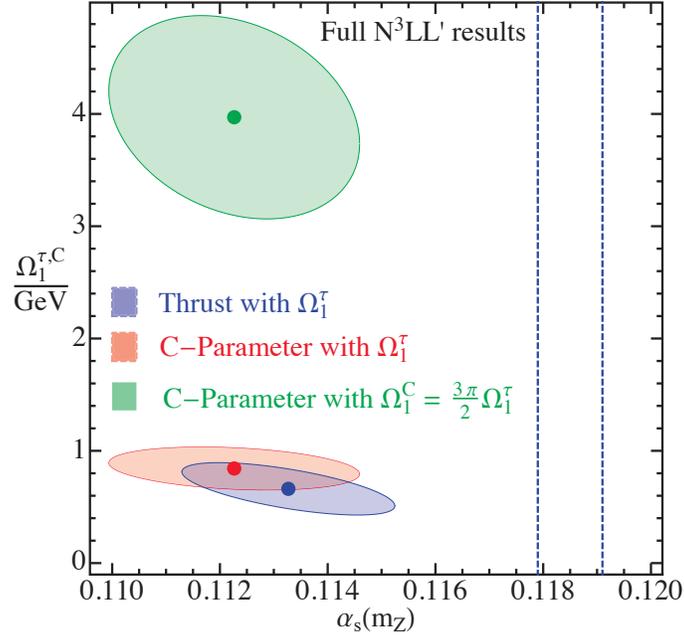

Figure 5: Comparison of $\alpha_s(m_Z)$ and $\Omega_1$ determinations from fits to the C-parameter with $\Omega_1^\tau$ (red), C-parameter with $\Omega_1^C$ (green), and thrust (blue) tail cross sections, at N$^3$LL$'$ with power corrections and in the Rgap scheme. The ellipses show the $\Delta\chi^2 = 2.3$ variations in the $\alpha_s(m_Z)$-$2\Omega_1$ plane, representing 1-$\sigma$ errors for two variables.

# Parton Showers since LEP

Peter Richardson[1,2]

[1] Theoretical Physics Department, CERN, Geneva, Switzerland
[2] IPPP, Department of Physics, Durham University

**Abstract:** We briefly discuss the development of Monte Carlo event generators over the last fifteen years. During this period there has been a revolutionary transformation in the accuracy of these programs as matching to higher-multiplicity matrix elements and next-to-leading order calculations has become standard with the first next-to-next-to-leading order processes now available. Finally the prospects for future improvements are discussed.

## Monte Carlo Simulations at LEP

Monte Carlo event generators came of age at LEP where for the first time a combination of better understanding of QCD and increased computing power provided simulated events which were in good quantitative agreement with the experimental results. These simulations used:

- a leading-order matrix element for $e^+e^- \to q\bar{q}$;

- a parton shower simulation for the evolution from the hard scale of the partonic collision to the infrared cut-off including the correct treatment of colour coherence;

- hadronization using either the non-perturbative string or cluster models.

The main programs used by the end of the LEP programme were PYTHIA 6[1] and HERWIG 6[2]. These simulations also included the matching of the hardest gluon emission for processes with a single colour line, for example $e^+e^- \to q\bar{q}$, Deep inelastic scattering and Drell-Yan, which effectively gave for $e^+e^- \to q\bar{q}$, apart from the trivial normalisation by a K-factor a next-to-leading order (NLO) simulation of the hard process.[*] The alternative dipole shower of ARIADNE (together with the string hadronization model) often provide the best agreement with the data [4].

## From LEP to the LHC

Starting in the early 2000's there was a major programme to develop better Monte Carlo simulations in order to describe the data from the energy frontier hadron colliders, first the Tevatron and now the LHC. This started with the development of the first viable approach allowing multiple hard emissions to be described correctly at leading order together with a parton shower simulation of soft and colinear radiation (CKKW)[5]. This was first used to describe the production of four jet events in $e^+e^-$ collisions where it gave quantitative improvements. However the main success of the approach was in hadron-hadron collisions where it allowed the accurate description of multiple jet production, for example in association with electroweak vector bosons, for the first time.

---

[*] In the case of $e^+e^- \to q\bar{q}$ the approach used by both HERWIG and PYTHIA is equivalent to more general POWHEG method [3] provided the total rate is normalised to the NLO value.

Together with the development of many variants of the original CKKW merging procedure Monte Carlo event generator development in the early 2000's focused on producing simulations that in addition to correctly treating hard emission (initially only the hardest emission) also had the correct NLO normalisation. All these approaches rely on rearranging the NLO cross section formula,

$$d\sigma = B(\Phi_B)d\Phi_B + (V(\Phi_B) + C(\Phi_B, \Phi_R)d\Phi_R)d\Phi_B + (R(\Phi_B, \Phi_R) - C(\Phi_B, \Phi_R))d\Phi_B d\Phi_R, \quad (1)$$

where $B(\Phi_B)$ is the leading-order contribution, $\Phi_B$ the N-body phase-space variables of the leading-order Born process whereas $\Phi_R$ are the radiative variables describing the phase space for the emission of an extra parton. The real contribution, $R(\Phi_B, \Phi_R)$, is the matrix element including the radiation of an additional parton multiplied by the relevant parton flux factors, and is regulated by subtracting the counter terms $C(\Phi_B, \Phi_R)$ which contain the same singularities as $R(\Phi_B, \Phi_R)$.

The first successful approach MC@NLO [6] chose to use the shower approximation for soft and colinear emission as the subtraction term

$$d\sigma = B(\Phi_B)d\Phi_B + (V(\Phi_B) + C_{\text{shower}}(\Phi_B, \Phi_R)d\Phi_R)d\Phi_B + (R(\Phi_B, \Phi_R) - C_{\text{shower}}(\Phi_B, \Phi_R))d\Phi_B d\Phi_R, \quad (2)$$

which allows a simulation to be constructed without double counting of radiation from the parton shower and hard real corrections. However, while correctly incorporating the resummation of the parton shower and the fixed NLO calculation this approach does lead to negative weighted events. Additionally as the subtraction term depends on the details of the specific parton-shower algorithm used it must be analytically recalculated for different approaches which can be complicated depending on the details of the parton-shower algorithm.

Later an alternative rearrangement POWHEG [3] was suggested

$$d\sigma = \overline{B}(\Phi_B)d\Phi_B \left[ \Delta_R^{(\text{NLO})}(0) + \Delta_R^{(\text{NLO})}(p_\perp)\frac{R(\Phi_B, \Phi_R)}{B(\Phi_B)}d\Phi_R \right], \quad (3)$$

where

$$\overline{B}(\Phi_B) = B(\Phi_B) + V(\Phi_B) + \int R(\Phi_B, \Phi_R)d\Phi_R, \quad (4)$$

$$\Delta_R^{(\text{NLO})}(p_\perp) = \exp\left[ -\int d\Phi_R \frac{R(\Phi_B, \Phi_R)}{B(\Phi_B)} \theta(k_\perp(\Phi_B, \Phi_R) - p_\perp) \right]. \quad (5)$$

While this looks more complicated it has the advantage that it is independent of the parton shower algorithm used to describe subsequent emissions and only generates positive weights. While a number of variants have been developed and different theoretical approaches suggested only these approaches, together with the more recent KrKNLO [7], have proved viable in practice.

Following the development of approaches for handing multiple emissions at leading order and one emission at NLO a number of approaches have now been developed to allow the merging of multiple emissions at NLO [8][9][10][11][12][13] together with the first processes at NNLO [14][15][16].

Together with the development of new approaches for the simulation of the hard processes with higher accuracy the last ten years has also seen the development of a number of new parton shower algorithms [17][18][19][20][21][22][23] primarily motivated by improving the matching to higher-order calculations.

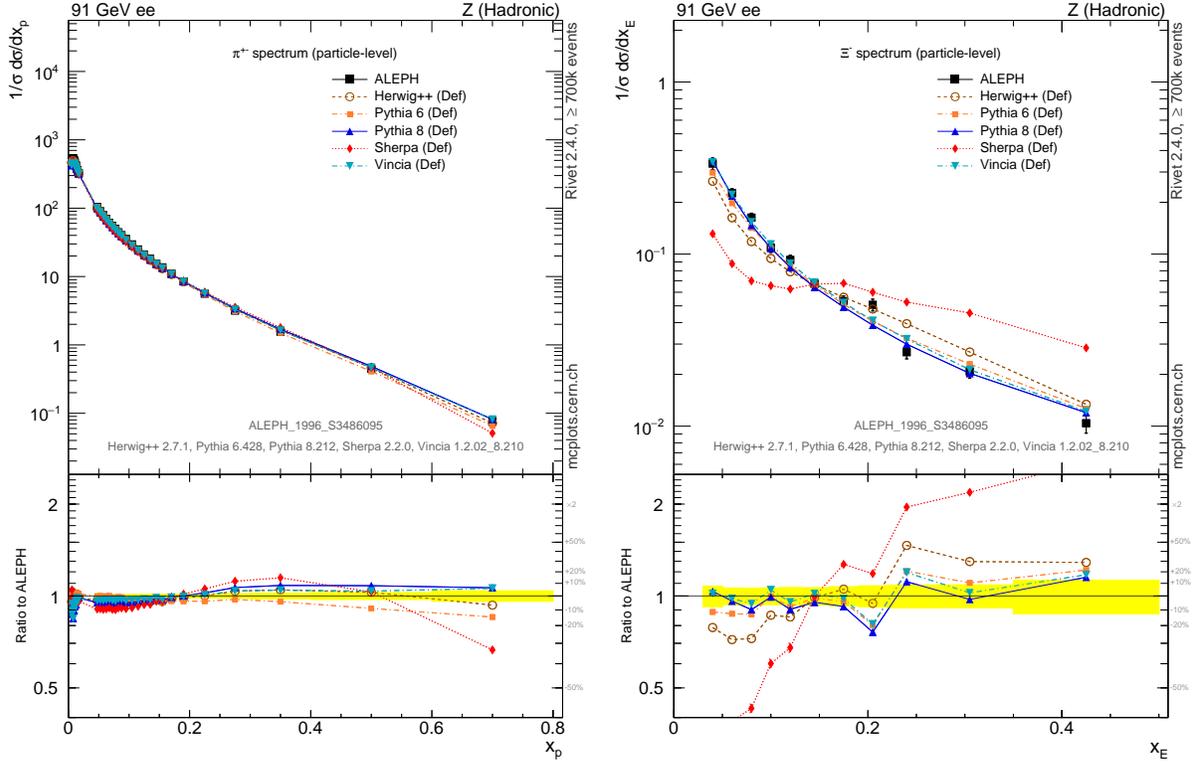

Figure 1: The $\pi^+$ and $\Xi^-$ spectra measured by the ALEPH experiment at LEP I compared to the predictions from taken from modern event generators. Plot from MCplots[28].

In the early 2000's it also became clear that the programming paradigm in particle physics was changing from procedural programming in FORTRAN to object oriented programming in C++. This led to the development of new programs to replace the successful HERWIG [24] and PYTHIA [25] simulation programs and the new SHERPA [26] program developed from scratch in C++. This new generation of event generators are now the workhorses for simulation at the LHC, together with specialised programs for the calculation of hard processes in the various merging schemes.

It is clear that there has been a dramatic development in Monte Carlo simulations over the last fifteen years motivated by the need to describe the unprecedented energy scale and accuracy of the LHC results. For a recent review of the current status of Monte Carlo event generators see [27]. However, describing LEP data is still important and all new shower algorithms are still developed, tested and tuned using data from $e^+e^-$ collisions. Many properties of the non-perturbative models, particularly relating to the production of specific hadrons are hard, if not impossible, to measure in the more complicated hadron–hadron environment, for example the $\Xi^-$ spectrum shown in Fig. 1. Some aspects of hadronization, particularly the production of baryons and hadrons containing strange quarks, remain poorly understood.

## LHC and the Future

The new generation of Modern Carlo event generators provide an impressive quantitative agreement with the LHC data, see for example [29] for the latest results on the production of a Z boson in association jets. While there has been dramatic progress in incorporating higher multiplicity matrix elements and higher-order virtual corrections there has been little progress improving either the underlying accuracy of the parton shower resummation or the non-perturbative modeling of the hadronization process.

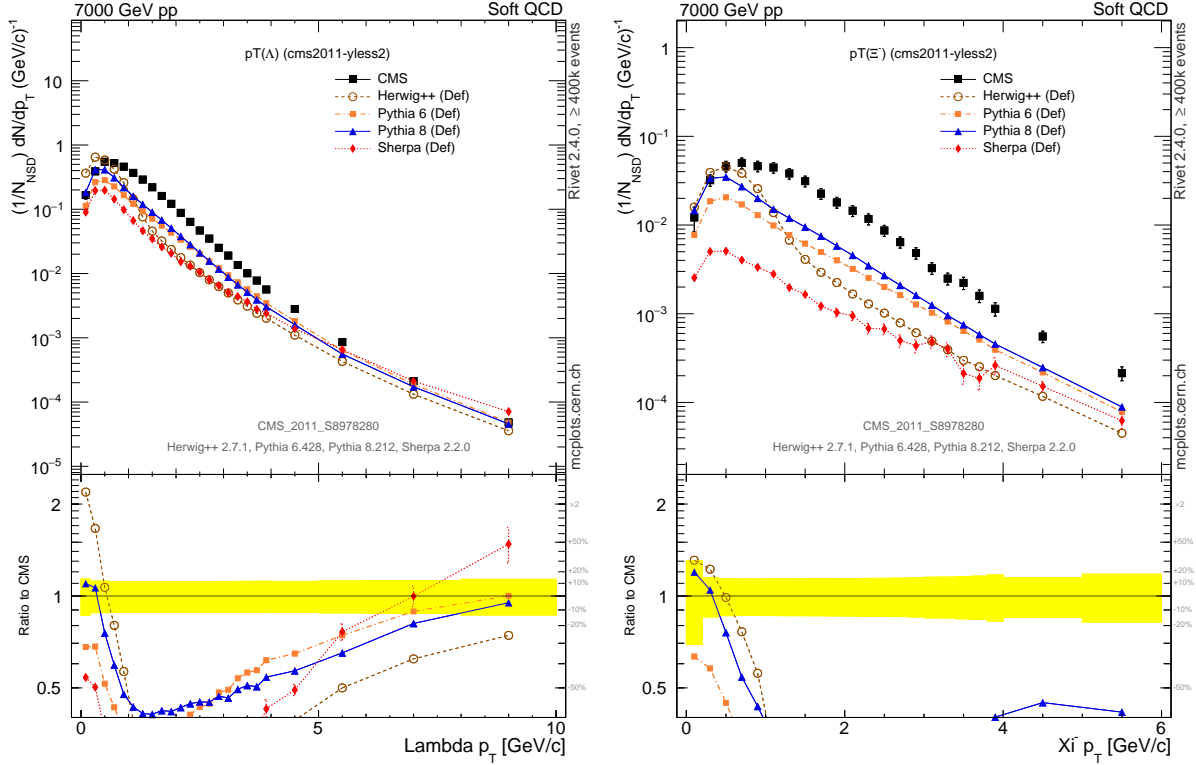

Figure 2: The $\Lambda^0$ and $\Xi^-$ spectra measured by the CMS experiment at the LHC compared to the predictions from taken from modern event generators. Plot from MCplots[28].

Recent work has started on improving the accuracy of the parton-shower algorithm which will be required as the accuracy of the matrix elements increases. This has looked at including subleading colour effects [30][31], and sub-leading colinear logarithms [32]. While this is the area where there is probably the greatest potential for improvement it remains to be seen if we can consistently improve the logarithmic accuracy particularly for all observables, or whether the accuracy can at least be improved for those classes of observables where higher-order analytic resummations are possible.

The standard assumption of universality was that we could develop the hadronization models using $e^+e^-$ data and then apply them in hadron–hadron collisions. Simulating hadronic collisions has always needed additional non-perturbative modeling of the underlying event and non-perturbative colour reconnection. However in the complicated environment of the LHC clearly other things are going on, or colour reconnection is much more complicated, and we need better modeling of non-perturbative effects, e.g. for strange hadron spectra (see Fig. 2). While there are some new ideas, e.g. [33], this is an area where clearly more work is required in order to describe the LHC results.

## Outlook

Given the massive progress in Monte Carlo event generation over the last 15 years it is impossible to say what the state-of-the-art in event simulation will be by the time of FCC-ee. Certainly there will continue to be significant developments over the course of the LHC high-luminosity programme to include higher order matrix elements, more accurate resummation and better non-perturbative modelling. While for the foreseeable this will be driven by the need to describe the results of the LHC given the much simple nature of leptonic collisions the results of LEP and earlier $e^+e^-$ colliders will continue to be used when developing new approaches. Given the decades of work we are looking forward to in order to fully exploit the LHC and improved understanding of QCD and better simulations that surely must follow it is impossible to say at this stage what, if any, further understanding would be obtained from the FCC-ee.

# Observables sensitive to Coherence in $e^+e^-$ Collisions

Simon Plätzer[1,2]

[1] IPPP, Department of Physics, Durham University
[2] Particle Physics Group, School of Physics and Astronomy, University of Manchester

**Abstract:** We describe coherence properties of QCD at electron-positron colliders. We emphasize that past and future measurements sensitive to this effect are vital to the development of reliable and precise parton shower event generators.

## Introduction

Coherence properties of QCD radiation govern a number of inter-jet and intra-jet properties in hadronic final states. The underlying interference effect for emission of a soft gluon can be interpreted within the ordering of emissions in a parton cascade and is a crucial ingredient to the design of contemporary parton shower algorithms.

Observables sensitive to coherence effects in $e^+e^-$ collisions have long been suggested [1]. With the advent of new parton shower algorithms which are either based on a new formulation of angular ordering [2], or which incorporate coherence effects through a dipole-type picture [3], interest has again grown in testing coherence properties at $e^+e^-$ colliders [4], using four-jet events as a testing ground.

## Coherence effects and their experimental manifestation

Coherence is governing multiple soft gluon emissions. It is an interference effect which is manifest at the level of scattering amplitudes, such as when *e.g.* a soft gluon is emitted off a pair of collinear partons. In this case the emission appears as if the gluon had been emitted by the splitting product's parent. This observation gave rise to the angular ordering property employed in parton shower event generators [5] to take into account single-logarithmically enhanced effects due large-angle gluon emission on top of the double-logarithmic structure encountered for soft-collinear emission.

A necessary ingredient to obtain the right description of large-angle soft gluon emission is the inclusion of the correct soft radiation pattern within a parton shower algorithm, at least at the leading power in the number of colours (see *e.g.* [6] for corrections to the radiation pattern). In the large-$N$ limit, colour connected dipoles of partons radiate coherently and approaches based on a dipole-type picture naturally allow for the correct soft radiation pattern as both legs of the emitting dipole are treated democratically [7][8][9]. However, also in approaches build on singling out one of the two legs to emit [2][10][11][12], an appropriate sharing of soft radiation either through disjoint phase space regions connected to angular ordering, or through partitioning within commonly accessible phase space, this goal can be achieved.

Coherence effects have been measured at LEP [13], as well as at hadron colliders [14][15], where a clean identification of the effect is much more challenging as many colour flow topologies contribute to jet production in these cases. In any case, specifically the CDF measurement [14] revealed the importance of coherent parton cascades.

**The quest for new observables**

As event generators have been subject to a tremendous development, specifically in the context of including higher order QCD corrections and the production of additional jets, recent work [4] has been focused on devising new observables sensitive to coherence effects and subsequent parton emission beyond strongly ordered limits. In order to have a clean handle on the effect[*], four jet topologies as the first non-trivial examples have been considered. In the limit of a soft fourth jet, and two collinear sub-leading ones recoiling against one leading jet, these topologies directly allow to expose the radiation pattern expected from coherence arguments. An anti-$k_\perp$ [16] type jet algorithm is desirable in these cases to not strip away soft particles into hard jets in the first place. At least some discriminative power has been found in a subsequent measurement based on OPAL data [17].

**Outlook**

Observables which test QCD dynamics, and in particular coherence effects, are a vital ingredient to the feedback loop of event generator development and validation. Specifically at $e^+e^-$ colliders much less ambiguities and complex dynamics are involved such that these algorithms can be subjected to data benchmarks in a most systematic way. An FCC ee option will provide unique opportunities to further elaborate on such observables and to carry out new measurements.

---

[*]With a clear statement that it is nowadays virtually impossible to construct parton shower algorithms which would not include at least the leading coherent evolution effects.

# QCD splitting-function dependence on evolution variable

S. Jadach[1], A. Kusina[2], W. Placzek[3] and M. Skrzypek[1]

[1] Institute of Nuclear Physics, Polish Academy of Sciences,
ul. Radzikowskiego 152, 31-342 Kraków, Poland
[2] Laboratoire de Physique Subatomique et de Cosmologie
53 Rue des Martyrs Grenoble, France
[3] Marian Smoluchowski Institute of Physics, Jagiellonian University,
ul. Łojasiewicza 11, 30-348 Kraków, Poland

**Abstract:** In this note we present the results of the study performed at IFJ PAN in Krakow in the context of the parton-shower-related research which suggests that QCD splitting function depends on the type of physical evolution variable. We sketch a method of including NLO corrections into a cascade (ladder) of a parton shower (PS). The method is based on the Curci-Furmanski-Petronzio classical paper and matching between PS and NLO distributions obtained from the Feynman diagrams. We note that one does not reproduce the classical CFP NLO kernels if the $k_T$ ordering is used for the calculation. This seems to be a limitation of the definition of the CFP procedure and requires further investigation.

## Parton-Shower-related activities at IFJ PAN

Let us give a short review of the results of the group at IFJ PAN related to the parton shower (PS) research.

(A) In the first step we developed a series of works on evolving parton distribution functions (PDFs) using Markovian Monte Carlo (MMC) programs. We solved exactly the LO and NLO DGLAP evolution equations using the MMC methods with long term aim of developing new types of PS MCs. a) The first MMCs were evolving quark and gluon PDFs according to the LO and NLO DGLAP iequations n $t = \ln \mu^2$ variable, with the running $\alpha_s(t)$ [1], [2]. b) Next we developed the MMC for CCFM (1-loop) gluonstrahlung with $t = \ln(k_T/(1-z))$ and $\alpha_s(k_T^2)$ [3]. c) Finally, the most sophisticated MMCs, with various choices of $t$ and an argument of $\alpha_s$, combining the full DGLAP and CCFM, together with the numerical results, were reported in [4]. d) A corresponding MMC program in C++ implementing variety of the above models was published in [5]. The PDFs $q(t,x)$ and $g(t,x)$ from MMCs agreed (within 3-digits) with non-MC calculations of the QCDnum16 and APCheb programs. e) Let us mention also a more recent study of the MCFM modelling exploiting our MMC done in Ref. [6].

(B) MMCs were designed to test numerically series of constrained MC (CMC) programs, implemented with the same evolution, but with a constrained predefined final $x$ variable (an alternative to the backward evolution in the PS MC) aiming at strict (NLO level) control of the distributions of LO PS MC. a) We began with CMC and MMC modules for single ladder/shower, without a hard process, with exclusive LO kernels, optionally inclusive NLO kernels. CMC programs tested with MMCs were published in [7], [8]. b) Next, we combined two CMCs with a hard process matrix element (ME) into PSMC for Drell-Yan (DY) process [9], [10]. Such an academic PS MC was instrumental in testing new ideas on the NLO corrections in the exclusive evolution included in all kernels in the initial state ladders/showers, and on the NLO corrections to the hard process (a simpler alternative to MC@NLO [11] and POWHEG [12] approaches) thanks to a perfect numerical and algebraical control over LO distributions.

(C) The next project concerns the NLO corrections in an exclusive form to PS MC (one ladder): a) The first solution, albeit limited to the non-singlet evolution, was proposed and tested numerically in [13], [14]. b) The exclusive NLO kernels had to be recalculated in the Curci-Furmanski-Petronzio (CFP) [15] framework from scratch. The non-singlet 2-real kernels were recalculated in [16] and the non-singlet 1-real-1-virtual ones in the PhD thesis [17]. Also the singlet evolution kernels are now almost complete (unpublished). c) Simplified and faster scheme was reported (numerical tests) in [18]. d) An even simpler and faster scheme of NLO-correcting was presented in [19]. e) A major problem was how to include consistently virtual corrections. The first solution has been formulated but not yet published. It exploits recalculated virtual corrections in the CFP scheme to the non-singlet kernels [20], [17].

D) Finally, the last project is the implementation of the NLO corrections to the hard process in a new way, named KrkNLO method. It provides a simpler alternative to MC@NLO and POWHEG methods: a) Methodology for DY process was defined (without numerical tests) in [21]. b) Numerical tests of KrkNLO with the help of Double-CMC PS was done in [10]. c) Further improvements and introduction of PDFs in the MC factorisation scheme can be found in [22], where the the MC implementation is still done with the help of the not-so realistic Double-CMC PS. d) Next, the KrkNLO method was implemented on top of the SHERPA [23] parton shower. Comparisons for the DY process with the fixed-order NLO calculations of MCFM [24] as well as with the NLO-PS matched calculations of MC@NLO and POWHEG were also done in [25] and in the conference note [26]. e) A complete definition of a new scheme for PDFs, called the MC scheme, necessary for the KrkNLO method, both for the DY processes and Higgs-boson production via gluon–gluon fusion was presented in [27] and in the conference note [28]. f) At last, the implementation of the KrkNLO method in the Herwig 7 MC generator [29] and a series of comparisons for Higgs production in gluon–gluon fusion with the programs: MCFM, MC@NLO, POWHEG and HNNLO [30] as well as with experimental data of the ATLAS Collaboration from the LHC Run 1 was published in [31].

In this note we briefly present the findings of Ref. [32]. This is however, a rather technical paper, so let us first explain the background and then summarise on the main results of Ref. [32]. Generally, the context is that of the N+NLO parton shower, with the NLO evolution kernels (PDFs) in the shower (ladder) and the NNLO hard process. As listed earlier, we have already defined and tested some important elements of the complete N+NLO parton shower.

### What is NLO Parton Shower?

Let us start with a simpler question: What is LO PS? It is built using the LO-class evolution kernels and the LO PDFs within a certain collinear factorisation scheme. LO PS MC implements the LO DGLAP evolution for semi-inclusive distributions (structure functions, etc.). If the hard process is corrected to the NLO level (N+LO), LO PS MC desirably encapsulates all collinear/soft singularities of the NLO hard process ME in the exclusive form. In the N+LO schemes *one* parton originally generated within LO PS MC gets promoted to the hard process.

Let us now get back to the NLO PS. It is built using the NLO-class evolution kernels and the NLO PDFs, and it implements the NLO DGLAP evolution for certain semi-inclusive distributions (structure functions etc). For the hard process corrected to the NNLO level (N+NLO), it is desirable that NLO PS MC encapsulates all collinear/soft singularities of the NNLO hard process ME in the exclusive form. In the N+NLO scheme up to *two* partons originally generated within the NLO PS MC get promoted to the hard process.

Introducing the complete NLO real and virtual corrections into PS MC in the exclusive form, in accordance with the collinear factorization theorem, is the main problem, theoretically and practically. To this end we proposed a theoretical framework based on the CFP [15] approach. We have formulated and tested three methods of practical implementation of the NLO corrections in the PS MC. One of them turned out to be quite promising.

Consistent matching of NLO PS with CFP is highly nontrivial. The procedure goes like this. One writes PS MC in $d = 4 + 2\epsilon$ space-time dimensions truncated to the 2nd order and applies to it the CFP subtraction procedure. In this step the following problem appears: PS MC has a Sudakov virtual correction with non-pole terms, while CFP has a wave-function renormalisation factor with pure poles. Therefore subtraction has to be done separately for the form factor and for the rest. Once PS MC distributions are processed through the CFP procedure, dummy functions inserted in the PS MC distributions are identified with their counterparts in the CFP distributions representing the complete NLO. There are two requirements of the procedure: 1) Both distributions, that of PS MC and that of the CFP calculations, have to be defined within *the same phase-space* limits. 2) PS MC has to implement exactly all collinear and soft limits of the NLO matrix element.

While constructing NLO PS we had to recalculate the NLO kernels in the exclusive form. We have done this in the CFP scheme, diagrammatically, in the axial gauge, both for real [16] and virtual [17] corrections. In the course of the calculation a technical improvement related to the use of principal value (PV) prescription was proposed [33].

### Problem with evolution variable

As sketched earlier, we successfully realised the matching of NLO PS with CFP for bremsstrahlung-type diagrams ($C_F^2$-type), shown in the right-hand-side plots of Fig. 1. We used various types of the evolution variable: $k_T$, $k_T/(1-z) \sim \theta$, $k^- = k_T/\sqrt{1-z}$. The situation changed when we included the graphs Vg and Vf, shown in the left-hand-side plots of Fig. 1. We encountered a problem when applying the $k_T$-ordering to the calculation of the Vg graph of Fig. 1. Namely, we found that in the CFP-based analytical calculation, with phase space as in PS MC and with $k_T$-ordering, the contribution of this graph to the NLO kernel is different than in the original CFP paper. This is, of course, rather a problem of CFP, not of PS MC, as according to the theorem proven in the CFP paper the dependence of the MS NLO kernel on the choice of the upper kinematic boundary $Q$ is forbidden. The problem is absent for the gluonstrahlung (non-singlet) kernels shown in the right-hand-side plots of Fig. 1 as well as for all single-pole diagrams. Next, we found that the choice of the type of the "evolution time", i.e. the $Q$-variable, is essential: for $Q =$ virtuality, $Q = |\vec{k}_{T1} + \vec{k}_{T2}|$ and $Q = |\vec{k}_{T1}| + |\vec{k}_{T2}|$ the problem is absent. It shows up only for the diagrams with the final-state radiation (FSR) gluon splitting into a gluon pair calculated for $Q = \max(k_{T1}, k_{T2})$, combined with the relevant virtual diagram counter-partner. The analytical calculation shows that source of the effect is a mismatch between the upper kinematic limit in the virtual and real diagram contributions. Such a mismatch is usually absent, except for the $Q = \max(k_{T1}, k_{T2})$ case.

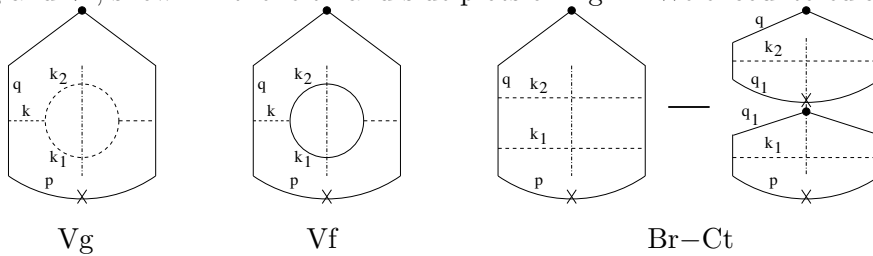

Figure 1: Real graphs with double poles contributing to the NLO non-singlet $P_{qq}$ kernel. The solid lines represent quarks and the dotted lines stand for gluons. Left plots: the FSR-type graphs, right plots: the gluonstrahlung-type graphs.

Let us discuss a list of possible solutions of this problem: A) The simplest explanation would be that the CFP scheme is not as general as expected, though it is rather unlikely. B) Is it then due to partial introduction of the dimensional regularisation in CFP, only for the collinear singularity, not for the light-cone variable? C) Perhaps it is due to the use of the axial gauge? D) Finally, maybe it cancels when combined with the NNLO hard process?

### Summary and outlook

We have reviewed a methodology of including NLO corrections along the parton shower MC ladder. It has been tested numerically. The method is based on matching between PS MC and CFP diagrammatic NLO distributions. We have spotted a problem of not reproducing NLO evolution kernels in the calculations with $k_T$-ordering kinematics. It is essentially the problem of the CFP scheme itself and requires better understanding.

# Jet reconstruction algorithms in $e^+e^-$ collisions

Javier Aparisi[1], Ignacio García[1], Martin Perelló[1], Ph. Roloff[2], Rosa Simoniello[2], **Marcel Vos**[1]

[1] *IFIC (UVEG/CSIC), València, Spain*
[2] *CERN, Geneva, Switzerland*

**Abstract:** Jet reconstruction is a key technique at future energy-frontier $e^+e^-$ colliders. Classical $e^+e^-$ algorithms are tested by several new challenges. In this contribution results are presented of studies into the jet reconstruction performance at high-energy $e^+e^-$ colliders.

## Introduction

A high-energy $e^+e^-$ collider can provide precise measurements of the interactions of the Higgs boson and the top quark. The Higgs-strahlung process ($e^+e^- \to ZH$), with a maximum cross section at $\sqrt{s} \sim 250$ GeV), is accessible with a large circular machine [1][2]. The largest rings envisaged (with a circumference of 100 km) can reach the top quark pair production threshold at $\sqrt{s} \sim 350$ GeV [3]. Processes at still higher energies (vector-boson fusion Higgs production, associated production of a top quark pair and a Higgs boson, di-Higgs boson production) are accessible at a linear collider. The ILC [4][5] project envisages operation at 250 GeV and 500 GeV, with the possibility of an upgrade to 1 TeV. The CLIC [6][7] project aims for multi-TeV operation, with an initial stage at 380 GeV.

An accurate reconstruction of hadronic final states is a prerequisite for a precise measurements of Higgs boson and top quark couplings [8][9]. The linear collider detector concepts [10] achieve excellent single-particle reconstruction with highly granular calorimeters [11][12], and particle-flow algorithms [13]. Excellent jet clustering is required to take full advantage of the potential of the machine and detectors.

## Challenges to jet clustering

Jet clustering at high-energy colliders differs in several respects from previous $e^+e^-$ colliders, such as LEP or SLC. The most important effects are listed below:

- **Multi-jet final states:** processes with many jets in the final state become more important. Key measurements at the lowest energy require an accurate reconstruction of four final-state jets ($e^+e^- \to ZH$, with hadronic Higgs and Z boson decays, $e^+e^- \to t\bar{t}$ in the lepton+jets channel). Processes with six-jet, eight-jet and even ten-jet final states open up. A correct clustering of the reconstructed particles into jets turns out to be far from trivial even if the inputs to the algorithm are accurately reconstructed. In analyses such as the extraction of the Higgs self-coupling clustering has a dominant contribution to the mass resolution [14].

- **Hard emissions:** the phase space for the emission of hard gluons opens up. In some cases the distance or energy scale of the emitted gluon is no longer small compared to the typical distance between the decay products of gauge bosons. The $n$ jets reconstructed with exclusive clustering using a sequential recombination algorithm (the standard procedure at previous lepton colliders and in benchmark studies of future $e^+e^-$ colliders) may not correspond to the $n$ final-state quarks. This problem may be circumvented in the pair production of very energetic objects with hadronic decays (i.e. boosted gauge or Higgs bosons, or top quarks), by reconstructing two *fat* jets that capture the energy flow of the boosted object [15][16]. For final states with a strong hierarchy between energy scales this effect leads to failures in jet reconstruction in a small fraction of events (i.e. di-Higgs production at very high energy, where the radiated Higgs boson remains rather soft).

- **Forward processes:** $t$-channel processes become increasing important. At high energy the final-state products of processes such as vector-boson-fusion Higgs boson production are strongly peaked in forward and backward directions [17]. Special care is needed in the detector design and in the development of jet clustering algorithms to ensure robust jet reconstruction performance over the full polar angle coverage of the experiment.

- **Background processes:** energy flow superposed on the *signal* event can affect jet reconstruction. Where such backgrounds could safely be ignored at previous $e^+e^-$ colliders, they may have a non-negligible effect at future installations. The $\gamma\gamma \rightarrow hadrons$ background renders the classical $e^+e^-$ algorithms inadequate for high-energy operation of linear colliders [7][18][19][14]. At circular colliders the rate of $\gamma\gamma \rightarrow hadrons$ is several orders of magnitude smaller. The effect of synchrotron radiation, and appropriate shielding measures, remain to be evaluated.

Higher energy also has some benificial effects. An additional boost collimates the jets, so that confusion due to clustering is reduced. The relative size of non-perturbative corrections diminishes strongly with increasing center-of-mass energy.

### Robust jet clustering in the presence of background

The impact of *pile-up* due to the energy flow of $\gamma\gamma \rightarrow hadrons$ was studied thoroughly as part of the CLIC conceptual design studies [7][18]. In multi-TeV operation a bunch train may deposit several TeV in the experiment. Timing cuts that select 1 ns around the *signal* bunch crossing reduce this contribution to the order of 100 GeV. In the classical approach (represented in most studies by exclusive clustering with the Durham algorithm [20]), all final-state particles are clustered into jets. The reconstructed jet properties are found to be strongly affected by the background energy flow [7][18][19][14]. At the highest energy classical $e^+e^-$ algorithms are inadequate.

Several alternatives have been considered to achieve more robust performance in the presence of background. The generalization of the $e^+e^-$ algorithms with a beam distance [21] yields jets with a limited area. Longitudinally invariant algorithms developed for hadron colliders [22][23] expose even less area in the forward and backward parts of the experiment, where the $\gamma\gamma \rightarrow hadrons$ background is most pronounced. In the VLC algorithm proposed in Ref. [19][14] this feature is combined with the traditional inter-particle distance of the Durham algorithm. The inter-particle and beam distance of these three classes of algorithms are given in Fig. 1, together with an indication of the jet area in the central and forward directions of the experiment.

The longitudinally invariant algorithms and VLC are found to be much more resilient than the classical $e^+e^-$ algorithms. The VLC algorithm outperforms the hadron collider algorithms in the most demanding environment [19][14].

At the ILC, with a much smaller $\gamma\gamma \rightarrow hadrons$ rate and a much larger bunch spacing, the effect of this background is much less pronounced. Still, a modest but non-negligible improvement of the performance can be achieved by adopting the VLC or longitudinally invariant algorithm. At circular colliders the rate of $\gamma\gamma \to hadrons$ is so low that its effect is expected to be negligible. The impact of synchrotron radiation has not been evaluated in detail.

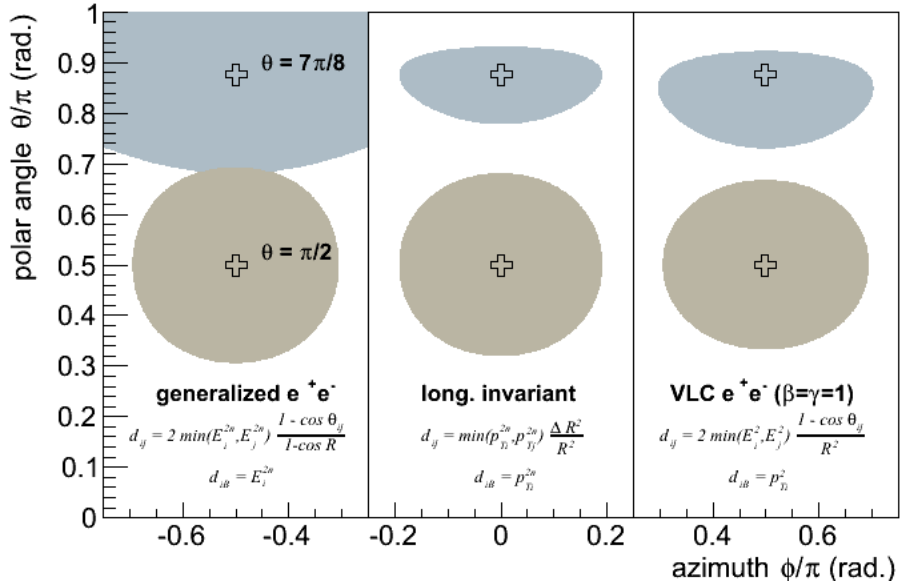

Figure 1: The area or *footprint* of jets reconstructed with a radius parameter $R = 0.5$, for the three major families of sequential recombination algorithms. The two shaded areas in each column correspond to a jet in the central detector ($\theta = \pi/2$) and to a forward jet ($\theta = 7\pi/8$). Reprinted from Ref. [14].

### Perturbative and non-perturbative corrections

Even in the ideal case of a perfect detector response the jet energy differs from that of the final-state parton due to a number of effects. The largest correction for jets with a finite size can be addressed in a perturbative calculation. A smaller, non-perturbative correction remains. Ref. [14] (following Ref. [24]) estimates both on simulated $e^+e^- \to q\bar{q}$ and $e^+e^- \to t\bar{t}$ from a Monte Carlo event generator (the MadGraph5_aMC@NLO package [25] interfaced to Pythia 8.180 [26]).

As expected, the relative perturbative energy correction - evaluated as the difference between the energy of the jet reconstructed on stable particles and the quark produced in the hard scatter - is roughly independent of center-of-mass energy and decreases as the catchment area of the jet increases. The correction is smallest for the classical $e^+e^-$ algorithms is smallest. For the longitudinally invariant and VLC algorithms with a radius parameter of 1.5, the average correction is less than 2%. A tail towards lower reconstructed jet energy remains, however, resulting in a median correction of approximately 5%.

Non-perturbative corrections are estimated as the difference between the energy of jets reconstructed on stable particles and at the parton-level. The results for $e^+e^- \to q\bar{q}$ production at $\sqrt{s} = 250$ GeV (left panel) and $e^+e^- \to t\bar{t}$ production at $\sqrt{s} = 3$ TeV (right panel) are shown in Fig. 2. The distributions are again asymmetric, with significant differences between the mean and median corrections (which can even have opposite signs). Non-perturbative corrections are reduced for large radius parameter, but do not vanish completely. The most striking effect is the strong reduction of the size of these corrections at high energy (the range of the Y-axis is reduced by a factor ten in the rightmost panel).

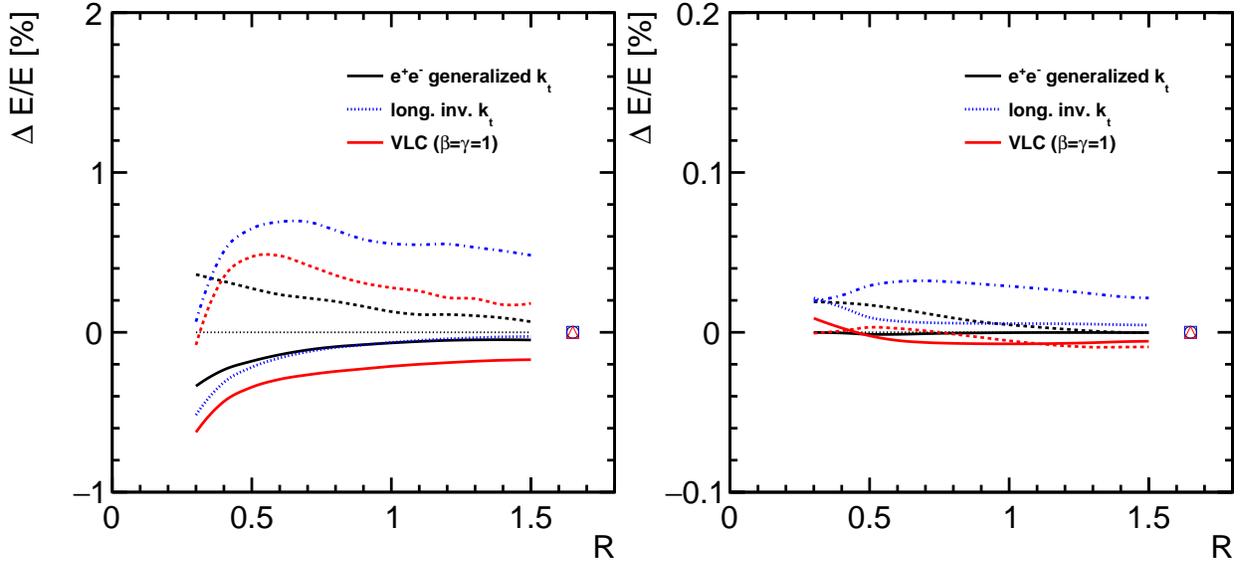

Figure 2: Non-perturbative jet energy corrections to the jet energy as a function of the jet radius parameter $R$ in $e^+e^- \to q\bar{q}$ production at $\sqrt{s} = 250$ GeV (left panel) and $e^+e^- \to t\bar{t}$ production at $\sqrt{s} = 3$ TeV (right panel). The continuous line corresponds to the median relative correction, the dashed line to the median. Results are shown for three algorithms: the generalized $e^+e^-$ algorithm, the longitudinally invariant $k_t$ algorithm and the VLC algorithm with $\beta = 1$. Reprinted from Ref. [14].

The non-perturbative contribution to the invariant mass of the jet is more important, with relative correction of several tens of % at low energy and a few % for $\sqrt{s} = 3$ TeV (for $R = 1$). Algorithms with the $e^+e^-$ inter-particle distance (generalized Durham and VLC) converge slightly faster than the longitudinally invariant algorithms.

## Conclusions

Jet clustering at future energy-frontier $e^+e^-$ facilities faces several challenges that are new to $e^+e^-$ colliders. Multi-jet final states and final states with very forward jets are much relevant than at LEP or SLC. Hard gluons emitted in events with relatively soft gauge bosons challenge exclusive jet reconstruction. Background such as $\gamma\gamma \to$ *hadrons* or synchrotron radiation may affect the jet reconstruction performance.

Detailed benchmark studies of the ILC and CLIC design study groups that jet clustering is the limiting factor in the analysis of complex multi-jet final states. Studies with realistic background levels show that for the most demanding environment in multi-TeV operation classical $e^+e^-$ algorithms are inadequate. Longitudinally invariant algorithms and the VLC algorithm proposed in Ref. [19] prove to be much more resilient. Classical $e^+e^-$ algorithms, on the other hand, show faster convergence of energy corrections with the radius parameter of the algorithm. The non-perturbative correction associated to hadronization decreases strongly with center-of-mass energy: from 1% at $\sqrt{s} = 250$ GeV to less than a per mil at $\sqrt{s} = 3$ TeV.

Given the relevance of jet clustering for the potential of high-energy $e^+e^-$ colliders we encourage new ideas and more exhaustive performance study for existing algorithms (e.g. those of Refs. [27][28]).

## Acknowledgement

The authors acknowledge the effort of the ILC and CLIC detector & physics groups in putting together the simulation infrastructure used to benchmark the algorithms. This work benefited from services provided by the ILC Virtual Organisation, supported by the national resource providers of the EGI Federation. The authors would like to thank Gavin Salam and Jesse Thaler for help with the VLC algorithm and the FastJet team for guidance creating the plugin code. The IFIC group is financially supported by the Spanish national programme for particle physics (FPA2015-65652-C4-3-R, MINECO/FEDER).

# Challenges in heavy-quark fragmentation

Gennaro Corcella[1]

[1] INFN, Laboratori Nazionali di Frascati, Via E. Fermi 40, I-00044, Frascati (RM)

**Abstract:** I discuss some open issues concerning the fragmentation of heavy quarks in $e^+e^-$ annihilation, from LEP/SLD to FCC-*ee*. In particular, I review the state of the art of resummed calculations and Monte Carlo event generators and underline some of the challenging objectives of FCC-*ee* in the heavy-flavour sector.

## Introduction

Heavy-quark (charm, bottom, top) phenomenology in different environments is one of the most challenging topics in high-energy physics, from both theoretical and experimental viewpoints, as it allows tests of QCD, parton model, factorization and power corrections. In fact, when calculating the cross section for heavy-quark production, the heavy-quark mass ($m$) regularizes the collinear singularity, but nevertheless differential distributions exhibit large logarithmic corrections, which need to be resummed to all orders, such as contributions $\sim \alpha_S \ln(m^2/Q^2)$, $Q$ being a typical scale. Calculations resumming such large logarithms are available [1] [2], as well as Monte Carlo parton shower algorithms, such as the HERWIG [3] [4] and PYTHIA [5] [6] codes, which have lately been matched to NLO calculations in the aMC@NLO [7] and POWHEG [8] frameworks, for a number of hard-scattering processes, including heavy-quark production. As for hadronization, resummed calculations typically use non-perturbative fragmentation functions depending on few parameters which are to be tuned to experimental data [9]; alternatively, one can model power corrections by including them in an effective [10] or frozen [11] strong coupling constant. On the other hand, Monte Carlo generators implement phenomenological models, such as the string [12] or cluster [13] models, to turn partons into hadrons, once a scale of the order of 1 GeV is reached in the shower. When fitting hadronization parameters and using the tuned models in other processes, it is essential describing the parton-level process always within the same framework, namely resummations or Monte Carlo showers.

In the following, as a case study for heavy-quark fragmentation, I shall concentrate on bottom production in $e^+e^-$ annihilation: I will discuss the state of art of perturbative calculations, at fixed order and resummed, review the current status of Monte Carlo generators, and finally comment on the perspectives at FCC-*ee* and make concluding remarks.

## Perturbative calculations for heavy-quark fragmentation

As discussed in the introduction, fixed-order and resummed calculations are available to describe bottom production in $e^+e^-$ annihilation and the fragmentation into $b$-flavoured mesons/baryons. Heavy-quark spectra are usually expressed in terms of $x$, the quark energy fraction in the centre-of-mass frame; the perturbative fragmentation approach, proposed in [1], factorizes the $x$-spectrum as the convolution of a massless, $\overline{\text{MS}}$-subtracted coefficient function and a massive perturbative fragmentation function. The initial condition of the perturbative fragmentation function is process-independent [2] and, by means of the DGLAP equations, for an evolution between scales of the order of the centre-of-mass energy $\sqrt{s}$ and $m$, one manages to resum the large logarithms $\alpha_S \ln(m^2/s)$ appearing in the NLO mass spectrum (collinear resummation).

Furthermore, both $\overline{\mathrm{MS}}$ coefficient function and initial condition contain terms which become large whenever the energy fraction $x$ gets close to unity, which corresponds to soft or collinear gluon radiation. It is therefore mandatory to resum even these contributions to all orders to obtain a reliable prediction (large-$x$ or threshold resummation).

The initial condition of the perturbative fragmentation function was computed at NLO in [1] and lately at NNLO in [14] and [15] for quark- and gluon-initiated contributions, respectively. The $\overline{\mathrm{MS}}$ coefficient function was instead calculated at NLO in [16] and at NNLO in [17]; Ref. [18] computes the NNLO corrections to the Altarelli–Parisi splitting functions, entering in the evolution of the perturbative fragmentation function. In fact, if the splitting functions are computed at NLO, the large mass logarithms $\alpha_S \ln(m^2/s)$ are resummed at NLL, whereas collinear resummation can be carried out at NNLL once even the NNLO corrections to the splitting functions are implemented.

As for threshold resummation, large-$x$ contribution to the coefficient function and initial condition of the perturbative fragmentation function can be resummed following standard techniques as in [19], where the calculation is carried out in the next-to-leading logarithmic approximation (NLL). By using the results in [20] and [21], one can extend threshold resummation to NNLL accuracy.

Although all the ingredients to calculate the heavy-quark spectra at NNLO, with the resummation of mass and threshold contributions to NNLL accuracy, are available, most phenomenological investigations have been carried out in the NLO+NLL approximation in [2] and [22] for $e^+e^-$ annihilation, in [23] for $b$-production in top decays $t \to bW$, in [24] for $H \to b\bar{b}$ processes, $H$ being the Standard Model Higgs boson. Extending the studies of heavy-quark fragmentation in $e^+e^-$ annihilation is nevertheless very useful, in order to further decrease the scale uncertainty on the predictions and improve the behaviour of the energy spectra at large $x$, which are instead unstable and oscillating whenever $x > 1 - \Lambda_{\mathrm{QCD}}/m$ [2].

As for the inclusion of hadronization corrections, the heavy-quark spectra yielded by resummed calculations are typically convoluted with non-perturbative fragmentation functions containing a few parameters which are to be tuned to experimental data, e.g., $B$-hadron production at LEP [25] [26] or SLD [27]. Details on the fitting of hadronization models can be found in [28], where the best-fitted models are also used to predict $B$-hadron energy distributions in top and Higgs decays.

Before concluding this section, it is worthwhile pointing out that most analyses on heavy-quark fragmentation are undertaken in the so-called non-singlet approximation and gluon splitting into heavy-quark pairs is neglected. Ref. [22] did include $g \to c\bar{c}(b\bar{b})$ splitting, but its contribution to charm/bottom fragmentation turned out to be small and not essential to fit the experimental data. In fact, LEP and SLD experiments measured the gluon branching fractions to heavy quarks, labelled as $g_{c\bar{c}}$ and $g_{b\bar{b}}$, and it was found $g_{c\bar{c}} \simeq 3 \times 10^{-2}$ [29] [30] and $g_{b\bar{b}} \simeq 2 \times 10^{-3}$ [31] [32] [33]. As will be commented later on, FCC will have a better sensitivity to $g \to c\bar{c}(b\bar{b})$ processes, thanks to expected higher statistics and more refined granularity of calorimeter and vertex detectors.

### Monte Carlo parton showers and heavy-quark fragmentation

As pointed out in the introduction, Monte Carlo event generators, implementing parton showers in the soft/collinear approximation, along with non-perturbative models for hadronization, are available tools to address heavy-quark fragmentation in $e^+e^-$ collisions. Ref. [28] discusses a tuning of HERWIG 6 [3] and PYTHIA 6 [5] event generators to LEP and SLD, taking particular care about fitting the Monte Carlo parameters which are directly related to the hadronization of the bottom quark. Those tunings were then used in [34] to estimate the Monte Carlo uncertainty on the top-quark mass due to the treatment of bottom fragmentation in top decays at the LHC. The overall result of these analyses is that it was necessary to retune both HERWIG and PYTHIA to get an acceptable description of B-hadron data at LEP and SLD. Indeed, one managed to tune PYTHIA to reproduce well the data, whereas HERWIG was only marginally consistent, although the fitting procedure improved the comparison pretty much.

Given the late progresses in Monte Carlo implementations, the results in Refs. [28] and [34] clearly need to be updated. On the one hand, both HERWIG and PYTHIA have new versions in C++, namely HERWIG 7 [4] and PYTHIA 8 [6]: Ref. [35] compares the PYTHIA 8 predictions with bottom-fragmentation data from LEP and SLD, showing that it is possible to tune PYTHIA 8 to reproduce such data. On the other hand, HERWIG 7 exhibits some discrepancies with respect to B-hadron data [36] and therefore a retuning is therefore mandatory.

As for the novel generation of NLO codes, such as aMC@NLO [7] and POWHEG [8], in principle the HERWIG and PYTHIA fragmentation parameters need to be retuned, once the hard scattering is implemented at NLO. However, HERWIG and PYTHIA standard showers are matched to NLO tree-level matrix elements, along the lines of [37] and [38], and the full virtual corrections, included in POWHEG and aMC@NLO, are relevant only at large energy fractions. Furthermore, up to power corrections $\mathcal{O}(m/\sqrt{s})^p$, the NLO $K$-factor for the total $e^+e^-$-annihilation cross section is small, being $K \simeq 1 + \alpha_S(s)/\pi$. Therefore, although a thorough investigation of bottom fragmentation using POWHEG and aMC@NLO is currently in progress [39] and should be very welcome, since such programs are heavily used even for $b$-quark production in top or Higgs decays at the LHC, one should expect very little differences in the best-fit parametrizations with respect to the standard tunings of PYTHIA and HERWIG.

**Prospects at FCC-*ee***

The perspectives of the FCC-*ee* program, with an integrated luminosity $\mathcal{L}_{\text{int}}=1$ ab$^{-1}$, are summarized in [40], where the authors also debate the challenging objectives at different centre-of-mass energies, namely the threshold for $Z$, Higgs, $WW$, $t\bar{t}$ and $HZ$ production.

In particular, the expected statistics will be $10^5$ larger than at LEP and therefore the statistical uncertainties will be reduced by a factor of 30. Also, because of the smaller beam-spot size and the new-generation vertex detectors, much more precise measurements of the $R_b$ ratio are foreseen. The current value is $R_b = 0.21629 \pm 0.00066$, whereas a precision about 2-5 $\times 10^{-5}$ is the goal of FCC-*ee* [41].

Furthermore, while the LEP and SLD fragmentation measurements were carried out essentially for inclusive spectra ($B$-mesons and possibly $\Lambda_b$ baryons) and chains like $B \to D^*\ell\nu$, $D^* \to D\pi$, $D \to K(n\pi)$, the high FCC-*ee* statistics will make it possible to distinguish $b$- and $c$-flavoured hadrons and separate fragmentation spectra (charged vs neutral, spin 1 vs spin 0, baryons vs mesons). In this way, one will be able to extract the non-perturbative fragmentation function very precisely, which is crucial to carry out any program of precision physics even in the Higgs and top-quark sectors. Moreover, according to the FCC project, one will be sensitive to rare $B$-decays, such as $B \to J/\psi X$, which, albeit the small branching ratio, can be easily discriminated from the backgrounds, after suitable cuts are set.

Finally, as commented above, the higher statistics and granularity of calorimeters and vertex detectors will allow one to disentangle the $g \to b\bar{b}$ and $g \to c\bar{c}$ splittings, through a double tagging of the jets originated from the $b(c)$ and $\bar{b}(\bar{c})$ quarks. Therefore, FCC-*ee* will be a unique environment to perform precise measurements of gluon-initiated contributions to heavy-quark fragmentation functions and thus perform further tests of QCD and factorization.

## Conclusions

I discussed heavy-quark fragmentation in $e^+e^-$ collisions, in the perspective of Future Circular Colliders, and briefly reviewed the state of the art of theoretical calculations and Monte Carlo generators. As for resummations, although phenomenological analyses have been so far carried out in the NLO+NLL approximation, all ingredients to promote them to NNLO+NNLL accuracy are available and such an extension will be certainly desirable in order to meet the precision goals of FCC-*ee*. Particular care will have to be taken to include in a consistent way non-perturbative corrections, once the higher-order corrections to the parton-level process are implemented. Futhermore, the large statistics and more refined detectors which are foreseen at FCC will allow more accurate determinations of the gluon branching fractions into heavy-quark pairs.

Thanks to lively activity in the latest years, much progress has been undertaken in the implementation of Monte Carlo generators. The new object-oriented versions of HERWIG and PYTHIA contain improved hadronization models and have been matched to NLO hard-scattering processes provided by POWHEG and aMC@NLO. A systematic investigation of heavy-quark/hadron production in $e^+e^-$ annihilation with the NLO+shower codes is currently in progress. Although NLO corrections to shapes and normalization of $e^+e^-$-annihilation cross section should be small, retuning non-perturbative models and studying heavy-hadron production with POWHEG and aMC@NLO will be nonetheless very interesting.

In summary, FCC-*ee* will be a great opportunity to study heavy-quark phenomenology with high precision, from the viewpoint of both perturbative and non-perturbative QCD. The late advances in QCD calculations and Monte Carlo implementations, as well as the ongoing work on heavy-quark phenomenology, should make the challenging objectives of the FCC-*ee* project reachable.

# $g \to b\bar{b}$ Studies at the LHC

Benjamin Nachman[1]

[1]Physics Division, Lawrence Berkeley National Lab, 1 Cyclotron Rd, Berkeley, CA, 94720, USA

**Abstract:** Gluon splitting to bottom quark pairs offers a unique probe of many aspects of the Standard Model. ATLAS and CMS have already started a program to measure this process, which will be significantly improved from efforts to advance boosted Higgs tagging. This note briefly describes what has been measured, what can/should be measured, what improvements have already been made, and what the outlook is to FCC-ee.

## Introduction

Gluon splitting to bottom quarks with small opening angles provides a rich and diverse probe of the Standard Model (SM). First of all, this process can be used to examine perturbative aspects of quantum chromodynamics (QCD). For example, it is essentially the only (nearly) direct measurement of a parton splitting function. Other measurements of jet structure can measure the energy distribution between prongs of a hard splitting but cannot unambiguously assign parton flavors. In addition, $g \to b\bar{b}$ offers a source of gluon jets that is unmatched with other event selections at a hadron collider[*]. Studying gluon splitting to bottom quarks also serves as a tool to measure subtle non-perturbative properties of jets. In particular, $g \to b\bar{b}$ is a relatively pure source of a color octet radiation pattern that can be directly compared with similar measurements of scalar or vector singlets in Higgs boson or W/Z boson decays - see Fig. 1 for an illustration. A third important reason for studying $g \to b\bar{b}$ is as a background process for $h \to b\bar{b}$. In most cases, current simulations are not trusted to make predictions for $g \to b\bar{b}$ but this situation may be improved with future measurements and new models inspired by and tested against the measurements.

This section is organized as follows. Section* 2 briefly reviews the $g \to b\bar{b}$ measurements from the ATLAS and CMS experiments conducted during the first run of the LHC. Then, Sec. 2 describes future measurements that would be useful for constraining $g \to b\bar{b}$ properties. Section* 2 describes ongoing efforts to construct boosted Higgs boson taggers that attempt to separate $g \to b\bar{b}$ from $h \to b\bar{b}$ and the section ends with conclusions and outlook in Sec. 2.

## LHC Run 1

Both ATLAS and CMS have performed measurements directly sensitive to $g \to b\bar{b}$ in the first part of the LHC Run 1[†]. In particular, there have been differential measurements for inclusive[‡] $g \to b\bar{b}$ [4,5], as well as in association with a $Z$ boson [1,2,3]. While these are very important measurements that lay the foundation for the future, they are limited in the range of $\Delta R$ between the $b$-jets/hadrons/quarks. In particular, the interesting regime $\Delta R \in [0, R]$, for $R \sim 0.4$-$0.5$ the jet radius parameter, is only a single bin of these measurements. Furthermore, the only existing measurements use only the $\sqrt{s} = 7$ TeV data, which was limited in size and thus the energy reach is small compared to the possibilities with the $\sqrt{s} = 8$ TeV Run 1 or the $\sqrt{s} = 13$ TeV Run 2 data. The Monte Carlo (MC) community has also made great strides since the $\sqrt{s} = 7$ TeV data, so unfortunately most of the comparisons with the existing measurements to simulation predictions use deprecated setups. Significant advances in close-by $b$-tagging by ATLAS and CMS since these initial measurements will allow for interesting measurements of the low $\Delta R$ region in the future. A discussion of some of these improvements appears in Sec. 2, where the efforts have been focused on developing powerful boosted $h \to b\bar{b}$ (and $t \to bjj$) tagging techniques. Before a discussion of those innovative algorithms, the next section documents some aspects of the small angle $g \to b\bar{b}$ that would be interesting to measure with more data, higher energy, smaller angles, and state-of-the-art simulations.

---

[*]Although, measurements on a jet containing $g \to b\bar{b}$ would be complicated by the $B$ hadrons that carry a significant fraction of the $b$ quark energy.

[†]There are also many interesting measurements from LEP, but they are very limited in statistics and energy. In addition, all of these measurements focus on the rate of $g \to b\bar{b}$ and not the differential kinematic properties.

[‡]Both collaborations also have measured $W+b(\bar{b})$ and $t\bar{t}+b(\bar{b})$, but there is little sensitivity in the public results to small angle $g \to b\bar{b}$; instead, the extra $b$ quarks are from hard splittings well described by fixed order matrix elements.

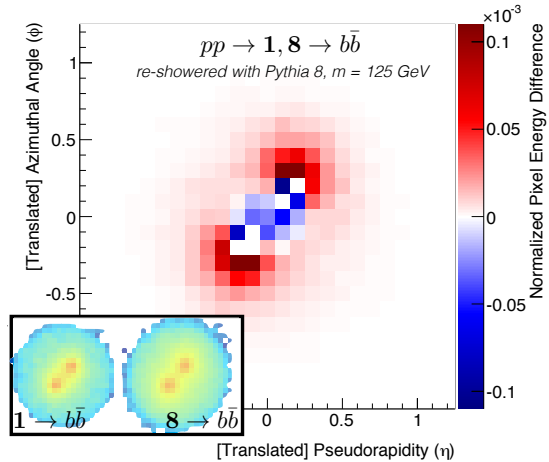

Figure 1: An illustration of the difference in radiation pattern between a jet formed by a color singlet (**1**) versus a color octet (**8**) scalar particle decay into quarks. The inset shows the average radiation pattern and the main plot is the difference between the two averages. The same hard decay is re-showered many times with Pythia 8 for both the singlet and octet cases.

## Future Measurements

In addition to the probing the octet radiation pattern as described in the introduction (and conceptually similar to a recent ATLAS measurement of colorflow in $W \to qq'$ [6]), there are several aspects of the $g \to b\bar{b}$ decay that would provide nearly direct measurements of the fragmentation function. Three key variables are $\Delta R(b, \bar{b})$, $\rho_{b\bar{b}} = m_{b\bar{b}}/p_{\mathrm{T},b\bar{b}}$, and $z_{b\bar{b}} = p_\mathrm{b}/p_{\mathrm{T,g}}$. None of the distributions of these quantities have been measured for $\Delta R(b, \bar{b}) < R$. Figure 2, shows the distribution of these three variables, as predicted by Pythia [7]. In addition to the nominal prediction, the sensitivity to variations in the modeling of fragmentation is illustrated by varying parameters in Pythia [8]. One of the key choices in this measurement is deciding to unfold to less observable quantities ($b$-quarks), more observable quantities ($b$-jets or $b$ track-jets), or somewhere in between ($B$-hadrons). The impact of this choice is illustrated in Fig. 3.

In addition to the QCD properties of the fragmentation, one can also make measurements of observables that are sensitive to the gluon spin, as in the angle between the gluon production and gluon 'decay' planes shown in Fig. 4.

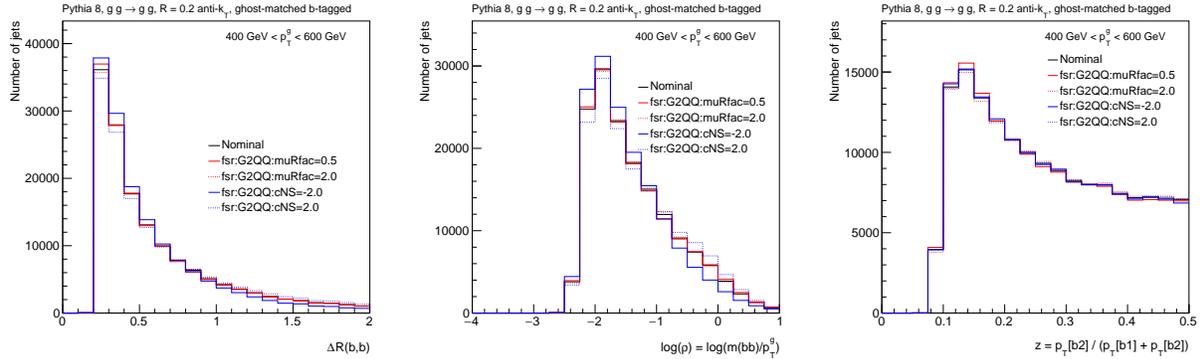

Figure 2: The distributions of $\Delta R(b,\bar{b})$, $\rho_{b\bar{b}} = m_{b\bar{b}}/p_{\mathrm{T},b\bar{b}}$, and $z_{b\bar{b}} = p_b/p_{\mathrm{T},g}$ in simulation along with a series of variations in the form of the fragmentation described by Pythia [8].

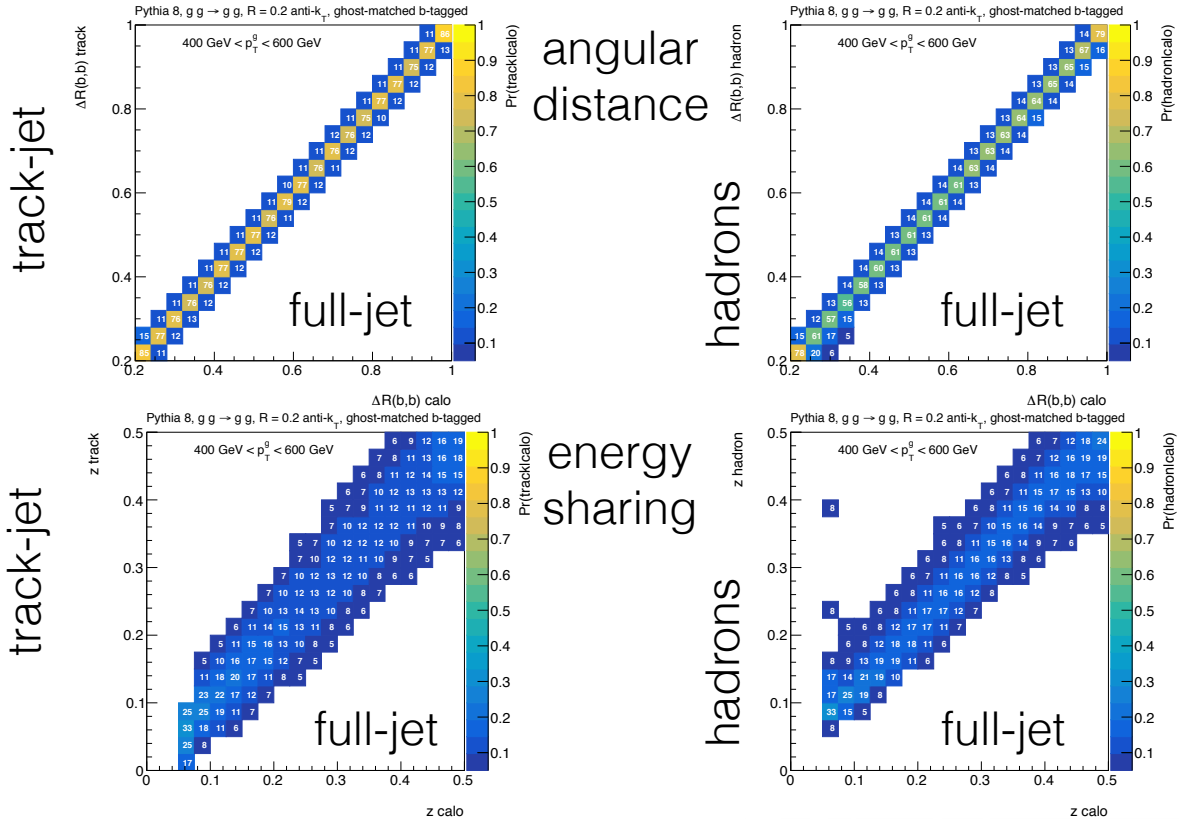

Figure 3: The two-dimensional distribution of two of the variables from Fig. 2, but using different definitions of 'b' (B-hadrons, b-jets, b-track-jets).

Overall, there is rich and largely unexplored region of phase space for $g \to b\bar{b}$ at low $\Delta R$ that will hopefully be an exciting part of the Run 2+ physics program at the LHC.

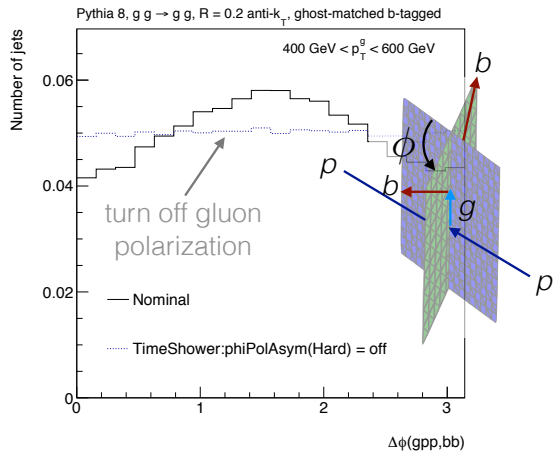

Figure 4: The angle $\phi$ that the $g \to b\bar{b}$ plane makes with respect to the $pp \to g$ plane. This distribution is sensitive to the gluon polarization.

## Higgs Tagging

Since early Run 1, most of the efforts related to gluon splitting have focused on discriminating boosted Higgs bosons from this QCD background process. Both ATLAS [10,11] and CMS [9] have developed sophisticated algorithms for improving $b$-tagging inside jets[§]. These algorithms make use of tracks and jet structure information in clever ways in order to maintain efficiency when the two $b$ quarks are close to each other, as it the case naturally for high $p_{T,\text{higgs}}$. There is no pure source of Higgs bosons to test these methods in data, but there are plenty of $g \to b\bar{b}$ events to assess the modeling of these new algorithms. ATLAS [12,13] and CMS [9,17] use template fits to study the efficiency of the taggers as well as the modeling of kinematic properties after selecting events enriched in $g \to b\bar{b}$. These results are public for both the $\sqrt{s} = 8$ TeV Run 1 data [13,17] and already for part of the Run 2 data [12,9].

In addition to improvements in (double) $b$-tagging inside jets, future measurements of $g \to b\bar{b}$ properties will be improved due to recent advances in the reconstruction of charged particle tracks inside the core of high $p_T$ jets. In this difficult region of phase space, pixel clusters merge, lowering the efficiency to reconstruct tracks. ATLAS [16] and CMS [18] have developed powerful techniques using pixel charge information to split clusters, recovering a significant amount of the efficiency loss even at high $p_T$.

## Conclusions and Outlook to FCC-ee

There is an exciting opportunity now for new measurements at the LHC that could provide powerful constraints on $g \to b\bar{b}$. Interesting in its own right as a fundamental QCD process, this is also a critical background for many searches and measurements involving Higgs bosons. Many of these analyses are forced to use data-driven techniques because the simulation is unreliable (or rather, unconstrained). If measurement-inspired simulation updates at the LHC and FCC-ee could change this situation, some of the most critical measurements at the LHC and FCC-hh such as $pp \to hh$ could be significantly improved.

---

[§]There is also an early Run 1 result from ATLAS for identifying single jets as containing a gluon splitting to $b\bar{b}$, but it has not used by any analysis.

In addition to providing a supporting role for Higgs and other measurements, $g \to b\bar{b}$ may also be useful for a direct comparison to $Z \to b\bar{b}$, and $h \to b\bar{b}$. In particular, these decays provide an opportunity to directly compare the impact of color and spin on the radiation pattern inside jets. This opportunity is only possible at an $e^+e^-$ machine that can directly produce Higgs bosons.

There is still a lot to learn about $g \to b\bar{b}$ at the LHC and there is a tremendous amount we could hope to study and apply at a FCC.

## Acknowledgments

I would like to thank Peter Skands and Zihao Jiang for many useful conversations.

# Colour Reconnections from LEP to Future Colliders

Torbjörn Sjöstrand[1]

[1] Department of Astronomy and Theoretical Physics, Lund University

**Abstract:** The phenomenon of colour reconnection (CR) is introduced, together with a selection of CR models and CR-related phenomena and observations.

## Introduction

Colour Reconnection (CR) was first discussed in the context of charmonium production [1],[2],[3], notably in weak $B$ decay to $J/\psi$, e.g. $\overline{B}^0 = b\overline{d} \to W^-c\overline{d} \to s\overline{c}c\overline{d} \to J/\psi \overline{K}^0$. In such decays the $c$ and $\overline{c}$ belong to two separate colour singlets, but ones that overlap in space–time, with the possibility of soft gluon exchange. Alternatively, colour algebra gives accidental $c\overline{c}$ colour singlets 1/9 of the time, but a dynamical principle would still be needed to override the original singlets.

The first large-scale application of CR was in the PYTHIA multiparton interaction (MPI) model of hadronic collisions [4], notably to explain the increasing mean transverse momentum $\langle p_\perp \rangle$ with increasing charged multiplicity $n_{\mathrm{ch}}$ observed at the $\mathrm{S}p\overline{p}\mathrm{S}$ [5]. If all MPIs draw out strings and fragment in the same manner, $\langle p_\perp \rangle(n_{\mathrm{ch}})$ would be essentially flat. CR was therefore introduced in such a way that the total string length is reduced. Each further MPI then on the average increases $n_{\mathrm{ch}}$ less than the previous one, while giving the same $p_\perp$ from (mini)jet production, resulting in an increasing $\langle p_\perp \rangle(n_{\mathrm{ch}})$.

The string length is conveniently described by the $\lambda$ measure [6], which is constructed such that $\lambda \propto \langle n_{\mathrm{hadrons}} \rangle \propto \langle n_{\mathrm{ch}} \rangle$ within the string model. For a simple $q\overline{q}$ string $\lambda = \ln(m_{q\overline{q}}^2/m_0^2)$, with $m_0 \approx 1$ GeV a measure of hadronic mass scale. The $\lambda$ measure becomes more difficult to evaluate for more complicated string topologies, and usually approximate expressions are used, like

$$\lambda \approx \sum_{i=0}^{n} \ln\left(1 + \frac{m_{i,i+1}^2}{m_0^2}\right) \; , \quad m_{i,i+1}^2 = (\epsilon_i p_i + \epsilon_{i+1} p_{i+1})^2 \; , \; \epsilon_q = 1 \; , \; \epsilon_g = \frac{1}{2} \; , \qquad (1)$$

for a string $q_0 g_1 g_2 \cdots g_n \overline{q}_{n+1}$, where $\epsilon_g = 1/2$ because gluon momenta are shared between two string pieces.

LEP 2 offered a good opportunity to search for CR effects. Specifically, in a process $e^+e^- \to W^+W^- \to q_1\overline{q}_2 q_3\overline{q}_4$, CR could lead to the formation of alternative "flipped" singlets $q_1\overline{q}_4$ and $q_3\overline{q}_2$, and correspondingly for more complicated string topologies. Such CR would be suppressed at the perturbative level, since it would force some $W$ propagators off the mass shell [7]. This suppression would not apply in the soft region, and a number of models were developed.

The main PYTHIA ones were scenarios I and II [7], which take their names from the analogy with type I and II superconductors. Strings are viewed as elongated bags in the former, and reconnection is proportional to the space–time overlap of these bags. In the latter, strings are instead imagined as vortex lines, and two cores need to cross each other for a reconnection to occur. In either case it is additionally possible to allow only reconnections that reduce $\lambda$, scenarios I' and II'.

Among other models, the ARIADNE ones were based on $\lambda$ reduction in combination with colour algebra restrictions [8],[9], whereas the HERWIG model acted to reduce the space–time size of clusters [10].

Based on a combination of results from all four LEP collaborations, the no-CR null hypothesis is excluded at 99.5% CL [11]. Within scenario I the best description is obtained for ∼50% of the 189 GeV $W^+W^-$ events being reconnected, in qualitative agreement with predictions.

As an aside, it was also proposed [12] that Bose–Einstein correlations between identical pions produced in the two $W$ systems could lead to further interconnection effects. Only $0.17 \pm 0.13$ of the predicted effect was observed [11], i.e. consistent with no effect at all, and at most giving a 7 MeV mass shift.

CR studies spread to HERA. Specifically, the Uppsala group described diffractive production in DIS, i.e. the presence of rapidity gaps in the hadronic system, in terms of CR [13]. This model was later extended also to other processes in $e^+e^-$ and $pp/p\bar{p}$, including rapidity gaps between jets and the production of gauge bosons. One main difference to the Lund approach is that minimization is imposed in terms of a string "area" $A \approx \sum m^2$ [14] rather than the $\lambda \approx \sum \ln m^2 = \ln \prod m^2$.

**CR models at the LHC**

The concept of CR has been well established also at the Tevatron and the LHC, e.g. by the same $\langle p_\perp \rangle (n_{ch})$ behaviour as at the S$p\bar{p}$S. Over the years, as the MPI modelling in PYTHIA has evolved, also new CR scenarios have been added. The detailed space–time picture of the LEP 2 models has been deemed too complicated and uncertain to apply to hadron collider events, so instead the reduction of the $\lambda$ measure has played a key role. In total PYTHIA 6 [15] came to contain twelve models, many of them involving annealing strategies to reduce $\lambda$.

In HERWIG++ [16] the default Plain CR considers all quark ends of clusters, and reconnects clusters $A$ and $B$ into $C$ and $D$ by a swap of the antiquark ends if $m_C + m_D < m_A + m_B$. If there are many possibilities open for cluster $A$, the one is picked which reduces the mass sum the most. The reconnection rate can be reduced by a probability $p_{\text{reco}}$ that an allowed reconnection is done. As an alternative, the Statistical CR minimizes the $\sum m_{\text{cluster}}^2$ by simulated annealing.

The current PYTHIA 8 [17] initially only contained one model. In it two MPIs can be merged with a probability $P = r^2 p_{\perp 0}^2 / (r^2 p_{\perp 0}^2 + p_{\perp \text{lower}}^2)$, where $r$ is a free parameter, $p_{\perp 0}$ is the standard dampening scale of MPIs, and $p_{\perp \text{lower}}$ is the scale of the lower-$p_\perp$ MPI. Each gluon of the latter MPI is put where it increases $\lambda$ the least for the higher-$p_\perp$ MPI. The procedure is applied iteratively, so for any MPI the probability of being reconnected is $P_{\text{tot}} = 1 - (1 - P)^{n_>}$, where $n_>$ is the number of MPIs with higher $p_\perp$.

A new QCD-based CR model [18] implemented a further range of reconnection possibilities, notably allowing the creation of junctions by the fusion of two or or three strings. A junction is a point where three string pieces come together, in a Y-shaped topology. The relative rate for different topologies is given by **SU(3)** colour rules in combination with a minimization of the $\lambda$ measure. The many junctions leads to an enhanced baryon production, although partly compensated by a shift towards strings with masses too low for baryon production. The model can explain some data but fails in other respects, see next presentation, by C. Bierlich.

Interestingly $t$, $Z^0$ and $W^\pm$ all have widths around 2 GeV, i.e. $c\tau \approx 0.1$ GeV. This means that their decays happen well after the (Lorentz-contracted) "pancakes" of the two incoming beams have passed through each other, and after the perturbative activity at scales above 2 GeV, but inside all the hadronization colour fields. The $t/Z/W$ decay products therefore have every chance of experiencing CR with the rest of the event.

Top mass determinations therefore have to take into account the uncertainty from our limited understanding of CR. As an example, the CMS measurement $m_t = 172.35 \pm 0.16 \pm 0.48$ GeV [19] involves an estimated systematic CR error of ±0.10 GeV, based on a comparison of the CR and noCR PYTHIA 6.4 Perugia 2011 tunes [20].

In order to provide an independent estimate, several new CR models were implemented in the PYTHIA 8 framework [21]. These fall in two classes. In the late $t$ decays one, ordinary CR is first carried out by the default description, with $t$ considered stable. After the subsequent $t$ and $W$ decays, the gluons from these can reconnect with the gluons from the rest of the event, using separate models. Some of these are intended to be straw-man ones, e.g. where random reconnections can occur, also when $\lambda$ increases. In the early decays class, the $t$ and $W$ decay products undergo CR on equal footing with the rest of the event. A gluon may be moved from one location to another, or two gluon chains may flip, i.e. reconnect with each other, or two gluons may be swapped. In either case a reduced $\lambda$ is required.

It is easy to shift the top mass downwards, by reconnecting the top decay products to particles outside the jet core and thereby broadening the jet profile, but more difficult to shift it upwards, since parton showers tend to select minimal $\lambda$ values from the onset. Extreme values can be excluded, however, since they would give too broad jet profiles and other problems. Restricting the models to acceptable parameter ranges, the resulting reconstructed mass range is around 0.5 GeV, i.e. ±0.25 GeV. This is in line with previous studies for the Tevatron [22], but now with a broader range of models.

### CR at future $e^+e^-$ colliders

The CR issues already noted for LEP 2 will reappear at any future high-energy $e^+e^-$ colliders. It will be especially relevant for the FCC-*ee*, with its high luminosities and resulting high precision. With the $W$ mass determined to better than 1 MeV by a threshold scan [23], the hadronic and semileptonic $WW$ channels can be used to probe the impact of CR. Some examples of how PYTHIA 8's CR scenarios shift the average reconstructed $W$ mass are shown in Table 1. A common trend is that effects are reasonably small near the threshold, then initially increase with energy, but eventually decrease as the $W$'s decay further apart. Most models also tend to shift the $W$ mass upwards, when away from the threshold region, but GM-I offers an interesting counterexample. The GM variants also nicely illustrate that different aspects of a CR model may go in opposite directions and partly cancel. The CS model, finally, is an example where mass shifts are tiny.

| $E_{\text{cm}}$ | $\langle \delta \overline{m}_{\text{W}} \rangle$ (MeV) | | | | | | |
|---|---|---|---|---|---|---|---|
| (GeV) | I | II | II$'$ | GM-I | GM-II | GM-III | CS |
| 170 | +18 | −14 | −6 | −41 | +49 | +2 | +7 |
| 240 | +95 | +29 | +25 | −74 | +400 | +104 | +9 |
| 350 | +72 | +18 | +16 | −50 | +369 | +60 | +4 |

Table 1: Reconstructed average $W$ mass shift for different CR models, relative to the no-CR baseline, at three different $e^+e^-$ CM energies [24]. The first three are the I, II and II$'$ models from LEP 2 days [7], the next three the "gluon move" model introduced for top mass studies [21], where I is only move, II is only flip and III is both, and finally CS is the new QCD-based model [18].

It should be stressed that this is for one (simple) mass reconstruction algorithm. Variations in the algorithm give somewhat different outcome, and thereby probe details of the models. Other measures can also be used, such as the particle flow between jets, or changes in the charged multiplicity as a function of topology. The prospects for pinning down the CR mechanism at the FCC-*ee* therefore are good.

An understanding of the CR not only is of interest in itself, but also for all kinds of precision studies. As an example we take the study of Higgs properties. In the Standard Model the 125 GeV Higgs is a pure $CP$-even state, but in various extensions there can be an $CP$-odd admixture, and an important task is to set stringent limits on this. One possibility is to study angular correlations in $H^0 \to W^+W^- \to q_1\bar{q}_2 q_3\bar{q}_4$ decays. The catch here is that CR also can shift jet directions, since the particle flow around a parton is biased in the direction towards its colour partner, by standard string effects. This can give rise to deviations that could be misinterpreted, unless CR is well understood [24]

## Summary and outlook

CR has been with us for 30 years, as a building block in the picture of multihadron production at high-energy colliders. Its existence has been convincingly demonstrated at LEP 2, but statistics was too small to allow any quantitative studies. The FCC-*ee* would allow detailed tests of the CR phenomenon, especially in the hadronic $WW$ channel, and the experience gained would help constrain the potential errors in other studies.

The picture is less clear for *pp* collisions, be it at the LHC or a future FCC-*pp*, where the busy environment not only allows much larger CR effects than in the relatively clean $e^+e^-$ setup, but also opens the way for many further poorly understood effects. Indeed, the LHC studies have revealed patterns more commonly associated with heavy-ion physics and quark-gluon-plasma formation, from the ridge effect [25] to the increase of strangeness production in high-multiplicity events [26]. These are not explained by the standard PYTHIA framework, with or without CR. A solution could be the fusing of several strings into colour ropes [27], as further described in the presentation by C. Bierlich. One consequence of the changing landscape is that what used to be considered a key proof of CR in *pp*, namely the rising of $\langle p_\perp \rangle(n_{\text{ch}})$, now could find alternative explanations, e.g. in terms of an increased string tension for closely-packed strings, or hadronic rescattering in a dense hadronic gas [28]. This does not mean that CR as such is in doubt, only that we may be faced with a cocktail of poorly understood effects, making further progress more challenging, but also invigorating the whole field of soft physics studies at hadronic collisions. To be continued ...

## Acknowledgements

Work supported in part by the Swedish Research Council, contract number 621-2013-4287, in part by the MCnetITN FP7 Marie Curie Initial Training Network, contract PITN-GA-2012-315877, and in part by the the European Research Council (ERC) under the European Union's Horizon 2020 research and innovation programme, grant agreement No 668679.

# Colour reconnections in pp collisions

Christian Bierlich[1]

[1] Lund University.

Work done in collaboration with Jesper Roy Christiansen, Gösta Gustafson, Leif Lönnblad and Andrey Tarasov.
Work supported in part by the MCnetITN FP7 Marie Curie Initial Training Network, contract
PITN-GA-2012-315877, and the Swedish Research Council (contracts 621-2012-2283 and 621-2013-4287)

**Abstract:** We discuss phenomenological results relating to Colour Reconnection, and review the Rope Hadronization formalism, including the ARIADNE/DIPSY final state swing. In this formalism, corrections to hadron flavour ratios in pp, have previously been calculated. We suggest new flavour ratio observables, based on event shapes, which are well suited for $e^+e^-$, and show that Colour Reconnection corrections can feasibly be measured at a new FCC-ee.

## Introduction

LHC have provided many interesting results related to Colour Reconnection (CR), here understood as an umbrella term for colour supressed corrections to final state parton shower and hadronization. We will discuss such phenomena in the context of the Lund string hadronization model [1][2]. In proton–proton collisions at the LHC, multiparton interactions (MPIs) are ubiquitous, leading to substantial space–time overlap of strings. At the same time effects, which in heavy ion collisions are linked to the formation of a Quark Gluon Plasma (QGP), have been observed in pp, such as enhanced production of strange hadrons [3] or ridges in the two–particle correlation function, linked to flow [4]. Common to these effects is that they are more pronounced in "central" collisions, *i.e.* collisions where a large number of particles in the forward direction are observed, since many produced particles correlates with many MPIs.

Corrections to the hadronization mechanism, with the aim of explaining QGP effects in pp, includes for example the core–corona model in EPOS [5], which imposes a macroscopic hydrodynamic model on dense events, QCD–based junction formation [6], a "microscopization" of hydrodynamics, by incorporating a thermal spectrum [7], as well as the Rope Hadronization model [8], the latter being the focus of these proceedings. Since the different models are based on quite different physical ideas, they can be reasonably expected to give different results for pp observables to which they are not tuned. In this way one can discriminate between the models. This was done *e.g.* for hadronic flavour ratios [9] and measured by ALICE [10], as well as several new underlying event observables [11], which are yet to be measured. It is clear from these studies that the models give different results, and that none of the models can yet fully describe soft pp collisions.

A future $e^+e^-$–collider could further illuminate the situation. Since CR effects require the presence of many strings in a small space–time volume to be visible, large effects are hard to come by in existing LEP data, due to limited statistics. With the large statistics expected from the new collider, one should be able to open up for new types of precision measurements.

## DIPSY and Rope hadronization

The DIPSY event generator [12] is built on an initial state cascade, based on Mueller's dipole formulation of QCD in impact parameter space and rapidity [13]. As such, full space–time information at the partonic level is accessible. After the initial state evolution and (multiple) interaction(s) between projectile and target, a final state shower is carried out by ARIADNE [14], where the space–time information is pertained. In figure 1, an example pp event (after final state shower) at $\sqrt{s} = 7$ TeV is shown in impact parameter space and rapidity. The coloured tubes are strings, and all string ends are quarks. Kinks on the strings are gluons. The event is finally hadronized by PYTHIA 8 [15], modified to include rope corrections.

When a number of string segments overlap, the colour charges of the string ends can act together coherently to form a "rope", which is conveniently described as an SU(3) multiplet, with two quantum numbers $p$ and $q$, given by the amount of overlapping segments with colour flow in each direction. Several possible multiplets can arise from combining strings. Following the work by Biro et al. [16], we let each individual rope form using a random walk procedure in colour space. Such a random walk can lead to three qualitatively different endpoints. Junction formation, singlet formation and the highest multiplet. We will only treat the two latter here, and refer to ref. [8] for a description of the junction treatment.

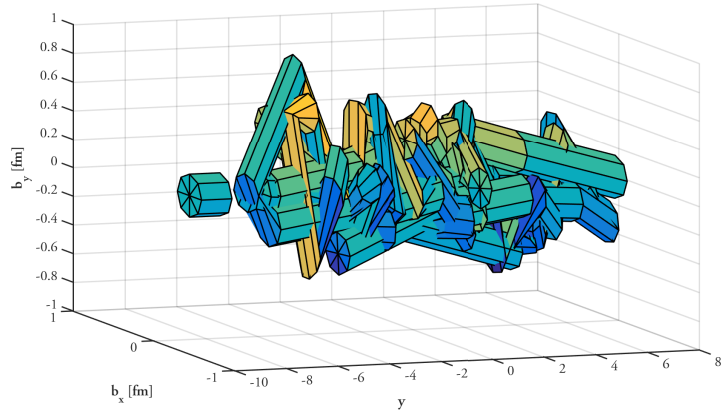

Figure 1: Example event from DIPSY in impact parameter space and rapidity, shoving sizable overlap of strings. Note that the transverse string radius in the picture is set to 0.2 fm, as opposed to 0.7 fm in the simulation, in order to improve readability of the figure.

**Singlet formation**

We address the possibility of singlet formation (*i.e.* the $\mathbf{1}$ in $\mathbf{3} \otimes \mathbf{\bar{3}} = \mathbf{8} \oplus \mathbf{1}$) using a final state swing already in the shower. In the shower, a dipole emits with a probability:

$$\frac{d\mathcal{P}_e}{d\ln(p_\perp^2)} \approx dy \frac{C_F \alpha_s}{2\pi}. \tag{1}$$

If there are colour compatible dipoles present in the event, they are allowed to recouple, competing with the emission, with a probability of:

$$\frac{d\mathcal{P}_r}{d\ln(p_\perp^2)} = \lambda \frac{(\vec{p_1} + \vec{p_2})^2 (\vec{p_3} + \vec{p_4})^2}{(\vec{p_1} + \vec{p_4})^2 (\vec{p_2} + \vec{p_3})^2}, \tag{2}$$

where $p_i$ are the parton momenta and $\lambda$ is a parameter. The swing greatly affects $\langle p_\perp \rangle (N_{ch})$, which in pp collisions is often linked to CR between different MPI systems (see also T. Sjöstrand, these proceedings). In figure 2 we show this observable as measured by ATLAS [17]. We see that the swing increases the $\langle p_\perp \rangle(N_{ch})$ dependence, as expected. For comparison, PYTHIA 8 with CR is included. In Pythia, which does not include any correlations between the MPIs from the initial state, CR is necessary to get any $\langle p_\perp \rangle(N_{ch})$ dependence. Since DIPSY already includes such correlations from the initial state model, it shows such a dependence even before including CR in the form of the swing.

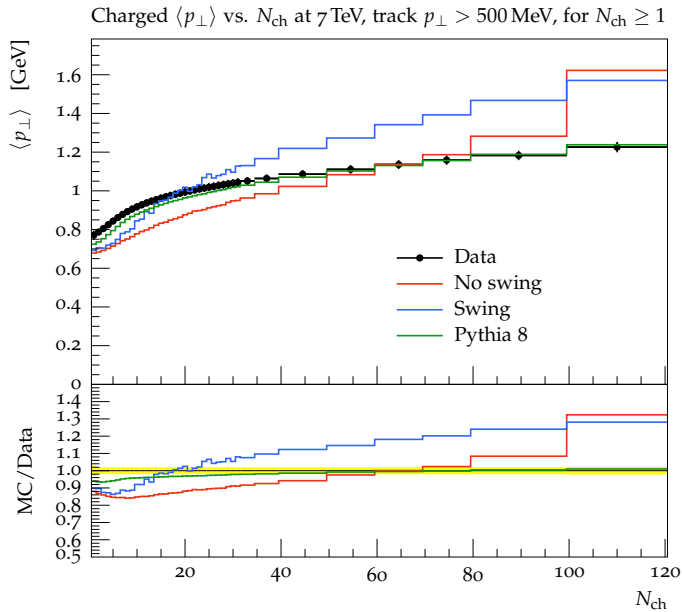

Figure 2: Average $p_\perp$ as function of $N_{ch}$, as measured by ATLAS, compared to DIPSY with and without swing, as well as Pythia 8.

**Highest multiplet**

The string tension of the rope has been calculated on the lattice [18], and is given by the secondary Casimir operator of the multiplet. The enhancement of string tension in a multiplet is thus:

$$\frac{\tilde{\kappa}}{\kappa} = \frac{C_2(p,q)}{1 \text{ GeV/fm}}. \tag{3}$$

In the string hadronization model, the string breaks by tunneling of a new $q\bar{q}$ pair with a tunneling probability of:

$$\frac{1}{\kappa}\frac{d\mathcal{P}_q}{d^2 p_\perp} \propto \exp(-\pi m_{\perp q}^2/\kappa) = \exp(-\pi p_\perp^2/\kappa)\exp(-\pi m_q^2/\kappa). \tag{4}$$

From the latter part, it is clear that production of heavier quarks will be suppressed with respect to lighter quarks. When the string tension is enhanced, the suppression vanishes. In the case of strange quarks with respect to up and down we call the suppression factor $\rho$, and one obtains directly for the modified suppression factor ($\tilde{\rho}$):

$$\tilde{\rho} = \exp\left(-\frac{\pi(m_s^2 - m_u^2)}{\tilde{\kappa}}\right) = \rho^{\frac{\kappa}{\tilde{\kappa}}}. \tag{5}$$

We therefore expect an increase in the amount of strange hadrons relative to non–strange with increasing string tension – and thus increasing event activity – which is also observed in pp data [10].

**New observables for $e^+e^-$**

It would be interesting to repeat the measurements of flavour ratios as function of event activity in $e^+e^-$. We focus here on $Z \to q\bar{q}$ events. Due to the absence of MPIs, a large number of produced particles in a $Z \to q\bar{q}$ event does not necessarily correspond to large overlap of strings, but can just as well correspond to a very long initial dipole. We suggest instead to look at event shape observables as a proxy for string overlap, also suggested for CR studies [19], not related to strangeness, in pp. The sphericity is defined as the linear combination $\frac{3}{2}(\lambda_2 + \lambda_3)$ of the second and third eigenvalues of the sphericity tensor. The sphericity approaches 1 for an isotropic event, which consists of many small gluon emissions, rather than two jets. With many small emissions, some strings are bound to be on top of each other, giving rise to rope hadronization effects.

In figure 3 we show the ratio of $K_s^0, \phi$ and $\Omega$ respectively to pions, in $e^+e^- \to Z \to q\bar{q}$, plus shower, with ARIADNE/DIPSY and Rope Hadronization. We ran $10^9$ events, roughly similar to the expected statistics of a new FCC-ee machine.

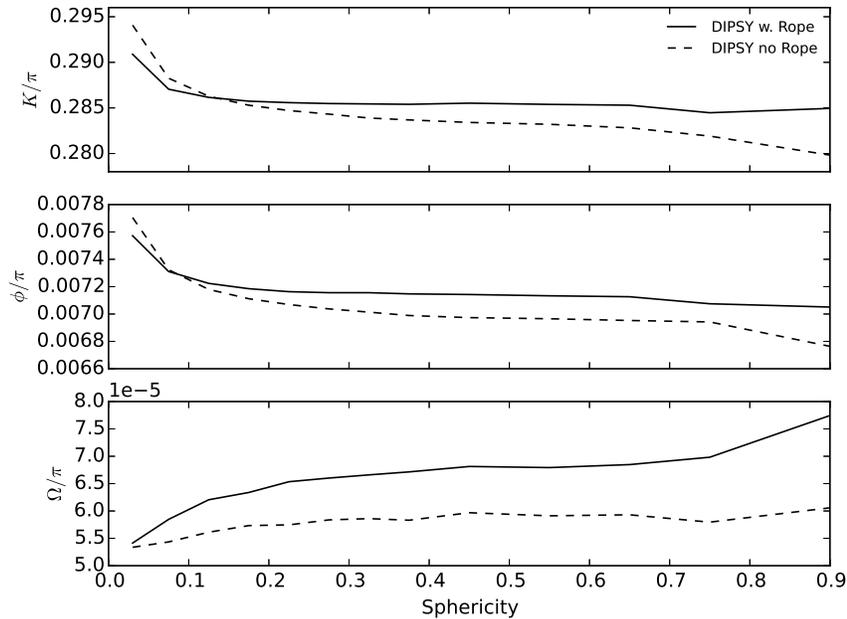

Figure 3: Hadron flavour ratios as function of sphericity in $e^+e^- \to Z \to q\bar{q}$. We show $K_s^0$ (single strange), $\phi$ (double strange) and $\Omega$ (triple strange) compared to pions.

It is seen that Rope Hadronization gives a sizeable effect in all three observables.

**Outlook**

We have briefly reviewed Rope Hadronization and the final state swing in DIPSY/ARIADNE, and how these mechanisms lead to increased $\langle p_\perp \rangle (N_{ch})$ and increased strangeness as function of $N_{ch}$ in pp collisions. We have argued that $N_{ch}$ cannot be directly translated to a proxy for string overlap in $e^+e^-$, and suggest the use of event shape observables instead. We show that effects on flavour composition from Rope Hadronization could feasibly be measured at an FCC-ee.

We expect that more interesting measurements relating to CR at an FCC-ee will surface, as CR is still an active, and not yet fully understood, topic in pp. New developments of the Rope Hadronization model [20] indicates that features similar to those observed in long–range near–side angular correlations (the ridge) can be obtained by allowing the excess energy in string overlaps generate a transverse pressure. If such a model can explain QGP–like features in systems as small as pp, it would not be far fetched to suggest a dedicated QGP programme at a new FCC-ee.

# Baryon Production and Correlations from LEP to FCC-ee

Stefan Kluth[1]

[1] MPI für Physik, München, Germany

**Abstract:** We report on two analyses of correlated baryon production by OPAL. The measurements of correlations in baryon production are sensitive to details of hadronisation models implemented in the Monte Carlo generators. The first analysis measures rates and kinematic distributions of correlated $\Lambda\overline{\Lambda}$ pairs. The second analysis measures fractions of correlated $\Sigma^-$ anti-hyperon pairs.

## Introduction

In this contribution we concentrate on OPAL analyses since these are known to the author in more detail than other compareable analyses. Following the suggestion of the conveners we first report on a study of parton framentation using rapidity differences of $\Lambda\overline{\Lambda}$ baryon pairs produced in hadronic final states of $e^+e^-$ annihilation. Then follows a presentation of the analysis of $\Sigma^-$ anti-hyperon correlations, which studies particle production only and is thus less model dependent compared to studies relying on kinematic quantities with corresponding modelling uncertainties.

## The OPAL experiment

The OPAL detector [1] operated from the year 1989 to 2000 as one of the four large experiments at the $e^+e^-$ collider LEP at centre-of-mass energies from around 91.2 GeV to the highest LEP 2 energies of 209 GeV. The analyses presented here use the large data set recorded at 91.2 GeV only. The OPAL detector had a large tracking chamber (jet chamber) in a solenoid magnetic field of 0.435 T complemented by Silicon micro-vertex and wire-based vertex detectors, and so-called Z-chambers to measure the coordinate parallel to the beam direction for tracks leaving the jet chamber. Tracks could be measured with $p_t > 150$ MeV.

The analyses use the complete data set of almost $4 \cdot 10^6$ hadronic Z decays recorded by OPAL during the LEP runs on the Z peak. For experimental corrections large samples of simulated events generated with JETSET 7.4 [2] or HERWIG 5.9 [3] and passed through a simulation of the OPAL detector [4] are used.

## $\Lambda\overline{\Lambda}$ correlations

The analysis [5] focuses on baryon number compensation in jet fragmentation using $\Lambda\overline{\Lambda}$ pairs. Hadronisation models describe the transition from the partons (quarks and gluons) of the hard interaction into the hadrons measured in the experiment. The model implemented in the Monte Carlo generator program JETSET (today replaced by Pythia) is based on a chainlike production with local compensation of baryon quantum numbers. In the HERWIG generator programs hadronisation is modelled using isotropic decays of clusters with conservation of baryon quantum numbers. This implies a correlation of kinematic quantities of corresponding baryons and anti-baryons as probes to check the valididy of the models.

In the analysis $\Lambda$ baryons are reconstructed in their pion-proton decay channel as a displaced opposite sign vertex. In the OPAL jet chamber measurements of the specific energy loss dE/dx of charged particles in the chamber gas are used to improve the purity of the sample.

As kinematic observable the decay angle $\theta^*$ of the $\Lambda$ in the rest frame of the $\Lambda\overline{\Lambda}$ system is measured. In addition the rapidities $y = 1/2 \ln((E+p_\parallel)/(E-p_\parallel))$ of the $\Lambda\overline{\Lambda}$ pair, where $p_\parallel$ is the momentum parallel to the thrust axis, and their difference $|\Delta y|$ are measured.

The numbers of selected events with $\Lambda\overline{\Lambda}$ pairs are corrected in bins of $|\Delta y|$ and $\cos\theta^*$ for background, efficiency and acceptance effects. The number of correlated pairs is derived from the total number of pairs by removing accidental pairs as $N^{cor.} = N_{\Lambda\overline{\Lambda}} - (N_{\Lambda\Lambda} + N_{\overline{\Lambda\Lambda}})$. In the sample of almost $4 \cdot 10^6$ hadronic Z decays 5262 correlated $\Lambda\overline{\Lambda}$ are estimated.

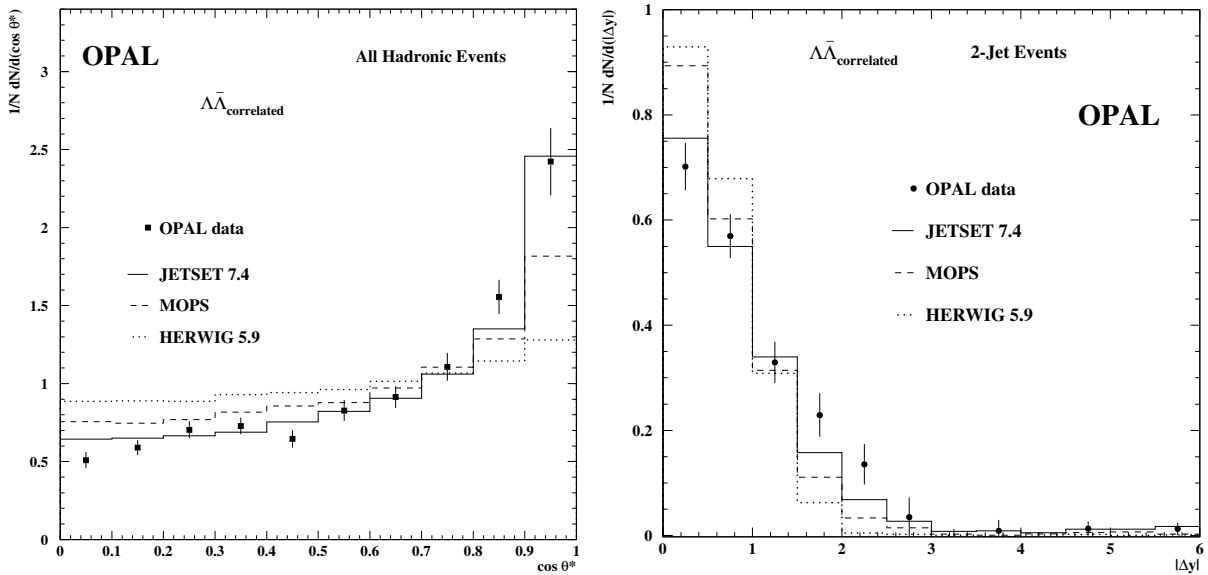

Figure 1: (left) Distribution of $\cos\theta*$ for correlated $\Lambda\overline{\Lambda}$ pairs. Superimposed are predictions by Monte Carlo generators as indicated. (right) Distribution of rapidity difference $|\Delta y|$ for correlated $\Lambda\overline{\Lambda}$ pairs in 2-jet events (see text). Both figures from [5].

Figure 1 (left) shows the $\cos\theta*$ distribution for correlated $\Lambda\overline{\Lambda}$ pairs compared with predictions by Monte Carlo generators. The JETSET 7.4 program gives the best description of the data. Figure 1 (right) presents the distribution of $|\Delta y|$ for correlated $\Lambda\overline{\Lambda}$ pairs in 2-jet events selected with $y_{cut} = 0.005$ in the Durham algorithm [6]. The JETSET 7.4 generator is in reasonable agreement with the data while the other models are not successful.

## $\Sigma^-$ anti-hyperon correlations

The analysis of $\Sigma^-$ anti-hyperon correlations by OPAL [7] uses $\Sigma^-$ baryons to tag events, since with strangeness and baryon number conservation a $\Sigma^-$ must be compensated by a anti-nucleon and a Kaon or a anti-hyperon. The analysis measures only rates and does not use kinematic quantities in order to stay as model-independent as possible.

The main production mechanism in the string hadronisation model [2] is the diquark model, where a pair of virtual quark-anti-quark pairs (diquark) together with single quark-anti-quark pairs produces a pair of baryons with compensating quantum numbers. In addition the popcorn mechanism [8] allows for a diquark with an intermediate quark-anti-quark pair to produce a baryon pair. The relative fraction of diquark and popcorn mechanism in baryon production is adjustable in the model.

In the analysis the fraction

$$F_{\overline{H}} = F_{\Sigma^-\overline{\Sigma}^-} + F_{\Sigma^-\overline{\Lambda}} + F_{\Sigma^-\overline{\Xi}^-} \qquad (1)$$

of $\Sigma^-$ in association with $\overline{\Sigma}^-$, $\overline{\Lambda}$ or $\overline{\Xi}^-$ is measured. These three correlations are accessible both by diquark and the popcorn mechanism and can be measured with sufficiently large efficiency.

The tag $\Sigma^-$ hyperons are reconstructed exclusively with a search for charged particle tracks with kinks assuming the decay $\Sigma^- \to n\pi^-$, where the neutron is not reconstructed. Two signatures for the correlated anti-hyperons are then searched for. The first signature is for $\overline{\Lambda}$ in the channel $\overline{\Lambda} \to \overline{p}\pi^+$ producing a secondary vertex. The flight direction displacement w.r.t. the beam direction of the $\overline{\Lambda}$ candidate allows to separate direct production from contributions of $\overline{\Xi}$ decays.

The second signature are charged pion tracks with significant displacement from the beam direction, since such pions are mostly produced in weak anti-hyperon decays. At this stage 9965 (10769) like-sign $\Sigma^-\pi^-$ and 11951 (13818) unlike-sign $\Sigma^-\pi^+$ pairs are found in the data (JETSET 7.4 based simulation). After normalising the simulation to the number of events in the data the difference between like-sign and unlike-sign pairs is larger in the simulation by $(28 \pm 7)\%$. This implies that the simulation contains too many correlated anti-hyperons. The further analysis quantifies this further by measuring the quantity $F_{\overline{H}}$.

The tag $\Sigma^-$ candidates are binned in their invariant mass and the decay angle of the pion in the $\Sigma^-$ restframe. Then the correlated particles, i.e. $\overline{\Lambda}$ or charged pion, are measured as function of their displacement w.r.t. beam direction. These measurements are compared with predictions from simulation of all relevant sources of anti-hyperon decays and the corresponding fractions are extracted.

The final results are

$$F_{\Sigma^-\overline{\Sigma}^-} = 0.17 \pm 0.11; \quad F_{\Sigma^-\overline{\Lambda}} = 0.057 \pm 0.056; \quad F_{\Sigma^-\overline{\Xi}^-} = 0.25 \pm 0.08; \quad F_{\overline{H}} = 0.48 \pm 0.10 \ . \qquad (2)$$

The error of the total fraction $F_{\overline{H}}$ is smaller than those of the components due to correlations.

These results are compared to predictions of the JETSET 7.4 generator with different configurations of the hadronisation model and in particular with or without the popcorn mechanism. A good decription of the data by the Monte Carlo generator can be obtained with for a large contribution of the popcorn mechanism to baryon production. Predictions without popcorn mechanism are inconsistent with the data.

### Conclusion and outlook

Since the analyses compared with predictions of now outdated Monte Carlo generators the question remains, how well modern Monte Carlo generator programs can describe the data of $\Lambda\overline{\Lambda}$ correlations and $\Sigma^-$ anti-hyperon correlations.

With much larger data samples of a future $e^+e^-$ collider the measurements could be improved significantly. The OPAL measurements of the $\Sigma^-$ anti-hyperon correlations are dominated by statistical uncertainties. With samples of about $4 \cdot 10^8$ hadronic Z decays the statistical uncertainties would match the OPAL systematic uncertainties corresponding to total errors reduced by a factor of about 10. The distributions measured with correlated $\Lambda\overline{\Lambda}$ pairs have small systematic uncertainties compared to the statistical errors and would thus profit directly from larger data samples. The total rates of $\Lambda\overline{\Lambda}$ pairs have systematic uncertainties of about the same size as the statistical errors and thus the impact of larger data samples is less direct. However, statistical errors of the data and simulation samples would be much smaller and in turn many systematic uncertainties can be studied in more detail probably leading to a reduction.

The requirements for detectors at future $e^+e^-$ colliders from the point of view of the analyses presented above can roughly be lined out as follows. The experiments should have particle identification capabilities at least at the level of the LEP experiments. The performance for charged particle momentum measurement and displaced vertex or kink reconstruction should be comparable or better than the LEP experiments. The data taking schedule should provide a sufficiently large sample of $O(10^8)$ Z decays.

# Bose-Einstein and Fermi-Dirac Correlations at LEP

Wesley J. Metzger

IMAPP, Radboud University, Nijmegen, Netherlands

**Abstract:** Results from LEP on Bose-Einstein and Fermi-Dirac correlations are briefly reviewed and suggestions are made of interesting questions that could be answered with the high statistics expected with the FCC-ee.

## Introduction

The interest in Bose-Einstein correlations (BEC) or Fermi-Dirac correlations (FDC), which are dependent on the relative momenta of identical bosons or fermions, respectively, is that they are related to the positions in space-time of the creation of the particles. However, this relationship is dependent on the functional form of the distribution of these points in space-time.

The correlation function is, in principle, a function of the particles' momenta. Since the particles are on the mass shell, we have, *e.g.*, that the two-particle correlation is given by

$$R_2 = \frac{\rho_2(\vec{p}_1, \vec{p}_2)}{\rho_1(\vec{p}_1)\rho_1(\vec{p}_2)} \quad (1)$$

where $\rho_n$ is the n-particle number density.

With the assumption that the particles are produced incoherently, the two-particle correlation function, $R_2$, is a function of the relative momentum, $\vec{Q} = |\vec{p}_1 - \vec{p}_2|$. Making the simplest assumption of a static source, *i.e.*, no time dependence, with a spherically symmetric distribution of creation points, the source distribution depends only on the radial coordinate: $S(\vec{r}, t) = S(r)$, and $R_2$ depends only on the invariant relative momentum $Q$:

$$R_2(Q) = 1 \pm \lambda |\tilde{S}(Q)|^2 \quad (2)$$

where $\tilde{S}$ is the Fourier transform of $S$. The plus sign holds for bosons, the minus sign for fermions. Although $\lambda$ is in principle unity, it is added *ad hoc* in an effort to account for the possibility that production is not entirely incoherent or for other effects that reduce the correlations, *e.g.*, contamination of non-identical particle pairs or multiple (distinguishable) sources such as long-lived resonances.

The source function is usually assumed to be Gaussian with radius $r$, leading to

$$R_2(Q) = 1 \pm \lambda \mathrm{e}^{-(Qr)^2}(1 + \epsilon Q) \: . \quad (3)$$

It is measured by

$$R_2(Q) = \frac{\rho_2(Q)}{\rho_0(Q)} \quad (4)$$

where $\rho_0$ is the two-particle density of a 'reference sample', which is constructed to be identical to $\rho_2$ except that there are no BEC (or FDC). Note the term $(1 + \epsilon Q)$ in Eq. 3, which is added to account for possible non-BEC (or non-FDC) not adequately removed by the reference sample.

**What have we learned from LEP?**

**1.** There is a large systematic dependence on the choice of reference sample. In the case of BEC in charged pions the two most common choices are opposite-sign pairs of particles and pairs of particles from different events (mixed events). The former suffers from the resonances present in $\pi^+\pi^-$ necessitating exclusion of a large range of $Q$ from the fits. For this reason the latter is generally preferable, but it too is not without problems. The two choices yield discrepant values of $\lambda$ and $r$, as shown in Fig. 1, where all the results with $r > 0.7$ fm were obtained using the unlike-sign reference sample while all those with smaller $r$ were obtained using the mixed event reference sample Also shown in this figure is a comparison of $r$ from BEC and FDC for charged pions, charged and uncharged kaons, (anti)protons and anti(lambdas). Given the discrepancies in values of $r$ for different reference samples, only values using the same type of reference sample should be compared. This indicates a value of about 0.55 fm for mesons and 0.1 fm for baryons rather than a $1/\sqrt{m}$ dependency as has been occasionally claimed [1].

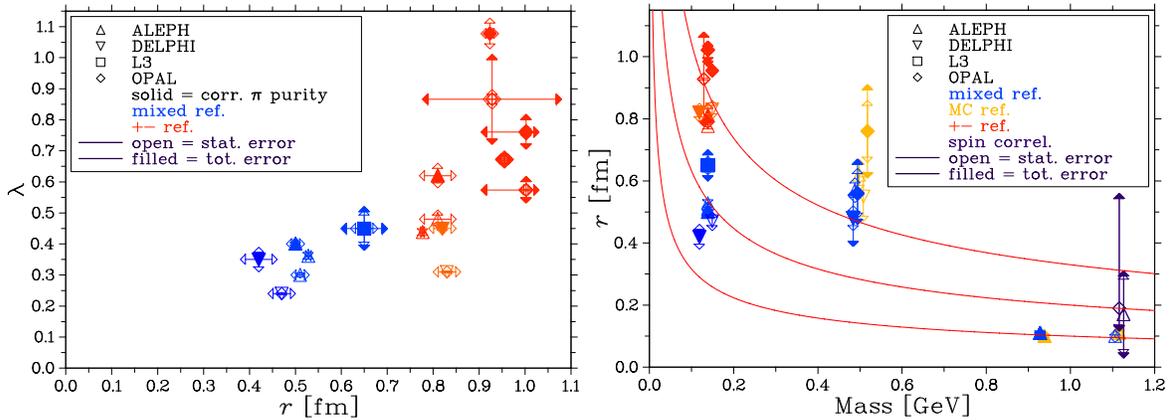

Figure 1: (left) $\lambda$ and $r$ for charged pions from 2-particle BEC at $\sqrt{s} = M_Z$ found in the LEP experiments [2,3,4,5,6,7,8]; (right) dependence of $r$ on the mass of the particle as determined at $\sqrt{s} = M_Z$ from 2-particle BEC for charged pions [2,3,4,5,6,7,8], charged kaons [9,10] and neutral kaons [11,9,12] and from Fermi-Dirac correlations for protons [11] and lambdas [13,14]. The curves illustrate a $1/\sqrt{m}$ dependence.

**2.** The simple picture sketched in the introduction is inadequate:

- $R_2$ is not Gaussian. An Edgeworth expansion[15] about a Gaussian or the assumption of a symmetric Lévy distribution[16] for the source provide (overwhelmingly) better fits to the BEC low-$Q$ peak.

- $R_2 \neq 1 + \lambda|\tilde{S}(Q)|$, i.e., $R_2$ is not equal to 1 + a positive-definite term. In the region $Q \approx 0.6$–1.3 GeV $R_2$ is less than unity, i.e., there is anti-correlation[16].

- $R_2 \neq R_2(Q)$, but rather $R_2(\vec{Q})$, as evidenced by unequal transverse and longitudinal radii in the LCMS[15,8,17,3]. This difference is smaller for 3-jet events than for 2-jet events (see Fig. 2).

- The BEC parameters, in particular $r$, depends on the jet structure of the event (see below).

**3.** Several aspects of BEC were investigated, but with large statistical uncertainty on their conclusions:

- Comparison with results from Petra/PEP and Tristan suggest that the value of $r$ in the Gaussian parametrization does not depend on $\sqrt{s}$ [18]. However, as seen in Fig. 1, the large disparity between the various measurements, which also exists at lower energies, prohibits a strong conclusion.

- The parameter $\lambda$ is less than unity if the particles are not produced completely incoherently. However, other effects also cause smaller values of $\lambda$ so that $\lambda < 1$ does not in itself imply any coherence. By also analyzing 3-$\pi$-BEC additional information on coherence is obtained. Genuine 3-$\pi$ BEC were found to exist, i.e., the amount of 3-$\pi$ BEC is greater than would be expected simply from the existence of 2-$\pi$ BEC [19,20,5]. Further, the results are compatible with complete incoherence, albeit with large statistical uncertaities[5].

- Naively, one might expect $r_{\pi^0\pi^0} \approx r_{\pi^\pm\pi^\pm}$. However, in models with local charge conservation two $\pi^0$s would be expected to be produced closer to each other than two charged pions of equal charge. On this question the results are contradictory [21,22], but both statistical and systematic uncertainties are large.

- The question of cross talk arose as a possible systematic effect in the determination of $M(\text{W}^\pm)$. If BEC exists between $\pi$'s from different jets, in particular between jets from different Ws, the apparent $M(\text{W})$ would be affected. The conclusion of the LEP experiments was that the data were compatible with no effect [23,24,25,26,27]. However this result had large statistical uncertainties as well as a large dependence on the available Monte Carlo models of BEC. Further, it depends on the selection criteria of the 4-jet events, which ensured well-separated jets. This minimizes the potential BEC, since it suppresses the possibility that particles from different jets have similar 3-momenta. Attempts to study this effect between different jets in 3-jet events proved inconclusive, as did studies of differences between BEC in gluon and quark jets[28,29].

### Selected topics in more detail

### Anticorrelation

The anti-correlation region mentioned above was first observed by L3 and found to be well-described by the $\tau$-model. In its simplified form $R_2$ is given by [16]

$$R_2(Q) = \gamma \left[ 1 + \lambda \cos\left((R_a Q)^{2\alpha}\right) \exp\left(-(RQ)^{2\alpha}\right) \right] (1 + \epsilon Q), \quad (5a)$$

$$\text{where} \quad R_a^{2\alpha} = \tan\left(\frac{\alpha\pi}{2}\right) R^{2\alpha}. \quad (5b)$$

Comparing this to the conventional parametrization, Eq. 3, with the Gaussian replaced by a symmetric Lévy distribution with index of stability $0 < \alpha \leq 2$,

$$R_2(Q) = \gamma \left[ 1 + \lambda \exp\left(-(rQ)^\alpha\right) \right] (1 + \epsilon Q), \quad (6)$$

we see that they essentially differ by the $\cos\left((R_a Q)^{2\alpha}\right)$ in the $\tau$-model expression. It is this factor which enables the anti-correlation.

Though similar in form, the interpretation of the parameters is very different. In the $\tau$-model it is the (proper) time distribution of particle creation which is described by a Lévy distribution rather than the spatial distribution as in the conventional interpretation; and consequently the parameters, $R$ and $r$, govern the 'width' of these distributions.

Not only LEP data, but also pp minimum bias data at the LHC are found to be described by Eq. 5a [30,31]. While Eq. 3 also describes the BEC peak, this parametrization (with $R_a$ a free parameter) describes the anti-correlation region as well.

In the $\tau$-model anti-correlation arises through the correlation of coordinate space and momentum space. Recently, another explanation has been proposed, namely the non-zero size of the pion [32,33]. The argument is basically that at separations less than the radius of a pion, one no longer has two pions, but only a bunch of quarks. Hence there is no Bose symmetry requirement. This introduces an anti-correlation between pions. Moreover, this anti-correlation occurs at approximately that value of $Q$ where it appears in the data.

This suggests, since Eq. 5a provides a good description of the data, that the cosine be regarded as an *ad hoc* factor. In this case, the parameter $R_a$ is regarded as a free parameter describing the anti-correlation, and $R$ and $\alpha$ are parameters of the spatial source distribution as in the conventional approach.

A good description of both the BEC peak and the anti-correlation region can also be achieved in a model-independent way by expanding the symmetric Lévy distribution in terms of Lévy polynomials [34,35]. However, this approach distributes the description of the anti-correlation region over two (or more) expansion coefficients rather than the single parameter $R_a$ of the $\tau$-model. The values of $R$ and $\alpha$ are also affected.

**Jet dependence**

In the conventional parametrization $r$ is found to increase with the jet multiplicity as well as with the track multiplicity [7]. This is also true of the parameter $R$ of the $\tau$-model [36]. However, since the track multiplicity increases with the jet multiplicity it is not completely clear which of the two is the primary source of the increase in size of the particle production region.

Recently the dependence of the size of the source on jet structure has been investigated more differentially. *e.g.*, in term of rapidity [37,38]. 'Jettiness' is defined in terms of $y_{23}$, the value of $y_{\text{cut}}$ in the JADE or Durham jet algorithm at which the classification of the event changes from 2-jet to 3-jet. Small values of $y_{23}$ correspond to narrow 2-jet events and large values to well-separated 3-jet events. The rapidity, $y_E$, is defined with respect to the thrust axis with the positive thrust axis chosen to be in the same hemisphere as the most energetic jet. Then $y_E > 1$ selects almost always tracks from the most energetic quark jet, and $y_E < -1$ selects mostly tracks from the other quark jet with the contribution of tracks from the gluon jet increasing as $y_{23}$ increases, *i.e.*, as the events become more 3-jetlike. The intermediate $y_E$ region contains tracks from the gluon jet and low-energy tracks from both quark jets. One observes, Fig. 2, that $R$ is roughly independent of $y_{23}$ for $y_E > 1$. This value of $R$ is also found for $y_E < -1$ in the case of two-jet events. These are the situations of 'pure' quark jets. As $y_{23}$ increases $R$ for $y_E < -1$ also increases, reflecting the increasing contribution of tracks from the gluon. The region $-1 < y_E < 1$ has a larger value of $R$ for two-jet events, and this value increases with $y_{23}$. For three-jet events, where the gluon contributes to both $-1 < y_E < 1$ and $y_E < -1$, the values of $R$ are approximately equal.

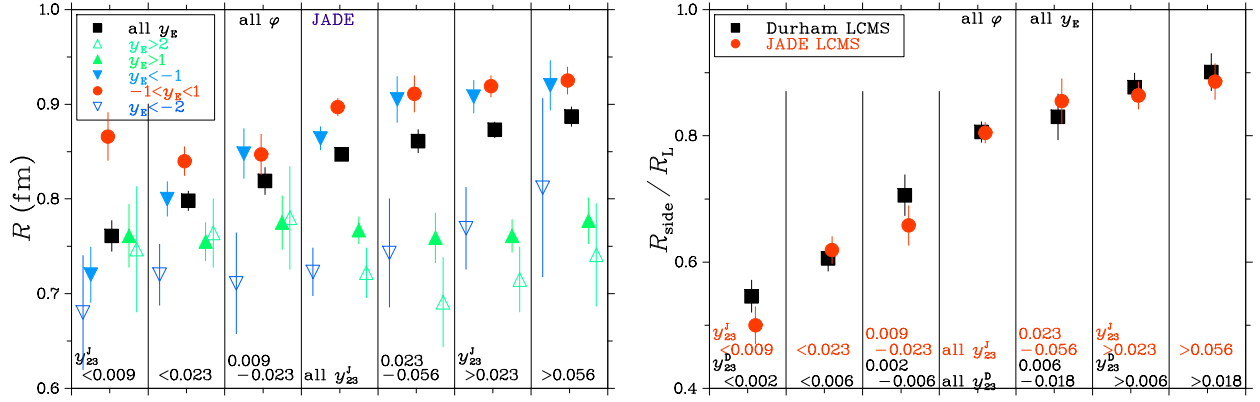

Figure 2: Left: $R$ obtained in fits of Eq. 5 for various rapidity and $y_{23}$ intervals (JADE jet algorithm). Right: Ratio $R_{\text{side}}/R_{\text{L}}$ in fits of a 3-D analog of Eq. 5 in the LCMS for various $y_{23}$ intervals

**Baryons**

As seen if Fig. 1, the FDC source radii found for baryons is very small, approximately 0.1 fm. This is an order of magnitude smaller than the radii of the baryons themselves, and thus seems unlikely. It has been pointed out [39,40] that the extraction of the FDC source radius is dependent on a realistic Monte Carlo event generator and that the generators are subject to large theoretical uncertainties, particularly in regard to dynamical correlations in baryon production. Anti-correlation effects, as seen in the pion case could further confound the measurement.

**Conclusions**

LEP has made a good start in investigating fragmentation with BEC.

- But statistics has limited the studies to 1-dimensional parmetrizations (in $Q$) or to very global 3-dimensional parametrizations (in $\vec{Q}$).

- A region ($Q \approx 0.6$–1.3 GeV of anti-correlation has been discovered. But its origin is unknown. Possibilities are a strong $x$-$p$ correlation (as in the $\tau$-model) or the size of the pion. Three-dimensional analyses for different topologies, *e.g.*, $y_{23}$, rapidity, transverse mass could help to solve this question.

- The correct parametrization is unknown. At present that of the $\tau$-model, or the model independent Lévy polynomial expansion appear to work best. But the $\tau$-model gives a parametrization dependent on $Q$, whereas the data show a difference of $r$ between longitudinal and transverse directions. Here too, three-dimensional analyses are required.

- The question whether $r$ depends on the mass of the particles remains open. The difference of $r$ between neutral and charged pion pairs and between mesons and baryons is relevant to the details of the hadronization model, as is any difference in BEC between quark and gluon jets.

**Desiderata for the FCC-ee**

As implied in the previous section, the first wish is a large increase in statistics to enable

1. narrower bins to better determine the parametrization;

2. more differential studies of event structure, *e.*g., measuring the components of $\vec{R}$ in the longitudinal, transverse in the event plane and out of the plane;

3. BEC dependence on the flavour of the quark jet; is BEC different for quarks and gluons?

LEP studies had 1–2 million Z events. To go from 1-D to 3-D analyses requires $N_{\text{bins}}^3$. For 100 bins this means $1\text{–}2 \cdot 10^{12}$ events, roughly corresponding to the expectation for the FCC-ee.

It is clear that the following are desired: $\pi/\text{K}/\text{p}$ identification; good acceptance; good track efficiency; good two-track resolution; good $\pi^0$ measurement; good $\text{K}^0$, $\Lambda$ measurement; and good b-tag efficiency.

# The Role of Quark Spin in Hadronisation

Hrayr H. Matevosyan[1]

[1] ARC Centre of Excellence for Particle Physics at the Tera-scale, and CSSM,
School of Chemistry and Physics, The University of Adelaide, Adelaide SA 5005, Australia

**Abstract:** The unpolarized and the Collins fragmentation functions (FF) quantitatively describe the hadronization of a polarized quark to unpolarized hadrons. They are needed for mapping the 3D structure of the nucleon from the semi-inclusive deep inelastic scattering experiments. We present our recent results in modeling the polarized quark hadronization in sequential hadron emission picture. Using the spin density matrix formalism, we describe the elementary $q \to q' + h$ process using the eight leading twist quark-to-quark transverse-momentum-dependent (TMD) elementary FFs. The unpolarized and the Collins FFs of light quarks to pions are then calculated using a Monte Carlo (MC) implementation of this model by utilizing the quark-jet framework for the sequential hadronization. We outline the the distinctive features of the resulting Collins FFs that reflect the underlying hadronization mechanism, such as the treatment of the intrinsic transverse momentum of the produced hadrons in the quark-jet framework. These polarized FFs can be precisely measured at FCC-ee, that would allow us to discriminate between different mechanisms for hadronisation, that in turn would provide a detailed description of various deep-inelastic hadron production processes.

## Introduction

One of the most challenging topics in high energy physics has been the description of the parton hadronization process because of its non-perturbative nature. The FFs that quantify the hadronization process are universal, in that according to the QCD factorization theorems they enter into the cross sections of various hard scattering processes with observed final state hadrons [1]. The so-called transverse-momentum-dependent (TMD) unpolarized FF can be interpreted as a number density for a quark to produce a hadron that carries a fraction of its light-cone momentum and a transverse momentum with respect to the momentum of the original quark. The modulations of this probability density for unpolarized hadrons produced by a transversely polarized quark is described by the so-called Collins FF [2]. The TMD FFs are needed to extract the TMD parton distribution functions, that encode the 3D structure of the nucleon in the momentum space, from semi-inclusive deep inelastic scattering experiments. One of the most widely used approach for describing hadronization is based on the Lund string model [3] and implemented in the Monte Carlo event generators PYTHIA [4]. Nevertheless, at present the polarized quark hadronization is not implemented in any MC event generator, and it is not possible to simulate the Collins effect.

Recently, we developed a self-consistent description of the polarized quark hadronization and a corresponding MC framework for calculating transversely polarized quark to pion FFs based on the extended quark-jet model [5],[6]. The quark-jet model describes the hadronization of a quark as a sequential emission of hadrons that do not interact with each other or re-interact with the remnant, as schematically depicted in Fig. 1. The quark to hadron fragmentation functions are then calculated as the corresponding number densities, either using integral equations or Monte Carlo techniques [7],[8],[9],[10],[11],[12],[13],[14]. Here we highlight the key findings of Refs. [5] and [6] in the perspective of future precise measurements of polarized FFs in FCC-ee experiment.

### Elementary $q \to q' + h$ process

We consider a quark hadronization mechanism where hadrons are produced one at a time in a sequential manner $q \to q_1 + h_1$, $q_1 \to q_2 + h_2$, etc. Thus, to describe this process we need to know both the probability density for the initial quark $q$ to produce a final quark $q_1$ of a given flavor and momentum, as well as how the polarization is transferred from $q$ to $q_1$. Let's denote the spin density matrices of $q$ and $q_1$ as $\rho_q$ and $\rho_{q_1}$ respectively, that are completely determined by the corresponding polarization vectors $\mathbf{s}_q$ and $\mathbf{s}_{q_1}$. The probability density for this transition can be expressed in terms of the respective density matrices $\rho_q$ and $\rho_{q_1}$,

$$f^{q \to q_1} = Tr[\rho_{q_1} A \rho_q \overline{A}], \tag{1}$$

where $A$ is some matrix describing the interaction with the other particles in this process. Then the probability density $f^{q \to q_1}$ should be a linear function in both $\mathbf{s}_q$ and $\mathbf{s}_{q_1}$,

$$f^{q \to q_1}(\mathbf{s}_q, \mathbf{s}_{q_1}) = \alpha_q + \beta_q \cdot \mathbf{s}_{q_1}, \tag{2}$$

where both $\alpha_q$ and $\beta_q$ are linear functions of $\mathbf{s}_q$ that also depend on the momenta of the quarks. We can express these coefficients in terms of the 8 leading-twist quark-to-quark TMD elementary FFs ( see Refs. [5],[6])

$$\alpha_q \equiv \hat{D}(z_1, p_{1\perp}^2) + \frac{(\mathbf{p}_{1\perp} \times \mathbf{s}_T) \cdot \hat{\mathbf{z}}}{z_1 \mathcal{M}} \hat{H}^\perp(z_1, p_{1\perp}^2), \tag{3}$$

$$\beta_{q\parallel} \equiv s_L \hat{G}_L(z_1, p_{1\perp}^2) - \frac{(\mathbf{p}_{1\perp} \cdot \mathbf{s}_T)}{z_1 \mathcal{M}} \hat{H}_L^\perp(z_1, p_{1\perp}^2), \tag{4}$$

$$\beta_{q\perp} \equiv \frac{\mathbf{p}'_{1\perp}}{z_1 \mathcal{M}} \hat{D}_T^\perp(z_1, p_{1\perp}^2) - \frac{\mathbf{p}_{1\perp}}{z_1 \mathcal{M}} s_L \hat{G}_T(z_1, p_{1\perp}^2) \tag{5}$$
$$+ \mathbf{s}_T \hat{H}_T(z_1, p_{1\perp}^2) + \frac{\mathbf{p}_{1\perp}(\mathbf{p}_{1\perp} \cdot \mathbf{s}_T)}{z_1^2 \mathcal{M}^2} \hat{H}_T^\perp(z_1, p_{1\perp}^2),$$

where $z_1$ and $\mathbf{p}_{1\perp}$ are the light-cone momentum fraction and the transverse momentum of $q_1$ with respect to $q$, while $\mathcal{M}$ is the mass of $q_1$. The momentum vector $\mathbf{p}'_{1\perp} \equiv (-p_{1,y}, p_{1,x})$. The unit vector $\hat{\mathbf{z}}$ denotes the direction of the 3-momentum of $q$, which also helps to define $\mathbf{s}_T$ and $s_L$ as the transverse and longitudinal components of $\mathbf{s}_q = (\mathbf{s}_T, s_L)$. In this work we use hats on TMD elementary FFs to distinguish them from the analogous TMD FFs .

Let us not that the quark $q_1$ is unobserved, then its polarization is completely determined by $\mathbf{s}_q$, $z_1$ and $\mathbf{p}_{1\perp}$. It can be expressed as $\mathbf{s}_{q_1} = \beta_q/\alpha_q$. The probability to produce quark $q_1$ with light-cone momentum fraction $z_1$ and transverse momentum $\mathbf{p}_{1\perp}$ is determined from Eq. (2), $\hat{f}^{q \to q_1}(z_1, \mathbf{p}_{1\perp}; \mathbf{s}_q) = \alpha_q$. The next fragmentation steps $q_1 \to q_2$ , can be treated in a completely analogous manner, where the results are expressed via light-cone momentum fraction $\eta_2$ and transverse momentum $\mathbf{p}_{2\perp}$ of quark $q_2$ relative to $q_1$. Nevertheless, since $\mathbf{s}_{q_1}$ itself is determined by $\mathbf{s}_q$, we can infer that $\mathbf{s}_{q_2}$ should also be completely determined by $\mathbf{s}_q$, as well as the light-cone momentum fraction $z_2$ and transverse momentum $\mathbf{P}_{2\perp}$ of quark $q_2$ with respect to $q$. Then, in the quark-jet framework, the probability of the $q \to q_2$ transition is given by

$$\hat{f}^{(2)}_{q \to q_2}(z_2, \mathbf{P}_{2\perp}; \mathbf{s}_q) = \hat{f}^{q \to q_1}(z_1, \mathbf{p}_{1\perp}; \mathbf{s}_q) \otimes \hat{f}^{q_1 \to q_2}(\eta_2, \mathbf{p}_{2\perp}; \mathbf{s}_{q_1}), \tag{6}$$

where the convolution $\otimes$ relates the corresponding relative momenta. We can then iterate this procedure for the subsequent fragmentation steps in a completely analogous manner.

## MC Implementation and Results

The iterative mechanism for the quark polarization transfer described in the previous section allows us to readily adapt the extended quark-jet framework for MC simulations of the polarized quark hadronization process with a finite number of produced hadrons, similar to our previous work in Refs. [9],[12], and [13]. The basic concept is to adapt the number density implementation of the FFs, which then can be calculated using Monte Carlo techniques as averages of these densities taken over a large number of quark hadronization event simulations. In the instance of polarized quark fragmentation into unpolarized hadrons, the corresponding number density is the following polarized fragmentation function:

$$D_{h/q^\uparrow}(z, p_\perp^2, \varphi) = D^{h/q}(z, p_\perp^2) - H^{\perp h/q}(z, p_\perp^2)\frac{p_\perp s_T}{z m_h}\sin(\varphi_C), \tag{7}$$

where $D^{h/q}(z, p_\perp^2)$ and $H^{\perp h/q}(z, p_\perp^2)$ denote the unpolarized and Collins fragmentation function, respectively. The variables $z$ and $p_\perp^2$ are the light-cone momentum fraction and the transverse momentum squared of the produced hadron with respect to the momentum of the initial fragmenting quark, and $m_h$ denotes its mass. Here, $s_T$ is the modulus of the transverse component of the quark's polarization. The Collins angle for the hadron $\varphi_C \equiv \varphi - \varphi_s$ is defined as the difference of the azimuthal angles of the produced hadron's transverse momentum $\varphi$ and the transverse polarization of the initial quark $\varphi_s$. We calculate $D_{h/q^\uparrow}(z, p_\perp^2, \varphi)$ by computing the average number of hadrons $h$ with given momenta produced in the hadronization chain of $q$. This can be accomplished by sampling the remnant quark's momentum according to the elementary quark-to-quark splitting functions, and calculating the type and the momentum of the produced hadron using flavor and momentum conservation. Then we calculate the polarization of the remnant quark using its momentum from the polarization of the fragmenting quark. We can continue the hadronization chain until we reach some predetermined termination condition, which we choose as a given number of produced hadrons $N_L$. The hadrons produced at the $n$th step in the hadronization chain are called rank-$n$ hadrons.

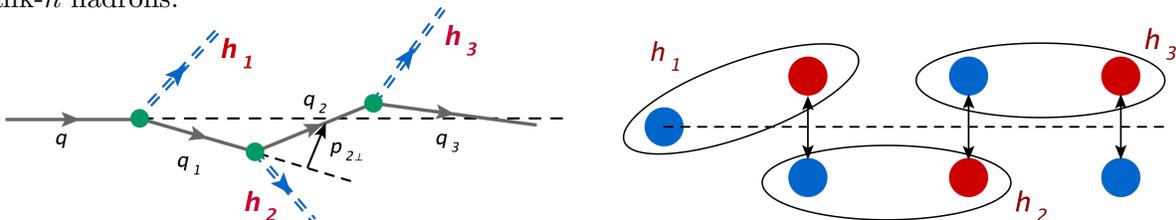

Figure 1: Schematic depiction of the transverse momentum generation for the extended quark-jet framework and the Lund model.

Here we point out a distinctive feature of the quark-jet model in describing the intrinsic transverse momentum of the produced hadrons, which is schematically depicted in the left panel of Fig. 1. In the $i$-th hadron emission step, we sample the light cone momentum fraction $\eta_i$ and the transverse momentum $\mathbf{p}_{i\perp}$ of the remnant quark $q_i$ with respect to the fragmenting quark $q_{i-1}$. The light-cone momentum fraction $z_i$ of the remnant quark with respect to the initial quark $q$ is given by the simple relation $z_i = \eta_1 \cdot ... \cdot \eta_i = z_{i-1} \cdot \eta_i$. To obtain the transverse momentum in the initial quark frame, we need to perform a Lorentz transform that preserves the light-cone momentum fraction. The resulting expression,

$$\mathbf{P}_{i\perp} = \eta_i\ \mathbf{p}_{i-1\perp} + \mathbf{p}_{i\perp}, \tag{8}$$

shows that the transverse momentum of the remnant quark (and the emitted hadron via momentum conservation), gets a contribution from the transverse momenta of the preceding quarks in the hadronization chain. The same is true for the produced hadrons, where the transverse momentum of the hadron emitted at a given step gets a contribution from the recoil of the transverse momenta of previously emitted hadrons. This is different in Lund model, where the different string breaks, that create the $q\bar{q}$ pairs, are causally disconnected. Thus, the direction of the string does not rotate after each quark pair creation, and only the transverse momenta of the hadrons of neighboring ranks can be correlated. For example, the transverse momenta of hadrons $h_1$ and $h_3$ in the left panel of Fig. 1 depicting the quark-jet model are correlated, while those in the right panel, schematically depicting the Lund model, are not.

The input for the MC calculations of the polarized FFs are the eight TMD elementary quark-to-quark FFs that describe the one step process and the polarization transfer. These elementary FFs have been modeled in Ref. [6]. Here we discuss the results for the calculations of the unpolarized and Collins functions of pions produced by an up quark. The isospin symmetry, assumed to be exact in the model, then can be used to extract the results for the down quark FFs. The plots in Figs. 2 show the unpolarized FFs, and the analyzing powers for an up quark fragmenting to pions. The results for the analyzing power of the Collins effect show the opposite sign for the favored and unfavored channels, and become equal in size at $z \simeq 0.2$. These are in agreement with recent the results by COMPASS, STAR and BELLE experiments that measured significant asymmetries at $z \simeq 0.2$ of opposite signs for the favored and unfavored FFs. In the future work, we can tune our results to best fit experimental data by changing the input elementary quark-to-quark FFs, as we have demonstrated in Ref. [6] a significant dependence of our results on these functions.

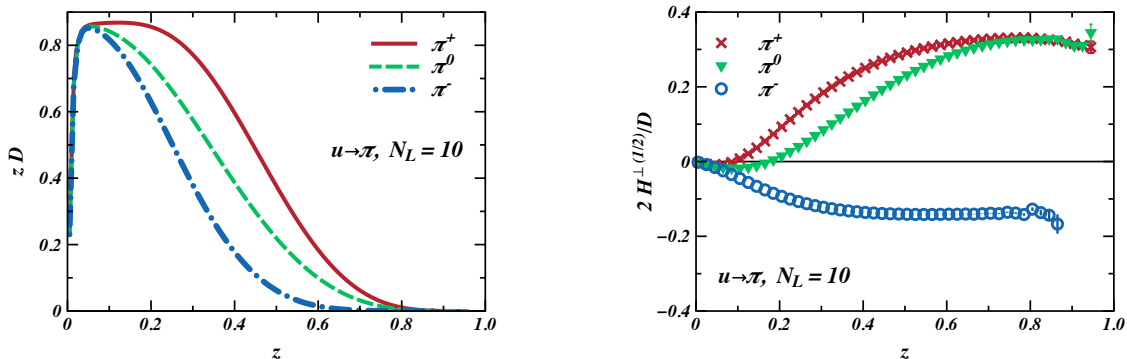

Figure 2: Fitted values of $zD$ (right panel), and $2H^{\perp(1/2)}/D$ (left panel) as a function of $z$ from Monte Carlo simulations for $u \to \pi$, with $N_L = 10$ emitted hadrons.

## Conclusions

The accurate description of the polarized quark hadronization process remains one of the most challenging aspects in the phenomenological description of deep inelastic scattering processes. For example, the treatment of the quark polarization and the corresponding correlations are, to date, not included in any of the well-known event generators, such as PYTHIA [4]. Here we presented a self-consistent model for polarized quark hadronization in an iterative setting, and the MC implementation of this model using the extended quark-jet hadronization framework, as first described in Refs. [5] and [6]. The MC approach was used to calculate the TMD polarized FFs of light quarks to pions, namely the unpolarized and Collins FFs. The results for the unpolarized FFs and the ratios of the 1/2 moments of the Collins functions to the unpolarized FFs were presented in Fig. 2. The analyzing powers demonstrated distinctive features: opposite sign for the large $z$ values for favored and unfavored channels. The results for the favored channel then fall off in magnitude more rapidly than the unfavored ones with decreasing $z$, and they cross the zero at some small $z$. These features reflect the underlying quark-jet hadronization mechanism, including the treatment of the hadron transverse momentum. It is also interesting to note that the shapes of the analyzing powers and the zero crossover points for the favored ones drastically depend on the forms of the input splitting functions [6]. The inclusion of the strange quarks and kaons, as well as the vector meson production and strong decays, will allow one to precisely describe a large range of phenomena that involve polarized quark hadronization. The computation of various polarized dihadron FFs will provide an improved set of predictions compared to our previous work [14] with a simplistic model. Further work on the model calculations of the input TMD FFs would give more predictive power to the framework which can be tested by precisely measuring the polarized FFs at FCC-ee in the future. At the same time, the polarization transfer mechanism used in this work can be readily adapted into the well-known MC event generators such as PYTHIA [4], with parametric forms for the input functions that can be tuned to best reproduce various experimental data.

**Acknowledgements** H.H.M. was supported by the Australian Research Council through the ARC Centre of Excellence for Particle Physics at the Terascale (CE110001104), by an ARC Australian Laureate Fellowship FL0992247 and Discovery Project No. DP151103101, and by the University of Adelaide.

# The Helix String

Šárka Todorova-Nová[1]

[1] LPSC, CNRS, Grenoble

**Abstract:** Helix string model has been originally proposed as an alternative to the one-dimensional string concept used by Lund fragmentation model in order to investigate gluon ordering of the end of parton shower cascade. A simple effective quantization of the helical string model is used to derive properties of QCD string from the mass spectrum of light mesons and to predict observable quantum effects in correlations between adjacent hadrons. The quantized fragmentation model is presented and compared with experimental observations.

## Introduction

The Lund string fragmentation [1], which is using a 1-dimensional string to model the QCD confinement, imposes a space-like distance between the string breakup vertices forming a hadron. The model relies on the concept of quantum tunneling to generate intrinsic transverse momenta of hadrons. The local charge and momentum conservation holds in the break-up vertex but there are no correlations, in the string transverse plane, between non-adjacent hadrons.

The situation changes substantially when the 1-dimensional string is replaced by a 3-dimensional string and the quantum tunneling turned into gluon splitting into quark-antiquark pair with a negligible momentum in the rest frame of the string. Fragmentation of such a string generates intrinsic transverse momenta which depends on the folding of the string and implies azimuthal correlations between hadrons. On the basis of angular properties of gluon emission under helicity conservation, it has been shown that the shape of the QCD string should be helix-like [2].

The helix-like shaped QCD string (or any 3-dimensional string) allows to develop fragmentation model where cross-talk between breakup vertices is possible (their distance is time-like). When the cross-talk (i.e. causal relation) between breakup vertices is imposed, the transverse shape of the string generates both transverse momentum and mass of the hadron. Quantization of the model then opens the way to build-up of hadron mass spectrum. It turns out that the causal constraint applied to the helical QCD field reveals a rather simple quantization pattern for particles with mass below 1 GeV. In particular, the pseudoscalar mesons $(\pi, \eta, \eta')$ can be regarded as string pieces fragmenting into (n=1,3,5) ground state hadrons (pions), with transverse mass ($E_T$) and momentum ($p_T$) of mesons defined by helical string properties [3]

$$E_T = \sqrt{m_n^2 + p_T^2} = n \; \kappa \; R \; \Delta\Phi, \qquad (1)$$
$$p_T = 2 \; \kappa \; R \; | \sin(n\Delta\Phi/2) |, \qquad (2)$$

where $R$ stands for the radius of the helix, $\kappa \sim 1$GeV/fm is the string tension, $\Delta\Phi$ is the quantized helix phase difference describing the minimal piece of string which can form a hadron, and $m_n$ is the (quantized) meson mass spectrum.

The mass spectrum of pseudoscalar mesons is used to extract the parameters of the helical QCD field and its quantum properties; the fit of the spectrum indicates a rather narrow radius of the helical string ($\kappa R$= 68 ±2 MeV) and the quantized phase difference $\Delta\Phi$ =2.82 ± 0.06 rad which translates into a quantized ground state transverse energy ($E_T|_{n=1} \sim 0.193$ GeV). It is the numerical value of $\Delta\Phi$ which suggests there should be a significant charge-combination asymmetry in the production of chains of ground state pions: while adjacent (opposite-sign) pairs of pions have to go apart in the transverse plane, the like-sign pion pairs with rank difference 2 should have a relatively small opening angle of $2.(\pi-2.8) \sim 0.7$ rad. Charge–combination asymmetry in the production of pairs of charged pions thus may be understood as a consequence of local charge conservation in (coherent) QCD string fragmentation into a quantized chain of ground state pions.

It is an interesting observation that the $\kappa R$ obtained from the fit of the mass spectrum of pseudoscalar mesons agrees with the estimate of the size of the QCD flux tube derived from the fit of mass spectrum of glueball states using topological constants of knotted flux tubes [4].

Neglecting the longitudinal momentum differences between adjacent hadrons ( i.e. assuming local homogeneity of the fragmenting QCD field ), the momentum difference between ground state pions can be written as function of their rank r:

$$Q(\text{r}) = \sqrt{-(p_i - p_{i+r})^2} \cong 2\, p_T^{thr} |\sin(\text{r}\,\Delta\Phi/2)|, \tag{3}$$

where $p_T^{thr}$ ($\sim 134$ MeV) stands for the intrinsic $p_T$ of ground state pions. The numerical values of the predicted momentum difference separating pairs of ground state hadrons with rank up to 5 are given in Table 1. The chain of n adjacent ground state pions has the mass

$$M_{nh} = \sqrt{n^2 m_\pi^2 + \sum_{i \neq j} Q_{ij}^2}, \tag{4}$$

where $m_\pi$ is the pion mass, and $Q_{ij}$ the momentum difference between pairs of hadrons forming the chain.

| Pair rank difference | 1 | 2 | 3 | 4 | 5 |
|---|---|---|---|---|---|
| Q expected [MeV] | 266 ± 8 | 91 ± 3 | 236 ± 7 | 171 ± 5 | 178 ± 5 |

Table 1: The expected momentum difference between ground state direct hadrons, in the quantized helix string model. The uncertainty (3%) is derived from the precision of the fit of the mass spectrum of light pseudoscalar mesons[3].

## Experimental data

The study of adjacent hadron pairs is complicated by the fact that the history of string fragmentation ( i.e. the exact hadron ordering along string ) is unknown. It is nevertheless possible to define observable in a way which allows an implicit study of adjacent hadron pairs. Assuming the local charge and momentum conservation in the string breakup, it is shown with help of MC samples that the subtraction of inclusive like-sign and opposite sign spectra agrees with the distribution of true adjacent charged hadron pairs ( obtained using the MC truth information ), up to the uncertainty related to the ordering of products of multibody decay of resonant states. Adjacent charged hadrons can be of opposite sign only due to the local charge conservation. The number of adjacent hadron pairs grows linearly with the number of charged particle in the sample which implies the integral of observable

$$\Delta(Q) = \frac{1}{N_{ch}}(N^{+-}(Q) - N^{++,--}(Q)) \qquad (5)$$

is an invariant (here $N_{ch}$ stands for number of charged tracks in the sample and $N^{+-}(N^{++,--})$ design the number of opposite-sign (like-sign) pairs (summed over events), while Q is the 4-momentum difference of the pair).

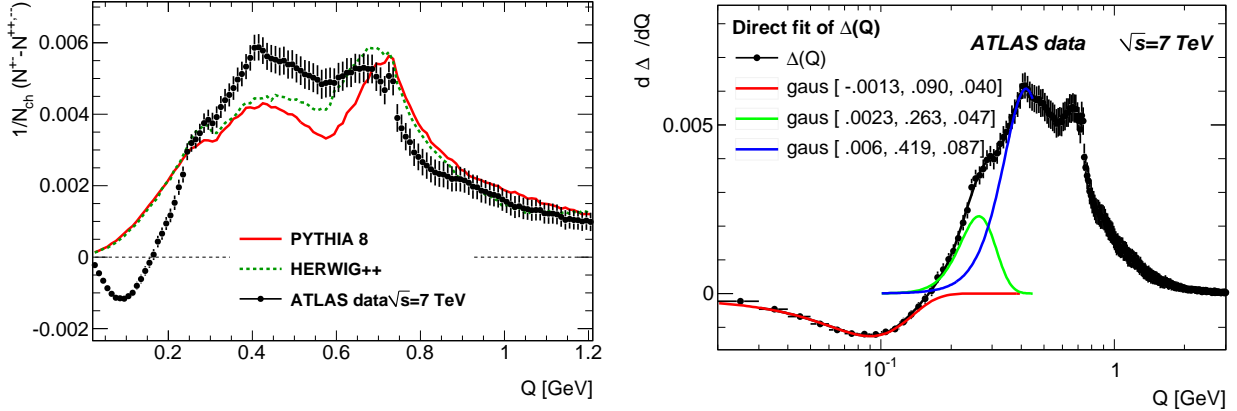

Figure 1: Left: Comparison of ATLAS minimum bias data [6] with MC prediction of Pythia and Herwig++. The inclusive $\Delta(Q)$ distribution is expected to reflect the distribution of adjacent pairs of charged hadrons. Quantum effects are not included in MC description. Right: Direct fit of the triple gaussian shape at low Q in the measured inclusive $\Delta(Q)$ distribution. The onset of adjacent pair production coincides with the position of (the first) $Q^{+-}$ peak at Q=0.26 GeV, the fit of the depleted area at low Q agrees with the expected momentum difference of pions with rank difference 2, the lowest rank difference possible for a pair of like-sign pions.

The comparison of MC predictions with the data published by ATLAS [6] is shown in Fig. 1 (left plot). A large difference is seen between the data and the MC predictions in low Q region : MC samples populate the whole phase space while in the data, the adjacent pairs seem to be absent at very low Q and in excess in the region 0.2 GeV < Q <0.6 GeV. The low Q region shows an excess of like-sign pairs which reaches maximum around $Q \sim 0.09$ GeV. This corresponds to the expected momentum difference between ground state hadrons with rank difference 2 in the quantized helix string fragmentation.

In order to get a rough estimate of the position of first local maximum for opposite-sign pairs, the measured $\Delta(Q)$ is fitted - in the region up to Q=0.4 GeV - by 3 gaussians describing the enhanced production of pairs of like-sign hadrons (seen as depletion at low Q), and the first 2 (positive) peaks for adjacent opposite-sign pairs of hadrons (Fig. 1, right plot). The position of the lowest peak for unlike-sign pairs agrees with the model prediction for the minimal momentum difference between a pair of adjacent ground state hadrons (0.26 GeV).

It should be noted that the enhanced production of like-sign hadron pairs is usually attributed to Bose-Einstein effect and that the model offers an alternative description in terms of fully coherent particle production.

## Prospects

The combination of causal and quantum constraints allows to develop model of QCD string fragmentation where the shape of QCD field translates into observable properties (mass, transverse momentum) of hadrons. Parameters of the QCD string are fitted using mass spectrum of pseudoscalar mesons. Due to the reduction of the number of free parameters, the model acquires considerable predictive power. The measured shape of 2-particle correlations (commonly attributed to Bose-Einstein effect) agrees with model predictions, which means the model successfully merges the description of previously disconnected aspects of hadronization (hadron mass spectra and hadron correlations). Implementation of quantum effects into hadronization models should therefore be among priorities for future modelling.

From the point of view of the model development, two directions seem to be particularly interesting: the model is based on causality relations which help to constrain the quantum effects. Since the data agree well with an effective quantization at the hadron level, it should be possible to develop a quantized model of gluon emission which would resolve the problem of soft collinear divergencies in QCD calculations (which presumably was on mind of the authors of the original helix string proposal). The other direction consists in study of topological properties of knotted helical strings which possibly generate a distinct particle properties [7].

# Workshop summary and FCC-ee prospects on parton radiation and fragmentation studies

P.Z. Skands[1]

[1] School of Physics and Astronomy, Monash University, Melbourne VIC-3800, Australia

**Abstract:** The contributions to the workshop on Parton Radiation and Fragmentation from LHC to FCC-ee are summarised. Emphasis is given to aspects that touch directly on the prospects for new measurements made possible by the large statistics and high energies accessible at the FCC-ee.

## Introduction

Quantum Chromodynamics is the only unbroken Yang-Mills theory that can be compared directly with experiment. It exhibits the unique and complicating feature of self-interactions among massless gauge bosons (moreover with a tree-level coupling that is relatively large), and related to that, the still fundamentally puzzling feature of confinement. Its rich structure is probed in a particularly clean way by the process $e^+e^- \to$ hadrons, in which quarks produced from the vacuum turn into (jets of) hadrons. At CM energies above a few GeV, this process can be described via a combination of quark and gluon bremsstrahlung showers, which may be treated perturbatively, and a hadronisation process, which must be parametrised or modeled nonperturbatively and constrained by fits to data. The two components correspond to short and long wavelengths, respectively, and may formally be factorised from each other. This is the basis of virtually all calculations in QCD.

A recurring theme at the workshop was the crucial legacy left in particular by the LEP (and SLD) experiments at and above the $Z$ pole, which continue to provide the first line of constraints on any new framework for high-energy jet fragmentation calculations. However, there are still important gaps in our knowledge, some of which will be highlighted below. Particularly exciting prospects for the FCC-ee will be to shed crucial new light on the detailed mechanism(s) of confinement, to measure the strong coupling constant with unprecedented accuracy, and to perform high-precision studies of the quasi-fractal patterns of jet (sub)structure in $e^+e^-$ (and $\gamma\gamma$) collisions. Furthermore, as new improved QCD calculations and models continue to be developed, pushing the envelope further in terms of both precision and detail, a constructive interplay — at the next order of precision — should be expected both with the other physics programs at the FCC-ee (for which a solid understanding of QCD corrections can translate into improved accuracy), but also crucially with the FCC-hh, in terms of a set of high-precision constraints on all aspects of final-state parton radiation and fragmentation, which can continue to be brought to bear during the FCC-hh era.

High-precision $\alpha_s$ measurements were the topic of a separate workshop held in 2015 [1], to which we refer for details on those prospects. Below, we focus on fragmentation functions, parton radiation and jet (sub)structure, and nonperturbative QCD effects and hadronisation, respectively.

## Fragmentation functions

The single-inclusive annihilation process (SIA), $e^+e^- \to h + X$, offers the cleanest environment to extract FFs both theoretically and experimentally. As emphasised in several talks, the FF field is now moving to NNLO accuracy [2][3][4][5] with improved resummation of low-$z$ dynamics [4][5], which implies, roughly, that a precision goal of order 1% or better can be envisioned already now.

Simultaneously, efforts are underway to reduce the effects of parametrisation bias in the fits [6] and improve the rigour of uncertainty estimates [7]. Currently, the most precise measurements come from the $B$ factories, with BaBar having collected $\sim 500$ fb$^{-1}$ [8] and Belle about twice that [2]. (Belle II will collect $\sim 50$ ab$^{-1}$.) Despite the huge data samples, the relatively low CM energy of $\sim 10.5$ GeV limits the reach to hadron momentum fractions greater than $z \sim 0.2$. Measurements at higher energies imply reduced hadron mass effects and can provide access to lower $z$. Furthermore, the access to the gluon FFs comes only from scaling violations which require a large lever am in $Q^2$. The LEP experiments collected about 0.2 fb$^{-1}$ at the $Z$ pole, while the FCC-ee would be able to deliver a sample similar in size to that of Belle II. This would set a completely new standard in particular for scaling studies, with two very large data samples separated by about two orders of magnitude in $Q^2$.

In the region of overlap in $z$, the LEP and SLD measurements were also limited to use bin sizes an order of magnitude larger than those of the $B$ factories, with steps of $\Delta z \geq 0.1$ being typical for $z > 0.3$ [2]. With an expected momentum resolution of 1%, FCC-ee measurements could potentially use a binning at least as fine as that of the $B$ factories, hence providing the optimal resolution on the scaling behaviour, bin by bin in $z$. In addition, some tensions exist among high energy SIA data sets at overlapping kinematics, which will require new high precision measurements of SIA at future facilities [7].

A point that will come back in the discussion of hadronisation studies is flavour separation. SIA offers the possibility to extract FFs of hyperons and other particles that are difficult to reconstruct in $pp$ and $ep$ environments [2]. So far, we are not aware of detailed phenomenological studies in this regard vis a vis the particle identification capabilities envisioned for the FCC-ee detectors. The need for accurate knowledge of FFs, especially for the production of kaons, was also highlighted [7] in the context of extracting the strange quark polarization in the nucleon.

Two theoretical contributions focused specifically on resummation effects at low $z$ [4][5]. On the experimental side, a requirement of 3 tracker hits could allow track reconstruction down to momentum values as low as 30-40 MeV, corresponding to a reach in $z$ of $\sim 10^{-3}$, which not only covers the logarithmic region but would also allow detailed studies of soft hadronisation effects, see below. It was pointed out [9] that even if nominal run conditions were to require a higher momentum cutoff, dedicated runs with a lower $B$-field could produce LEP-sized samples in minutes.

Further prospects include heavy-quark FFs for which the current state of the art is dominated by LEP+SLD [2] with NLO+NLL theory accuracy (although in principle all pieces required for NNLO+NNLL are now available [10]), $p_T$ dependence in hadron+jet production [2], and extraction of polarisation dependent FFs [2][11]. Such measurements would also all place relevant constraints on hadronisation models, especially with the addition of particle correlations, discussed below. All these expected experimental and theoretical advances will allow for $\alpha_\text{s}$ extractions at NNLO accuracy (or beyond) from global fits to the data with sub-percent uncertainty [5].

### Parton radiation and jet (sub)structure

As in any (massless) gauge theory, gluon and quark bremsstrahlung build up radiation patterns that exhibit scale invariance before renormalisation, on top of which the running coupling imparts a scaling violation. The structure of these patterns are both of fundamental interest from a theoretical-mathematical point of view and also underpin the application of any jet-based observables in both phenomenological studies and experimental measurements.

Among the central questions discussed at the workshop, the ability to discriminate between quark-like and gluon-like jets was highlighted by several speakers. A theoretical requirement for any application beyond the classical level is that taggers (of any kind) be based solely on final-state observable quantities [12]. Thus we move away from the quantum mechanically ill-defined concept of distinct (LO) partons that occasionally appeared in discussions of LEP-era analyses [13] to classifying jets as quark-like or gluon-like purely according to observable-based taggers, which may then be optimised on selected reference processes with 'known' properties, such as $Z \to q\bar{q} \to$ hadrons, $H \to gg \to$ hadrons, etc. Generalised angularities were used as an example of a convenient set of observables (which include e.g. track multiplicity, jet mass, and several other commonly used observables as special cases) both at the workshop [12][14][15] and in a recent Les Houches study [16]. To first approximation, one expects simple ratios like $C_A/C_F \sim 2.25$ between, e.g., the track multiplicities in gluon-like vs quark-like jets, but even at LO this could only really hold in the collinear limit of jets; at larger angles, QCD coherence implies that there is no 'independent fragmentation' in particular for soft wide-angle emissions, a facet that was also explicitly studied at LEP, albeit with limited precision [13][17]. From both experiment [18][19] and theory [12][15], it was emphasised that widely used MC event generators differ significantly in their predictions of q/g separation power, an issue it will be important to resolve in coming years. An interesting development in this regard was the development of a standalone shower model that can interpolate smoothly between a PYTHIA-like and a HERWIG-like evolution [15].

For a 125-GeV SM Higgs, the branching fraction for $H \to gg$ is about 8%. Hence at the $ZH$ threshold, a unique sample of $\mathcal{O}\left(2 \times 10^5\right)$ $H \to gg$ events will become available in which the fragmentation of a high-mass purely gluonic system can be studied for the first time [12]. In this context, $H \to b\bar{b}$ constitutes a background six times larger, and hence some of the signal sample will have to be sacrificed for purity; thus, it will be relevant to establish what precision can actually be achieved in the end. Other possibilities already in use at LEP include contrasting the fragmentation of $Z \to b\bar{b}g$, $Z \to q\bar{q}g$, and $Z \to q\bar{q}\gamma$ events [13].

One of the major advances in jet technology has been the emergence of algorithms based on particle flow. Combined with improved detectors at FCC-ee (relative to LEP) and a reliable theoretical modeling of QCD jet substructure, these are expected to lead to substantial gains in the achievable jet calibration systematics. The first level of jet substructure, captured by the $y_{23}$ jet resolution variable, is currently known at NNLL+NNLO precision, with a residual QCD scale uncertainty below 5% [20], with explicit numerical calculations provided by the ARES (NNLL) and EERAD3 (NNLO) codes [20][21]. Likewise, the state of the art for events shapes is generally [14][20], with the $C$ parameter even known to N$^3$LL [21]. These calculations are instrumental in the quest for precision determinations of $\alpha_s$ [1]. It should be noted that the non-perturbative power corrections can be significant, and that there are generally discrepancies between the size of such corrections extracted from Monte Carlo event generators and analytical fits, with MC generators tending to predict larger corrections. One possible reason is that the evolution equations of MC generators are formally less accurate, and that the hadronisation modeling may be 'overcompensating' for deficiencies in the perturbative modeling; it is therefore interesting to note that, for the first time in many years, new work is appearing on the fundamental accuracy of the parton-shower algorithms [22][23]. The advent of higher-order showers would not only improve the precision on the perturbative part of the MC calculation itself, but should also enable a reevaluation of the constraints on hadronisation models, at a formal level more consistent with that of the inclusive resummation calculations.

At deeper levels of jet substructure, the inclusive $y_{ij}$ jet clustering scales remain relevant but they are only partially informative; one method to extract information about the (coherent) emission of a soft parton from a hard $n$-jet state is to isolate specific event topologies (via cuts) and measure one or more complementary observables designed to be sensitive to coherence or other aspects beyond collinear factorisation [17]. Collinear structure may also be further targeted by small-radius (sub)jets, for which resummation in the jet radius parameter can become relevant [24].

The importance of scaling studies (with the CM energy of the hadronically decaying system) was already emphasised in the discussion of FFs. It is worth noting that radiative events, with hard initial-state photon bremsstrahlung, can provide access to lower $Q^2$ values as well. While not competitive with the $B$ factories at 10 GeV, FCC-ee could presumably outperform all previous experiments in the intermediate region between 10 GeV and $m_Z$, depending only on how far forward the instrumented region goes (and with what resolution).

The question of resilience of jet algorithms to backgrounds (e.g., from $\gamma\gamma$ pileup) has been studied closely for linear facilities like ILC and CLIC, where longitudinally invariant algorithms can mitigate the effects [25]. At a circular collider, however, the effects are orders of magnitude smaller and are expected to be negligible, modulo that the effect of synchroton radiation, and appropriate shielding measures, remain to be evaluated.

Efforts at the LHC to measure $g \to b\bar{b}$ splittings in gluon jets were summarised [26]. On the theory side, this is described by the least singular splitting kernel in QCD (aside from $g \to t\bar{t}$) and consequently ranks among the most poorly constrained aspects of jet evolution [10]. Moreover, tagged $g \to b\bar{b}$ splittings can also provide a relatively pure source of colour-octet $b$-quark pairs (as opposed to the singlet pairs in e.g. $\gamma^*/Z/W/H \to b\bar{b}$), the radiation patterns of which are therefore also interesting to study. At FCC-ee, the higher statistics and granularity of calorimeters and vertex detectors should allow to disentangle the $g \to b\bar{b}$ and $g \to c\bar{c}$ splittings, through a double tagging of the jets from the $b(c)$ and $\bar{b}(\bar{c})$ quarks [10], providing a unique environment for precise measurements of gluon-initiated heavy-quark fragmentation.

Heavy-quark radiation patterns themselves exhibit the so-called dead-cone effect which suppresses collinear radiation at angles $\theta < m/E$, translating to an angular resolution scale of 0.1 for $Z \to b\bar{b}$ and 0.03 for $Z \to c\bar{c}$. For slower massive quarks (e.g., from $g \to b\bar{b}$) the corresponding angles are larger. A complicating factor in attempts to directly measure the dead-cone effect is that this same region of phase space is filled in by the decay products of the massive D, B hadrons [27].

### Nonperturbative QCD effects and hadronisation

Confinement remains one of the fundamental unsolved problems in quantum field theory. For infrared safe observables its effects may be minimised and cast in terms of factorised and universal long-distance corrections (power corrections) and/or functions (FFs, PDFs). Monte Carlo models (such as the string or cluster ones) attempt to go further and provide an explicit non-perturbative modelling of the hadronisation process. The latter allows for direct comparisons with data at the particle level. Moreover, confronting models based on qualitatively different assumptions with data enables us (if done carefully) to extract knowledge about the properties of the nonperturbative dynamics, a famous example being the discovery of the 'string effect' in the early eighties [28]. In both cases, there are free parameters that must be extracted from fits (or 'tunes', in the MC context) to data. In this context, the clean environment of $e^+e^-$ collisions is the sine qua non.

Particle identification is crucial for many aspects of hadronisation studies. To give just one example, leading baryons in jets are an interesting discriminator since it is not possible for a gluon to fragment directly into a baryon in cluster models while this is allowed in string-based models. In more generality, it is fair to say that baryon production is the least well-understood component of the fragmentation process in both string and cluster models [22], making measurements of baryon production rates, spectra, and correlations particularly interesting. With regard to correlations, baryons are obviously produced in pairs due to baryon-number conservation. An information-rich question is then: how far away are such pairs allowed to be (e.g., in rapidity) from one another?

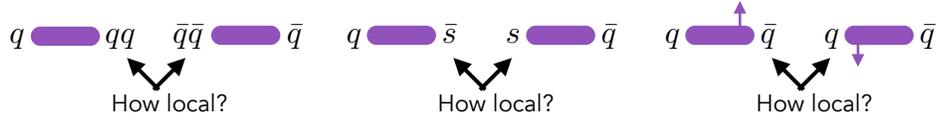

Figure 1: Illustration of string breaking, highlighting the conservation of (from left to right) baryon number, strangeness, and transverse momentum, respectively.

How 'locally' does baryon-number conservation act, see illustration in Fig. 1? It was noted that both of the OPAL studies of baryon-baryon rapidity correlations were statistics-limited and would only have reached the OPAL systematics uncertainties with $10^8$ Z decays [9]. Since the FCC-ee detectors will exhibit further reduced systematics uncertainties, these measurements could likely be improved by more than an order of magnitude at FCC-ee. Note that this determination obviously benefits from being able to identify as many types of baryons as possible, with high fidelity, so that as few as possible 'slip through the net'.

For identical baryons, another expected effects is that of Fermi-Dirac (FD) correlations [29]; again due to baryon-number conservation, the production of two identical baryons requires two antibaryons to be produced as well, thus the FD correlations probe a combined hadronisation involving at least four particles. It was noted that the radius extracted from these measurements at LEP (across multiple experiments and multiple baryon species) was smaller than the proton radius, a puzzle which remains unresolved [29], although MC modeling dependence may have played a role. The capability of the FCC-ee to perform a much cleaner determination of the underlying two-baryon correlations, and the consequent improved constraints on MC models of baryon production, would presumably be the ideal prerequisite to performing a new measurement of FD correlations, settling this question. On the topic of Bose-Einstein correlations between identical mesons, it was noted that these studies were limited (by statistics) at LEP to 1D distributions. To go from 1D to 3D analyses requires $N^3$ bins. Thus, the increased statistics at FCC-ee should allow for fully 3D distributions with roughly 100 bins [29].

Strangeness likewise represents a clearly identifiable quantum number. This has recently been highlighted in new measurements from the LHC [30]. In particular, ALICE has observed (in $pp$ collisions) a significant increase in strangeness fractions with event multiplicity [31]. This would appear to indicate that fundamental properties of the hadronisation process, such as strangeness suppression, can depend on global properties of the event. There is a lively activity in the theory community now, discussing whether the origin of these (and other, e.g., "ridge-like") observations should be sought in physics that is ultimately thermal, and/or in interaction effects between multiple hadronising systems. For at least the author of this summary, the ALICE measurement ranks as the discovery of the year for 2016 and indicates a clear breakdown of the way jet universality has been imposed at least in the conventional event generators such as PYTHIA. In any case, analogous measurements in $e^+e^-$ collisions (both in hadronic Z decays and in fully hadronic $WW$ decays) would serve as crucial and theoretically much cleaner reference cases. It is also worth emphasising that the strongest effect in ALICE is seen in $\Omega$ baryons, which only occur at a rate of less than 1 per-mille in hadronic Z decays, making them a rare sight indeed at LEP but abundant in the (hundreds of) millions at FCC-ee. It would be particularly interesting to repeat the measurements of flavour ratios as function of event activity in $e^+e^-$ [30].

Measurements at LEP convincingly demonstrated that colour reconnections (e.g., between the two decay systems in hadronic $WW$ events) exist in nature, ruling out the no-CR scenario at 99.5% CL [28]. The effect became the topic of significant interest in particular since 'string drag' had a noticeable effect on extractions of the $W$ mass, but apart from the conclusion that it exists and the exclusion of some of the most aggressive (and possibly unrealistic) scenarios, not much differential information is available to constrain new generations of models. The issue has reappeared with potentially enhanced effects in hadron collisions, where it can affect the top mass extraction and significant effects may be indicated by soft-QCD distributions in both minimum-bias and underlying-event measurements. High-statistics $e^+e^-$ measurements are needed to tell the other side of the story [28]. In particular, it was suggested [28] that with thousands of times more $WW$ events at FCC-ee than at LEP, the $W$ mass issue could be turned around, using the huge sample of semi-leptonic events (in which no CR can occur between the two decay systems) to determine $M_W$ with high precision and then use that extracted value as a constraint to measure CR in the fully hadronic ones. Other measures include particle flow between jets, or changes in the charged multiplicity as a function of topology. Complementary studies should of course also be done in $e^+e^- \to t\bar{t}$ and in multi-jet final states of hadronic $Z$ decays. The prospects for pinning down the CR mechanism at the FCC-ee are therefore good.

In addition, we emphasise that measurements of especially the soft components of particle spectra, with $|p| < \Lambda_{\text{QCD}}$, are instrumental to reveal the genuinely non-perturbative soft physics. At LEP, charged-kaon spectra were measured down to $p_K \sim 250$ MeV, with about 10% uncertainties for the lowest bins (see, e.g., [32] and references therein). This would be relevant to extend to softer momenta and increased precision, to obtain solid constraints on the full soft part of the fragmentation. Ideally this would be accompanied by a division into longitudinal and transverse components along a jet or event-shape axis, similar to that required for measurements of $p_T$-dependent FFs discussed above. Finally, strangeness and $p_T$ correlations are similarly revealing as baryon correlations about the degree of locality of quantum number (and momentum) conservation in the hadronisation process, see Fig. 1. The extraction of polarised FFs with high precision at FCC-ee should allow further discrimination between different mechanisms for hadronisation [11]. Measurements of spin/polarisation correlations are in principle also informative [33] though also more challenging.

A further more speculative prospect would be to search for imprints on the final state of sphalerons or instantons between QCD vacuum states with different winding numbers, which the high statistics for especially high-multiplicity $e^+e^-$ annihilation at the FCC-ee could potentially make accessible for the first time [2].

**Acknowledgments:** PS is the recipient of an Australian Research Council Future Fellowship, FT130100744: "Virtual Colliders: high-accuracy models for high energy physics".